\documentclass[preprint,prd,aps,showpacs,showkeys,nofootinbib]{revtex4}
\usepackage{graphicx}
\begin{document}

%\preprint{hep-ph/0412147}

\title{Spontaneous R-parity violation in the minimal gauged $(B-L)$ supersymmetry with a $125\;{\rm GeV}$ Higgs}

\author{Chao-Hsi Chang$^{a,b,c}$\footnote{email:zhangzx@itp.ac.cn},Tai-Fu
Feng$^{b,c}$\footnote{email:fengtf@hbu.edu.cn}, Yu-Li
Yan$^{b,c}$, Hai-Bin Zhang$^{c,d}$, Shu-Min
Zhao$^{b,c}$\footnote{email:zhaosm@hbu.edu.cn}}

\affiliation{$^a$ CCAST (World Laboratory),P.O.Box 8730, Beijing 100190, China\\
$^b$ State Key Laboratory of Theoretical Physics (KLTP),\\
Institute of theoretical
Physics, Chinese Academy of Sciences, Beijing, 100190, China\\
$^c$ Department of Physics, Hebei University, Baoding, 071002, China\\
$^d$ Department of Physics, Dalian University of Technology, Dalian,
116024, China}

\begin{abstract}

We precisely derive the mass squared matrices for charged and neutral
(CP-odd and CP-even) Higgs, as well as the mass matrices for
neutrino-neutralino and charged lepton-chargino in the minimal
R-parity violating supersymmetry with local $U(1)_{B-L}$ symmetry.
In the framework the nonzero TeV scale vacuum expectations of right-handed
sneutrinos induce the heavy mass of neutral $U(1)_{B-L}$ gauge boson,
and result in relatively large mixing between the lightest CP-even Higgs
and three generation right-handed sneutrinos when we include the one-loop
corrections to the scalar potential. We numerically show
that there is parameter space of the considered model
to accommodate experimental data on the newly ones of Higgs signal from
LHC and experimental observations on the neutrino oscillation simultaneously.
\end{abstract}

\keywords{supersymmetry, Higgs, neutrino}
\pacs{12.60.Jv, 14.60.St, 14.80.Cp}

\maketitle

\section{Introduction\label{sec1}}
\indent\indent
A main destination of the Large Hadron Collider (LHC)
is to understand the origin of the electroweak symmetry breaking,
and to study the properties of neutral Higgs predicted by the Standard Model (SM)
and its various extensions. In the year of 2012, ATLAS and CMS reported
significantly excess events in a few channels which are interpreted
as the neutral Higgs with mass $m_{h_0}\sim 124-126\;{\rm
GeV}$\cite{CMS,ATLAS}, and CP properties and couplings of the particle are also being
established\cite{CMS1,ATLAS1,CMS2,ATLAS2} recently. It implies that the
Higgs mechanism to break electroweak symmetry has an
experimental cornerstone now. Another important progress of particle
physics in the last year is that nonzero experimental observation on the neutrino mixing
angle $\theta_{13}$ is obtained with high precision\cite{theta13-3},
which opens several prospects for neutrino physics. In this work,
we investigate the constraints on parameter space of the minimal
R-parity violating supersymmetry with local $U(1)_{B-L}$ symmetry
from the updated experimental data mentioned above.

R-parity, as a discrete symmetry, is defined through
$R=(-1)^{3(B-L)+2S}$, where $B$, $L$ and $S$ are baryon number,
lepton number and spin respectively for a concerned
field\cite{R-parity}. When $B-L$ is violated by an even amount,
R-parity conservation is guaranteed. However, breaking $B-L$ via
nonzero vacuum expectation values (VEVs) of neutral scalar fields with odd
$U(1)_{B-L}$ charges will induce the R-parity violation
simultaneously. In the minimal supersymmetric extension of SM (MSSM) with
local $U(1)_{B-L}$ symmetry, R-parity is spontaneously
broken when left- and right-handed sneutrinos acquire nonzero
VEVs\cite{Perez1,Perez2,Perez3,Perez4}.
Actually, both spontaneously violated R-parity and broken local $U(1)_{B-L}$ symmetry
replicate the MSSM with conserving baryon number but violating lepton number.
The authors of Ref.\cite{Perez5} further propose an extension
of the MSSM, which includes right-handed neutrino superfields and
two additional superfields $\hat{X},\;\hat{X^\prime}$ with even
$U(1)_{B-L}$ charges. When sneutrinos and scalar components of
$\hat{X},\;\hat{X^\prime}$ acquire non zero VEVs simultaneously,
local $U(1)_{B-L}$ symmetry and R-parity are broken spontaneously. To account for
the neutrino oscillation experiment, tiny neutrino masses are
generated through an extended seesaw mechanism in the framework
proposed in Ref.\cite{Perez1,Perez2,Perez3,Perez4,Perez5}.
Furthermore, the neutral Higgs fields $H_{_u}^0,\;H_{_d}^0$
mix with the scalar components of neutrino superfields and
$\hat{X},\;\hat{X}^\prime$ superfields after the electroweak symmetry is broken in those models.
Assuming that the scalar components of $\hat{X},\;\hat{X^\prime}$
and neutral Higgs fields $H_{_u}^0,\;H_{_d}^0$ acquire nonzero VEVs,
Ref.\cite{Porod1} studies mass spectrum in the model proposed in
Ref.\cite{Perez5}.

Here we study the constraints from the
observed Higgs signal and neutrino oscillation experimental data on
parameter space of the MSSM with local $U(1)_{B-L}$ symmetry in the scenarios where
sneutrinos obtain nonzero VEVs\cite{Perez1,Perez2,Perez3,Perez4}.
%%%%%%%%%%%%%%%%%%%%%%%%%%%%%%%%%%%%%%%%%BEGIN REVISION%%%%%%%%%%%%%%%%%%%%%%%%%
Since the tree level mixing between the lightest CP-even Higgs and right-handed
sneutrinos is suppressed by the tiny neutrino masses, we include the one-loop
corrections to the mixing which are mainly originated from the third generation
fermions and their supersymmetric partners.
%%%%%%%%%%%%%%%%%%%%%%%%%%%%%%%%%%%%%%%%%%END REVISION%%%%%%%%%%%%%%%%%%%%%%%%%%
Numerically the MSSM with local $U(1)_{B-L}$ symmetry
accommodates naturally the experimental data on the Higgs particle
from ATLAS/CMS collaborations and the updated experimental observations on
the neutrino oscillation simultaneously. In addition, the model also
predicts two sterile neutrinos with sub-eV
masses\cite{Perez6,Senjanovic1}, which are favored by the Big-bang
nucleosynthesis (BBN) in cosmology\cite{Hamann}.

Certainly the deviation from unitarity of the leptonic mixing matrix
intervening in charged currents might induce a tree-level
enhancement of $R_{_P}=\Gamma(P^+\rightarrow
e^+\nu)/\Gamma(P^+\rightarrow \mu^+\nu)\;(P^+=K^+,\;\pi^+)$
\cite{Abadaa} because of additional mixings between the active
neutrinos and the sub-eV sterile states.
Ignoring the difference between hadronic matrix elements in $P^+\rightarrow e^+\nu$
and that in $P^+\rightarrow \mu^+\nu$, one finds that the experimental observations
on $R_{_P}$ also constrain the parameter space of considered model.
Furthermore, the experimental data on $Z$ invisible width\cite{PDG}
also constrain the mixings between the active neutrinos and the
sub-eV sterile ones. We will address the constraints on the
mixings between the active neutrinos and the sub-eV sterile ones
from lepton flavor universality (LFU) and $Z$ invisible width
elsewhere \cite{Chang-Feng}.

Our presentation is organized as follows. In section \ref{sec2}, we
briefly summarize the main ingredients of the MSSM with
local $U(1)_{B-L}$ symmetry, then present the mass
squared matrices for CP-odd and charged Higgs sectors, respectively.
We analyze the loop corrections on the mass squared matrix of CP-even
Higgs in section \ref{sec3}, and present the mass matrices
for neutrino-neutralino and charged lepton-chargino in section
\ref{sec4} and section \ref{sec5}, respectively. Furthermore, we
also present the decay widths for
$h^0\rightarrow\gamma\gamma,\;VV^*,\;(V=Z,\;W)$ in section \ref{sec6}. The
numerical analyses are given in section \ref{sec7}, and our
conclusions are summarized in section \ref{sec8}.

\section{The MSSM with local $U(1)_{B-L}$ symmetry\label{sec2}}
\indent\indent
When $U(1)_{B-L}$ is a local gauge symmetry, one can
enlarge the local gauge group of the SM to $SU(3)_{_C}\otimes
SU(2)_{_L}\otimes U(1)_{_Y}\otimes U(1)_{_{(B-L)}}$. In the
model proposed in Ref.\cite{Perez1,Perez2,Perez3,Perez4}, the
exotic superfields are three generation right-handed neutrinos
$\hat{N}_{_i}^c\sim(1,\;1,\;0,\;1)$. Meanwhile, quantum numbers of
the matter chiral superfields for quarks and leptons are given by
\begin{eqnarray}
&&\hat{Q}_{_I}=\left(\begin{array}{l}\hat{U}_{_I}\\
\hat{D}_{_I}\end{array}\right)\sim(3,\;2,\;{1\over3},\;{1\over3})\;,\;\;
\hat{L}_{_I}=\left(\begin{array}{l}\hat{\nu}_{_I}\\
\hat{E}_{_I}\end{array}\right)\sim(1,\;2,\;-1,\;-1)\;,
\nonumber\\
&&\hat{U}_{_I}^c\sim(3,\;1,\;-{4\over3},\;-{1\over3})\;,\;\;
\hat{D}_{_I}^c\sim(3,\;1,\;{2\over3},\;-{1\over3})\;,\;\;
\hat{E}_{_I}^c\sim(1,\;1,\;2,\;1)\;,
\label{quantum-number1}
\end{eqnarray}
with $I=1,\;2,\;3$ denoting the index of generation. In addition,
the quantum numbers of two Higgs doublets
are assigned as
\begin{eqnarray}
&&\hat{H}_{_u}=\left(\begin{array}{l}\hat{H}_{_u}^+\\
\hat{H}_{_u}^0\end{array}\right)\sim(1,\;2,\;1,\;0)\;,\;\;
\hat{H}_{_d}=\left(\begin{array}{l}\hat{H}_{_d}^0\\
\hat{H}_{_d}^-\end{array}\right)\sim(1,\;2,\;-1,\;0)\;.
\label{quantum-number2}
\end{eqnarray}
The superpotential of the MSSM with local $U(1)_{B-L}$ symmetry is written as
\begin{eqnarray}
&&{\cal W}={\cal W}_{_{MSSM}}+{\cal W}_{_{(B-L)}}^{(1)}\;.
\label{superpotential1}
\end{eqnarray}
Here ${\cal W}_{_{MSSM}}$ is superpotential of the MSSM, and
\begin{eqnarray}
&&{\cal W}_{_{(B-L)}}^{(1)}=\Big(Y_{_N}\Big)_{_{IJ}}\hat{H}_{_u}^Ti\sigma_2\hat{L}_{_I}\hat{N}_{_J}^c\;.
\label{superpotential-BL}
\end{eqnarray}
Correspondingly, the soft breaking terms for the MSSM with local $U(1)_{B-L}$
symmetry are generally given as
\begin{eqnarray}
&&{\cal L}_{_{soft}}={\cal L}_{_{soft}}^{MSSM}+{\cal L}_{_{soft}}^{(1)}\;.
\label{soft-breaking1}
\end{eqnarray}
Here ${\cal L}_{_{soft}}^{MSSM}$ is soft breaking terms of the MSSM, and
\begin{eqnarray}
&&{\cal L}_{_{soft}}^{(1)}=
-(m_{_{\tilde{N}^c}}^2)_{_{IJ}}\tilde{N}_{_I}^{c*}\tilde{N}_{_J}^c
-\Big(m_{_{BL}}\lambda_{_{BL}}\lambda_{_{BL}}+h.c.\Big)
+\Big\{\Big(A_{_N}\Big)_{_{IJ}}H_{_u}^Ti\sigma_2\tilde{L}_{_I}\tilde{N}_{_J}^c+h.c.\Big\}\;,
\label{soft-breaking3}
\end{eqnarray}
with $\lambda_{_{BL}}$ denoting the gaugino of $U(1)_{_{B-L}}$.
After the $SU(2)_L$ doublets $H_{_u},\;H_{_d},\;\tilde{L}_{_I}$ and $SU(2)_L$ singlets $\tilde{N}_{_I}^c$
acquire the nonzero VEVs,
\begin{eqnarray}
&&H_{_u}=\left(\begin{array}{c}H_{_u}^+\\{1\over\sqrt{2}}\Big(\upsilon_{_u}+H_{_u}^0+iP_{_u}\Big)\end{array}\right)\;,
\nonumber\\
&&H_{_d}=\left(\begin{array}{c}{1\over\sqrt{2}}\Big(\upsilon_{_d}+H_{_d}^0+iP_{_d}\Big)\\H_{_d}^-\end{array}\right)\;,
\nonumber\\
&&\tilde{L}_{_I}=\left(\begin{array}{c}{1\over\sqrt{2}}\Big(\upsilon_{_{L_I}}+\tilde{\nu}_{_{L_I}}+iP_{_{L_I}}\Big)\\
\tilde{L}_{_I}^-\end{array}\right)\;,\nonumber\\
&&\tilde{N}_{_I}^c={1\over\sqrt{2}}\Big(\upsilon_{_{N_I}}+\tilde{\nu}_{_{R_I}}+iP_{_{N_I}}\Big)\;,
\label{VEVs}
\end{eqnarray}
the R-parity is broken spontaneously, and the local gauge symmetry $SU(2)_{_L}\otimes U(1)_{_Y}\otimes U(1)_{_{(B-L)}}$
is broken down to the electromagnetic symmetry $U(1)_{_e}$. Assuming that all parameters are real, we obtain
the minimization conditions at one-loop level in the model considered here
\begin{eqnarray}
&&T_{_u}^0+\Delta T_{_u}\upsilon_{_u}=0\;,
\nonumber\\
&&T_{_d}^0+\Delta T_{_d}\upsilon_{_d}=0\;,
\nonumber\\
&&T_{_{\tilde{L}_I}}^0+\Delta T_{_{\tilde{L}}}\upsilon_{_{L_I}}=0\;,
\nonumber\\
&&T_{_{\tilde{N}_I}}^0+\Delta T_{_{\tilde{N}}}\upsilon_{_{N_I}}=0\;,
\label{minimizations}
\end{eqnarray}
where $T_{_u}^0,\;T_{_d}^0,\;T_{_{\tilde{L}_I}}^0$, $T_{_{\tilde{N}_I}}^0$
denote the tree level tadpole conditions, and $\Delta T_{_u}$, $\Delta T_{_d}$,
$\Delta T_{_{\tilde{L}}}$ as well as $\Delta T_{_{\tilde{N}}}$
are the one-loop radiative corrections to the minimization conditions from
top, bottom, tau and their supersymmetric partners respectively, their concrete
expressions are given in the appendix.\ref{app1}.
After the local gauge group $SU(2)_{_L}\otimes U(1)_{_Y}\otimes U(1)_{_{(B-L)}}$
is broken down to the electromagnetic symmetry $U(1)_e$, the masses of neutral and
charged gauge bosons are respectively formulated as
\begin{eqnarray}
&&m_{_{\rm Z}}^2={1\over4}(g_1^2+g_2^2)\upsilon_{_{\rm EW}}^2\;,\nonumber\\
&&m_{_{\rm W}}^2={1\over4}g_2^2\upsilon_{_{\rm EW}}^2\;,\nonumber\\
&&m_{_{Z_{BL}}}^2=g_{_{BL}}^2\Big(\upsilon_{_N}^2+\upsilon_{_{\rm EW}}^2-\upsilon_{_{\rm SM}}^2\Big)\;.
\label{gauge-masses}
\end{eqnarray}
Where $\upsilon_{_{\rm SM}}^2=\upsilon_{_u}^2+\upsilon_{_d}^2$,
$\upsilon_{_{\rm EW}}^2=\upsilon_{_u}^2+\upsilon_{_d}^2+\sum\limits_{\alpha=1}^3\upsilon_{_{L_\alpha}}^2$,
$\upsilon_{_N}^2=\sum\limits_{\alpha=1}^3\upsilon_{_{N_\alpha}}^2$,
and $g_2,\;g_1$, $g_{_{BL}}$ denote the gauge couplings of
$SU(2)_{_L},\;\;U(1)_{_Y}$ and $U(1)_{_{(B-L)}}$, respectively.

%%%%%%%%%%%%%%%%%%%%%%%%%%%%Begin 1st Revision%%%%%%%%%%%%%%
To satisfy present electroweak precision observations we assume
the mass of neutral $U(1)_{_{(B-L)}}$ gauge boson $m_{_{Z_{BL}}}>1\;{\rm TeV}$ which implies
$\upsilon_{_N}>1\;{\rm TeV}$ when $g_{_{BL}}<1$, then we derive $\max((Y_{_{N}})_{ij})\le10^{-6}$
and $\max(\upsilon_{_{L_I}})\le10^{-3}\;{\rm GeV}$\cite{Perez4}
to explain experimental data on neutrino oscillation.
Ignoring the small terms and assuming that the $3\times3$ matrices $m_{_{\tilde L}}^2,\;
m_{_{{\tilde N}^c}}^2$ are real, we simplify the minimization conditions in
Eq.(\ref{minimizations}) as
\begin{eqnarray}
&&\upsilon_{_u}\Big\{\mu^2+m_{_{H_u}}^2+{g_1^2+g_2^2\over8}\Big(2\upsilon_{_u}^2
-\upsilon_{_{\rm EW}}^2\Big)+\Delta T_{_u}\Big\}+B\mu\upsilon_{_d}\simeq0
\;,\nonumber\\
%%%%%%%%%%%%%%%%%%%%%%%%%%%%%%%%%%%%%%%%%%%%%%%%%%%
&&\upsilon_{_d}\Big\{\mu^2+m_{_{H_d}}^2-{g_1^2+g_2^2\over8}\Big(2\upsilon_{_u}^2
-\upsilon_{_{\rm EW}}^2\Big)+\Delta T_{_d}\Big\}+B\mu\upsilon_{_u}\simeq0
\;,\nonumber\\
%%%%%%%%%%%%%%%%%%%%%%%%%%%%%%%%%%%%%%%%%%%%%%%%%%%
&&\sum\limits_{\alpha=1}^3\Big[\Big(m_{_{\tilde L}}^2\Big)_{I\alpha}
+\Delta T_{_{\tilde L}}\delta_{I\alpha}\Big]\upsilon_{_{L_\alpha}}
+{\upsilon_{_u}\over\sqrt{2}}\sum\limits_{\alpha=1}^3\Big(A_{_N}\Big)_{I\alpha}\upsilon_{_{N_\alpha}}
+{\mu\upsilon_{_d}\over\sqrt{2}}\zeta_{_I}
\nonumber\\
&&\hspace{0.0cm}
-\upsilon_{_{L_I}}\Big\{{g_1^2+g_2^2\over8}\Big(2\upsilon_{_u}^2-\upsilon_{_{\rm EW}}^2\Big)
+{m_{_{Z_{BL}}}^2\over2}\Big\}\simeq0
\;,\nonumber\\
%%%%%%%%%%%%%%%%%%%%%%%%%%%%%%%%%%%%%%%%%%%%%%%%%%%
&&\sum\limits_{\alpha=1}^3\Big[\Big(m_{_{{\tilde N}^c}}^2\Big)_{I\alpha}
+\Delta T_{_{\tilde N}}\delta_{I\alpha}\Big]\upsilon_{_{N_\alpha}}
+{m_{_{Z_{BL}}}^2\over2}\upsilon_{_{N_I}}\simeq0\;,
\label{minimizations-simplification}
\end{eqnarray}
with $\zeta_{_I}=\sum\limits_{\alpha=1}^3\Big(Y_{_N}\Big)_{I\alpha}\upsilon_{_{N_\alpha}}$.
Note here that the first two minimization conditions respectively for $H_{_u}^0,\;H_{_d}^0$
are not greatly modified from that in the MSSM, the third condition keeps
the linear terms of $\upsilon_{_{L_I}}$ or $Y_{_N}$, and the last equation implies that the vector
$(\upsilon_{_{N_1}},\;\upsilon_{_{N_2}},\;\upsilon_{_{N_3}})$  is an eigenvector of $3\times3$
mass squared matrix $m_{_{{\tilde N}^c}}^2$ with eigenvalue
$-m_{_{Z_{BL}}}^2/2-\Delta T_{_{\tilde N}}$.
A possible symmetric $3\times3$ matrix satisfying the last equation in
Eq.(\ref{minimizations-simplification}) is written as
\begin{eqnarray}
&&m_{_{{\tilde N}^c}}^2\simeq\left(\begin{array}{ccc}
\Lambda_{_{\tilde{N}_1^c}}^2-\Lambda_{_{BL}}^2\;,\;\;&0\;,\;\;
&-{\upsilon_{_{N_1}}\over\upsilon_{_{N_3}}}\Lambda_{_{\tilde{N}_1^c}}^2\\
0\;,\;\;&\Lambda_{_{\tilde{N}_2^c}}^2-\Lambda_{_{BL}}^2\;,\;\;
&-{\upsilon_{_{N_2}}\over\upsilon_{_{N_3}}}\Lambda_{_{\tilde{N}_2^c}}^2\\
-{\upsilon_{_{N_1}}\over\upsilon_{_{N_3}}}\Lambda_{_{\tilde{N}_1^c}}^2\;,\;\;
&-{\upsilon_{_{N_2}}\over\upsilon_{_{N_3}}}\Lambda_{_{\tilde{N}_2^c}}^2\;,\;\;
&{\upsilon_{_{N_1}}^2\Lambda_{_{\tilde{N}_1^c}}^2+\upsilon_{_{N_2}}^2\Lambda_{_{\tilde{N}_2^c}}^2
\over\upsilon_{_{N_3}}^2}-\Lambda_{_{BL}}^2
\end{array}\right)
\label{R-sneutrino-mass}
\end{eqnarray}
with $\Lambda_{_{BL}}^2=m_{_{Z_{BL}}}^2/2+\Delta T_{_{\tilde N}}$.
In order to make our final results transparently, we further assume in our
following discussion
\begin{eqnarray}
&&\Big(m_{_{\tilde L}}^2\Big)_{IJ}\simeq m_{_{{\tilde L}_I}}^2\delta_{IJ}\;,
\;\;(I,\;J=1,\;2,\;3)\;,
\label{L-sneutrino-mass}
\end{eqnarray}
then we obtain
\begin{eqnarray}
&&\upsilon_{_{L_I}}\simeq-{4\sqrt{2}\Big[\upsilon_{_u}\sum\limits_{\alpha=1}^3\Big(A_{_N}\Big)_{I\alpha}
\upsilon_{_{N_\alpha}}+\mu\upsilon_{_d}\zeta_{_I}\Big]\over8(m_{_{{\tilde L}_I}}^2+\Delta T_{_{\tilde L}})
-(g_1^2+g_2^2)\Big(\upsilon_{_u}^2-\upsilon_{_d}^2\Big)-4m_{_{Z_{BL}}}^2}\;.
\label{L-sneutrino-vacuum}
\end{eqnarray}
As $m_{_{{\tilde L}_I}}\sim1\;{\rm TeV}$, the condition $\max(\upsilon_{_{L_I}})\le10^{-2}\;{\rm GeV}$
requires $A_{_N}\sim0.01\;{\rm GeV}$. This implies that tree level contributions to the mixing
between the lightest CP-even Higgs and right-handed sneutrinos can be ignored, leading
contributions to the mixing are mainly originated from one-loop radiative corrections.
%%%%%%%%%%%%%%%%%%%%%%%End 1st Revision%%%%%%%%%%%%%%%%%%%%%%%%%%

\subsection{The mass squared matrix for charged Higgs}
\indent\indent
Using those minimization conditions, we derive the $8\times8$ mass squared matrix for charged Higgs
\begin{eqnarray}
&&\left(\begin{array}{cc}\Big[{\cal M}_{_{CH}}^2\Big]_{2\times2}\;\;\;&\Big[A_{_{CH}}\Big]_{2\times6}\\
\Big[A_{_{CH}}^T\Big]_{6\times2}\;\;\;&\Big[M_{_{\tilde E}}^2\Big]_{6\times6}\end{array}\right)\;,
\label{CH1}
\end{eqnarray}
in the interaction eigenstates $H_{_{CH}}^T=(H_{_u}^-,\;H_{_d}^-,\;\tilde{L}_{_I}^-,\;\tilde{E}_{_J}^{c*}),\;\;(I,\;J=1,\;2,\;3)$.
Here, elements of the $2\times2$ matrix ${\cal M}_{_{CH}}^2$ are given as
\begin{eqnarray}
&&\Big[{\cal M}_{_{CH}}^2\Big]_{11}=\Big(B\mu+\Delta_{odd}\Big){\upsilon_{_d}\over\upsilon_{_u}}
-{g_2^2\over4}\Big(\upsilon_{_{\rm EW}}^2-\upsilon_{_u}^2\Big)
+{1\over\sqrt{2}\upsilon_{_u}}\sum\limits_{\alpha,\beta}^3\upsilon_{_{L_\alpha}}
\Big(A_{_N}\Big)_{\alpha\beta}\upsilon_{_{N_\beta}}
\nonumber\\
&&\hspace{2.2cm}
+{1\over2}\sum\limits_{\alpha,\beta}^3\upsilon_{_{L_\alpha}}
\Big(Y_{_N}Y_{_N}^\dagger\Big)_{\alpha\beta}\upsilon_{_{L_\beta}}
\;,\nonumber\\
%%%%%%%%%%%%%%%%%%%%%%%%%%%%%%%%%%%%%%%%%%%%%%%%%%%%
&&\Big[{\cal M}_{_{CH}}^2\Big]_{12}=\Big(B\mu+\Delta_{odd}\Big)-{g_2^2\over4}\upsilon_{_u}\upsilon_{_d}
\;,\nonumber\\
%%%%%%%%%%%%%%%%%%%%%%%%%%%%%%%%%%%%%%%%%%%%%%%%%%%%
&&\Big[{\cal M}_{_{CH}}^2\Big]_{22}=\Big(B\mu+\Delta_{odd}\Big){\upsilon_{_u}\over\upsilon_{_d}}
+{g_2^2\over4}\Big(\upsilon_{_{\rm EW}}^2-\upsilon_{_{\rm SM}}^2-\upsilon_{_u}^2\Big)
+{\mu\varepsilon_{_N}^2\over\sqrt{2}\upsilon_{_d}}
\nonumber\\
&&\hspace{2.2cm}
-{1\over2}\sum\limits_{\alpha,\beta=1}^3\upsilon_{_{L_\alpha}}
\Big(Y_{_E}Y_{_E}^T\Big)_{\alpha\beta}\upsilon_{_{L_\beta}}\;,
\label{CH1-2}
\end{eqnarray}
with $\varepsilon_{_N}^2=\sum\limits_{\alpha,\beta=1}^3\upsilon_{_{L_\alpha}}
\Big(Y_{_N}\Big)_{\alpha\beta}\upsilon_{_{N_\beta}}$.
Additionally the $3\times3$ matrix $Y_{_E}$ is Yukawa couplings in charged lepton sector,
and the one-loop radiative correction is written as
\begin{eqnarray}
&&\Delta_{odd}={3g_{_2}^2\over32\pi^2\sin^2\beta}{m_{_t}^2A_{_t}\mu\over m_{_{\rm W}}^2}
{f(m_{_{{\tilde t}_1}}^2)-f(m_{_{{\tilde t}_2}}^2)\over m_{_{{\tilde t}_1}}^2-m_{_{{\tilde t}_2}}^2}
\nonumber\\
&&\hspace{1.5cm}
+{3g_{_2}^2\over32\pi^2\cos^2\beta}{m_{_b}^2A_{_b}\mu\over m_{_{\rm W}}^2}
{f(m_{_{{\tilde b}_1}}^2)-f(m_{_{{\tilde b}_2}}^2)\over m_{_{{\tilde b}_1}}^2-m_{_{{\tilde b}_2}}^2}
\nonumber\\
&&\hspace{1.5cm}
+{g_{_2}^2\over32\pi^2\cos^2\beta}{m_{_\tau}^2A_{_\tau}\mu\over m_{_{\rm W}}^2}
{f(m_{_{{\tilde\tau}_1}}^2)-f(m_{_{{\tilde\tau}_2}}^2)\over
m_{_{{\tilde\tau}_1}}^2-m_{_{{\tilde\tau}_2}}^2}\;.
\label{CP-odd2a}
\end{eqnarray}
Here $m_{_{\tilde{t}_{1,2}}}^2$, $m_{_{\tilde{b}_{1,2}}}^2$ and
$m_{_{\tilde{\tau}_{1,2}}}^2$ are the eigenvalues of the $\tilde{t}$,
$\tilde{b}$ and $\tilde{\tau}$ mass-squared matrices, the form factor
$f(m^2)=m^2(\ln(m^2/\Lambda^2)-1)$ with $\Lambda$ denoting renormalization scale.
Additionally, the concrete expressions for the symmetric matrix $M_{_{\tilde E}}^2$ and $A_{_{CH}}$ can
be found in appendix.\ref{app1}. Actually, the symmetric matrix in Eq.(\ref{CH1}) contains an eigenvector
with zero eigenvalue
\begin{eqnarray}
&&G^\pm={\upsilon_{_u}\over \upsilon_{_{\rm EW}}}H_{_u}^\pm-{\upsilon_{_d}\over\upsilon_{_{\rm EW}}}H_{_d}^\pm
-\sum\limits_{\alpha=1}^3{\upsilon_{_{L_\alpha}}\over\upsilon_{_{\rm EW}}}\tilde{L}_{_\alpha}^\pm\;,
\label{CH-Goldstone}
\end{eqnarray}
which corresponds to the charged Goldstone eaten by charged gauge boson
as electroweak symmetry broken spontaneously.  Applying the $8\times8$ orthogonal matrix
\begin{eqnarray}
&&{\cal Z}_{_{CH}}^{(0)}=Z_{_{CH}}^{(0)}\bigoplus1_{_{3\times3}}\;,
\label{C-H2}
\end{eqnarray}
we separate the charged Goldstone boson from the physical states:
\begin{eqnarray}
&&{\cal Z}_{_{CH}}^{(0)T}\cdot\left(\begin{array}{cc}\Big[{\cal M}_{_{CH}}^2\Big]_{2\times2}\;\;\;&\Big[A_{_{CH}}\Big]_{2\times6}\\
\Big[A_{_{CH}}^T\Big]_{6\times2}\;\;\;&\Big[M_{_{\tilde E}}^2\Big]_{6\times6}\end{array}\right)
\cdot {\cal Z}_{_{CH}}^{(0)}=
\left(\begin{array}{cc}0&0_{_{1\times7}}\\
0_{_{7\times1}}&M_{_{H^\pm}}^2\end{array}\right)\;.
\label{C-H3}
\end{eqnarray}
Where the $5\times5$ orthogonal matrix $Z_{_{CH}}^{(0)}$ is given as
\begin{eqnarray}
&&Z_{_{CH}}^{(0)}=\left(\begin{array}{ccc}
{\upsilon_{_u}\over\upsilon_{_{\rm EW}}},&{\upsilon_{_d}\over\upsilon_{_{\rm SM}}},&
\Big({\upsilon_{_u}\upsilon_{_{L_K}}\over\upsilon_{_{\rm SM}}\upsilon_{_{\rm EW}}}\Big)_{_{1\times3}}\\
-{\upsilon_{_d}\over\upsilon_{_{\rm EW}}},&{\upsilon_{_u}\over\upsilon_{_{\rm SM}}},&
\Big(-{\upsilon_{_d}\upsilon_{_{L_K}}\over\upsilon_{_{\rm SM}}\upsilon_{_{\rm EW}}}\Big)_{_{1\times3}}\\
\Big(-{\upsilon_{_{L_I}}\over\upsilon_{_{\rm EW}}}\Big)_{_{3\times1}},&0_{_{3\times1}},&\Big({\upsilon_{_{\rm SM}}\over\upsilon_{_{\rm EW}}}
\delta_{IK}+\sum\limits_{\alpha=1}^3\varepsilon_{IK\alpha}{\upsilon_{_{L_\alpha}}\over\upsilon_{_{\rm EW}}}\Big)_{_{3\times3}}
\end{array}\right)\;.
\label{C-H4}
\end{eqnarray}
Finally, we give the $8\times8$ mixing matrix $Z_{_{CH}}$ in charged Higgs sector as
\begin{eqnarray}
&&Z_{_{CH}}={\cal Z}_{_{CH}}^{(0)}\cdot\left(\begin{array}{cc}1\;\;&0_{1\times7}\\0_{7\times1}\;\;&
\Big[Z_{_{H^\pm}}\Big]_{7\times7}\end{array}\right)
\label{C-H5}
\end{eqnarray}
with $Z_{_{H^\pm}}^\dagger\cdot M_{_{H^\pm}}^2\cdot Z_{_{H^\pm}}=diag(m_{_{H_2^\pm}}^2,\cdots,m_{_{H_8^\pm}}^2)$.

\subsection{The mass squared matrix for CP-odd Higgs}
\indent\indent
In the interaction basis $P^{0,T}=(P_{_u}^0,\;P_{_d}^0,\;P_{_{\tilde{L}_I}}^0,\;P_{_{\tilde{N}_J}}^0),\;\;(I,\;J=1,\;2,\;3)$,
the $8\times8$ mass matrix for neutral CP-odd scalars is
\begin{eqnarray}
&&\left(\begin{array}{cc}\Big[{\cal M}_{_{CPO}}^2\Big]_{2\times2}\;\;\;&\Big[A_{_{CPO}}^{(0)}\Big]_{2\times6}\\
\Big[A_{_{CPO}}^{(0)T}\Big]_{6\times2}\;\;\;&\Big[M_{_{P}}^2\Big]_{6\times6}
\end{array}\right)\;,
\label{CP-odd1}
\end{eqnarray}
the elements of $2\times2$ mass squared matrix are
\begin{eqnarray}
&&\Big[{\cal M}_{_{CPO}}^2\Big]_{11}=\Big(B\mu+\Delta_{odd}\Big){\upsilon_{_d}\over\upsilon_{_u}}
+{1\over\sqrt{2}\upsilon_{_u}}\sum\limits_{\alpha,\beta}^3\upsilon_{_{L_\alpha}}
\Big(A_{_N}\Big)_{\alpha\beta}\upsilon_{_{N_\beta}}
\;,\nonumber\\
%%%%%%%%%%%%%%%%%%%%%%%%%%%%%%%%%%%%%%%%%%%%%%%%%%%%
&&\Big[{\cal M}_{_{CPO}}^2\Big]_{12}=B\mu+\Delta_{odd}
\;,\nonumber\\
%%%%%%%%%%%%%%%%%%%%%%%%%%%%%%%%%%%%%%%%%%%%%%%%%%%%
&&\Big[{\cal M}_{_{CPO}}^2\Big]_{22}=\Big(B\mu+\Delta_{odd}\Big){\upsilon_{_u}\over\upsilon_{_d}}
+{\mu\varepsilon_{_N}^2\over\sqrt{2}\upsilon_{_d}}\;.
\label{CP-odd2}
\end{eqnarray}
As we assign the VEVs of left-handed sneutrinos to zero, the expressions in Eq.(\ref{CP-odd2}) recover
the elements of mass-squared matrix for CP-odd Higgs in the MSSM. Additionally,
the concrete expressions for elements of the matrix $A_{_{CPO}}^{(0)}$ can be found in appendix.\ref{app1}.
Similarly, the symmetric matrix in Eq.(\ref{CP-odd1}) contains two massless eigenstates
which correspond to the neutral Goldstones swallowed by neutral gauge bosons $Z,\;Z_{_{BL}}$
after the symmetry $SU(2)\times U_{_Y}(1)\times U_{_{(B-L)}}$  is broken down to the
electromagnetic symmetry $U_e(1)$:
\begin{eqnarray}
&&G^0={\upsilon_{_u}\over \upsilon_{_{\rm EW}}}P_{_u}^0-{\upsilon_{_d}\over\upsilon_{_{\rm EW}}}P_{_d}^0
-\sum\limits_{\alpha=1}^3{\upsilon_{_{L_\alpha}}\over\upsilon_{_{\rm EW}}}P_{_{\tilde{L}_\alpha}}^0
\;,\nonumber\\
&&G_{_{(B-L)}}^0=\eta{\upsilon_{_u}\over\upsilon_t}P_{_u}^0-\eta{\upsilon_{_d}\over\upsilon_t}P_{_d}^0
+(1-\eta)\sum\limits_{\alpha=1}^3{\upsilon_{_{L_\alpha}}\over\upsilon_t}
P_{_{\tilde{L}_\alpha}}^0-\sum\limits_{\alpha=1}^3{\upsilon_{_{N_\alpha}}\over\upsilon_t}P_{_{\tilde{N}_\alpha}}^0\;,
\label{neutral-Goldstones}
\end{eqnarray}
with $\eta=1-{\upsilon_{_{\rm SM}}^2\over\upsilon_{_{\rm EW}}^2}$, and
$\upsilon_t^2=\upsilon_{_{\rm N}}^2+\eta\upsilon_{_{\rm SM}}^2$.
To separate neutral Goldstones from physical states, we define the $8\times8$
orthogonal matrix
\begin{eqnarray}
&&{\cal Z}_{_P}^{(0)}=\left\{Z_{_{CH}}^{(0)}\bigoplus{\cal Z}_{_{\tilde{N}^c}}^P
\right\}
\nonumber\\
&&\hspace{1.2cm}\times
\left\{1_{_{2\times2}}\bigoplus\left(\begin{array}{cccc}
-{\upsilon_{_{\rm SM}}\upsilon_{_{L_1}}\over\upsilon_{_{\rm EW}}\upsilon_t},
&{\upsilon_{_N}\over\upsilon_t},
&{\upsilon_{_{\rm SM}}\upsilon_{_{L_3}}\over\upsilon_{_{\rm EW}}\upsilon_t},
&-{\upsilon_{_{\rm SM}}\upsilon_{_{L_2}}\over\upsilon_{_{\rm EW}}\upsilon_t}\\
-{\upsilon_{_{\rm SM}}\upsilon_{_{L_2}}\over\upsilon_{_{\rm EW}}\upsilon_t},
&-{\upsilon_{_{\rm SM}}\upsilon_{_{L_3}}\over\upsilon_{_{\rm EW}}\upsilon_t},
&{\upsilon_{_N}\over\upsilon_t},
&{\upsilon_{_{\rm SM}}\upsilon_{_{L_1}}\over\upsilon_{_{\rm EW}}\upsilon_t}\\
-{\upsilon_{_{\rm SM}}\upsilon_{_{L_3}}\over\upsilon_{_{\rm EW}}\upsilon_t},
&{\upsilon_{_{\rm SM}}\upsilon_{_{L_2}}\over\upsilon_{_{\rm EW}}\upsilon_t},
&-{\upsilon_{_{\rm SM}}\upsilon_{_{L_1}}\over\upsilon_{_{\rm EW}}\upsilon_t},
&{\upsilon_{_N}\over\upsilon_t}\\
{\upsilon_{_N}\over\upsilon_t},
&{\upsilon_{_{\rm SM}}\upsilon_{_{L_1}}\over\upsilon_{_{\rm EW}}\upsilon_t},
&{\upsilon_{_{\rm SM}}\upsilon_{_{L_2}}\over\upsilon_{_{\rm EW}}\upsilon_t},
&{\upsilon_{_{\rm SM}}\upsilon_{_{L_3}}\over\upsilon_{_{\rm EW}}\upsilon_t}\\
\end{array}\right)\bigoplus 1_{2\times2}\right\}
\nonumber\\
&&\hspace{1.2cm}\times
\left\{\left(\begin{array}{ccc}1,&0,&0\\0,&0,&1\\0,&1,&0\end{array}\right)
\bigoplus1_{5\times5}\right\}
\;,
\label{ZP0}
\end{eqnarray}
then we have
\begin{eqnarray}
&&{\cal Z}_{_P}^{(0)T}\cdot\left(\begin{array}{cc}\Big[{\cal M}_{_{CPO}}^2\Big]_{2\times2}\;\;\;&\Big[A_{_{CPO}}\Big]_{2\times6}\\
\Big[A_{_{CPO}}^T\Big]_{6\times2}\;\;\;&\Big[M_{_{P}}^2\Big]_{6\times6}\end{array}\right)\cdot {\cal Z}_{_P}^{(0)}=
\left(\begin{array}{cc}0_{_{2\times2}}&0_{_{2\times6}}
\;,\nonumber\\
0_{_{6\times2}}&\Big[M_{_{P^0}}^2\Big]_{6\times6}\end{array}\right)\;.
\label{ZPO1}
\end{eqnarray}
Finally, the $8\times8$ mixing matrix $Z_{_{A^0}}$ in CP-odd Higgs sector is written as
\begin{eqnarray}
&&Z_{_{A^0}}= {\cal Z}_{_P}^{(0)}\cdot\left(\begin{array}{cc}1_{2\times2}&0_{2\times6}\\
0_{6\times2}&\Big(Z_{_P}\Big)_{6\times6}
\end{array}\right)
\label{Z-CPO2}
\end{eqnarray}
with $Z_{_P}^\dagger\cdot M_{_{P^0}}^2\cdot Z_{_P}={\rm diag}(m_{_{A_3^0}}^2,\cdots,m_{_{A_8^0}}^2)$.

%%%%%%%%%%%%%%%%%%%%%%%%%%%%%%%%%%%%%BEGIN REVISION%%%%%%%%%%%%%%%%%%
Where
\begin{eqnarray}
&&{\cal Z}_{_{\tilde{N}^c}}^{P,T} m_{_{{\tilde N}^c}}^2
{\cal Z}_{_{\tilde{N}^c}}^P
={\rm diag}(0,{\omega_{_A}-\omega_{_B}\over2\upsilon_{_{N_3}}^2},\;
{\omega_{_A}+\omega_{_B}\over2\upsilon_{_{N_3}}^2})\;,
\label{I-sneutrino-mass}
\end{eqnarray}
and the concrete expressions for $\omega_{_{A,B}}$ are
\begin{eqnarray}
&&\omega_{_A}=\Lambda_{_{\tilde{N}_1^c}}^2\Big(\upsilon_{_N}^2-\upsilon_{_{N_2}}^2\Big)
+\Lambda_{_{\tilde{N}_2^c}}^2\Big(\upsilon_{_N}^2-\upsilon_{_{N_1}}^2\Big)\;,
\nonumber\\
&&\omega_{_B}^2=\omega_{_A}^2-4\Lambda_{_{\tilde{N}_1^c}}^2\Lambda_{_{\tilde{N}_2^c}}^2
\upsilon_{_N}^2\upsilon_{_{N_3}}^2\;.
\label{I-sneutrino-mass1}
\end{eqnarray}
Additionally the orthogonal $3\times3$ rotation is written as
\begin{eqnarray}
&&\left(\begin{array}{c}\Big({\cal Z}_{_{\tilde{N}^c}}^P\Big)_{11}\\
\Big({\cal Z}_{_{\tilde{N}^c}}^P\Big)_{21}\\\Big({\cal Z}_{_{\tilde{N}^c}}^P\Big)_{31}
\end{array}\right)={1\over\upsilon_{_N}}\left(\begin{array}{c}
\upsilon_{_{N_1}}\\\upsilon_{_{N_2}}\\\upsilon_{_{N_3}}\end{array}\right)\;,
\nonumber\\
&&\left(\begin{array}{c}\Big({\cal Z}_{_{\tilde{N}^c}}^P\Big)_{12}\\
\Big({\cal Z}_{_{\tilde{N}^c}}^P\Big)_{22}\\\Big({\cal Z}_{_{\tilde{N}^c}}^P\Big)_{32}
\end{array}\right)={1\over\sqrt{|x_-|^2+|y|^2+|z_-|^2}}\left(\begin{array}{c}
x_-\\y\\z_-\end{array}\right)\;,
\nonumber\\
&&\left(\begin{array}{c}\Big({\cal Z}_{_{\tilde{N}^c}}^P\Big)_{13}\\
\Big({\cal Z}_{_{\tilde{N}^c}}^P\Big)_{23}\\\Big({\cal Z}_{_{\tilde{N}^c}}^P\Big)_{33}
\end{array}\right)={1\over\sqrt{|x_+|^2+|y|^2+|z_+|^2}}\left(\begin{array}{c}
x_+\\y\\z_+\end{array}\right)\;,
\label{I-sneutrino-mass2}
\end{eqnarray}
with
\begin{eqnarray}
%%%%%%%%%%%%%%%%%%%%%%%%%%%%%%%%%%%%%%%%%%%%%%%%%%%%%%%%%%%%%%%%%
&&x_\mp=-{\Lambda_{_{\tilde{N}_2^c}}^4\upsilon_{_{N_2}}^2\over\Lambda_{_{\tilde{N}_1^c}}^2
\upsilon_{_{N_1}}\upsilon_{_{N_3}}}+\Big[\Lambda_{_{\tilde{N}_2^c}}^2
-{\omega_{_A}\mp\omega_{_B}\over2\upsilon_{_{N_3}}^2}\Big]\Big\{{\upsilon_{_{N_1}}\over
\upsilon_{_{N_3}}}
\nonumber\\
&&\hspace{1.2cm}
+{\Lambda_{_{\tilde{N}_2^c}}^2\upsilon_{_{N_2}}^2\over
\Lambda_{_{\tilde{N}_1^c}}^2\upsilon_{_{N_1}}\upsilon_{_{N_3}}}
-{\omega_{_A}\mp\omega_{_B}\over2\Lambda_{_{\tilde{N}_1^c}}^2\upsilon_{_{N_1}}\upsilon_{_{N_3}}}\Big\}
\;,\nonumber\\
%%%%%%%%%%%%%%%%%%%%%%%%%%%%%%%%%%%%%%%%%%%%%%%%%%%%%%%%%%%%%%%%%
&&y={\Lambda_{_{\tilde{N}_2^c}}^2\upsilon_{_{N_2}}\over\upsilon_{_{N_3}}}
\;,\nonumber\\
%%%%%%%%%%%%%%%%%%%%%%%%%%%%%%%%%%%%%%%%%%%%%%%%%%%%%%%%%%%%%%%%%%
&&z_\mp=\Lambda_{_{\tilde{N}_2^c}}^2-{\omega_{_A}\mp\omega_{_B}\over2\upsilon_{_{N_3}}^2}\;.
%%%%%%%%%%%%%%%%%%%%%%%%%%%%%%%%%%%%%%%%%%%%%%%%%%%%%%%%%%%%%%%%%%
\label{I-sneutrino-mass3}
\end{eqnarray}
%%%%%%%%%%%%%%%%%%%%%%%%%%%%%%%%%%%%%%END REVISION%%%%%%%%%%%%%%%%%%%%
Since one-loop effective potential does not induce corrections to the mixing
between $P_{_u},\;P_{_d}$ and $P_{_{L_I}},\;P_{_{N_I}}$, the mixing is dominated
by the $2\times6$ matrix $A_{_{CPO}}^{(0)}$ originating from tree level contributions.
Considering the constraints from neutrino oscillation, we derive the correction
to mass of the lightest CP-odd neutral Higgs $\sim0.01\;{\rm GeV}$ from the mixing
between $P_{_u},\;P_{_d}$ and $P_{_{L_I}},\;P_{_{N_I}}$ as
$m_{_{{\tilde L}_I}}\simeq\Lambda_{_{\tilde{N}_{1,2}^c}}\simeq m_{_{Z_{BL}}}\sim1\;{\rm TeV}$.
This fact implies the mass of the lightest CP-odd Higgs
\begin{eqnarray}
%%%%%%%%%%%%%%%%%%%%%%%%%%%%%%%%%%%%%%%%%%%%%%%%%%%%%%%%%%%%%%%%%%%%%%%%%%%%%%%%%%%%%
&&m_{_{A_3^0}}^2\simeq{B\mu+\Delta_{odd}\over\sin2\beta}\;,
%%%%%%%%%%%%%%%%%%%%%%%%%%%%%%%%%%%%%%%%%%%%%%%%%%%%%%%%%%%%%%%%%%%%%%%%%%%%%%%%%%%%%
\label{MA0}
\end{eqnarray}
here we adopt the definition
\begin{eqnarray}
&&\tan\beta=\upsilon_{_u}/\sqrt{\upsilon_{_d}^2+\sum\limits_{\alpha=1}^3\upsilon_{_{L_\alpha}}^2}\;.
\label{tanbeta}
\end{eqnarray}
Accordingly the masses of other CP-odd scalars are formulated as
\begin{eqnarray}
%%%%%%%%%%%%%%%%%%%%%%%%%%%%%%%%%%%%%%%%%%%%%%%%%%%%%%%%%%%%%%%%%%%
&&m_{_{A_{(3+i)}^0}}^2\simeq
m_{_{{\tilde L}_i}}^2+\Delta T_{_{\tilde L}}+{1\over2}\Big(m_{_{\rm Z}}^2\cos2\beta-m_{_{Z_{BL}}}^2\Big)\;,\;\;\;(i=1,\;2,\;3)
\;,\nonumber\\
&&m_{_{A_7^0}}^2\simeq{\omega_{_A}-\omega_{_B}\over2\upsilon_{_{N_3}}^2}
\;,\nonumber\\
&&m_{_{A_8^0}}^2\simeq {\omega_{_A}+\omega_{_B}\over2\upsilon_{_{N_3}}^2}\;,
%%%%%%%%%%%%%%%%%%%%%%%%%%%%%%%%%%%%%%%%%%%%%%%%%%%%%%%%%%%%%%%%%%%
\label{MA4-8}
\end{eqnarray}
and the $6\times6$ mixing matrix is approximated as
\begin{eqnarray}
%%%%%%%%%%%%%%%%%%%%%%%%%%%%%%%%%%%%%%%%%%%%%%%%%%%%%%%%%%%%%%%%%%%
&&Z_{_P}\simeq\left(\begin{array}{ccc}
1&\Big[{(\delta^2m_{_{HL}}^{odd})_{I}\over m_{_{A_{(3+I)}^0}}^2-m_{_{A_3^0}}^2}\Big]_{1\times3}&0_{1\times2}\\
-\Big[{(\delta^2m_{_{HL}}^{odd})_{I}\over m_{_{A_{(3+I)}^0}}^2-m_{_{A_3^0}}^2}\Big]_{3\times1}&1_{3\times3}&0_{3\times2}\\
0_{2\times1}&0_{2\times3}&1_{2\times2}\end{array}\right)\;.
%%%%%%%%%%%%%%%%%%%%%%%%%%%%%%%%%%%%%%%%%%%%%%%%%%%%%%%%%%%%%%%%%%%
\label{mixing-CP-odd1}
\end{eqnarray}
Where
\begin{eqnarray}
%%%%%%%%%%%%%%%%%%%%%%%%%%%%%%%%%%%%%%%%%%%%%%%%%%%%%%%%%%%%%%%%%%%
&&(\delta^2m_{_{HL}}^{odd})_{I}\simeq
-{\mu\upsilon_{_{EW}}\over\sqrt{2}\upsilon_{_u}}\zeta_{_I}+\cos\beta\Big[m_{_{\tilde{L}_I}}^2+{1\over2}m_{_{Z_{BL}}}^2
+{g_1^2+g_2^2\over8}(\upsilon_{_u}^2-\upsilon_{_d}^2)\Big]{\upsilon_{_{L_I}}\over\upsilon_{_u}}
%%%%%%%%%%%%%%%%%%%%%%%%%%%%%%%%%%%%%%%%%%%%%%%%%%%%%%%%%%%%%%%%%%%
\label{mixing-CP-odd2}
\end{eqnarray}

\section{The lightest CP-even Higgs mass matrix\label{sec3}}
\indent\indent
It is well known for quite long time that radiative
corrections modify the tree level mass squared matrix of neutral
Higgs substantially in supersymmetry, and the main effect in those radiative
contributions originates from Feynman loops
involving the third generation fermions and their supersymmetric
partners\cite{Haber1}. In order to obtain mass of the lightest
neutral CP-even Higgs reasonably, we should also include the one-loop
corrections from those fermions and corresponding supersymmetric
partner in the MSSM with local $U(1)_{B-L}$ symmetry. In the interaction basis
$H^{0,T}=(H_{_u}^0,\;H_{_d}^0,\;\tilde{\nu}_{_{L_I}},\;\tilde{\nu}_{_{R_J}})\;\;(I,\;J=1,\;2,\;3)$,
the $8\times8$ symmetric mass squared matrix is written as
\begin{eqnarray}
&&\left(\begin{array}{cc}\Big[M_{_{H^0}}^2\Big]_{2\times2}\;\;\;&\Big[A_{_{CPE}}\Big]_{2\times6}\\
\Big[A_{_{CPE}}^T\Big]_{6\times2}\;\;\;&\Big[M_{_{S}}^2\Big]_{6\times6}
\end{array}\right)\;,
\label{M-CPE0}
\end{eqnarray}
here the $2\times2$ mass squared matrix $M_{_{H^0}}^2$ is
\begin{eqnarray}
&&M_{_{H^0}}^2=\left(\begin{array}{ll}\Big[{\cal M}^2_{_{CPE}}\Big]_{11}+\Delta_{11}&
\Big[{\cal M}^2_{_{CPE}}\Big]_{12}+\Delta_{12}\\
\Big[{\cal M}^2_{_{CPE}}\Big]_{12}+\Delta_{12}&\Big[{\cal M}^2_{_{CPE}}\Big]_{22}+\Delta_{22}\end{array}\right)\;,
\label{M-CPE}
\end{eqnarray}
with
\begin{eqnarray}
&&\Big[{\cal M}^2_{_{CPE}}\Big]_{11}=\Big(B\mu+\Delta_{odd}\Big){\upsilon_{_d}\over\upsilon_{_u}}
+m_{_{\rm Z}}^2\sin^2\beta
\;,\nonumber\\
&&\Big[{\cal M}^2_{_{CPE}}\Big]_{12}=-\Big(B\mu+\Delta_{odd}\Big)-m_{_{\rm Z}}^2\sin\beta\cos\beta\;,
\nonumber\\
&&\Big[{\cal M}^2_{_{CPE}}\Big]_{22}=\Big(B\mu+\Delta_{odd}\Big){\upsilon_{_u}\over\upsilon_{_d}}
+m_{_{\rm Z}}^2\cos^2\beta\;.
\label{M-CPE1}
\end{eqnarray}
Where
\begin{eqnarray}
&&\Delta_{11}=\Delta_{11}^T+\Delta_{11}^{B}+\Delta_{11}^{L}\;,
\nonumber\\
&&\Delta_{12}=\Delta_{12}^T+\Delta_{12}^{B}+\Delta_{12}^{L}\;,
\nonumber\\
&&\Delta_{22}=\Delta_{22}^T+\Delta_{22}^{B}+\Delta_{22}^{L}\;,
\nonumber\\
\label{M-CPE2}
\end{eqnarray}
and $\Delta_{11}^T,\;\Delta_{12}^T,\;\Delta_{22}^T$ represent the tree level corrections to CP-even
Higgs mass squared matrix from sneutrinos after electroweak symmetry is broken:
\begin{eqnarray}
&&\Delta_{11}^T={1\over\sqrt{2}\upsilon_{_u}}\sum\limits_{\alpha,\beta}^3\upsilon_{_{L_\alpha}}
\Big(A_{_N}\Big)_{\alpha\beta}\upsilon_{_{N_\beta}}\;,
\nonumber\\
&&\Delta_{12}^T=m_{_{\rm Z}}^2\sin\beta\cos\beta\Big\{1
-{\upsilon_{_d}\over\sqrt{\upsilon_{_d}^2+\upsilon_{_{\rm EW}}^2-\upsilon_{_{\rm SM}}^2}}\Big\}
\;,\nonumber\\
&&\Delta_{22}^T={g_1^2+g_2^2\over4}(\upsilon_{_{\rm EW}}^2-\upsilon_{_{\rm SM}}^2)
+{\mu\varepsilon_{_N}^2\over\sqrt{2}\upsilon_{_d}}\;.
\nonumber\\
\label{M-CPE2a}
\end{eqnarray}
In fact $\Delta_{11}^T=\Delta_{12}^T=\Delta_{22}^T=0$ when the VEVs of left-handed sneutrinos
are assigned to zero.
The concrete expressions for the radiative corrections from quark sector
$\Delta_{ij}^{B}\;\;(i,j=1,\;2)$ up to
two-loop level can be found in literature\cite{2loop-HiggsM}, and the one-loop
corrections from lepton sectors can also be found in \cite{1loop-HiggsM}
within framework of the MSSM. Obviously radiative corrections modify the mass spectrum
of neutral Higgs drastically, and two-loop corrections decrease that from one-loop
in most of the MSSM parameter space. Here it is sufficient to include the one-loop
corrections and the leading terms of two-loop corrections to the mass matrix of CP-even Higgs, and the expressions for
$\Delta_{ij}^{B},\;\Delta_{ij}^{L}$ are given in appendix.\ref{app1a}.

Furthermore, the $2\times6$ matrix $A_{_{CPE}}$ is
\begin{eqnarray}
&&A_{_{CPE}}=A_{_{CPE}}^{(0)}+\Delta A_{_{CPE}}\;,
\label{M-CPE2b}
\end{eqnarray}
where the tree level contribution $A_{_{CPE}}^{(0)}$ is given in appendix.\ref{app1},
and nontrivial one-loop corrections $\Delta A_{_{CPE}}$ are
\begin{eqnarray}
%%%%%%%%%%%%%%%%%%%%%%%%%%%%%%%%%%%%%%%%%%%%%%%%%%%%%%%
&&\Big(\Delta A_{_{CPE}}\Big)_{1(3+I)}={G_{_F}m_{_t}^2\over2\sqrt{2}\pi^2}
{g_{_{BL}}^2\upsilon_{_u}\upsilon_{_{N_I}}\over\sin^2\beta}\Big(m_{_{\tilde{t}_L}}^2-m_{_{\tilde{t}_R}}^2\Big)
\Big\{{\ln m_{_{\tilde{t}_1}}^2-\ln m_{_{\tilde{t}_2}}^2\over m_{_{\tilde{t}_1}}^2-m_{_{\tilde{t}_2}}^2}
\nonumber\\
&&\hspace{3.2cm}
+{A_{_t}(A_{_t}-\mu\cot\beta)\over(m_{_{\tilde{t}_1}}^2-m_{_{\tilde{t}_2}}^2)^2}
g(m_{_{\tilde{t}_1}}^2,m_{_{\tilde{t}_2}}^2)\Big\}
\nonumber\\
&&\hspace{3.2cm}
-{G_{_F}m_{_b}^2\over2\sqrt{2}\pi^2}
{g_{_{BL}}^2\upsilon_{_d}\upsilon_{_{N_I}}\over\cos^2\beta}\Big(m_{_{\tilde{b}_L}}^2-m_{_{\tilde{b}_R}}^2\Big)
{\mu(A_{_b}-\mu\tan\beta)\over(m_{_{\tilde{b}_1}}^2-m_{_{\tilde{b}_2}}^2)^2}
g(m_{_{\tilde{b}_1}}^2,m_{_{\tilde{b}_2}}^2)
\nonumber\\
&&\hspace{3.2cm}
-{G_{_F}m_{_\tau}^2\over2\sqrt{2}\pi^2}
{g_{_{BL}}^2\upsilon_{_d}\upsilon_{_{N_I}}\over\cos^2\beta}\Big(m_{_{\tilde{\tau}_L}}^2-m_{_{\tilde{\tau}_R}}^2\Big)
{\mu(A_{_\tau}-\mu\tan\beta)\over(m_{_{\tilde{\tau}_1}}^2-m_{_{\tilde{\tau}_2}}^2)^2}
g(m_{_{\tilde{\tau}_1}}^2,m_{_{\tilde{\tau}_2}}^2)
\;,\nonumber\\
%%%%%%%%%%%%%%%%%%%%%%%%%%%%%%%%%%%%%%%%%%%%%%%%%%%%%%%%%%
&&\Big(\Delta A_{_{CPE}}\Big)_{2(3+I)}=-{G_{_F}m_{_t}^2\over2\sqrt{2}\pi^2}
{g_{_{BL}}^2\upsilon_{_u}\upsilon_{_{N_I}}\over\sin^2\beta}\Big(m_{_{\tilde{t}_L}}^2-m_{_{\tilde{t}_R}}^2\Big)
{\mu(A_{_t}-\mu\cot\beta)\over(m_{_{\tilde{t}_1}}^2-m_{_{\tilde{t}_2}}^2)^2}
g(m_{_{\tilde{t}_1}}^2,m_{_{\tilde{t}_2}}^2)
\nonumber\\
&&\hspace{3.2cm}
+{G_{_F}m_{_b}^2\over2\sqrt{2}\pi^2}
{g_{_{BL}}^2\upsilon_{_d}\upsilon_{_{N_I}}\over\cos^2\beta}\Big(m_{_{\tilde{b}_L}}^2-m_{_{\tilde{b}_R}}^2\Big)
\Big\{{\ln m_{_{\tilde{b}_1}}^2-\ln m_{_{\tilde{b}_2}}^2\over m_{_{\tilde{b}_1}}^2-m_{_{\tilde{b}_2}}^2}
\nonumber\\
&&\hspace{3.2cm}
+{A_{_b}(A_{_b}-\mu\tan\beta)\over(m_{_{\tilde{b}_1}}^2-m_{_{\tilde{b}_2}}^2)^2}
g(m_{_{\tilde{b}_1}}^2,m_{_{\tilde{b}_2}}^2)\Big\}
\nonumber\\
&&\hspace{3.2cm}
+{G_{_F}m_{_\tau}^2\over2\sqrt{2}\pi^2}
{g_{_{BL}}^2\upsilon_{_d}\upsilon_{_{N_I}}\over\cos^2\beta}\Big(m_{_{\tilde{\tau}_L}}^2-m_{_{\tilde{\tau}_R}}^2\Big)
\Big\{{\ln m_{_{\tilde{\tau}_1}}^2-\ln m_{_{\tilde{\tau}_2}}^2\over m_{_{\tilde{\tau}_1}}^2-m_{_{\tilde{\tau}_2}}^2}
\nonumber\\
&&\hspace{3.2cm}
+{A_{_\tau}(A_{_\tau}-\mu\tan\beta)\over(m_{_{\tilde{\tau}_1}}^2-m_{_{\tilde{\tau}_2}}^2)^2}
g(m_{_{\tilde{\tau}_1}}^2,m_{_{\tilde{\tau}_2}}^2)\Big\}\;,
\;\;\;(I=1,\;2,\;3)\;,
%%%%%%%%%%%%%%%%%%%%%%%%%%%%%%%%%%%%%%%%%%%%%%%%%%%%%%%%%%%%
\label{M-CPE2c}
\end{eqnarray}
with the concrete expression of $g(x,y)$ presented in appendix.\ref{app1a}.
Meanwhile the radiative corrections to $\Big(\Delta A_{_{CPE}}\Big)_{1I},\;\Big(\Delta A_{_{CPE}}\Big)_{2I}$
are proportional to $\upsilon_{_{L_I}}$, and can be neglected safely here.
Note that $\Big(\Delta A_{_{CPE}}\Big)_{1(3+I)},\;\Big(\Delta A_{_{CPE}}\Big)_{2(3+I)}$
are independent of the renormalization scale $\Lambda$, as they should be.

At the tree level, i.e. $\Delta_{_{ij}}^{B}=\Delta_{_{ij}}^L=0\;\;(i,j=1,\;2)$
and $\Delta A_{_{CPE}}=0$,
there are relations between the CP-even and CP-odd Higgs masses\cite{Chang-Feng1}
\begin{eqnarray}
&&\sum\limits_{i=1}^8m_{_{H_i^0}}^2=m_{_{\rm Z}}^2+m_{_{Z_{BL}}}^2+\sum\limits_{i=1}^6m_{_{A_{(2+i)}}}^2
\;,\nonumber\\
&&\prod\limits_{i=1}^8m_{_{H_i^0}}^2=\Big({\upsilon_{_{\rm EW}}^2-2\upsilon_{_u}^2
\over\upsilon_{_{\rm EW}}^2}\Big)^2
\Big({\upsilon_{_{\rm N}}^2-\upsilon_{_{\rm EW}}^2+\upsilon_{_{\rm SM}}^2\over
\upsilon_{_{\rm N}}^2+\upsilon_{_{\rm EW}}^2-\upsilon_{_{\rm SM}}^2}\Big)^2
m_{_{\rm Z}}^2m_{_{Z_{BL}}}^2\prod\limits_{i=1}^6m_{_{A_{(2+i)}}}^2\;.
\label{neutral-Relations}
\end{eqnarray}
Certainly, radiative corrections to the neutral Higgs mass squared matrices
destroy the relations in Eq.(\ref{neutral-Relations}) strongly.

Considering the constraints from neutrino oscillation on parameter space
of the model considered here, we find that the radiative correction from right-handed
neutrinos/sneutrinos on the lightest CP-even Higgs mass is
negligible. This conclusion coincides with that presented in
Ref\cite{Haber2}.

%%%%%%%%%%%%%%%%%%%%%%%%%%%%%%%%%%%%%%%BEGIN REVISION%%%%%%%%%%%%%%%%%%%%%%%%
Applying above equations, one finds that
the mass squared matrices for real part of sneutrinos is
approximately approached as
\begin{eqnarray}
&&M_{_S}^2\simeq\left(\begin{array}{cc}
\left[\begin{array}{l}[\delta^2m_{_{LL}}^{even}]_{IJ}+(m_{_{{\tilde L}_I}}^2+\Delta T_{_{\tilde L}})\delta_{IJ}\\
+\Big({1\over2}m_{_{\rm Z}}^2\cos2\beta
-{m_{_{Z_{BL}}}^2\over2}\Big)\delta_{IJ}\end{array}\right]_{3\times3}\;,
&\Big[\Big(\delta^2m_{_{LR}}^{even}\Big)_{IJ^\prime}\Big]_{3\times3}\\
\Big[\Big(\delta^2m_{_{LR}}^{even}\Big)_{I^\prime J}\Big]_{3\times3}\;,
&\Big[\Big(M_{_{{\tilde N}^c}}^2\Big)_{I^\prime J^\prime}
+\Big(\delta^2m_{_{RR}}^{even}\Big)_{IJ^\prime}\Big]_{3\times3}\end{array}\right)\;,
\label{sneutrino-mass-squared matrices}
\end{eqnarray}
where the $3\times3$ mass squared matrix $M_{_{{\tilde N}^c}}^2$ is
\begin{eqnarray}
&&M_{_{{\tilde N}^c}}^2\simeq\left(\begin{array}{ccc}
\Lambda_{_{\tilde{N}_1^c}}^2+{g_{_{BL}}^2\over2}\upsilon_{_{N_1}}^2\;,\;\;
&{g_{_{BL}}^2\over2}\upsilon_{_{N_1}}\upsilon_{_{N_2}}\;,\;\;
&{g_{_{BL}}^2\over2}\upsilon_{_{N_1}}\upsilon_{_{N_3}}
-{\upsilon_{_{N_1}}\over\upsilon_{_{N_3}}}\Lambda_{_{\tilde{N}_1^c}}^2\\
{g_{_{BL}}^2\over2}\upsilon_{_{N_1}}\upsilon_{_{N_2}}\;,\;\;&
\Lambda_{_{\tilde{N}_2^c}}^2+{g_{_{BL}}^2\over2}\upsilon_{_{N_2}}^2\;,\;\;
&{g_{_{BL}}^2\over2}\upsilon_{_{N_2}}\upsilon_{_{N_3}}
-{\upsilon_{_{N_2}}\over\upsilon_{_{N_3}}}\Lambda_{_{\tilde{N}_2^c}}^2\\
{g_{_{BL}}^2\over2}\upsilon_{_{N_1}}\upsilon_{_{N_3}}
-{\upsilon_{_{N_1}}\over\upsilon_{_{N_3}}}\Lambda_{_{\tilde{N}_1^c}}^2
\;,\;\;&{g_{_{BL}}^2\over2}\upsilon_{_{N_2}}\upsilon_{_{N_3}}
-{\upsilon_{_{N_2}}\over\upsilon_{_{N_3}}}\Lambda_{_{\tilde{N}_2^c}}^2\;,\;\;
&{g_{_{BL}}^2\over2}\upsilon_{_{N_3}}^2+{\upsilon_{_{N_1}}^2\Lambda_{_{\tilde{N}_1^c}}^2
+\upsilon_{_{N_2}}^2\Lambda_{_{\tilde{N}_2^c}}^2\over\upsilon_{_{N_3}}^2}
\end{array}\right)\;,
\label{R-sneutrino-mass1}
\end{eqnarray}
and the concrete expressions for $\delta^2m_{_{LL}}^{even},\;\delta^2m_{_{LR}}^{even}$ and
$\delta^2m_{_{RR}}^{even}$ can be found in the appendix.\ref{app1}.
Defining the orthogonal $3\times3$ rotation ${\cal Z}_{_{\tilde{N}^c}}$, we get
\begin{eqnarray}
&&{\cal Z}_{_{\tilde{N}^c}}^TM_{_{{\tilde N}^c}}^2{\cal Z}_{_{\tilde{N}^c}}={\rm diag}
(M_{_{\tilde{\nu}_{_R}^1}}^2,\;M_{_{\tilde{\nu}_{_R}^2}}^2,\;M_{_{\tilde{\nu}_{_R}^3}}^2)\;,
\label{R-sneutrino-mass2}
\end{eqnarray}
where
\begin{eqnarray}
&&M_{_{\tilde{\nu}_{_R}^1}}^2={1\over2}m_{_{Z_{BL}}}^2\;,
\nonumber\\
&&M_{_{\tilde{\nu}_{_R}^2}}^2={\omega_{_A}-\omega_{_B}\over2\upsilon_{_{N_3}}^2}\;,
\nonumber\\
&&M_{_{\tilde{\nu}_{_R}^3}}^2={\omega_{_A}+\omega_{_B}\over2\upsilon_{_{N_3}}^2}\;.
\label{R-sneutrino-mass3}
\end{eqnarray}
Additionally the orthogonal $3\times3$ rotation is written as
\begin{eqnarray}
&&\left(\begin{array}{c}\Big({\cal Z}_{_{\tilde{N}^c}}\Big)_{11}\\
\Big({\cal Z}_{_{\tilde{N}^c}}\Big)_{21}\\\Big({\cal Z}_{_{\tilde{N}^c}}\Big)_{31}
\end{array}\right)={1\over\upsilon_{_N}}\left(\begin{array}{c}
\upsilon_{_{N_1}}\\\upsilon_{_{N_2}}\\\upsilon_{_{N_3}}\end{array}\right)\;,
\nonumber\\
&&\left(\begin{array}{c}\Big({\cal Z}_{_{\tilde{N}^c}}\Big)_{12}\\
\Big({\cal Z}_{_{\tilde{N}^c}}\Big)_{22}\\\Big({\cal Z}_{_{\tilde{N}^c}}\Big)_{32}
\end{array}\right)={1\over\sqrt{|X_-|^2+|Y_-|^2+|Z_-|^2}}\left(\begin{array}{c}
X_-\\Y_-\\Z_-\end{array}\right)\;,
\nonumber\\
&&\left(\begin{array}{c}\Big({\cal Z}_{_{\tilde{N}^c}}\Big)_{13}\\
\Big({\cal Z}_{_{\tilde{N}^c}}\Big)_{23}\\\Big({\cal Z}_{_{\tilde{N}^c}}\Big)_{33}
\end{array}\right)={1\over\sqrt{|X_+|^2+|Y_+|^2+|Z_+|^2}}\left(\begin{array}{c}
X_+\\Y_+\\Z_+\end{array}\right)\;,
\label{R-sneutrino-mass5}
\end{eqnarray}
with
\begin{eqnarray}
&&X_\mp={\upsilon_{_{N_3}}\over\upsilon_{_{N_1}}}\Big\{\Big[2\Lambda_{_{\tilde{N}_1^c}}^2
-g_{_{BL}}^2(\upsilon_{_N}^2-\upsilon_{_{N_1}}^2)\Big]\Big[\Lambda_{_{\tilde{N}_2^c}}^2
(\upsilon_{_N}^2-\upsilon_{_{N_1}}^2)\pm\omega_{_B}\Big]
\nonumber\\
&&\hspace{1.2cm}
-2\Lambda_{_{\tilde{N}_1^c}}^4(\upsilon_{_N}^2-\upsilon_{_{N_2}}^2)
-g_{_{BL}}^2\Lambda_{_{\tilde{N}_1^c}}^2\upsilon_{_{N_1}}^2\upsilon_{_{N_2}}^2
+g_{_{BL}}^2\Lambda_{_{\tilde{N}_1^c}}^2\upsilon_{_N}^2\upsilon_{_{N_3}}^2\Big\}
\;,\nonumber\\
&&Y_\mp=\upsilon_{_{N_2}}\upsilon_{_{N_3}}\Big\{g_{_{BL}}^2(\omega_{_A}\pm\omega_{_B})
-4\Lambda_{_{\tilde{N}_1^c}}^2\Lambda_{_{\tilde{N}_2^c}}^2\Big\}
\;,\nonumber\\
&&Z_\mp=2\Lambda_{_{\tilde{N}_1^c}}^4(\upsilon_{_N}^2-\upsilon_{_{N_2}}^2)
+2\Lambda_{_{\tilde{N}_1^c}}^2\Lambda_{_{\tilde{N}_2^c}}^2(\upsilon_{_{N_2}}^2-\upsilon_{_{N_3}}^2)
-g_{_{BL}}^2\Lambda_{_{\tilde{N}_1^c}}^2(\upsilon_{_N}^2+\upsilon_{_{N_2}}^2)\upsilon_{_{N_3}}^2
\nonumber\\
&&\hspace{1.2cm}
+g_{_{BL}}^2\Lambda_{_{\tilde{N}_2^c}}^2(\upsilon_{_N}^2-\upsilon_{_{N_1}}^2)\upsilon_{_{N_3}}^2
\mp(2\Lambda_{_{\tilde{N}_1^c}}^2-g_{_{BL}}^2\upsilon_{_{N_3}}^2)\omega_{_B}\;.
\label{R-sneutrino-mass6}
\end{eqnarray}

To continue with our analysis on the mass spectrum and mixing in the neutral scalar sector,
we assume $\max(|\upsilon_{_u}\Big(A_{_N}\Big)_{IJ}|,\;|\mu\upsilon_{_d}(Y_{_N})_{_{IJ}}|,\;
g_{_{BL}}^2\upsilon_{_{L_I}}\upsilon_{_{N_J}})\ll\min(|m_{_{{\tilde L}_I}}^2
+\Delta T_{_{\tilde L}}+\Big({1\over2}m_{_{\rm Z}}^2\cos2\beta
-{m_{_{Z_{BL}}}^2\over2}\Big)-M_{_{\tilde{\nu}_{_R}^J}}^2|)$, then obtain
\begin{eqnarray}
&&m_{_{S_i}}^2\simeq m_{_{{\tilde L}_i}}^2+\Delta T_{_{\tilde L}}+{1\over2}\Big(m_{_{\rm Z}}^2\cos2\beta
-m_{_{Z_{BL}}}^2\Big)
\nonumber\\
&&\hspace{1.2cm}
+\sum\limits_{\alpha=1}^3{2M_{_{\tilde{\nu}_{_R}^\alpha}}^2
((\delta^2m_{_{LR}}^{even}){\cal Z}_{_{{\tilde N}^c}})_{i\alpha}\over(2M_{_{\tilde{\nu}_{_R}^\alpha}}^2
-2m_{_{{\tilde L}_i}}^2-2\Delta T_{_{\tilde L}}-m_{_{\rm Z}}^2\cos2\beta+m_{_{Z_{BL}}}^2)^2}
\;,\nonumber\\
&&m_{_{S_{(3+i)}}}^2\simeq M_{_{\tilde{\nu}_{_R}^i}}^2
+\sum\limits_{\alpha=1}^3{(2m_{_{{\tilde L}_\alpha}}^2+2\Delta T_{_{\tilde L}}
+m_{_{\rm Z}}^2\cos2\beta-2m_{_{Z_{BL}}}^2)
((\delta^2m_{_{LR}}^{even}){\cal Z}_{_{{\tilde N}^c}})_{i\alpha}\over(2M_{_{\tilde{\nu}_{_R}^i}}^2
-2m_{_{{\tilde L}_\alpha}}^2-2\Delta T_{_{\tilde L}}
-m_{_{\rm Z}}^2\cos2\beta+m_{_{Z_{BL}}}^2)^2}\;.
\label{sneutrino-mass1}
\end{eqnarray}
Meanwhile, the mixing matrix $Z_{_S}$ is
\begin{eqnarray}
&&Z_{_S}\simeq{\tiny \left(\begin{array}{cc}
[Z_{_{\tilde L}}]_{3\times3}\;,\;\;\;&\Big[{2((\delta^2m_{_{LR}}^{even}){\cal Z}_{_{{\tilde N}^c}})_{ij\prime}\over
2M_{_{\tilde{\nu}_{_R}^{j\prime}}}^2-2m_{_{{\tilde L}_{i}}}^2-2\Delta T_{_{\tilde L}}-m_{_{\rm Z}}^2\cos2\beta
+m_{_{Z_{BL}}}^2}\Big]_{3\times3}\\
-\Big[{{2((\delta^2m_{_{LR}}^{even}){\cal Z}_{_{{\tilde N}^c}})_{i^\prime j}\over
2M_{_{\tilde{\nu}_{_R}^\alpha}}^2-2m_{_{{\tilde L}_j}}^2-2\Delta T_{_{\tilde L}}-m_{_{\rm Z}}^2\cos2\beta
+m_{_{Z_{BL}}}^2}}\Big]_{3\times3}\;,\;\;\;&\Big[({\cal Z}_{_{\tilde{N}^c}})_{i\prime j\prime}\Big]_{3\times3}
\end{array}\right)}\;,
\label{sneutrino-mass3}
\end{eqnarray}
and the $8\times8$ mixing matrix is written as
\begin{eqnarray}
&&Z_{_{H_0}}=\left(\begin{array}{cc}
\Big[(Z_{_R})_{ij}\Big]_{2\times2}\;,\;\;\;&\Big[\sum\limits_{\alpha=1}^2{(Z_{_R})_{i\alpha}
(Z_{_R}^TA_{_{CPE}}Z_{_S})_{\alpha b}\over m_{_{S_b}}^2-\lambda_\alpha}\Big]_{2\times6}\\
-\Big[\sum\limits_{\alpha=1}^6{(Z_{_S})_{a\alpha}(Z_{_S}^TA_{_{CPE}}^TZ_{_R})_{\alpha j}
\over m_{_{S_\alpha}}^2-\lambda_j}\Big]_{6\times2}\;,\;\;\;&\Big[(Z_{_S})_{ab}\Big]_{6\times6}
\end{array}\right)\;.
\label{sneutrino-mass4}
\end{eqnarray}
Correspondingly the expression for $Z_{_{\tilde L}}$ can be found in appendix .\ref{app1}.
Then we formulate the mass squared for those CP-even neutral scalars as
\begin{eqnarray}
&&m_{_{H_1^0}}^2\simeq\lambda_1+\sum\limits_{\alpha=1}^6{\Big((Z_{_R}^TA_{_{CPE}}Z_{_S})_{1\alpha}
\Big)^2\over(m_{_{S_\alpha}}^2-\lambda_1)^2}m_{_{S_\alpha}}^2
\;,\nonumber\\
&&m_{_{H_2^0}}^2\simeq\lambda_2+\sum\limits_{\alpha=1}^6{\Big((Z_{_R}^TA_{_{CPE}}Z_{_S})_{2\alpha}
\Big)^2\over(m_{_{S_\alpha}}^2-\lambda_2)^2}m_{_{S_\alpha}}^2
\;,\nonumber\\
&&m_{_{H_{(2+i)}^0}}^2\simeq m_{_{S_i}}^2+{\Big((Z_{_R}^TA_{_{CPE}}Z_{_S})_{1\alpha}
\Big)^2\over(m_{_{S_\alpha}}^2-\lambda_1)^2}\lambda_1+{\Big((Z_{_R}^TA_{_{CPE}}Z_{_S})_{2\alpha}
\Big)^2\over(m_{_{S_\alpha}}^2-\lambda_2)^2}\lambda_2\;,
\label{sneutrino-mass5}
\end{eqnarray}
with
\begin{eqnarray}
&&Z_{_R}=\left(\begin{array}{cc}\cos\alpha_{_H}&-\sin\alpha_{_H}\\
\sin\alpha_{_H}&\cos\alpha_{_H}\end{array}\right)
\;,\nonumber\\
&&\tan2\alpha_{_H}={2\Big(M_{_{H^0}}^2\Big)_{12}\over
\Big(M_{_{H^0}}^2\Big)_{11}-\Big(M_{_{H^0}}^2\Big)_{22}}
\;,\nonumber\\
&&\lambda_{1,2}={1\over2}\Big[\Big(M_{_{H^0}}^2\Big)_{11}+\Big(M_{_{H^0}}^2\Big)_{22}\Big]
\mp\sqrt{{1\over4}\Big[\Big(M_{_{H^0}}^2\Big)_{11}-\Big(M_{_{H^0}}^2\Big)_{22}\Big]^2
+\Big(M_{_{H^0}}^2\Big)_{12}^2}\;.
\label{sneutrino-mass6}
\end{eqnarray}
%%%%%%%%%%%%%%%%%%%%%%%%%%%%%%%%%%%%%%%END REVISION%%%%%%%%%%%%%%%%%%%%%%%%

One most stringent constraint on parameter space of the model is that the mass squared
matrix in Eq.(\ref{M-CPE0}) should produce an eigenvalue
around $(125\;{\rm GeV})^2$ as mass squared of the lightest neutral CP-even Higgs.
The current combination of the ATLAS and CMS data gives:
\begin{eqnarray}
&&m_{_{h^0}}=125.9\pm2.1\;{\rm GeV}\;,
\label{M-h0}
\end{eqnarray}
this fact constrains parameter space of extensions of the SM strongly.
%%%%%%%%%%%%%%%%%%%Begin modification%%%%%%%%%%%%%%%%%%%%%%%%%%%%
In the MSSM, the SM-liked Higgs satisfying the condition (\ref{M-h0}) demands
both relatively light scalar top quarks with large mixing, or a large mass
hierarchy between two scalar top quarks in the case that one scalar top quark
is light \cite{Carena1}.
%%%%%%%%%%%%%%%%%%%%%End modification%%%%%%%%%%%%%%%%%%%%%%%%%%%
Considering the constraints from neutrino oscillation experimental data,
the mixing between $H_{_u}^0,\;H_{_d}^0$ and real part of right-handed sneutrinos
is only about $10^{-6}$ at tree level in the MSSM with local $U(1)_{B-L}$ symmetry.
Nevertheless one-loop radiative corrections
can enhance the mixing drastically when $A_{_t},\;\upsilon_{_N}\ge1\;{\rm TeV}$
and $\tan\beta>1$, and the corrections to mass of the lightest CP-even Higgs
from this mixture increase the corresponding MSSM radiative contributions.

\section{The mass matrix for neutralinos and neutrinos\label{sec4}}
\indent\indent
After the local gauge symmetry $SU(2)_{_L}\otimes U(1)_{_Y}\otimes U(1)_{_{(B-L)}}$
is broken down, the nonzero VEVs of left- and right-handed sneutrinos induce the mixing between neutralinos (charginos)
and neutrinos (charged leptons). As mentioned above, the MSSM with  local $U(1)_{B-L}$
symmetry naturally predicates two sterile neutrinos with sub-eV masses\cite{Perez6,Senjanovic1}.
In the basis $(\nu_{_{L_I}},\;N_{_J}^c,\;i\lambda_{_{BL}},\;i\lambda_{_B},\;i\lambda_{_A}^3,\;
\psi_{_{H_d}}^1,\;\psi_{_{H_u}}^2)$, the mass matrix for neutralino-neutrino is formulated as
\begin{eqnarray}
&&M_{_N}=\left(\begin{array}{ccc}
0_{3\times3}\;\;&\Big({\cal A}_{_N}^{(1)}\Big)_{3\times4}\;\;&\Big({\cal A}_{_N}^{(2)}\Big)_{3\times4}\;\;\\
\Big({\cal A}_{_N}^{(1)T}\Big)_{4\times3}\;\;&\Big({\cal M}_{_N}^{(0)}\Big)_{4\times4}\;\;
&\Big({\cal A}_{_N}^{(3)}\Big)_{4\times4}\;\;\\
\Big({\cal A}_{_N}^{(2)T}\Big)_{4\times3}\;\;&\Big({\cal A}_{_N}^{(3)T}\Big)_{4\times4}\;\;
&\Big({\cal M}_{_N}\Big)_{4\times4}\;\;
\end{array}\right)\;,
\label{neutralino-neutrino1}
\end{eqnarray}
where ${\cal M}_{_N}$ denotes the $4\times4$ mass matrix for neutralinos in the MSSM,
the concrete expressions for ${\cal M}_{_N}^{(0)}$, ${\cal A}_{_N}^{(1)}$, ${\cal A}_{_N}^{(2)}$
and ${\cal A}_{_N}^{(3)}$ are
\begin{eqnarray}
&&{\cal M}_{_N}^{(0)}=\left(\begin{array}{cc}
0_{3\times3}\;\;\;&\Big(g_{_{BL}}\upsilon_{_{N_{J^\prime}}}\Big)_{3\times1}\;\;\;\\
\Big(g_{_{BL}}\upsilon_{_{N_J}}\Big)_{1\times3}\;\;\;&2m_{_{BL}}\;\;\;
\end{array}\right)
\;,\nonumber\\\nonumber\\
&&{\cal A}_{_N}^{(1)}=\left(\begin{array}{cc}\Big({\upsilon_{_u}\over\sqrt{2}}(Y_{_N})_{I^\prime J}\Big)_{3\times3}\;\;&
\Big(-g_{_{BL}}\upsilon_{_{L_{I^\prime}}}\Big)_{3\times1}\end{array}\right)
\;,\nonumber\\\nonumber\\
&&{\cal A}_{_N}^{(2)}=\left(\begin{array}{cccc}\Big(-{g_1\over2}\upsilon_{_{L_{I^\prime}}}\Big)_{3\times1}\;\;&
\Big({g_2\over2}\upsilon_{_{L_{I^\prime}}}\Big)_{3\times1}\;\;&0_{3\times1}\;\;
&\Big({\zeta_{_{I^\prime}}\over\sqrt{2}}\Big)_{3\times1}
\end{array}\right)
\nonumber\\
&&\hspace{1.3cm}
+\Big[\Delta {\cal A}_{_N}^{(2)}\Big]_{3\times4}
\;,\nonumber\\\nonumber\\
&&{\cal A}_{_N}^{(3)}=\left(\begin{array}{cccc}0_{3\times1}\;\;\;&0_{3\times1}\;\;\;&0_{3\times1}\;\;\;&
\Big({1\over\sqrt{2}}\sum\limits_{\alpha=1}^3\upsilon_{_{L_\alpha}}(Y_{_N})_{\alpha J}\Big)_{3\times1}\;\;\\
0\;\;\;&0\;\;\;&0\;\;\;&0\;\;\;
\end{array}\right)\;,
\label{neutralino-neutrino2}
\end{eqnarray}
with the row indices of matrix $I^\prime,\;J^\prime=1,\;2,\;3$, and the column
indices of matrix $I,\;J=1,\;2,\;3$, respectively.
%%%%%%%%%%%%%%%%%%%%BEGIN PRD VERSION%%%%%%%%%%%%%%%%%%%
In addition, the $3\times4$ matrix $\Delta {\cal A}_{_N}^{(2)}$ represents the one-loop radiative corrections to
neutrino mass matrix from virtual sneutrino-neutralino loop with relatively large $A_{_N}$ in the
soft breaking terms\cite{RneutrinoMass}:
\begin{eqnarray}
&&\Big[\Delta {\cal A}_{_N}^{(2)}\Big]_i=
\sum\limits_{\beta=1}^4\Big[{\cal N}_{_F}^{(2)}\Big]_{i,\beta}\left(\begin{array}{cccc}
s_{_{\rm W}}(U_{\chi^0})_{4\beta},\;&
-c_{_{\rm W}}(U_{\chi^0})_{4\beta},\;&
0,\;&\Big[s_{_{\rm W}}(U_{\chi^0})_{1\beta}
-c_{_{\rm W}}(U_{\chi^0})_{2\beta}\Big]
\end{array}\right),
\label{radiative-corrections-neutrino1}
\end{eqnarray}
with
\begin{eqnarray}
&&\Big[{\cal N}_{_F}^{(2)}\Big]_{i,\beta}={\alpha\over8\sqrt{2}\pi s_{_{\rm W}}^2c_{_{\rm W}}^2}
\Big[s_{_{\rm W}}(U_{\chi^0})_{1\beta}-c_{_{\rm W}}(U_{\chi^0})_{2\beta}\Big]
m_{\chi_\beta^0}(A_{_N}\upsilon_{_N})_i
\nonumber\\
&&\hspace{2.0cm}\times
\Big\{\sum\limits_{k=1}^2(Z_{_R})_{1k}\varrho_{_{1,1}}(m_{\chi_\beta^0}^2,\;m_{_{H_k^0}}^2,\;m_{_{{\tilde L}_i}}^2)
+\sin\beta\varrho_{_{0,1}}(m_{\chi_\beta^0}^2,\;m_{_{{\tilde L}_i}}^2)
\nonumber\\
&&\hspace{2.0cm}
+\cos\beta\varrho_{_{1,1}}(m_{\chi_\beta^0}^2,\;m_{_{A_3^0}}^2,\;m_{_{{\tilde L}_i}}^2)\Big\}
\label{radiative-corrections-neutrino2}
\end{eqnarray}
where the functions $\varrho_{_{m,n}}(x_1,x_2,\cdots,\;x_{_N})$ are defined by
\begin{eqnarray}
&&\varrho_{_{m,n}}(x_1,x_2,\cdots,\;x_{_N})=\sum\limits_{i=1}^N{x_i^m\ln^nx_i
\over\prod\limits_{j\neq i}(x_i-x_j)}\;.
\label{varrho-functions}
\end{eqnarray}
To estimate magnitudes of the radiative corrections on neutrino masses,
we obtain
\begin{eqnarray}
&&\Big[{\cal N}_{_F}^{(2)}\Big]_{i,\alpha}\sim{\alpha\Lambda_{_{EW}}(A_{_N}\upsilon_{_N})_i
\over8\sqrt{2}\pi s_{_{\rm W}}^2c_{_{\rm W}}^2m_{_{{\tilde L}_i}}^2}\ln{
m_{_{{\tilde L}_i}}\over\Lambda_{_{EW}}}
\Big[s_{_{\rm W}}(U_{\chi^0})_{1\alpha}-c_{_{\rm W}}(U_{\chi^0})_{2\alpha}\Big]
\nonumber\\
&&\hspace{2.0cm}\times
\Big\{\sum\limits_{k=1}^2(Z_{_R})_{1k}
+\sin\beta+\cos\beta\Big\}
\label{radiative-corrections-neutrino2a}
\end{eqnarray}
as $m_{_{{\tilde L}_i}}\gg m_{\chi_\beta^0}\simeq m_{_{H_k^0}}\simeq m_{_{A_3^0}}
\simeq\Lambda_{_{EW}}$,where $\Lambda_{_{EW}}$ denotes the electroweak energy scale.
Taking $m_{_{{\tilde L}_i}}=\upsilon_{_{N_i}}\sim1\;{\rm TeV}$, $\Lambda_{_{EW}}\sim100$ GeV,
$A_{_{N_i}}\sim0.01\;{\rm GeV}$, one acquires $\Big[{\cal N}_{_F}^{(2)}\Big]_{i,\alpha}
\sim10^{-6}$ GeV which is negligible comparing with the tree level contributions
to the $3\times4$ matrix ${\cal A}_{_N}^{(2)}\sim10^{-4}$ GeV as
$\upsilon_{_{L_i}}\sim10^{-3}\;{\rm GeV}$ and $Y_{_N}\sim10^{-7}$.
Nevertheless, this radiative pieces account for tiny masses of the lightest active
and two sterile neutrinos naturally because only two left-handed neutrinos
acquire nonzero masses at tree level \cite{Perez6}.
%%%%%%%%%%%%%%%%%%%%%END PRD VERSION%%%%%%%%%%%%%%%%%%%

Defining the $4\times4$ orthogonal matrix
\begin{eqnarray}
&&{\cal Z}_{_N}^{(0)}=\left(\begin{array}{cccc}
-{\upsilon_{_{N_3}}\over\sqrt{\upsilon_{_{N_1}}^2+\upsilon_{_{N_3}}^2}}\;\;\;&
-{\upsilon_{_{N_2}}\over\sqrt{\upsilon_{_{N_1}}^2+\upsilon_{_{N_2}}^2}}\;\;\;&
-{ig_{_{BL}}\upsilon_{_{N_1}}\over\sqrt{2}\Delta_{_{BL}}\eta_{_{BL}}^-}\;\;\;&
{g_{_{BL}}\upsilon_{_{N_1}}\over\sqrt{2}\Delta_{_{BL}}\eta_{_{BL}}^+}\;\;\;\\
0\;\;\;&{\upsilon_{_{N_1}}\over\sqrt{\upsilon_{_{N_1}}^2+\upsilon_{_{N_2}}^2}}\;\;\;&
-{ig_{_{BL}}\upsilon_{_{N_2}}\over\sqrt{2}\Delta_{_{BL}}\eta_{_{BL}}^-}\;\;\;&
{g_{_{BL}}\upsilon_{_{N_2}}\over\sqrt{2}\Delta_{_{BL}}\eta_{_{BL}}^+}\;\;\;\\
{\upsilon_{_{N_1}}\over\sqrt{\upsilon_{_{N_1}}^2+\upsilon_{_{N_3}}^2}}\;\;\;&
0\;\;\;&-{ig_{_{BL}}\upsilon_{_{N_3}}\over\sqrt{2}\Delta_{_{BL}}\eta_{_{BL}}^-}\;\;\;&
{g_{_{BL}}\upsilon_{_{N_3}}\over\sqrt{2}\Delta_{_{BL}}\eta_{_{BL}}^+}\;\;\;\\
0\;\;\;&0\;\;\;&{i\over\sqrt{2}}\eta_{_{BL}}^-\;\;\;&{1\over\sqrt{2}}\eta_{_{BL}}^+\;\;\;
\end{array}\right)\;,
\label{neutralino-neutrino3}
\end{eqnarray}
one obtains
\begin{eqnarray}
&&\left(\begin{array}{ccc}
1_{3\times3},\;\;\;&0_{3\times4}\;\;\;&0_{3\times4}\;\;\;\\
0_{4\times3}\;\;\;&\Big({\cal Z}_{_N}^{(0)T}\Big)_{4\times4}\;\;\;&0_{4\times4}\;\;\;\\
0_{4\times3}\;\;\;&0_{4\times4}\;\;\;&1_{4\times4}\end{array}\right)\cdot M_{_N}\cdot
\left(\begin{array}{ccc}
1_{3\times3},\;\;\;&0_{3\times4}\;\;\;&0_{3\times4}\;\;\;\\
0_{4\times3}\;\;\;&\Big({\cal Z}_{_N}^{(0)}\Big)_{4\times4}\;\;\;&0_{4\times4}\;\;\;\\
0_{4\times3}\;\;\;&0_{4\times4}\;\;\;&1_{4\times4}\end{array}\right)
\nonumber\\&&
=\left(\begin{array}{ccc}
0_{3\times3}\;\;&\Big({\cal A}_{_N}^{(1)}{\cal Z}_{_N}^{(0)}\Big)_{3\times4}\;\;&
\Big({\cal A}_{_N}^{(2)}\Big)_{3\times4}\;\;\\
\Big({\cal Z}_{_N}^{(0)T}{\cal A}_{_N}^{(1)T}\Big)_{4\times3}\;\;&
\Big({\cal Z}_{_N}^{(0)T}{\cal M}_{_N}^{(0)}{\cal Z}_{_N}^{(0)}\Big)_{4\times4}\;\;
&\Big({\cal Z}_{_N}^{(0)T}{\cal A}_{_N}^{(3)}\Big)_{4\times4}\;\;\\
\Big({\cal A}_{_N}^{(2)T}\Big)_{4\times3}\;\;&\Big({\cal A}_{_N}^{(3)T}{\cal Z}_{_N}^{(0)}\Big)_{4\times4}\;\;
&\Big({\cal M}_{_N}\Big)_{4\times4}\;\;
\end{array}\right)
\nonumber\\&&
=\left(\begin{array}{cc}
\Big(m_\nu\Big)_{5\times5}\;\;\;&\Big(m_{_D}\Big)_{5\times6}\;\;\;\\
\Big(m_{_D}^T\Big)_{6\times5}\;\;\;&\Big({\cal M}\Big)_{6\times6}
\end{array}\right)\;.
\label{neutralino-neutrino4}
\end{eqnarray}
Where $\Delta_{_{BL}}=\sqrt{m_{_{BL}}^2+g_{_{BL}}^2\upsilon_{_N}^2}$,
$\eta_{_{BL}}^\pm=\sqrt{1\pm{m_{_{BL}}\over\Delta_{_{BL}}}}$, and
${\cal Z}_{_N}^{(0)T}{\cal M}_{_N}^{(0)}{\cal Z}_{_N}^{(0)}=diag(0,\;\;0,\;\;\Delta_{_{BL}}-m_{_{BL}},\;\;
\Delta_{_{BL}}+m_{_{BL}})$, respectively.

Using Eq.(\ref{neutralino-neutrino1}) and Eq.(\ref{neutralino-neutrino3}),
we formulate the submatrices in Eq.(\ref{neutralino-neutrino4}) respectively as
\begin{eqnarray}
&&{\cal M}=\left(\begin{array}{cccccc}
\Delta_{_{BL}}-m_{_{BL}}\;\;\;&0\;\;\;&0\;\;\;&0\;\;\;&0\;\;\;&-i\varepsilon_-\;\;\;\\
0\;\;\;&\Delta_{_{BL}}+m_{_{BL}}\;\;\;&0\;\;\;&0\;\;\;&0\;\;\;&\varepsilon_+\;\;\;\\
0\;\;\;&0\;\;\;&2m_1\;\;\;&0\;\;\;&-{g_1\upsilon_d\over2}\;\;\;&{g_1\upsilon_u\over2}\;\;\;\\
0\;\;\;&0\;\;\;&0\;\;\;&2m_2\;\;\;&{g_2\upsilon_d\over2}\;\;\;&-{g_2\upsilon_u\over2}\;\;\;\\
0\;\;\;&0\;\;\;&-{g_1\upsilon_d\over2}\;\;\;&{g_2\upsilon_d\over2}\;\;\;&0\;\;\;&\mu\;\;\;\\
-i\varepsilon_-\;\;\;&\varepsilon_+\;\;\;&{g_1\upsilon_u\over2}\;\;\;&-{g_2\upsilon_u\over2}\;\;\;
&\mu\;\;\;&0\;\;\;
\end{array}\right)
\;,\nonumber\\\nonumber\\
&&m_{_D}=\left(\begin{array}{cccccc}
-i\delta_1^-\;\;\;&\delta_1^+\;\;\;&\Big[{\cal A}_{_N}^{(2)}\Big]_{1,1}
\;\;\;&\Big[{\cal A}_{_N}^{(2)}\Big]_{1,2}\;\;\;
&0\;\;\;&\Big[{\cal A}_{_N}^{(2)}\Big]_{1,4}\\
-i\delta_2^-\;\;\;&\delta_2^+\;\;\;&\Big[{\cal A}_{_N}^{(2)}\Big]_{2,1}\;\;\;&\Big[{\cal A}_{_N}^{(2)}\Big]_{2,2}\;\;\;
&0\;\;\;&\Big[{\cal A}_{_N}^{(2)}\Big]_{2,4}\\
-i\delta_3^-\;\;\;&\delta_3^+\;\;\;&\Big[{\cal A}_{_N}^{(2)}\Big]_{3,1}\;\;\;&\Big[{\cal A}_{_N}^{(2)}\Big]_{3,2}\;\;\;
&0\;\;\;&\Big[{\cal A}_{_N}^{(2)}\Big]_{3,4}\\
0\;\;\;&0\;\;\;&0\;\;\;&0\;\;\;&0\;\;\;&\varepsilon_{13}\\
0\;\;\;&0\;\;\;&0\;\;\;&0\;\;\;&0\;\;\;&\varepsilon_{12}
\end{array}\right)
\;,\nonumber\\\nonumber\\
&&m_\nu=\left(\begin{array}{ccccc}
0\;\;\;&0\;\;\;&0\;\;\;&\delta_{13}\;\;\;&\delta_{12}\;\;\;\\
0\;\;\;&0\;\;\;&0\;\;\;&\delta_{23}\;\;\;&\delta_{22}\;\;\;\\
0\;\;\;&0\;\;\;&0\;\;\;&\delta_{33}\;\;\;&\delta_{32}\;\;\;\\
\delta_{13}\;\;\;&\delta_{23}\;\;\;&\delta_{33}\;\;\;&0\;\;\;&0\;\;\;\\
\delta_{12}\;\;\;&\delta_{22}\;\;\;&\delta_{32}\;\;\;&0\;\;\;&0\;\;\;
\end{array}\right)\;,
\label{neutralino-neutrino5}
\end{eqnarray}
where the abbreviations are
\begin{eqnarray}
&&\varepsilon_\pm={g_{_{BL}}\varepsilon_{_N}^2\over2\Delta_{_{BL}}\eta_{_{BL}}^\pm}
\;,\nonumber\\
&&\delta_{i2}={\upsilon_{_u}\over\sqrt{2(\upsilon_{_{N_1}}^2+\upsilon_{_{N_2}}^2)}}
\Big[-\Big(Y_{_N}\Big)_{i1}\upsilon_{_{N_2}}+\Big(Y_{_N}\Big)_{i2}\upsilon_{_{N_1}}\Big]
\;,\nonumber\\
&&\delta_{i3}={\upsilon_{_u}\over\sqrt{2(\upsilon_{_{N_1}}^2+\upsilon_{_{N_3}}^2)}}
\Big[-\Big(Y_{_N}\Big)_{i1}\upsilon_{_{N_3}}+\Big(Y_{_N}\Big)_{i3}\upsilon_{_{N_1}}\Big]
\;,\nonumber\\
&&\varepsilon_{12}={1\over\upsilon_{_u}}\sum\limits_{\alpha=1}^3
\upsilon_{_{L_\alpha}}\delta_{\alpha2}
\;,\nonumber\\
&&\varepsilon_{13}={1\over\upsilon_{_u}}\sum\limits_{\alpha=1}^3
\upsilon_{_{L_\alpha}}\delta_{\alpha3}
\;,\nonumber\\
&&\delta_i^\pm={g_{_{BL}}\upsilon_{_u}\over\sqrt{2}\Delta_{_{BL}}\eta_{_{BL}}^\pm}\zeta_i
\mp{g_{_{BL}}\eta_{_{BL}}^\pm\over\sqrt{2}}\upsilon_{_{L_i}}\;.
\label{neutralino-neutrino6}
\end{eqnarray}

Defining a $11\times11$ approximated orthogonal transformation matrix ${\cal Z}_{_N}$
\begin{eqnarray}
&&{\cal Z}_{_N}=\left(\begin{array}{cc}\Big[1-{1\over2}m_{_D}\cdot {\cal M}^{-2}\cdot m_{_D}^T\Big]_{5\times5}
&\Big[m_{_D}\cdot {\cal M}^{-1}+m_\nu\cdot m_{_D}\cdot{\cal M}^{-2}\Big]_{5\times6}\\
-\Big[{\cal M}^{-1}\cdot m_{_D}^T+{\cal M}^{-2}\cdot m_{_D}^T\cdot m_\nu\Big]_{6\times5}&
\Big[1-{1\over2}{\cal M}^{-1}\cdot m_{_D}^T\cdot m_{_D}\cdot{\cal M}^{-1}\Big]_{6\times6}
\end{array}\right),
\label{neutralino-neutrino7}
\end{eqnarray}
we finally write the effective mass matrix for five light neutrinos (three active and two sterile) as
\begin{eqnarray}
&&m_\nu^{eff}\simeq m_\nu-m_{_D}\cdot{\cal M}^{-1}\cdot m_{_D}^T-{1\over2}m_\nu\cdot
m_{_D}\cdot {\cal M}^{-2}\cdot m_{_D}^T-{1\over2}m_{_D}\cdot {\cal M}^{-2}\cdot m_{_D}^T\cdot m_\nu
\nonumber\\
&&\hspace{1.0cm}
\simeq\left(\begin{array}{cc}
\Big[M_\nu^{LL}\Big]_{3\times3}\;\;\;&\Big[M_\nu^{LR}\Big]_{3\times2}\;\;\;\\
\Big[M_\nu^{LR,T}\Big]_{2\times3}\;\;\;&\Big[M_\nu^{RR}\Big]_{2\times2}\;\;\;
\end{array}\right)\;.
\label{neutralino-neutrino8}
\end{eqnarray}

In order to accommodate naturally the experimental data on neutrino oscillation and $Z$ invisible
decay width in this framework, we find that only one possibility $M_\nu^{LL}\gg M_\nu^{LR},\;M_\nu^{RR}$
is reasonable \cite{Perez6}. In fact, this point implies
\begin{eqnarray}
%%%%%%%%%%%%%%%%%%%%%%%%%%%%%%%%%%%%%%%%%%%%%%%%%%%%%%%%%%%%%%%%%%%%%%%%%%
&&\delta_{i2},\;\delta_{i3}\ll
\Big[M_\nu^{LL}\Big]_{ii}\;.
%%%%%%%%%%%%%%%%%%%%%%%%%%%%%%%%%%%%%%%%%%%%%%%%%%%%%%%%%%%%%%%%%%%%%%%%%%
\label{neutralino-neutrino10a}
\end{eqnarray}
%%%%%%%%%%%%%%%%%%%%%%%%%%%%%%%%%%Begin 1st Revision%%%%%%%%%%%%%%%%%%%%%%%%%%%%%%%%%%%%
Three active neutrinos with sub-eV masses require $\max(\upsilon_{_{L_i}},\zeta_j)< 10^{-2}$GeV,
where $i,\;j=1,\;2,\;3$ are the generation indices.

%%%%%%%%%%%%%%%%%%%%%%%%%%%%%%%%%%%%End 1st Revision%%%%%%%%%%%%%%%%%%%%%%%%%%%%%%%%%%%%

To guarantee decoupling two light sterile neutrinos from the active ones,
we choose the Yukawa couplings for right-handed neutrinos as
\begin{eqnarray}
%%%%%%%%%%%%%%%%%%%%%%%%%%%%%%%%%%%%%%%%%%%%%%%%%%%%%%%%%%%%%%%%%%%%%%%%%%
&&Y_{_N}={1\over\upsilon_{_N}}
\left(\begin{array}{ccc}
\upsilon_{_{N_1}}Y_{_1},&\upsilon_{_{N_2}}Y_{_1},&\upsilon_{_{N_3}}Y_{_1}\\
\upsilon_{_{N_1}}Y_{_2},&\upsilon_{_{N_2}}Y_{_2},&\upsilon_{_{N_3}}Y_{_2}\\
\upsilon_{_{N_1}}Y_{_3},&\upsilon_{_{N_2}}Y_{_3},&\upsilon_{_{N_3}}Y_{_3}\end{array}\right)
={1\over\upsilon_{_N}}\left(\begin{array}{ccc}
Y_{_1},&0,&0\\0,&Y_{_2},&0\\0,&0,&Y_{_3}\end{array}\right)
\left(\begin{array}{ccc}
\upsilon_{_{N_1}},&\upsilon_{_{N_2}},&\upsilon_{_{N_3}}\\
\upsilon_{_{N_1}},&\upsilon_{_{N_2}},&\upsilon_{_{N_3}}\\
\upsilon_{_{N_1}},&\upsilon_{_{N_2}},&\upsilon_{_{N_3}}\end{array}\right)\;,
%%%%%%%%%%%%%%%%%%%%%%%%%%%%%%%%%%%%%%%%%%%%%%%%%%%%%%%%%%%%%%%%%%%%%%%%%%
\label{Yukawa-right-neutrinos}
\end{eqnarray}
then get
\begin{eqnarray}
%%%%%%%%%%%%%%%%%%%%%%%%%%%%%%%%%%%%%%%%%%%%%%%%%%%%%%%%%%%%%%%%%%%%%%%%%%
&&\zeta_{_i}= Y_{_i}\upsilon_{_N}\;,\;
\delta_{i2}=\delta_{i3}=0\;,\;\;\;(i=1,\;2,\;3)\;.
%%%%%%%%%%%%%%%%%%%%%%%%%%%%%%%%%%%%%%%%%%%%%%%%%%%%%%%%%%%%%%%%%%%%%%%%%%
\label{Yukawa-right-neutrinos1}
\end{eqnarray}
Adopting the assumption on relevant parameter space, only two left-handed neutrinos
acquire nonzero masses at tree level. Two sterile neutrinos and
another active left-handed neutrino acquire their physical masses after we consider
radiative corrections to the neutrino mass matrix in Eq.(\ref{neutralino-neutrino8}).

In the case of three active neutrino mixing, so far the available
measurements on the neutrino oscillations can determine the neutrino
mass spectrum up to two possible solutions:
\begin{itemize}
\item the normal ordering (NO) spectrum:
\begin{eqnarray}
&&m_{\nu_1}<m_{\nu_2}<m_{\nu_3},\;\Delta m_{_A}^2=\Delta m_{_{31}}^2,\;
\Delta m_{_\odot}^2=\Delta m_{_{21}}^2;
\label{NH1}
\end{eqnarray}

\item the inverted ordering (IO) spectrum:
\begin{eqnarray}
&&m_{\nu_3}<m_{\nu_1}<m_{\nu_2},\;\Delta m_{_A}^2=\Delta m_{_{23}}^2,\;
\Delta m_{_\odot}^2=\Delta m_{_{13}}^2.
\label{IH1}
\end{eqnarray}
\end{itemize}
The flavor neutrinos are mixed into massive neutrinos $\nu_{1,2,3}$ during their
flight, and the mixings are described by the Pontecorvo-Maki-Nakagawa-Sakata matrix
$U_{_{PMNS}}$~\cite{neutrino-oscillations,neutrino-oscillations1}:
\begin{eqnarray}
%%%%%%%%%%%%%%%%%%%%%%%%%%%%%%%%%%%%%%%%%%%%%%%%%%%%%%%%%%%%%%%%%%%%%%%%%%
&&\sin\theta_{13}=\Big|\Big(U_{_{PMNS}}\Big)_{13}\Big|,\;\;\;
\cos\theta_{13}=\sqrt{1-\Big|\Big(U_{_{PMNS}}\Big)_{13}\Big|^2}
\;,\nonumber\\
%%%%%%%%%%%%%%%%%%%%%%%%%%%%%%%%%%%%%%%%%%%%%%%%%%%%%%%%%%%%%%%%%%%%%%%%%%
&&\sin\theta_{23}={\Big|\Big(U_{_{PMNS}}\Big)_{23}\Big|\over\sqrt{1-\Big|\Big(U_{_{PMNS}}\Big)_{13}\Big|^2}}
\;,\;\;\;
\cos\theta_{23}={\Big|\Big(U_{_{PMNS}}\Big)_{33}\Big|\over\sqrt{1-\Big|\Big(U_{_{PMNS}}\Big)_{13}\Big|^2}}
\;,\nonumber\\
%%%%%%%%%%%%%%%%%%%%%%%%%%%%%%%%%%%%%%%%%%%%%%%%%%%%%%%%%%%%%%%%%%%%%%%%%%
&&\sin\theta_{12}={\Big|\Big(U_{_{PMNS}}\Big)_{12}\Big|\over\sqrt{1-\Big|\Big(U_{_{PMNS}}\Big)_{13}\Big|^2}}
\;,\;\;\;
\cos\theta_{12}={\Big|\Big(U_{_{PMNS}}\Big)_{11}\Big|\over\sqrt{1-\Big|\Big(U_{_{PMNS}}\Big)_{13}\Big|^2}}\;.
%%%%%%%%%%%%%%%%%%%%%%%%%%%%%%%%%%%%%%%%%%%%%%%%%%%%%%%%%%%%%%%%%%%%%%%%%%
\label{neutralino-neutrino15}
\end{eqnarray}

Through several recent reactor oscillation experiments~\cite{theta13,theta13-1,theta13-2,theta13-3,theta13-4},
the mixing angle $\theta_{13}$ is now precisely known. The global fit of $\theta_{13}$
gives~\cite{Garcia}
\begin{eqnarray}
\sin^2\theta_{13}=0.023\pm 0.0023\;,
\label{neutrino-oscillations1}
\end{eqnarray}
and other experimental observations relating solar and atmospheric neutrino oscillations
are shown as\cite{PDG}
\begin{eqnarray}
&&\Delta m_{\odot}^2=7.58_{-0.26}^{+0.22}\times 10^{-5} {\rm eV}^2,\nonumber\\
&&\Delta m_{_A}^2=2.35_{-0.09}^{+0.12}\times 10^{-3} {\rm eV}^2,\nonumber\\
&&\sin^2\theta_{\odot}=0.306_{-0.015}^{+0.018}\;,\qquad \sin^2\theta_{_A}=0.42_{-0.03}^{+0.08}\;.
\label{neutrino-oscillations2}
\end{eqnarray}

%%%%%%%%%%%%%%%%%%%%%%%%%%%%%%%Begin Modification%%%%%%%%%%%%%%%%%%%%%%%%%%%%%%%%
In addition, the cosmological observations also constrain the light neutrino masses
strongly. The WMAP collaboration gets an upper bound on the sum of neutrino masses
as $\sum m_\nu\le0.44$ eV from nine-year Wilkinson microwave anisotropy probe data
\cite{WMAP}, and the Planck collaboration derives a more stringent bound on the sum
of neutrino masses as $\sum m_\nu\le0.23$ eV through the measurements of the cosmic
microwave background temperature and lensing-potential power spectra \cite{Planck}.
In our analysis, we consider the constraint from the Planck collaboration on the sum
of neutrino masses.

Considering extra minus in definition of $\mu$-parameter and only including the leading terms
in  Eq.(\ref{neutralino-neutrino8}), our result is consistent with Eq.(4.5) in literature \cite{Perez4}.
One easily finds that the parameters $Y_i\;(i=1,\;2,\;3)$ are the Yukawa couplings $Y_\nu^{i3}$,
and $\delta_{i2},\;\delta_{i3}$ are the  Yukawa couplings $Y_\nu^{i\alpha}\;(\alpha=1,\;2)$  in
literature \cite{Perez4} where only the third generation right-handed scalar neutrino acquires
nonzero VEV. Choosing the Yukawa couplings for right-handed neutrinos in Eq.(\ref{Yukawa-right-neutrinos}),
we derive that there is no mixings between three active and two light sterile neutrinos since $\delta_{i2}\simeq0,\;\delta_{i3}\simeq0$.
In order to satisfy the experimental bounds on active neutrino masses, one also numerically
finds that the allowed values for $Y_i$ in the range $(10^{-7},\;10^{-5})$ as the VEVs of left-handed
scalar neutrinos change in the region $(10^{-5},\;10^{-2})$ GeV. This conclusion also coincides
with the corresponding result quantitatively in literature \cite{Perez4}.  Assuming that the parameters
$\mu,\;\upsilon_{_N},\;m_{_{Z_{BL}}},\;m_{_{BL}}$ are all exceed TeV scale and taking $m_1=200\;{\rm GeV},\;
m_2=400$ GeV, one finds $\tilde{m}\simeq129$ GeV and
$1/\Lambda_\upsilon\simeq-2\times10^{-4}\;{\rm GeV}^{-1}$.  Choosing $\upsilon_{_L}\sim10^{-3}$ GeV,
we derive that the magnitude to neutrino mass from the terms $\upsilon_{_{L_i}}\upsilon_{_{L_j}}/\Lambda_\upsilon$
is about 0.1 eV.

%%%%%%%%%%%%%%%%%%%%%%%%%%%%End Modification%%%%%%%%%%%%%%

\section{The mass matrix for charginos and charged leptons\label{sec5}}
\indent\indent
The mass terms of charginos are written as
\begin{eqnarray}
&&-{\cal L}_{_{charginos}}=\left(\begin{array}{ccc}e_{_{L_I}}^-,&i\lambda_{_A}^-,&\psi_{_{H_d}}^2
\end{array}\right){\cal M}_c\left(\begin{array}{c}e_{_{R_J}}^+\\ i\lambda_{_A}^+\\\psi_{_{H_u}}^1\end{array}\right)
+h.c.
\label{chargino-lepton1}
\end{eqnarray}
with
\begin{eqnarray}
&&{\cal M}_c=\left(\begin{array}{cc}\Big[{1\over\sqrt{2}}
\Big(Y_{_E}\Big)_{IJ}\upsilon_{_d}\Big]_{3\times3},&E_{3\times2}\\
E^\prime_{2\times3},&{\cal M}_\pm^{(0)}
\end{array}\right)\;.
\label{chargino-lepton2}
\end{eqnarray}
Here ${\cal M}_\pm^{(0)}$ denotes the chargino mass matrix in the MSSM
\begin{eqnarray}
&&{\cal M}_\pm^{(0)}=\left(\begin{array}{cc}2m_2,&{e\upsilon_{_u}\over\sqrt{2}s_{_{\rm W}}}\\
{e\upsilon_{_d}\over\sqrt{2}s_{_{\rm W}}},&-\mu
\end{array}\right)\;,
\label{chargino-lepton2a}
\end{eqnarray}
and
\begin{eqnarray}
&&E=\left(\begin{array}{cc}
{g_2\over\sqrt{2}}\upsilon_{_{L_1}},&-{1\over\sqrt{2}}\zeta_{_1}\\
{g_2\over\sqrt{2}}\upsilon_{_{L_2}},&-{1\over\sqrt{2}}\zeta_{_2}\\
{g_2\over\sqrt{2}}\upsilon_{_{L_3}},&-{1\over\sqrt{2}}\zeta_{_3}
\end{array}\right)\;,
\nonumber\\
%%%%%%%%%%%%%%%%%%%%%%%%%%%%%%%%%%%%%%%%%%%%%%%%%%%%%%%%%%%%%%%%
&&E^\prime=\left(\begin{array}{ccc}
0&0&0\\
-\sum\limits_{\alpha=1}^3{\upsilon_{_{L_\alpha}}\over\sqrt{2}}\Big(Y_{_E}\Big)_{\alpha 1}&
-\sum\limits_{\alpha=1}^3{\upsilon_{_{L_\alpha}}\over\sqrt{2}}\Big(Y_{_E}\Big)_{\alpha 2}&
-\sum\limits_{\alpha=1}^3{\upsilon_{_{L_\alpha}}\over\sqrt{2}}\Big(Y_{_E}\Big)_{\alpha 3}
\end{array}\right)
\label{chargino-lepton2b}
\end{eqnarray}
Because $\max\Big[{1\over\sqrt{2}}(Y_{_E})_{ij}\upsilon_{_d},\;E_{i,\alpha},\;
E^\prime_{\beta,j}\Big]\ll \min\Big[{\cal M}_\pm^{(0)}\Big]\;\;(i.j=1,\;2,\;3,
\;\;\alpha,\;\beta=1,\;2)$, we diagonalize ${\cal M}_c$ through two unitary
matrices\cite{Hirsch-Valle}
\begin{eqnarray}
&&{\cal Z}_-\simeq\left(\begin{array}{cc}\Big[1-{1\over2}\xi_{_L}^T\xi_{_L}\Big]_{3\times3},
&\Big[-\xi_{_L}^T\Big]_{3\times2}\\
\Big[\xi_{_L}\Big]_{2\times3},&\Big[1-{1\over2}\xi_{_L}\xi_{_L}^T\Big]_{2\times2}
\end{array}\right)
\left(\begin{array}{cc}V_{_L}&0\\0&U_-\end{array}\right)\;,
\label{chargino-lepton3}
\end{eqnarray}
and
\begin{eqnarray}
&&{\cal Z}_+\simeq\left(\begin{array}{cc}\Big[1-{1\over2}\xi_{_R}^T\xi_{_R}\Big]_{3\times3},
&\Big[-\xi_{_R}^T\Big]_{3\times2}\\
\Big[\xi_{_R}\Big]_{2\times3},&\Big[1-{1\over2}\xi_{_R}\xi_{_R}^T\Big]_{2\times2}
\end{array}\right)
\left(\begin{array}{cc}V_{_R}&0\\0&U_+\end{array}\right)\;.
\label{chargino-lepton4}
\end{eqnarray}
Expanding the corresponding matrices in powers of $\upsilon_{_{L_i}}/\upsilon_{_{EW}}$ etc.
to the second order, we have
\begin{eqnarray}
&&\xi_{_L}\simeq-\Big({\cal M}_\pm^{(0),T}\Big)^{-1}\cdot E^T-{\upsilon_{_d}\over\sqrt{2}}
\Big({\cal M}_\pm^{(0),T}\Big)^{-1}\cdot\Big({\cal M}_\pm^{(0)}\Big)^{-1}\cdot E^\prime\cdot Y_{_E}
\;,\nonumber\\
%%%%%%%%%%%%%%%%%%%%%%%%%%%%%%%%%%%%%%%%%%%%%%%%%%%%%%%%%%%%%%%%%%%%%%%%%%%%%%
&&\xi_{_R}\simeq-\Big({\cal M}_\pm^{(0)}\Big)^{-1}\cdot E^\prime
-{\upsilon_{_d}\over\sqrt{2}}\Big({\cal M}_\pm^{(0)}\Big)^{-1}
\cdot\Big({\cal M}_\pm^{(0),T}\Big)^{-1}\cdot E^T\cdot Y_{_E}\;.
\label{chargino-lepton4a}
\end{eqnarray}
Using above equations, one can check that there is no mixing between charged leptons and
charginos in the matrix ${\cal Z}_-^T\cdot{\cal M}_c\cdot{\cal Z}_+$
to the second order of $\upsilon_{_{L_i}}/\upsilon_{_{EW}}$ etc.
For convenience, we write the elements of $\xi_{_{L,R}}$ explicitly as
\begin{eqnarray}
&&\Big(\xi_{_L}\Big)_{1I}=-{1\over\Delta_c}\Big\{
{e\mu\upsilon_{_{L_I}}\over\sqrt{2}s_{_{\rm W}}}-{e\upsilon_{_d}\over2s_{_{\rm W}}}\zeta_{_I}
\nonumber\\
&&\hspace{1.8cm}
+{e\upsilon_{_d}(2m_2\upsilon_{_u}-\mu\upsilon_{_d})\over2\sqrt{2}s_{_{\rm W}}\Delta_c}
\sum\limits_{\alpha=1}^3\upsilon_{_{L_\alpha}}\Big(Y_{_E}Y_{_E}^T\Big)_{\alpha I}\Big\}\;,
\nonumber\\
&&\Big(\xi_{_L}\Big)_{2I}=-{1\over\Delta_c}\Big\{
{e^2\upsilon_{_u}\upsilon_{_{L_I}}\over2s_{_{\rm W}}^2}+\sqrt{2}m_2\zeta_{_I}
\nonumber\\
&&\hspace{1.8cm}
-{\upsilon_{_d}(8s_{_{\rm W}}m_2+e^2\upsilon_{_u}^2)\over4s_{_{\rm W}}^2\Delta_c}
\sum\limits_{\alpha=1}^3\upsilon_{_{L_\alpha}}\Big(Y_{_E}Y_{_E}^T\Big)_{\alpha I}\Big\}\;,
\label{chargino-lepton5}
\end{eqnarray}
and
\begin{eqnarray}
&&\Big(\xi_{_R}\Big)_{1I}={1\over\Delta_c}\Big\{\Big[{e\upsilon_{_u}\over2s_{_{\rm W}}}
-{e\upsilon_{_d}(2s_{_{\rm W}}^2\mu^2+e^2\upsilon_{_u}^2)\over4s_{_{\rm W}}^2\Delta_c}\Big]
\sum\limits_{\alpha=1}^3\upsilon_{_{L_\alpha}}\Big(Y_{_E}\Big)_{\alpha I}
\nonumber\\
&&\hspace{1.8cm}
-{e\upsilon_{_d}(2s_{_{\rm W}}^2\mu^2+e^2\upsilon_{_u}^2)\over4s_{_{\rm W}}^2\Delta_c}
\sum\limits_{\alpha=1}^3\upsilon_{_{N_\alpha}}\Big(Y_{_N}^TY_{_E}\Big)_{\alpha I}\Big\}\;,
\nonumber\\
&&\Big(\xi_{_R}\Big)_{2I}=-{1\over\Delta_c}\Big\{\Big[\sqrt{2}m_2
-{e^2\upsilon_{_d}(2m_2\upsilon_{_u}-\mu\upsilon_{_d})\over2\sqrt{2}s_{_{\rm W}}^2\Delta_c}\Big]
\sum\limits_{\alpha=1}^3\upsilon_{_{L_\alpha}}\Big(Y_{_E}\Big)_{\alpha I}
\nonumber\\
&&\hspace{1.8cm}
-{\upsilon_{_d}(8s_{_{\rm W}}^2m_2^2+e^2\upsilon_{_d}^2)\over4s_{_{\rm W}}^2\Delta_c}
\sum\limits_{\alpha=1}^3\upsilon_{_{N_\alpha}}\Big(Y_{_N}^TY_{_E}\Big)_{\alpha I}\Big\}\;.
\label{chargino-lepton6}
\end{eqnarray}
Here, $I=1,\;2,\;3$ and $\Delta_c=2m_2\mu+e^2\upsilon_{_u}\upsilon_{_d}/(2s_{_{\rm W}}^2)$.
Considering corrections from the mixing between charged leptons and charginos, we approximate
the elements of $3\times3$ mass matrix for charged leptons as
\begin{eqnarray}
&&\Big(m_{_E}\Big)_{IJ}={1\over\sqrt{2}}\Big(Y_{_E}\Big)_{IJ}\upsilon_{_d}
-{e^2\upsilon_{_u}\over2\sqrt{2}s_{_{\rm W}}^2\Delta_c}\upsilon_{_{L_I}}
\sum\limits_{\alpha=1}^3\upsilon_{_{L_\alpha}}\Big(Y_{_E}\Big)_{\alpha J}
\nonumber\\
&&\hspace{1.8cm}
-{m_2\over\Delta_c}\zeta_{_I}\sum\limits_{\alpha=1}^3\upsilon_{_{L_\alpha}}\Big(Y_{_E}\Big)_{\alpha J}\;.
\label{chargino-lepton7}
\end{eqnarray}
Similarly, the elements of $2\times2$ mass matrix for charginos are approached as
\begin{eqnarray}
&&\Big({\cal M}_\pm\Big)_{11}=2m_2+{e^2\mu\over4s_{_{\rm W}}^2\Delta_c}(\upsilon_{_{\rm EW}}^2
-\upsilon_{_{\rm SM}}^2)-{e^2\upsilon_{_d}\varepsilon_{_N}^2\over4\sqrt{2}s_{_{\rm W}}^2\Delta_c}
\;,\nonumber\\
&&\Big({\cal M}_\pm\Big)_{12}={e\upsilon_{_u}\over\sqrt{2}s_{_{\rm W}}}
-{e\mu\varepsilon_{_N}^2\over4s_{_{\rm W}}\Delta_c}
-{e\upsilon_{_d}\zeta^2\over4\sqrt{2}s_{_{\rm W}}\Delta_c}
\;,\nonumber\\
&&\Big({\cal M}_\pm\Big)_{21}={e\upsilon_{_d}\over\sqrt{2}s_{_{\rm W}}}
+{e^3\upsilon_{_u}\over4\sqrt{2}s_{_{\rm W}}^3\Delta_c}(\upsilon_{_{\rm EW}}^2-\upsilon_{_{\rm SM}}^2)
+{em_2\varepsilon_{_N}^2\over4s_{_{\rm W}}\Delta_c}
\nonumber\\
&&\hspace{2.0cm}
+{e\upsilon_{_u}\over4\sqrt{2}s_{_{\rm W}}\Delta_c}
\sum\limits_{\alpha,\beta}^3\upsilon_{_{L_\alpha}}\Big(Y_{_E}Y_{_E}^T\Big)_{\alpha\beta}\upsilon_{_{L_\beta}}
\;,\nonumber\\
&&\Big({\cal M}_\pm\Big)_{22}=-\mu-{e^2\upsilon_{_u}\varepsilon_{_N}^2\over4\sqrt{2}s_{_{\rm W}}^2\Delta_c}
-{m_2\zeta^2\over2\Delta_c}
-{m_2\over2\Delta_c}\sum\limits_{\alpha,\beta}^3
\upsilon_{_{L_\alpha}}\Big(Y_{_E}Y_{_E}^T\Big)_{\alpha\beta}\upsilon_{_{L_\beta}}\;.
\label{chargino-lepton8}
\end{eqnarray}
Furthermore, the submatrices $V_{_{L,R}}$ and $U_\pm$ respectively diagonalize $m_{_E}$
and ${\cal M}_\pm$ in the following manner
\begin{eqnarray}
&&V_{_L}^T\cdot m_{_E}\cdot V_{_R}=diag\Big(m_e,\;m_\mu,\;m_\tau\Big)\;,
\nonumber\\
&&U_{_-}^T\cdot {\cal M}_\pm\cdot U_+=diag\Big(m_{_{\chi_1^\pm}},\;m_{_{\chi_2^\pm}}\Big)\;.
\label{chargino-lepton9}
\end{eqnarray}
In numerical analyses we choose $Y_{_E}={\sqrt{2}\over\upsilon_{_d}}{\rm diag}(m_e,\;m_\mu,\;m_\tau)$
for simplicity, then get $V_{_L}\simeq V_{_R}\simeq1_{3\times3}$. Taking $m_2\sim\mu\sim100$ GeV,
$\upsilon_{_{N_i}}\sim1$ TeV, $Y_{_N}\sim10^{-6}$ and $\upsilon_{_{L_i}}\sim10^{-3}$ GeV,
one acquires $\Big(m_{_E}\Big)_{11}\simeq m_e(1-10^{-10}\tan\beta)$,
$\Big(m_{_E}\Big)_{12}\simeq -10^{-10}m_\mu\tan\beta$,
$\Big(m_{_E}\Big)_{13}\simeq -10^{-10}m_\tau\tan\beta$ etc.
In other words the corrections to lepton masses from next to leading terms are
negligible.

%%%%%%%%%%%%%%%%%%%%%%%BEGIN PRD MODIFICATION%%%%%%%%%%%%%%%%%%%%%%%
Certainly the experimental observables on lepton flavor violation processes
such as $\mu\rightarrow e\gamma$ constrain the R-parity violating parameters
strongly. Generally the decay width of  $\mu\rightarrow e\gamma$
is written as \cite{Hisano}
\begin{eqnarray}
&&\Gamma(e_{_I}\rightarrow e_{_J}\gamma)={\alpha\over4}m_{e_I}^3\Big[|(A_{_L})_{_{IJ}}|^2+|(A_{_R})_{_{IJ}}|^2\Big]\;,
\label{mu-e-gamma-width}
\end{eqnarray}
where the Wilson coefficients $(A_{_L})_{_{IJ}},\;(A_{_R})_{_{IJ}}$ are extracted from the effective Lagrangian
\begin{eqnarray}
&&{\cal L}_{eff}^{e_{_I}\rightarrow e_{_J}\gamma}={e\over2}\bar{e}_{_J}\sigma_{\mu\nu}F^{\mu\nu}\Big[(A_{_L})_{_{IJ}}\omega_-
+(A_{_R})_{_{IJ}}\omega_+\Big]e_{_I}\;,
\label{mu-e-gamma-Lag}
\end{eqnarray}
with $\omega_\mp=(1\mp\gamma_5)/2$.

For simplicity we assume that those relevant soft breaking parameters
$m_{_{\tilde L}}^2,\;m_{_{\tilde E}}^2,\;A_{_E},\;A_{_N}$ are flavor conservation,
i.e. $(m_{_{\tilde L}}^2)_{_{IJ}}=(m_{_{\tilde E}}^2)_{_{IJ}}=(A_{_E})_{_{IJ}}
=(A_{_N})_{_{IJ}}=0$ as $I\neq J$. Under this assumption, the lepton flavor violating
processes are only induced by the R-parity violating couplings in this model.
In mass basis we expand relevant
couplings in powers of $\upsilon_{_{L_i}}/\upsilon_{_{EW}}$ etc. to the second order,
and present the concrete expressions of interactions in appendix.\ref{app2}
in the t'Hooft-Feynman gauge with $\xi=1$.
Then the transition $\mu\rightarrow e\gamma$ is evoked by the following pieces.
\begin{itemize}
\item   The interactions between CP-even Higgs and charginos/charged leptons,
the corresponding Wilson coefficients are written as

\begin{eqnarray}
%%%%%%%%%%%%%%%%%%%%%%%%%%%%%%%%%%%%%%%%%%%%%%%%%%%%%%%%%%%%
&&(A_{_L}^1)_{_{IJ}}={m_{_{e_I}}\over(4\pi)^2\Lambda^2}\sum\limits_{i=1}^2\sum\limits_{\alpha=1}^2
[\delta\xi_1^{m}]_{_{iJ\alpha}}[\delta\xi_1^{m}]_{_{iI\alpha}}^*f_1(x_{_{H_i^0}},x_{_{\chi_\alpha^\pm}})
\nonumber\\
&&\hspace{1.8cm}
+{1\over(4\pi)^2\Lambda^2}\sum\limits_{i=1}^2\sum\limits_{\alpha=1}^2m_{_{\chi_\alpha^\pm}}
[\delta\xi_1^{m}]_{_{iJ\alpha}}[\delta\xi_2^{m}]_{_{i\alpha I}}f_2(x_{_{H_i^0}},x_{_{\chi_\alpha^\pm}})
\nonumber\\
&&\hspace{1.8cm}
+{m_{_{e_I}}\over\sqrt{2}(4\pi)^2\Lambda^2}\sum\limits_{K=1}^3\sum\limits_{\alpha=1}^2
\Big[(U_-)_{2\alpha}(Y_{_E}V_{_R})_{_{KJ}}[\delta^2\xi_1^{m}]_{_{(2+K)I\alpha}}^*
\nonumber\\
&&\hspace{1.8cm}
+(U_-)_{2\alpha}^*(Y_{_E}V_{_R})_{_{KI}}^*[\delta^2\xi_1^{m}]_{_{(2+K)J\alpha}}\Big]
f_1(x_{_{H_{2+K}^0}},x_{_{\chi_\alpha^\pm}})
\nonumber\\
&&\hspace{1.8cm}
+{1\over\sqrt{2}(4\pi)^2\Lambda^2}\sum\limits_{K=1}^3\sum\limits_{\alpha=1}^2m_{_{\chi_\alpha^\pm}}
\Big[(U_-)_{2\alpha}(Y_{_E}V_{_R})_{_{KJ}}[\delta^2\xi_2^{m}]_{_{(2+K)\alpha I}}
\nonumber\\
&&\hspace{1.8cm}
+{e\over s_{_{\rm W}}}(U_+)_{1\alpha}(V_{_L})_{_{KI}}[\delta^2\xi_1^{m}]_{_{(2+K)J\alpha}}\Big]
f_2(x_{_{H_{2+K}^0}},x_{_{\chi_\alpha^\pm}})
\nonumber\\
&&\hspace{1.8cm}
-{\sqrt{2}m_{_{e_I}}^2\over(4\pi)^2\Lambda^2\upsilon_{_d}}\sum\limits_{i=1}^2(Z_{_R})_{2i}\Big[
[\delta^2\xi_1^{e}]_{_{iJI}}^*f_1(x_{_{H_i^0}},0)+[\delta^2\xi_1^{e}]_{_{iJI}}f_2(x_{_{H_i^0}},0)\Big]
\nonumber\\
&&\hspace{1.8cm}
+{1\over(4\pi)^2\Lambda^2}\sum\limits_{K=1}^3\sum\limits_{I^\prime=1}^3\Big[m_{_{e_I}}
[\delta\xi_1^{e}]_{_{(2+K)II^\prime}}[\delta\xi_1^{e}]_{_{(2+K)JI^\prime}}^*f_1(x_{_{H_{2+K}^0}},0)
\nonumber\\
&&\hspace{1.8cm}
+m_{_{e_{I^\prime}}}[\delta\xi_1^{e}]_{_{(2+K)II^\prime}}[\delta\xi_1^{e}]_{_{(2+K)I^\prime J}}f_2(x_{_{H_{2+K}^0}},0)\Big]
\;,\nonumber\\
%%%%%%%%%%%%%%%%%%%%%%%%%%%%%%%%%%%%%%%%%%%%%%%%%%%%%%%%%%%%
&&(A_{_R}^1)_{_{IJ}}={m_{_{e_I}}\over(4\pi)^2\Lambda^2}\sum\limits_{i=1}^2\sum\limits_{\alpha=1}^2
[\delta\xi_2^{m}]_{_{i\alpha J}}^*[\delta\xi_2^{m}]_{_{i\alpha I}}f_1(x_{_{H_i^0}},x_{_{\chi_\alpha^\pm}})
\nonumber\\
&&\hspace{1.8cm}
+{1\over(4\pi)^2\Lambda^2}\sum\limits_{i=1}^2\sum\limits_{\alpha=1}^2m_{_{\chi_\alpha^\pm}}
[\delta\xi_2^{m}]_{_{i\alpha J}}^*[\delta\xi_1^{m}]_{_{iI\alpha}}^*f_2(x_{_{H_i^0}},x_{_{\chi_\alpha^\pm}})
\nonumber\\
&&\hspace{1.8cm}
+{em_{_{e_I}}\over\sqrt{2}(4\pi)^2s_{_{\rm W}}\Lambda^2}\sum\limits_{K=1}^3\sum\limits_{\alpha=1}^2
\Big[(U_+)_{1\alpha}^*(V_{_L})_{_{KJ}}^*[\delta^2\xi_2^{m}]_{_{(2+K)\alpha I}}^*
\nonumber\\
&&\hspace{1.8cm}
+(U_+)_{1\alpha}(V_{_L})_{_{KI}}[\delta^2\xi_2^{m}]_{_{(2+K)\alpha J}}^*\Big]
f_1(x_{_{H_{2+K}^0}},x_{_{\chi_\alpha^\pm}})
\nonumber\\
&&\hspace{1.8cm}
+{1\over\sqrt{2}(4\pi)^2\Lambda^2}\sum\limits_{K=1}^3\sum\limits_{\alpha=1}^2m_{_{\chi_\alpha^\pm}}
\Big[(U_-)_{2\alpha}^*(Y_{_E}V_{_R})_{_{KI}}^*[\delta^2\xi_2^{m}]_{_{(2+K)\alpha J}}
\nonumber\\
&&\hspace{1.8cm}
+{e\over s_{_{\rm W}}}(U_+)_{1\alpha}^*(V_{_L})_{_{KJ}}^*[\delta^2\xi_1^{m}]_{_{(2+K)I\alpha}}^*\Big]
f_2(x_{_{H_{2+K}^0}},x_{_{\chi_\alpha^\pm}})
\nonumber\\
&&\hspace{1.8cm}
+{m_{_{e_I}}\over(4\pi)^2\Lambda^2}\sum\limits_{K=1}^3\sum\limits_{\alpha=1}^2
[\delta\xi_2^{m}]_{_{(5+K)\alpha J}}^*[\delta\xi_2^{m}]_{_{(5+K)\alpha I}}f_1(x_{_{H_{5+K}^0}},x_{_{\chi_\alpha^\pm}})
\nonumber\\
&&\hspace{1.8cm}
-{\sqrt{2}m_{_{e_I}}^2\over(4\pi)^2\Lambda^2\upsilon_{_d}}\sum\limits_{i=1}^2(Z_{_R})_{2i}^*\Big[
[\delta^2\xi_1^{e}]_{_{iJI}}f_1(x_{_{H_i^0}},0)+[\delta^2\xi_1^{e}]_{_{iJI}}^*f_2(x_{_{H_i^0}},0)\Big]
\nonumber\\
&&\hspace{1.8cm}
+{1\over(4\pi)^2\Lambda^2}\sum\limits_{K=1}^3\sum\limits_{I^\prime=1}^3\Big[m_{_{e_I}}
[\delta\xi_1^{e}]_{_{(2+K)I^\prime I}}^*[\delta\xi_1^{e}]_{_{(2+K)I^\prime J}}f_1(x_{_{H_{2+K}^0}},0)
\nonumber\\
&&\hspace{1.8cm}
+m_{_{e_{I^\prime}}}[\delta\xi_1^{e}]_{_{(2+K)I^\prime I}}^*[\delta\xi_1^{e}]_{_{(2+K)JI^\prime }}^*f_2(x_{_{H_{2+K}^0}},0)\Big]
\;,\nonumber\\
%%%%%%%%%%%%%%%%%%%%%%%%%%%%%%%%%%%%%%%%%%%%%%%%%%%%%%%%%%%%
\label{eI-eJ+gamma1}
\end{eqnarray}
where $\Lambda$ denotes the matching scale between full theory and effective one,
$x_i=m_i^2/\Lambda^2$, and
\begin{eqnarray}
%%%%%%%%%%%%%%%%%%%%%%%%%%%%%%%%%%%%%%%%%%%%%%%%%%%%%%%%%%%%
&&f_1(x,y)=\Big[{1\over12}{\partial^3\varrho_{_{3,1}}\over\partial x^3}-{1\over2}{\partial^2\varrho_{_{2,1}}\over\partial x^2}
+{1\over2}{\partial\varrho_{_{1,1}}\over\partial x}\Big](x,y)
\;,\nonumber\\
%%%%%%%%%%%%%%%%%%%%%%%%%%%%%%%%%%%%%%%%%%%%%%%%%%%%%%%%%%%%
&&f_2(x,y)=\Big[{1\over2}{\partial^2\varrho_{_{2,1}}\over\partial x^2}-{\partial\varrho_{_{1,1}}\over\partial x}
+{\partial\varrho_{_{1,1}}\over\partial y}\Big](x,y)\;.
%%%%%%%%%%%%%%%%%%%%%%%%%%%%%%%%%%%%%%%%%%%%%%%%%%%%%%%%%%%%
\label{FormFactor-ee+gamma1}
\end{eqnarray}
Those concrete expressions for lepton number violating couplings are
collected in appendix.(\ref{app2}).

\item The interactions between CP-odd Higgs and charginos/charged leptons,
the corresponding Wilson coefficients are formulated as
\begin{eqnarray}
%%%%%%%%%%%%%%%%%%%%%%%%%%%%%%%%%%%%%%%%%%%%%%%%%%%%%%%%%%%%
&&(A_{_L}^2)_{_{IJ}}={m_{_{e_I}}\over(4\pi)^2\Lambda^2}\sum\limits_{i=1,3}\sum\limits_{\alpha=1}^2
[\delta\eta_1^{m}]_{_{iJ\alpha}}[\delta\eta_1^{m}]_{_{iI\alpha}}^*f_1(x_{_{A_i^0}},x_{_{\chi_\alpha^\pm}})
\nonumber\\
&&\hspace{1.8cm}
-{1\over(4\pi)^2\Lambda^2}\sum\limits_{i=1,3}\sum\limits_{\alpha=1}^2m_{_{\chi_\alpha^\pm}}
[\delta\eta_1^{m}]_{_{iJ\alpha}}[\delta\eta_2^{m}]_{_{i\alpha I}}f_2(x_{_{A_i^0}},x_{_{\chi_\alpha^\pm}})
\nonumber\\
&&\hspace{1.8cm}
+{m_{_{e_I}}\over\sqrt{2}(4\pi)^2\Lambda^2}\sum\limits_{K=1}^3\sum\limits_{\alpha=1}^2
\Big[(U_-)_{2\alpha}(Y_{_E}V_{_R})_{_{KJ}}[\delta^2\eta_1^{m}]_{_{(3+K)I\alpha}}^*
\nonumber\\
&&\hspace{1.8cm}
+(U_-)_{2\alpha}^*(Y_{_E}V_{_R})_{_{KI}}^*[\delta^2\eta_1^{m}]_{_{(3+K)J\alpha}}\Big]
f_1(x_{_{A_{3+K}^0}},x_{_{\chi_\alpha^\pm}})
\nonumber\\
&&\hspace{1.8cm}
-{1\over\sqrt{2}(4\pi)^2\Lambda^2}\sum\limits_{K=1}^3\sum\limits_{\alpha=1}^2m_{_{\chi_\alpha^\pm}}
\Big[(U_-)_{2\alpha}(Y_{_E}V_{_R})_{_{KJ}}[\delta^2\eta_2^{m}]_{_{(3+K)\alpha I}}
\nonumber\\
&&\hspace{1.8cm}
+{e\over s_{_{\rm W}}}(U_+)_{1\alpha}(V_{_L})_{_{KI}}[\delta^2\eta_1^{m}]_{_{(3+K)J\alpha}}\Big]
f_2(x_{_{A_{3+K}^0}},x_{_{\chi_\alpha^\pm}})
\nonumber\\
&&\hspace{1.8cm}
+{m_{_{e_I}}\over(4\pi)^2\Lambda^2}\sum\limits_{i=1}^2\sum\limits_{\alpha=1}^2
[\delta\eta_1^{m}]_{_{(6+i)J\alpha}}[\delta\eta_1^{m}]_{_{(6+i)I\alpha}}f_1(x_{_{A_{6+i}^0}},x_{_{\chi_\alpha^\pm}})
\nonumber\\
&&\hspace{1.8cm}
-{1\over(4\pi)^2\Lambda^2}\sum\limits_{i=1}^2\sum\limits_{\alpha=1}^2m_{_{\chi_\alpha^\pm}}
[\delta\eta_1^{m}]_{_{(6+i)J\alpha}}[\delta\eta_2^{m}]_{_{(6+i)\alpha I}}f_2(x_{_{A_{6+i}^0}},x_{_{\chi_\alpha^\pm}})
\nonumber\\
&&\hspace{1.8cm}
+{\sqrt{2}m_{_{e_I}}^2\cos\beta\over(4\pi)^2\Lambda^2\upsilon_{_d}}\Big\{[\delta^2\eta^{e}]_{_{1JI}}f_1(x_{_Z},0)
-[\delta^2\eta^{e}]_{_{1IJ}}^*f_2(x_{_Z},0)\Big\}
\nonumber\\
&&\hspace{1.8cm}
-{\sqrt{2}m_{_{e_I}}^2\sin\beta\over(4\pi)^2\Lambda^2\upsilon_{_d}}\Big\{[\delta^2\eta^{e}]_{_{3JI}}f_1(x_{_{A_3^0}},0)
-[\delta^2\eta^{e}]_{_{3IJ}}^*f_2(x_{_{A_3^0}},0)\Big\}
\;,\nonumber\\
%%%%%%%%%%%%%%%%%%%%%%%%%%%%%%%%%%%%%%%%%%%%%%%%%%%%%%%%%%%%
&&(A_{_R}^2)_{_{IJ}}={m_{_{e_I}}\over(4\pi)^2\Lambda^2}\sum\limits_{i=1}^3\sum\limits_{\alpha=1}^2
[\delta\eta_2^{m}]_{_{i\alpha J}}^*[\delta\eta_2^{m}]_{_{i\alpha I}}f_1(x_{_{A_i^0}},x_{_{\chi_\alpha^\pm}})
\nonumber\\
&&\hspace{1.8cm}
-{1\over(4\pi)^2\Lambda^2}\sum\limits_{i=1,3}\sum\limits_{\alpha=1}^2m_{_{\chi_\alpha^\pm}}
[\delta\eta_2^{m}]_{_{i\alpha J}}^*[\delta\eta_1^{m}]_{_{iI\alpha}}^*f_2(x_{_{A_i^0}},x_{_{\chi_\alpha^\pm}})
\nonumber\\
&&\hspace{1.8cm}
+{em_{_{e_I}}\over\sqrt{2}(4\pi)^2s_{_{\rm W}}\Lambda^2}\sum\limits_{K=1}^3\sum\limits_{\alpha=1}^2
\Big[(U_+)_{1\alpha}^*(V_{_L})_{_{KJ}}^*[\delta^2\eta_2^{m}]_{_{(3+K)\alpha I}}^*
\nonumber\\
&&\hspace{1.8cm}
+(U_+)_{1\alpha}(V_{_L})_{_{KI}}[\delta^2\eta_2^{m}]_{_{(3+K)\alpha J}}^*\Big]
f_1(x_{_{A_{3+K}^0}},x_{_{\chi_\alpha^\pm}})
\nonumber\\
&&\hspace{1.8cm}
-{1\over\sqrt{2}(4\pi)^2\Lambda^2}\sum\limits_{K=1}^3\sum\limits_{\alpha=1}^2m_{_{\chi_\alpha^\pm}}
\Big[(U_-)_{2\alpha}^*(Y_{_E}V_{_R})_{_{KI}}^*[\delta^2\eta_2^{m}]_{_{(3+K)\alpha J}}
\nonumber\\
&&\hspace{1.8cm}
+{e\over s_{_{\rm W}}}(U_+)_{1\alpha}^*(V_{_L})_{_{KJ}}^*[\delta^2\eta_1^{m}]_{_{(3+K)I\alpha}}^*\Big]
f_2(x_{_{A_{3+K}^0}},x_{_{\chi_\alpha^\pm}})
\nonumber\\
&&\hspace{1.8cm}
+{m_{_{e_I}}\over(4\pi)^2\Lambda^2}\sum\limits_{i=1}^2\sum\limits_{\alpha=1}^2
[\delta\eta_2^{m}]_{_{(6+i)\alpha J}}^*[\delta\eta_2^{m}]_{_{(6+i)\alpha I}}f_1(x_{_{A_{6+i}^0}},x_{_{\chi_\alpha^\pm}})
\nonumber\\
&&\hspace{1.8cm}
-{1\over(4\pi)^2\Lambda^2}\sum\limits_{i=1}^2\sum\limits_{\alpha=1}^2m_{_{\chi_\alpha^\pm}}
[\delta\eta_2^{m}]_{_{(6+i)\alpha J}}^*[\delta\eta_1^{m}]_{_{(6+i)I\alpha}}f_2(x_{_{A_{6+i}^0}},x_{_{\chi_\alpha^\pm}})
\nonumber\\
&&\hspace{1.8cm}
+{\sqrt{2}m_{_{e_I}}^2\cos\beta\over(4\pi)^2\Lambda^2\upsilon_{_d}}\Big\{[\delta^2\eta^{e}]_{_{1IJ}}^*f_1(x_{_Z},0)
-[\delta^2\eta^{e}]_{_{1JI}}f_2(x_{_Z},0)\Big\}
\nonumber\\
&&\hspace{1.8cm}
-{\sqrt{2}m_{_{e_I}}^2\sin\beta\over(4\pi)^2\Lambda^2\upsilon_{_d}}\Big\{[\delta^2\eta^{e}]_{_{3IJ}}^*f_1(x_{_{A_3^0}},0)
-[\delta^2\eta^{e}]_{_{3JI}}f_2(x_{_{A_3^0}},0)\Big\}
\;,\nonumber\\
%%%%%%%%%%%%%%%%%%%%%%%%%%%%%%%%%%%%%%%%%%%%%%%%%%%%%%%%%%%%
\label{eI-eJ+gamma2}
\end{eqnarray}
the concrete expressions for lepton number violating couplings are
collected in appendix.(\ref{app2}).

\item The interactions between charged Higgs and neutralinos/neutrinos,
we write the relevant Wilson coefficients as
\begin{eqnarray}
%%%%%%%%%%%%%%%%%%%%%%%%%%%%%%%%%%%%%%%%%%%%%%%%%%%%%%%%%%%%
&&(A_{_L}^3)_{_{IJ}}={m_{_{e_I}}\over(4\pi)^2\Lambda^2}\sum\limits_{K=1}^3\sum\limits_{b=1}^5
[\delta\zeta_{\tilde L}^L]_{_{KJb}}[\delta\zeta_{\tilde L}^L]_{_{KIb}}^*f_3(x_{_{H_{2+K}^\pm}},0)
\nonumber\\
&&\hspace{1.8cm}
+{m_{_{e_I}}\over(4\pi)^2\Lambda^2}\sum\limits_{K=1}^3\sum\limits_{b=1}^5
[\delta\zeta_{\tilde R}^L]_{_{KJb}}[\delta\zeta_{\tilde R}^L]_{_{KIb}}^*f_3(x_{_{H_{5+K}^\pm}},0)
\nonumber\\
&&\hspace{1.8cm}
+{1\over(4\pi)^2\Lambda^2}\sum\limits_{\alpha=1}^4m_{_{e_I}}[\delta\zeta_G^L]_{_{1J\chi_\alpha^0}}
[\delta\zeta_G^L]_{_{1I\chi_\alpha^0}}^*f_3(x_{_{\rm W}},x_{_{\chi_\alpha^0}})
\nonumber\\
&&\hspace{1.8cm}
+{1\over(4\pi)^2\Lambda^2}\sum\limits_{\alpha=1}^4m_{_{\chi_\alpha^0}}[\delta\zeta_G^L]_{_{1J\chi_\alpha^0}}
[\delta\zeta_G^R]_{_{1I\chi_\alpha^0}}^*f_4(x_{_{\rm W}},x_{_{\chi_\alpha^0}})
\nonumber\\
&&\hspace{1.8cm}
+{1\over(4\pi)^2\Lambda^2}\sum\limits_{\alpha=1}^4m_{_{e_I}}[\delta\zeta_H^L]_{_{2J\chi_\alpha^0}}
[\delta\zeta_H^L]_{_{2I\chi_\alpha^0}}^*f_3(x_{_{H_2^\pm}},x_{_{\chi_\alpha^0}})
\nonumber\\
&&\hspace{1.8cm}
+{1\over(4\pi)^2\Lambda^2}\sum\limits_{\alpha=1}^4m_{_{\chi_\alpha^0}}[\delta\zeta_H^L]_{_{2J\chi_\alpha^0}}
[\delta\zeta_H^R]_{_{2I\chi_\alpha^0}}^*f_4(x_{_{H_2^\pm}},x_{_{\chi_\alpha^0}})
\nonumber\\
&&\hspace{1.8cm}
+{m_{_{e_I}}\over(4\pi)^2\Lambda^2}\sum\limits_{K=1}^3\sum\limits_{\alpha=1}^4\Big\{
\Big[{e\sqrt{2}\over c_{_{\rm W}}}(Z_{_{\tilde{E}_K}})_{21}\Big(V_{_R}\Big)_{KJ}(U_\chi)_{1\alpha}
\nonumber\\
&&\hspace{1.8cm}
+(Y_{_E}V_{_R})_{_{KJ}}(Z_{_{\tilde{E}_K}})_{11}(U_\chi)_{3\alpha}\Big]
[\delta^2\zeta_{_{\tilde L}}^L]_{_{KI\chi_\alpha^0}}^*
+\Big[{e\sqrt{2}\over c_{_{\rm W}}}(Z_{_{\tilde{E}_K}})_{21}^*\Big(V_{_R}\Big)_{KI}^*(U_\chi)_{1\alpha}^*
\nonumber\\
&&\hspace{1.8cm}
+(Y_{_E}V_{_R})_{_{KI}}^*(Z_{_{\tilde{E}_K}})_{11}^*(U_\chi)_{3\alpha}^*\Big]
[\delta^2\zeta_{_{\tilde L}}^L]_{_{KJ\chi_\alpha^0}}\Big\}f_3(x_{_{H_{2+K}^\pm}},x_{_{\chi_\alpha^0}})
\nonumber\\
&&\hspace{1.8cm}
+{1\over(4\pi)^2\Lambda^2}\sum\limits_{K=1}^3\sum\limits_{\alpha=1}^4m_{_{\chi_\alpha^0}}\Big\{
\Big[{e\sqrt{2}\over c_{_{\rm W}}}(Z_{_{\tilde{E}_K}})_{21}\Big(V_{_R}\Big)_{KJ}(U_\chi)_{1\alpha}
\nonumber\\
&&\hspace{1.8cm}
+(Y_{_E}V_{_R})_{_{KJ}}(Z_{_{\tilde{E}_K}})_{11}(U_\chi)_{3\alpha}\Big]
[\delta^2\zeta_{_{\tilde L}}^R]_{_{KI\chi_\alpha^0}}^*
\nonumber\\
&&\hspace{1.8cm}
+\Big[-{e\over\sqrt{2}s_{_{\rm W}}c_{_{\rm W}}}(Z_{_{\tilde{E}_K}})_{11}^*\Big(V_{_L}\Big)_{KI}\Big(
c_{_{\rm W}}(U_\chi)_{2\alpha}+s_{_{\rm W}}(U_\chi)_{1\alpha}\Big)
\nonumber\\
&&\hspace{1.8cm}
+(Z_{_{\tilde{E}_K}})_{21}^*(V_{_L}^\dagger Y_{_E})_{_{IK}}^*(U_\chi)_{3\alpha}\Big]
[\delta^2\zeta_{_{\tilde L}}^L]_{_{KJ\chi_\alpha^0}}\Big\}f_4(x_{_{H_{2+K}^\pm}},x_{_{\chi_\alpha^0}})
\nonumber\\
&&\hspace{1.8cm}
+{m_{_{e_I}}\over(4\pi)^2\Lambda^2}\sum\limits_{K=1}^3\sum\limits_{\alpha=1}^4\Big\{
\Big[{e\sqrt{2}\over c_{_{\rm W}}}(Z_{_{\tilde{E}_K}})_{22}\Big(V_{_R}\Big)_{KJ}(U_\chi)_{1\alpha}
\nonumber\\
&&\hspace{1.8cm}
+(Y_{_E}V_{_R})_{_{KJ}}(Z_{_{\tilde{E}_K}})_{12}(U_\chi)_{3\alpha}\Big]
[\delta^2\zeta_{_{\tilde R}}^L]_{_{KI\chi_\alpha^0}}^*
+\Big[{e\sqrt{2}\over c_{_{\rm W}}}(Z_{_{\tilde{E}_K}})_{22}^*\Big(V_{_R}\Big)_{KI}^*(U_\chi)_{1\alpha}^*
\nonumber\\
&&\hspace{1.8cm}
+(Y_{_E}V_{_R})_{_{KI}}^*(Z_{_{\tilde{E}_K}})_{12}^*(U_\chi)_{3\alpha}^*\Big]
[\delta^2\zeta_{_{\tilde R}}^L]_{_{KJ\chi_\alpha^0}}\Big\}f_3(x_{_{H_{5+K}^\pm}},x_{_{\chi_\alpha^0}})
\nonumber\\
&&\hspace{1.8cm}
+{1\over(4\pi)^2\Lambda^2}\sum\limits_{K=1}^3\sum\limits_{\alpha=1}^4m_{_{\chi_\alpha^0}}\Big\{
\Big[{e\sqrt{2}\over c_{_{\rm W}}}(Z_{_{\tilde{E}_K}})_{22}\Big(V_{_R}\Big)_{KJ}(U_\chi)_{1\alpha}
\nonumber\\
&&\hspace{1.8cm}
+(Y_{_E}V_{_R})_{_{KJ}}(Z_{_{\tilde{E}_K}})_{12}(U_\chi)_{3\alpha}\Big]
[\delta^2\zeta_{_{\tilde R}}^R]_{_{KI\chi_\alpha^0}}^*
\nonumber\\
&&\hspace{1.8cm}
+\Big[-{e\over\sqrt{2}s_{_{\rm W}}c_{_{\rm W}}}(Z_{_{\tilde{E}_K}})_{12}^*\Big(V_{_L}\Big)_{KI}\Big(
c_{_{\rm W}}(U_\chi)_{2\alpha}+s_{_{\rm W}}(U_\chi)_{1\alpha}\Big)
\nonumber\\
&&\hspace{1.8cm}
+(Z_{_{\tilde{E}_K}})_{22}^*(V_{_L}^\dagger Y_{_E})_{_{IK}}^*(U_\chi)_{3\alpha}\Big]
[\delta^2\zeta_{_{\tilde R}}^L]_{_{KJ\chi_\alpha^0}}\Big\}f_4(x_{_{H_{5+K}^\pm}},x_{_{\chi_\alpha^0}})
\;,\nonumber\\
%%%%%%%%%%%%%%%%%%%%%%%%%%%%%%%%%%%%%%%%%%%%%%%%%%%%%%%%%%%%
&&(A_{_R}^3)_{_{IJ}}={\cos\beta m_{_{e_I}}\over(4\pi)^2\Lambda^2}\sum\limits_{b=1}^3\Big\{[\delta^2\zeta_G^L]_{_{1Ib}}
(Y_{_E}V_{_R})_{bJ}^*+[\delta^2\zeta_G^L]_{_{1Jb}}^*(Y_{_E}V_{_R})_{bI}\Big\}f_3(x_{_{\rm W}},0)
\nonumber\\
&&\hspace{1.8cm}
-{\sin\beta m_{_{e_I}}\over(4\pi)^2\Lambda^2}\sum\limits_{b=1}^5\Big\{[\delta^2\zeta_H^L]_{_{2Ib}}(Y_{_E}V_{_R})_{bJ}^*
+[\delta^2\zeta_H^L]_{_{2Jb}}^*(Y_{_E}V_{_R})_{bI}\Big\}f_3(x_{_{H_2^\pm}},0)
\nonumber\\
&&\hspace{1.8cm}
+{m_{_{e_I}}\over(4\pi)^2\Lambda^2}\sum\limits_{K=1}^3\sum\limits_{b=1}^5
[\delta\zeta_{\tilde L}^R]_{_{KJb}}[\delta\zeta_{\tilde L}^R]_{_{KIb}}^*f_3(x_{_{H_{2+K}^\pm}},0)
\nonumber\\
&&\hspace{1.8cm}
+{m_{_{e_I}}\over(4\pi)^2\Lambda^2}\sum\limits_{K=1}^3\sum\limits_{b=1}^5
[\delta\zeta_{\tilde R}^R]_{_{KJb}}[\delta\zeta_{\tilde R}^R]_{_{KIb}}^*f_3(x_{_{H_{5+K}^\pm}},0)
\nonumber\\
&&\hspace{1.8cm}
+{1\over(4\pi)^2\Lambda^2}\sum\limits_{\alpha=1}^4m_{_{e_I}}[\delta\zeta_G^R]_{_{1J\chi_\alpha^0}}
[\delta\zeta_G^R]_{_{1I\chi_\alpha^0}}^*f_3(x_{_{\rm W}},x_{_{\chi_\alpha^0}})
\nonumber\\
&&\hspace{1.8cm}
+{1\over(4\pi)^2\Lambda^2}\sum\limits_{\alpha=1}^4m_{_{\chi_\alpha^0}}[\delta\zeta_G^R]_{_{1J\chi_\alpha^0}}
[\delta\zeta_G^L]_{_{1I\chi_\alpha^0}}^*f_4(x_{_{\rm W}},x_{_{\chi_\alpha^0}})
\nonumber\\
&&\hspace{1.8cm}
+{1\over(4\pi)^2\Lambda^2}\sum\limits_{\alpha=1}^4m_{_{e_I}}[\delta\zeta_H^R]_{_{2J\chi_\alpha^0}}
[\delta\zeta_H^R]_{_{2I\chi_\alpha^0}}^*f_3(x_{_{H_2^\pm}},x_{_{\chi_\alpha^0}})
\nonumber\\
&&\hspace{1.8cm}
+{1\over(4\pi)^2\Lambda^2}\sum\limits_{\alpha=1}^4m_{_{\chi_\alpha^0}}[\delta\zeta_H^R]_{_{2J\chi_\alpha^0}}
[\delta\zeta_H^L]_{_{2I\chi_\alpha^0}}^*f_4(x_{_{H_2^\pm}},x_{_{\chi_\alpha^0}})
\nonumber\\
&&\hspace{1.8cm}
+{m_{_{e_I}}\over(4\pi)^2\Lambda^2}\sum\limits_{K=1}^3\sum\limits_{\alpha=1}^4\Big\{
\Big[-{e\over\sqrt{2}s_{_{\rm W}}c_{_{\rm W}}}(Z_{_{\tilde{E}_K}})_{11}\Big(V_{_L}\Big)_{KJ}^*\Big(
c_{_{\rm W}}(U_\chi)_{2\alpha}^*+s_{_{\rm W}}(U_\chi)_{1\alpha}^*\Big)
\nonumber\\
&&\hspace{1.8cm}
+(Z_{_{\tilde{E}_K}})_{21}(V_{_L}^\dagger Y_{_E})_{_{JK}}(U_\chi)_{3\alpha}^*\Big]
[\delta^2\zeta_{_{\tilde L}}^R]_{_{KI\chi_\alpha^0}}^*
\nonumber\\
&&\hspace{1.8cm}
+\Big[-{e\over\sqrt{2}s_{_{\rm W}}c_{_{\rm W}}}(Z_{_{\tilde{E}_K}})_{11}^*\Big(V_{_L}\Big)_{KI}\Big(
c_{_{\rm W}}(U_\chi)_{2\alpha}+s_{_{\rm W}}(U_\chi)_{1\alpha}\Big)
\nonumber\\
&&\hspace{1.8cm}
+(Z_{_{\tilde{E}_K}})_{21}^*(V_{_L}^\dagger Y_{_E})_{_{IK}}^*(U_\chi)_{3\alpha}\Big]
[\delta^2\zeta_{_{\tilde L}}^R]_{_{KJ\chi_\alpha^0}}\Big\}f_3(x_{_{H_{2+K}^\pm}},x_{_{\chi_\alpha^0}})
\nonumber\\
&&\hspace{1.8cm}
+{1\over(4\pi)^2\Lambda^2}\sum\limits_{K=1}^3\sum\limits_{\alpha=1}^4m_{_{\chi_\alpha^0}}\Big\{
\Big[{e\sqrt{2}\over c_{_{\rm W}}}(Z_{_{\tilde{E}_K}})_{21}^*\Big(V_{_R}\Big)_{KI}(U_\chi)_{1\alpha}^*
\nonumber\\
&&\hspace{1.8cm}
+(Y_{_E}V_{_R})_{_{KI}}^*(Z_{_{\tilde{E}_K}})_{11}^*(U_\chi)_{3\alpha}^*\Big]
[\delta^2\zeta_{_{\tilde L}}^R]_{_{KJ\chi_\alpha^0}}
\nonumber\\
&&\hspace{1.8cm}
+\Big[-{e\over\sqrt{2}s_{_{\rm W}}c_{_{\rm W}}}(Z_{_{\tilde{E}_K}})_{11}\Big(V_{_L}\Big)_{KJ}\Big(
c_{_{\rm W}}(U_\chi)_{2\alpha}^*+s_{_{\rm W}}(U_\chi)_{1\alpha}^*\Big)
\nonumber\\
&&\hspace{1.8cm}
+(Z_{_{\tilde{E}_K}})_{21}(V_{_L}^\dagger Y_{_E})_{_{JK}}(U_\chi)_{3\alpha}^*\Big]
[\delta^2\zeta_{_{\tilde L}}^L]_{_{KI\chi_\alpha^0}}^*\Big\}f_4(x_{_{H_{2+K}^\pm}},x_{_{\chi_\alpha^0}})
\nonumber\\
&&\hspace{1.8cm}
+{m_{_{e_I}}\over(4\pi)^2\Lambda^2}\sum\limits_{K=1}^3\sum\limits_{\alpha=1}^4\Big\{
\Big[-{e\over\sqrt{2}s_{_{\rm W}}c_{_{\rm W}}}(Z_{_{\tilde{E}_K}})_{12}\Big(V_{_L}\Big)_{KJ}^*\Big(
c_{_{\rm W}}(U_\chi)_{2\alpha}^*+s_{_{\rm W}}(U_\chi)_{1\alpha}^*\Big)
\nonumber\\
&&\hspace{1.8cm}
+(Z_{_{\tilde{E}_K}})_{22}(V_{_L}^\dagger Y_{_E})_{_{JK}}(U_\chi)_{3\alpha}^*\Big]
[\delta^2\zeta_{_{\tilde R}}^R]_{_{KI\chi_\alpha^0}}^*
\nonumber\\
&&\hspace{1.8cm}
+\Big[-{e\over\sqrt{2}s_{_{\rm W}}c_{_{\rm W}}}(Z_{_{\tilde{E}_K}})_{12}^*\Big(V_{_L}\Big)_{KI}\Big(
c_{_{\rm W}}(U_\chi)_{2\alpha}+s_{_{\rm W}}(U_\chi)_{1\alpha}\Big)
\nonumber\\
&&\hspace{1.8cm}
+(Z_{_{\tilde{E}_K}})_{22}^*(V_{_L}^\dagger Y_{_E})_{_{IK}}^*(U_\chi)_{3\alpha}\Big]
[\delta^2\zeta_{_{\tilde R}}^R]_{_{KJ\chi_\alpha^0}}\Big\}f_3(x_{_{H_{5+K}^\pm}},x_{_{\chi_\alpha^0}})
\nonumber\\
&&\hspace{1.8cm}
+{1\over(4\pi)^2\Lambda^2}\sum\limits_{K=1}^3\sum\limits_{\alpha=1}^4m_{_{\chi_\alpha^0}}\Big\{
\Big[{e\sqrt{2}\over c_{_{\rm W}}}(Z_{_{\tilde{E}_K}})_{22}^*\Big(V_{_R}\Big)_{KI}(U_\chi)_{1\alpha}^*
\nonumber\\
&&\hspace{1.8cm}
+(Y_{_E}V_{_R})_{_{KI}}^*(Z_{_{\tilde{E}_K}})_{12}^*(U_\chi)_{3\alpha}^*\Big]
[\delta^2\zeta_{_{\tilde R}}^R]_{_{KJ\chi_\alpha^0}}
\nonumber\\
&&\hspace{1.8cm}
+\Big[-{e\over\sqrt{2}s_{_{\rm W}}c_{_{\rm W}}}(Z_{_{\tilde{E}_K}})_{12}\Big(V_{_L}\Big)_{KJ}\Big(
c_{_{\rm W}}(U_\chi)_{2\alpha}^*+s_{_{\rm W}}(U_\chi)_{1\alpha}^*\Big)
\nonumber\\
&&\hspace{1.8cm}
+(Z_{_{\tilde{E}_K}})_{22}(V_{_L}^\dagger Y_{_E})_{_{JK}}(U_\chi)_{3\alpha}^*\Big]
[\delta^2\zeta_{_{\tilde R}}^L]_{_{KI\chi_\alpha^0}}^*\Big\}f_4(x_{_{H_{5+K}^\pm}},x_{_{\chi_\alpha^0}})
\;,\nonumber\\
%%%%%%%%%%%%%%%%%%%%%%%%%%%%%%%%%%%%%%%%%%%%%%%%%%%%%%%%%%%%
\label{eI-eJ+gamma3}
\end{eqnarray}
with
\begin{eqnarray}
%%%%%%%%%%%%%%%%%%%%%%%%%%%%%%%%%%%%%%%%%%%%%%%%%%%%%%%%%%%%
&&f_3(x,y)=\Big[-{1\over12}{\partial^3\varrho_{_{3,1}}\over\partial x^3}+{1\over4}{\partial^2\varrho_{_{2,1}}\over\partial x^2}
\Big](x,y)
\;,\nonumber\\
%%%%%%%%%%%%%%%%%%%%%%%%%%%%%%%%%%%%%%%%%%%%%%%%%%%%%%%%%%%%
&&f_4(x,y)=\Big[-{1\over2}{\partial^2\varrho_{_{2,1}}\over\partial x^2}+{\partial\varrho_{_{1,1}}\over\partial x}
\Big](x,y)\;,
%%%%%%%%%%%%%%%%%%%%%%%%%%%%%%%%%%%%%%%%%%%%%%%%%%%%%%%%%%%%
\label{FormFactor-ee+gamma2}
\end{eqnarray}
and the explicit expressions for lepton number violating couplings are
collected in appendix.(\ref{app2}).

\item Neutral (charged) gauge bosons and chargino/charged leptons (neutralinos/neutrinos),
the Wilson coefficients are formulated as
\begin{eqnarray}
%%%%%%%%%%%%%%%%%%%%%%%%%%%%%%%%%%%%%%%%%%%%%%%%%%%%%%%%%%%%
&&(A_{_L}^4)_{_{IJ}}={e^2\over2(4\pi)^2c_{_{\rm W}}^2\Lambda^2}m_{_{e_I}}\Big\{
2[\delta^2{\cal C}^R]_{_{JI}}f_5(x_{_{\rm Z}},0)+[\delta^2{\cal C}^L]_{_{JI}}f_6(x_{_{\rm Z}},0)\Big\}
\nonumber\\
&&\hspace{1.8cm}
+{e^2\over4(4\pi)^2s_{_{\rm W}}^2c_{_{\rm W}}^2\Lambda^2}m_{_{e_I}}\sum\limits_{\alpha=1}^2[\delta{\cal C}^R]_{_{J\chi_\alpha^-}}
[\delta{\cal C}^R]_{_{I\chi_\alpha^-}}^*f_5(x_{_{\rm Z}},x_{_{\chi_\alpha^-}})
\nonumber\\
&&\hspace{1.8cm}
+{e^2\over4(4\pi)^2s_{_{\rm W}}^2c_{_{\rm W}}^2\Lambda^2}\sum\limits_{\alpha=1}^2m_{_{\chi_\alpha^-}}[\delta{\cal C}^L]_{_{J\chi_\alpha^-}}
[\delta{\cal C}^R]_{_{I\chi_\alpha^-}}^*f_6(x_{_{\rm Z}},x_{_{\chi_\alpha^-}})
\nonumber\\
&&\hspace{1.8cm}
+{g_{_{BL}}^2\over(4\pi)^2\Lambda^2}m_{_{e_I}}\Big\{
2[\delta^2{\cal B}^R]_{_{JI}}f_5(x_{_{Z_{BL}}},0)+[\delta^2{\cal B}^L]_{_{JI}}f_6(x_{_{Z_{BL}}},0)\Big\}
\nonumber\\
&&\hspace{1.8cm}
+{g_{_{BL}}^2\over(4\pi)^2\Lambda^2}m_{_{e_I}}\sum\limits_{\alpha=1}^2[\delta{\cal B}^R]_{_{J\chi_\alpha^-}}
[\delta{\cal B}^R]_{_{I\chi_\alpha^-}}^*f_5(x_{_{Z_{BL}}},x_{_{\chi_\alpha^-}})
\nonumber\\
&&\hspace{1.8cm}
+{g_{_{BL}}^2\over(4\pi)^2\Lambda^2}\sum\limits_{\alpha=1}^2m_{_{\chi_\alpha^-}}[\delta{\cal B}^L]_{_{J\chi_\alpha^-}}
[\delta{\cal B}^R]_{_{I\chi_\alpha^-}}^*f_6(x_{_{Z_{BL}}},x_{_{\chi_\alpha^-}})
\nonumber\\
&&\hspace{1.8cm}
+{e^2\over(4\pi)^2s_{_{\rm W}}^2\Lambda^2}m_{_{e_I}}\sum\limits_{\beta=1}^4[\delta{\cal V}^R]_{_{J\chi_\beta^0}}
[\delta{\cal V}^R]_{_{I\chi_\beta^0}}^*f_7(x_{_{\rm W}},x_{_{\chi_\beta^0}})
\nonumber\\
&&\hspace{1.8cm}
+{e^2\over(4\pi)^2s_{_{\rm W}}^2\Lambda^2}\sum\limits_{\beta=1}^4m_{_{\chi_\beta^0}}[\delta{\cal V}^R]_{_{J\chi_\beta^0}}
[\delta{\cal V}^L]_{_{I\chi_\beta^0}}^*f_8(x_{_{\rm W}},x_{_{\chi_\beta^0}})
\;,\nonumber\\
%%%%%%%%%%%%%%%%%%%%%%%%%%%%%%%%%%%%%%%%%%%%%%%%%%%%%%%%%%%%
&&(A_{_R}^4)_{_{IJ}}={e^2(2s_{_{\rm W}}^2-1)\over4(4\pi)^2s_{_{\rm W}}^2c_{_{\rm W}}^2\Lambda^2}m_{_{e_I}}\Big\{
2[\delta^2{\cal C}^L]_{_{JI}}f_5(x_{_{\rm Z}},0)+[\delta^2{\cal C}^R]_{_{JI}}f_6(x_{_{\rm Z}},0)\Big\}
\nonumber\\
&&\hspace{1.8cm}
+{e^2\over4(4\pi)^2s_{_{\rm W}}^2c_{_{\rm W}}^2\Lambda^2}m_{_{e_I}}\sum\limits_{\alpha=1}^2[\delta{\cal C}^L]_{_{J\chi_\alpha^-}}
[\delta{\cal C}^L]_{_{I\chi_\alpha^-}}^*f_5(x_{_{\rm Z}},x_{_{\chi_\alpha^-}})
\nonumber\\
&&\hspace{1.8cm}
+{e^2\over4(4\pi)^2s_{_{\rm W}}^2c_{_{\rm W}}^2\Lambda^2}\sum\limits_{\alpha=1}^2m_{_{\chi_\alpha^-}}[\delta{\cal C}^R]_{_{J\chi_\alpha^-}}
[\delta{\cal C}^L]_{_{I\chi_\alpha^-}}^*f_6(x_{_{\rm Z}},x_{_{\chi_\alpha^-}})
\nonumber\\
&&\hspace{1.8cm}
+{g_{_{BL}}^2\over(4\pi)^2\Lambda^2}m_{_{e_I}}\Big\{
2[\delta^2{\cal B}^L]_{_{JI}}f_5(x_{_{Z_{BL}}},0)+[\delta^2{\cal B}^R]_{_{JI}}f_6(x_{_{Z_{BL}}},0)\Big\}
\nonumber\\
&&\hspace{1.8cm}
+{g_{_{BL}}^2\over(4\pi)^2\Lambda^2}m_{_{e_I}}\sum\limits_{\alpha=1}^2[\delta{\cal B}^L]_{_{J\chi_\alpha^-}}
[\delta{\cal B}^L]_{_{I\chi_\alpha^-}}^*f_5(x_{_{Z_{BL}}},x_{_{\chi_\alpha^-}})
\nonumber\\
&&\hspace{1.8cm}
+{g_{_{BL}}^2\over(4\pi)^2\Lambda^2}\sum\limits_{\alpha=1}^2m_{_{\chi_\alpha^-}}[\delta{\cal B}^R]_{_{J\chi_\alpha^-}}
[\delta{\cal B}^L]_{_{I\chi_\alpha^-}}^*f_6(x_{_{Z_{BL}}},x_{_{\chi_\alpha^-}})
\nonumber\\
&&\hspace{1.8cm}
-{e^2\over\sqrt{2}(4\pi)^2s_{_{\rm W}}^2\Lambda^2}m_{_{e_I}}\sum\limits_{I^\prime=1}^3\Big\{[\delta^2{\cal V}^L]_{II^\prime}^*
(V_{_L})_{I^\prime J}^*+[\delta^2{\cal V}^L]_{JI^\prime}(V_{_L})_{I^\prime I}\Big\}f_7(x_{_{\rm W}},0)
\nonumber\\
&&\hspace{1.8cm}
+{e^2\over(4\pi)^2s_{_{\rm W}}^2\Lambda^2}m_{_{e_I}}\sum\limits_{\beta=1}^4[\delta{\cal V}^L]_{_{J\chi_\beta^0}}
[\delta{\cal V}^L]_{_{I\chi_\beta^0}}^*f_7(x_{_{\rm W}},x_{_{\chi_\beta^0}})
\nonumber\\
&&\hspace{1.8cm}
+{e^2\over(4\pi)^2s_{_{\rm W}}^2\Lambda^2}\sum\limits_{\beta=1}^4m_{_{\chi_\beta^0}}[\delta{\cal V}^L]_{_{J\chi_\beta^0}}
[\delta{\cal V}^R]_{_{I\chi_\beta^0}}^*f_8(x_{_{\rm W}},x_{_{\chi_\beta^0}})
\;,\nonumber\\
%%%%%%%%%%%%%%%%%%%%%%%%%%%%%%%%%%%%%%%%%%%%%%%%%%%%%%%%%%%%
\label{eI-eJ+gamma4}
\end{eqnarray}
with
\begin{eqnarray}
%%%%%%%%%%%%%%%%%%%%%%%%%%%%%%%%%%%%%%%%%%%%%%%%%%%%%%%%%%%%
&&f_5(x,y)=\Big[{1\over6}{\partial^3\varrho_{_{3,1}}\over\partial x^3}-{\partial\varrho_{_{1,1}}\over\partial x}\Big](x,y)
\;,\nonumber\\
%%%%%%%%%%%%%%%%%%%%%%%%%%%%%%%%%%%%%%%%%%%%%%%%%%%%%%%%%%%%
&&f_6(x,y)=\Big[-2{\partial^2\varrho_{_{2,1}}\over\partial x^2}+4{\partial\varrho_{_{1,1}}\over\partial x}\Big](x,y)
\;,\nonumber\\
%%%%%%%%%%%%%%%%%%%%%%%%%%%%%%%%%%%%%%%%%%%%%%%%%%%%%%%%%%%%
&&f_7(x,y)=\Big[-{1\over6}{\partial^3\varrho_{_{3,1}}\over\partial x^3}-{1\over2}{\partial^2\varrho_{_{2,1}}\over\partial x^2}\Big](x,y)
\;,\nonumber\\
%%%%%%%%%%%%%%%%%%%%%%%%%%%%%%%%%%%%%%%%%%%%%%%%%%%%%%%%%%%%
&&f_8(x,y)=2{\partial^2\varrho_{_{2,1}}\over\partial x^2}(x,y)\;.
%%%%%%%%%%%%%%%%%%%%%%%%%%%%%%%%%%%%%%%%%%%%%%%%%%%%%%%%%%%%
\label{FormFactor-ee+gamma3}
\end{eqnarray}
\end{itemize}
Actually the coefficients in above equations are not depend on the concrete choice
of matching energy scale $\Lambda$, and we choose $\Lambda=m_{_{\rm W}}$ in numerical analysis.
When we ignore the mixing between left- and right-handed sleptons,
the above results are consistent with that from mass insertion approach.
At 90\% confidence level the upper bound on the branching ratio of
$\mu\rightarrow e\gamma$ is \cite{MEG}
\begin{eqnarray}
%%%%%%%%%%%%%%%%%%%%%%%%%%%%%%%%%%%%%%%%%%%%%%%%%%%%%%%%%%%%
&&B(\mu\rightarrow e\gamma)\le B_{_{\mu\rightarrow e\gamma}}^{ub}=5.7\times10^{-13}\;.
%%%%%%%%%%%%%%%%%%%%%%%%%%%%%%%%%%%%%%%%%%%%%%%%%%%%%%%%%%%%
\label{exp-mutoegamma}
\end{eqnarray}
To investigate the constraint on the parameter space from above experimental data,
we define the ratios between the theoretical evaluation $B_{_{\mu\rightarrow e\gamma}}^{th}$
on the branching ratio of $\mu\rightarrow e\gamma$ and corresponding experimental upper bound
$B_{_{\mu\rightarrow e\gamma}}^{ub}$ as
\begin{eqnarray}
%%%%%%%%%%%%%%%%%%%%%%%%%%%%%%%%%%%%%%%%%%%%%%%%%%%%%%%%%%%%
&&R_{_{\mu\rightarrow e\gamma}}={B_{_{\mu\rightarrow e\gamma}}^{th}\over
B_{_{\mu\rightarrow e\gamma}}^{ub}}\;,
%%%%%%%%%%%%%%%%%%%%%%%%%%%%%%%%%%%%%%%%%%%%%%%%%%%%%%%%%%%%
\label{R-mutoegamma}
\end{eqnarray}
and $R_{_{\mu\rightarrow e\gamma}}<1$ implies theoretical evaluation
satisfying the experimental bound.

Similarly the lepton flavor violating process $\mu\rightarrow3e$ is evoked by
neutral gauge bosons $Z,\;Z_{_{BL}}$ at tree level, the corresponding decay
width is \cite{Hisano}
\begin{eqnarray}
%%%%%%%%%%%%%%%%%%%%%%%%%%%%%%%%%%%%%%%%%%%%%%%%%%%%%%%%%%%%
&&\Gamma_{\mu\rightarrow3e}={m_\mu^5\over1536\pi^3}\Big\{2|F_{_{LL}}|^2
+2|F_{_{RR}}|^2+|F_{_{LR}}|^2+|F_{_{RL}}|^2\Big\}\;,
%%%%%%%%%%%%%%%%%%%%%%%%%%%%%%%%%%%%%%%%%%%%%%%%%%%%%%%%%%%%
\label{muto3e-width}
\end{eqnarray}
with
\begin{eqnarray}
%%%%%%%%%%%%%%%%%%%%%%%%%%%%%%%%%%%%%%%%%%%%%%%%%%%%%%%%%%%%
&&F_{_{LL}}={e^2(2s_{_{\rm W}}^2-1)\over4s_{_{\rm W}}^2c_{_{\rm W}}^2m_{_{\rm Z}}^2}
[\delta^2{\cal C}^L]_{_{21}}+{g_{_{BL}}^2\over m_{_{Z_{BL}}}^2}[\delta^2{\cal B}^L]_{_{21}}
\;,\nonumber\\
%%%%%%%%%%%%%%%%%%%%%%%%%%%%%%%%%%%%%%%%%%%%%%%%%%%%%%%%%%%%
&&F_{_{RR}}={e^2\over2c_{_{\rm W}}^2m_{_{\rm Z}}^2}
[\delta^2{\cal C}^R]_{_{21}}+{g_{_{BL}}^2\over m_{_{Z_{BL}}}^2}[\delta^2{\cal B}^R]_{_{21}}
\;,\nonumber\\
%%%%%%%%%%%%%%%%%%%%%%%%%%%%%%%%%%%%%%%%%%%%%%%%%%%%%%%%%%%%
&&F_{_{LR}}={e^2\over2c_{_{\rm W}}^2m_{_{\rm Z}}^2}
[\delta^2{\cal C}^L]_{_{21}}+{g_{_{BL}}^2\over m_{_{Z_{BL}}}^2}[\delta^2{\cal B}^L]_{_{21}}
\;,\nonumber\\
%%%%%%%%%%%%%%%%%%%%%%%%%%%%%%%%%%%%%%%%%%%%%%%%%%%%%%%%%%%%
&&F_{_{RL}}={e^2(2s_{_{\rm W}}^2-1)\over4s_{_{\rm W}}^2c_{_{\rm W}}^2m_{_{\rm Z}}^2}
[\delta^2{\cal C}^R]_{_{21}}+{g_{_{BL}}^2\over m_{_{Z_{BL}}}^2}[\delta^2{\cal B}^R]_{_{21}}\;.
%%%%%%%%%%%%%%%%%%%%%%%%%%%%%%%%%%%%%%%%%%%%%%%%%%%%%%%%%%%%
\label{muto3e-width1}
\end{eqnarray}
The current bound on the $\mu\rightarrow3e$ decay has been set by the SINDRUM experiment
at PSI \cite{SINDRUM}:
\begin{eqnarray}
%%%%%%%%%%%%%%%%%%%%%%%%%%%%%%%%%%%%%%%%%%%%%%%%%%%%%%%%%%%%
&&B(\mu\rightarrow3e)\le B_{_{\mu\rightarrow3e}}^{ub}=10^{-12}\;.
%%%%%%%%%%%%%%%%%%%%%%%%%%%%%%%%%%%%%%%%%%%%%%%%%%%%%%%%%%%%
\label{exp-muto3e}
\end{eqnarray}
%%%%%%%%%%%%%%%%%%%%%%%END PRD MODIFICATION%%%%%%%%%%%%%%%%%%%%%%%%

\section{$gg\rightarrow h^0$ and $h^0\rightarrow\gamma\gamma,\;ZZ^*,\;WW^*$\label{sec6}}
\indent\indent
The Higgs is produced chiefly through the gluon fusion at the LHC.  In the SM, the leading order (LO) contributions originate
from the one-loop diagram which involves virtual top quarks. The cross section for this process is known to
the next-to-next-to-leading order (NNLO)\cite{NNLO} which can enhance the LO result by 80-100\%. Furthermore, any new particle
which strongly couples with the Higgs can significantly modify this cross section. In extension of the SM considered here,
the LO decay width for the process $h^0\rightarrow gg$ is given by (see Ref.\cite{Gamma1} and references therein)
\begin{eqnarray}
&&\Gamma_{_{NP}}(h^0\rightarrow gg)={G_{_F}\alpha_s^2m_{_{h^0}}^3\over64\sqrt{2}\pi^3}
\Big|\sum\limits_qg_{_{h^0qq}}A_{1/2}(x_q)
+\sum\limits_{\tilde q}g_{_{h^0\tilde{q}\tilde{q}}}{m_{_{\rm Z}}^2\over m_{_{\tilde q}}^2}
A_{0}(x_{_{\tilde{q}}})\Big|^2\;,
\label{hgg}
\end{eqnarray}
with $x_a=m_{_{h^0}}^2/(4m_a^2)$. In the sum above, $q=t,\;b$
and $\tilde{q}=\tilde{\cal U}_{_i},\;\tilde{\cal D}_{_i},\;(i=1,\;\cdots,\;6)$.
The concrete expressions for $g_{_{h^0tt}},\;g_{_{h^0bb}},\;g_{_{h^0\tilde{\cal U}_i\tilde{\cal U}_i}}
,\;g_{_{h^0\tilde{\cal D}_i\tilde{\cal D}_i}},\;(i=1,\;6)$ are
\begin{eqnarray}
&&g_{_{h^0tt}}={1\over\sin\beta}\Big(Z_{_{H_0}}\Big)_{11}
\;,\nonumber\\
&&g_{_{h^0bb}}=-{1\over\cos\beta}\Big(Z_{_{H_0}}\Big)_{21}\sqrt{1+\sum\limits_{\alpha=1}^3
{\upsilon_{_{L_\alpha}}^2\over\upsilon_{_d}^2}}
\;,\nonumber\\
&&g_{_{h^0\tilde{\cal U}_i\tilde{\cal U}_i}}=-{s_{_{\rm W}}c_{_{\rm W}}\over em_{_{\rm Z}}}
\xi_{1ii}^{\tilde{\cal U}}\;,\;\;(i=1,\;\cdots, 6)
\;,\nonumber\\
&&g_{_{h^0\tilde{\cal D}_i\tilde{\cal D}_i}}=-{s_{_{\rm W}}c_{_{\rm W}}\over em_{_{\rm Z}}}
\xi_{1ii}^{\tilde{\cal D}}\;,\;\;(i=1,\;\cdots, 6)\;.
\label{g-coupling1}
\end{eqnarray}
Here, we adopt the abbreviation $s_{_{\rm W}}=\sin\theta_{_{\rm W}}$ with $\theta_{_{\rm W}}$
denoting the Weinberg angle. Furthermore, $e$ is the electromagnetic coupling constant, and
the concrete expressions of $\xi_{1ii}^{\tilde{\cal U}},\;\xi_{1ii}^{\tilde{\cal D}}$ can be
found in appendix \ref{app3}.  The form factors $A_{1/2},\;A_0$ in Eq.(\ref{hgg}) are defined as
\begin{eqnarray}
&&A_{1/2}(x)=2\Big[x+(x-1)g(x)\Big]/x^2\;,\nonumber\\
&&A_0(x)=-(x-g(x))/x^2\;,
\label{loop-function1}
\end{eqnarray}
with
\begin{eqnarray}
&&g(x)=\left\{\begin{array}{l}\arcsin^2\sqrt{x},\;x\le1\\
-{1\over4}\Big[\ln{1+\sqrt{1-1/x}\over1-\sqrt{1-1/x}}-i\pi\Big]^2,\;x>1\end{array}\right.
\label{g-function}
\end{eqnarray}

The Higgs to diphoton decay is also obtained from loop diagrams, the LO contributions are derived from the
one-loop diagrams containing virtual charged gauge boson $W^\pm$ or virtual top quarks
in the SM. In this model, the charged Higgs together with corresponding supersymmetric
partner, and the supersymmetric partners of charged standard particles also contribute the corrections to the decay width
of Higgs to diphoton at LO, the corresponding correction is written as
\begin{eqnarray}
&&\Gamma_{_{NP}}(h^0\rightarrow\gamma\gamma)={G_{_F}\alpha^2m_{_{h^0}}^3\over128\sqrt{2}\pi^3}
\Big|\sum\limits_fN_cQ_{_f}^2g_{_{h^0ff}}A_{1/2}(x_f)+g_{_{h^0WW}}A_1(x_{_{\rm W}})
\nonumber\\
&&\hspace{3.2cm}
+\sum\limits_{i=2}^8g_{_{h^0H_i^+H_i^-}}{m_{_{\rm Z}}^2\over m_{_{H_i^\pm}}^2}A_0(x_{_{H_i^\pm}})
+\sum\limits_{i=1}^2g_{_{h^0\chi_i^+\chi_i^-}}{m_{_{\rm W}}\over m_{_{\chi_i}}}A_{1/2}(x_{_{\chi_i}})
\nonumber\\
&&\hspace{3.2cm}
+\sum\limits_{\tilde q}N_cQ_{_f}^2g_{_{h^0\tilde{q}\tilde{q}}}{m_{_{\rm Z}}^2\over m_{_{\tilde q}}^2}
A_{0}(x_{_{\tilde{q}}})\Big|^2\;,
\label{hpp}
\end{eqnarray}
where the concrete expression for the loop functions $A_1$ is
\begin{eqnarray}
&&A_1(x)=-\Big[2x^2+3x+3(2x-1)g(x)\Big]/x^2\;.\nonumber\\
\label{loop-function2}
\end{eqnarray}
In addition, the couplings $g_{_{h^0\tau\tau}},\;g_{_{h^0WW}}$,and $g_{_{h^0H^+H^-}}$ are
expressed as
\begin{eqnarray}
&&g_{_{h^0\tau\tau}}\simeq-{1\over\cos\beta}\Big(Z_{_{H_0}}\Big)_{21}\sqrt{1+\sum\limits_{\alpha=1}^3
{\upsilon_{_{L_\alpha}}^2\over\upsilon_{_d}^2}}
\;,\nonumber\\
&&g_{_{h^0WW}}=\sin\beta\Big(Z_{_{H_0}}\Big)_{11}+{\cos\beta\over
\sqrt{\upsilon_{_d}^2+\sum\limits_{\alpha=1}^3\upsilon_{_{L_\alpha}}^2}}\Big\{
\upsilon_{_d}\Big(Z_{_{H_0}}\Big)_{21}
+\sum\limits_{I=1}^3\upsilon_{_{L_I}}\Big(Z_{_{H_0}}\Big)_{(2+I)1}\Big\}
\;,\nonumber\\
&&g_{_{h^0H_i^-H_i^+}}=-{s_{_{\rm W}}c_{_{\rm W}}\over em_{_{\rm Z}}}
\xi_{1ii}^{H^\pm}\;,\;\;(i=2,\;\cdots, 8)\;,
\label{g-coupling2}
\end{eqnarray}
and the couplings between the lightest neutral CP-even Higgs and charginos
$g_{_{h^0\chi_i^+\chi_i^-}}$ are
\begin{eqnarray}
&&g_{_{h^0\chi_i^+\chi_i^-}}=-{2\over e}\Re\Big[\xi_{_{1ii}}^{\chi^\pm}\Big]\;,\;\;(i=1,\;2)\;.
\label{g-coupling3}
\end{eqnarray}

The lightest neutral CP-even Higgs with $125\;{\rm GeV}$ mass can also decay through the modes
$h^0\rightarrow ZZ^*,\;h^0\rightarrow WW^*$, where $Z^*/W^*$ denote the off-shell neutral/charged
electroweak gauge bosons. Summing over all channels available to $W^*$ or $Z^*$, one writes
the widths as\cite{Keung1}
\begin{eqnarray}
&&\Gamma(h^0\rightarrow WW^*)={3e^4m_{_{h^0}}\over512\pi^3s_{_{\rm W}}^4}|g_{_{h^0WW}}|^2
F({m_{_{\rm W}}\over m_{_{h^0}}}),\;\nonumber\\
&&\Gamma(h^0\rightarrow ZZ^*)={e^4m_{_{h^0}}\over2048\pi^3s_{_{\rm W}}^4c_{_{\rm W}}^4}|g_{_{h^0ZZ}}|^2
\Big(7-{40\over3}s_{_{\rm W}}^2+{160\over9}s_{_{\rm W}}^4\Big)F({m_{_{\rm Z}}\over m_{_{h^0}}}),\;\nonumber\\
\label{h-WW*ZZ*}
\end{eqnarray}
with $g_{_{h^0ZZ}}=g_{_{h^0WW}}$ and the abbreviation $c_{_{\rm W}}=\cos\theta_{_{\rm W}}$.
The form factor $F(x)$ is given as
\begin{eqnarray}
&&F(x)=-(1-x^2)\Big({47\over2}x^2-{13\over2}+{1\over x^2}\Big)-3(1-6x^2+4x^4)\ln x
\nonumber\\
&&\hspace{1.5cm}
+{3(1-8x^2+20x^4)\over\sqrt{4x^2-1}}\cos^{-1}\Big({3x^2-1\over2x^3}\Big)\;.\nonumber\\
\label{form-factor1}
\end{eqnarray}

Besides the Higgs discovery the ATLAS and CMS experiments have both observed an excess
in Higgs production and decay into diphoton channel which slightly differs from
the SM expectations. The observed signals for the Higgs decaying channels are quantified by the ratio
\begin{eqnarray}
&&R_{\gamma\gamma}={\sigma_{_{NP}}(h_0\rightarrow gg)\;{\rm BR}_{_{NP}}(h_0\rightarrow\gamma\gamma)\over
\sigma_{_{SM}}(h_0\rightarrow gg)\;{\rm BR}_{_{SM}}(h_0\rightarrow\gamma\gamma)}
\nonumber\\
&&\hspace{0.8cm}\simeq
{\Gamma_{_{NP}}(h_0\rightarrow gg)\;{\rm BR}_{_{NP}}(h_0\rightarrow\gamma\gamma)\over
\Gamma_{_{SM}}(h_0\rightarrow gg)\;{\rm BR}_{_{SM}}(h_0\rightarrow\gamma\gamma)}
\;,\nonumber\\
&&R_{VV^*}={\sigma_{_{NP}}(h_0\rightarrow gg)\;{\rm BR}_{_{NP}}(h_0\rightarrow VV^*)\over
\sigma_{_{SM}}(h_0\rightarrow gg)\;{\rm BR}_{_{SM}}(h_0\rightarrow VV^*)}
\nonumber\\
&&\hspace{1.0cm}\simeq
{\Gamma_{_{NP}}(h_0\rightarrow gg)\;{\rm BR}_{_{NP}}(h_0\rightarrow VV^*)\over
\Gamma_{_{SM}}(h_0\rightarrow gg)\;{\rm BR}_{_{SM}}(h_0\rightarrow VV^*)}
\;,\;\;(V=Z,\;W)\;.
\label{signal}
\end{eqnarray}
To obtain the Higgs production cross sections normalised to the SM values,
we adopt
\begin{eqnarray}
&&{\sigma_{_{NP}}(h_0\rightarrow gg)\over\sigma_{_{SM}}(h_0\rightarrow gg)}
\simeq{\Gamma_{_{NP}}^{h_0}\;{\rm BR}_{_{NP}}(h_0\rightarrow gg)\over
\Gamma_{_{SM}}^{h_0}\;{\rm BR}_{_{SM}}(h_0\rightarrow gg)}
\label{signal1}
\end{eqnarray}
where $\Gamma_{_{SM}}^{h_0}$ denotes the total decay width of Higgs in the SM,
and $\Gamma_{_{NP}}^{h_0}$ denotes the total decay width of the lightest Higgs $h_0$ in
the supersymmetry with local $U(1)_{B-L}$ symmetry, respectively.
To accommodate the observed Higgs signals at CMS and ATLAS, we require the theoretical
predictions on $R_{\gamma\gamma},\;R_{VV^*}$ satisfying
\begin{eqnarray}
&&0.9 < R_{\gamma\gamma}<2.2\;,
\nonumber\\
&&0.2<R_{VV^*}<1.4\;,(V=Z,\;W)\;.
\label{signal-exp}
\end{eqnarray}
The lower bounds of the ranges originate from the lower limits of $95\%$ C.L. range
for observed Higgs strength\cite{CMS1,ATLAS1}, and the upper bounds of the ranges
originate from $95\%$ C.L. exclusion from Higgs searching\cite{CMS2,ATLAS2}.

%%%%%%%%%%%%%%%%%%%%%%%%%%%%%%%%Begin Modification%%%%%%%%%%%%%%%%%%%%%%%%%%%%%
In the SM the main contribution to $R_{\gamma\gamma}$ originates from charged gauge bosons
and is partially compensated by the top quark contribution. Within framework of the MSSM,
the branching ratio is enhanced by light scalar quarks originating from large mixing
between left- and right-hand scalar partners. However this effect is generally
overcompensated by a suppression of the gluon fusion production rate, and the evaluations
on a Higgs gluon fusion production times photon decay rate is slightly lower than
the corresponding one of the SM in the parameter region consistent with a $125$ GeV
Higgs \cite{Dermisek}. In most cases the corrections to $R_{\gamma\gamma}$
from charginos also tend to reduce corresponding evaluation of the SM. The
possible correction to enhance theoretical evaluations originates from scalar tau leptons,
in which large mixing between left- and right-handed scalar tau leptons is evoked
by large values of $\mu$ parameter and $\tan\beta$ \cite{Carena2}. Because mixing
between the lightest Higgs and real components of left- and right-handed sneutrinos
is below $0.02$ in the parameter space consisting with present experimental data,
the main source of corrections to $R_{\gamma\gamma}$ in the considered model here is similar
to that in the MSSM.
%%%%%%%%%%%%%%%%%%%%%%%%%%%%%%%%%End Modification%%%%%%%%%%%%%%%%%%%%%%%%%%%%%%

\section{Numerical analyses\label{sec7}}
\indent\indent

%%%%%%%%%%%%%%%%%%%%%%%%%%%%%%%%%%%%%%%%%%%%%%%%%%%%%
\begin{figure}[h]
\setlength{\unitlength}{1mm}
\centering
\includegraphics[width=3.0in]{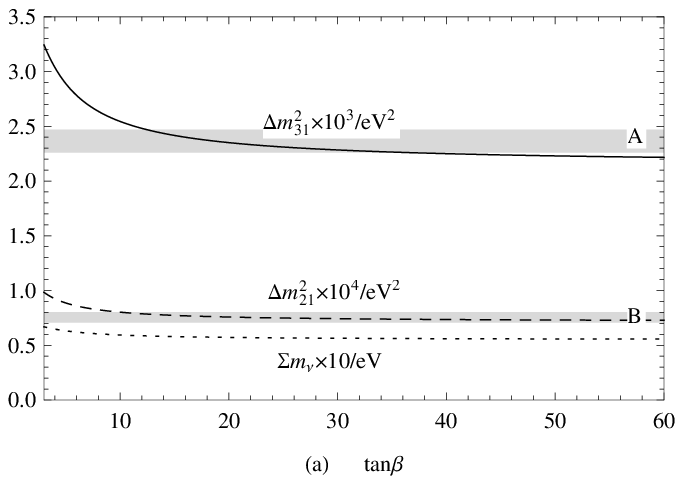}
\vspace{0.5cm}
\includegraphics[width=3.0in]{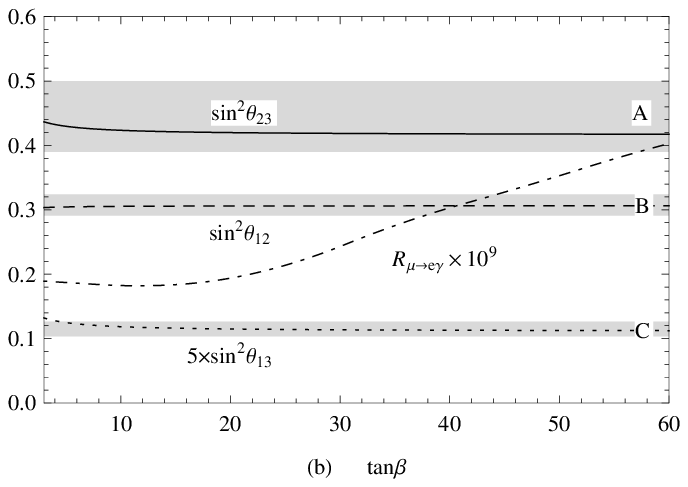}
\vspace{0cm}
\caption[]{Assuming neutrino mass spectrum with NO and taking $\mu=2$ TeV,
$m_{_{BL}}=1$ TeV, we plot the mass squared differences and mixing angles
of neutrinos versus $\tan\beta$. Where (a) the solid line stands $\Delta m_{_{31}}^2$ versus $\tan\beta$,
the dashed line stands $\Delta m_{_{21}}^2$ versus $\tan\beta$, the dotted line denotes $\sum m_{\nu}$ varying with $\tan\beta$,
together with the gray band A represents the points which deviate the experimental central value on
$\Delta m_{_A}^2=\Delta m_{_{31}}^2$ within 1 standard deviation,
the gray band B represents the points which deviate the experimental central value on
$\Delta m_{_\odot}^2=\Delta m_{_{21}}^2$ within 1 standard deviation;
and (b) solid line stands $\sin^2\theta_{_{23}}$
versus $\tan\beta$, the dashed line stands $\sin^2\theta_{_{12}}$ versus $\tan\beta$, the dotted
line stands $\sin^2\theta_{_{13}}$ versus $\tan\beta$,
the dashed-dotted line stands $R_{\mu\rightarrow e\gamma}\times10^9$ versus $\tan\beta$,
as well as the gray band A represents the points which deviate the experimental central value on
$\sin^2\theta_{_{23}}$ within 1 standard deviation,
the gray band B represents the points which deviate the experimental central value on
$\sin^2\theta_{_{12}}$ within 1 standard deviation,
the gray band C represents the points which deviate the experimental central value on
$\sin^2\theta_{_{13}}$ within 1 standard deviation,
respectively.}
\label{fig1}
\end{figure}
%%%%%%%%%%%%%%%%%%%%%%%%%%%%%%%%%%%%%%%%%%%%%%%%%%%%%

As mentioned above, the most stringent constraint on the parameter space is that the
$8\times8$ mass squared matrix in Eq.(\ref{M-CPE}) should produce the lightest eigenstate
with a mass $m_{_{h_0}}\simeq125.9\;{\rm GeV}$. Furthermore, the neutrino oscillation experimental data
and cosmological observations from Planck collaboration also constrain relevant parameter space strongly.
In numerical analysis, we choose the mass of the lightest CP-odd Higgs $m_{_{A_3^0}}$ as an input.
%%%%%%%%%%%%%%%%%%%%%%%%%%%%%%%%%%%%%%%%%%%%%%%%%%%%%
\begin{figure}[h]
\setlength{\unitlength}{1mm}
\centering
\includegraphics[width=3.0in]{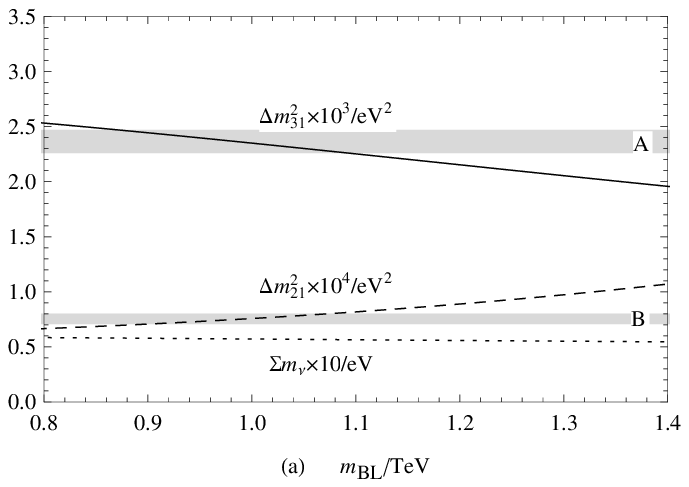}
\vspace{0.5cm}
\includegraphics[width=3.0in]{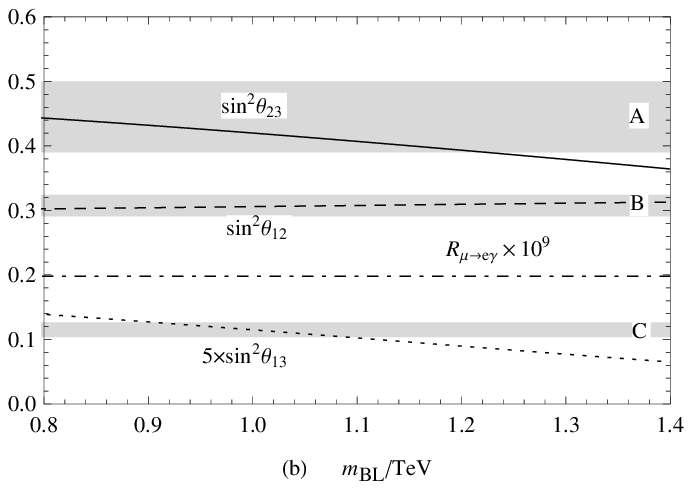}
\vspace{0cm}
\caption[]{Assuming neutrino mass spectrum with NO and taking
$\tan\beta=20$, $\mu=2$ TeV, we plot the mass squared differences and mixing angles
of neutrinos versus $U(1)_{_{(B-L)}}$ gauging mass $m_{_{BL}}$. Where (a) the solid line denotes
$\Delta m_{_{31}}^2$ versus $m_{_{BL}}$, the dashed line denotes $\Delta m_{_{21}}^2$ versus
$m_{_{BL}}$,  and the dotted line denotes $\sum m_\nu$ versus $m_{_{BL}}$, together with
the gray band A represents the points which deviate the experimental central value on
$\Delta m_{_A}^2=\Delta m_{_{31}}^2$ within 1 standard deviation,
the gray band B represents the points which deviate the experimental central value on
$\Delta m_{_\odot}^2=\Delta m_{_{21}}^2$ within 1 standard deviation;
and (b) solid line denotes $\sin^2\theta_{_{23}}$
versus $m_{_{BL}}$, the dashed line denotes $\sin^2\theta_{_{12}}$ versus $m_{_{BL}}$, the dotted
line denotes $\sin^2\theta_{_{13}}$ versus $m_{_{BL}}$,
the dashed-dotted line denotes $R_{\mu\rightarrow e\gamma}\times10^9$ versus $m_{_{BL}}$,
as well as the gray band A represents the points which deviate the experimental central value on
$\sin^2\theta_{_{23}}$ within 1 standard deviation,
the gray band B represents the points which deviate the experimental central value on
$\sin^2\theta_{_{12}}$ within 1 standard deviation,
the gray band C represents the points which deviate the experimental central value on
$\sin^2\theta_{_{13}}$ within 1 standard deviation,
respectively.}
\label{fig2}
\end{figure}
%%%%%%%%%%%%%%%%%%%%%%%%%%%%%%%%%%%%%%%%%%%%%%%%%%%%%
In addition, we include the radiative corrections to
the neutrino mass matrix in Eq.(\ref{radiative-corrections-neutrino1}),
and adopt the ansatz on the parameter space
\begin{eqnarray}
&&m_{_{\tilde{U}_3}}=1\;{\rm TeV},\;\;
m_{_{\tilde{D}_3}}=2\;{\rm TeV}\;,
\nonumber\\
&&\Lambda_{_{\tilde{N}_1^c}}=\Lambda_{_{\tilde{N}_2^c}}=3\;{\rm TeV}\;,\nonumber\\
&&m_{_1}=200\;{\rm GeV}\;,\;\;m_{_2}=400\;{\rm GeV}\;,\nonumber\\
&&\upsilon_{_{N_1}}=\upsilon_{_{N_2}}=\upsilon_{_{N_3}}=2\;{\rm TeV}
\;,\nonumber\\
&&m_{_{A_3^0}}=A_{_b}=1\;{\rm TeV}
\;,\nonumber\\
&&A_{_\tau}=1.5\;{\rm TeV}
\;\nonumber\\
&&m_{_{Z_{BL}}}=2.4\;{\rm TeV}
\label{assumption1}
\end{eqnarray}
to reduce the number of free parameters in the model considered here.
For relevant parameters in the SM, we choose\cite{PDG}
\begin{eqnarray}
&&\alpha_s(m_{_{\rm Z}})=0.118\;,\;\;\alpha(m_{_{\rm Z}})=1/128\;,\;\;
s_{_{\rm W}}^2(m_{_{\rm Z}})=0.23\;,
\nonumber\\
&&m_t=174.2\;{\rm GeV}\;,\;\;
m_b=4.18\;{\rm GeV}\;,\;\;m_{_{\rm W}}=80.4\;{\rm GeV}\;.
\label{PDG-SM}
\end{eqnarray}

%%%%%%%%%%%%%%%%%%%%%%%%%%%%%%%%%%%%%%%%%%%%%%%%%%%%%
\begin{figure}[h]
\setlength{\unitlength}{1mm}
\centering
\includegraphics[width=3.0in]{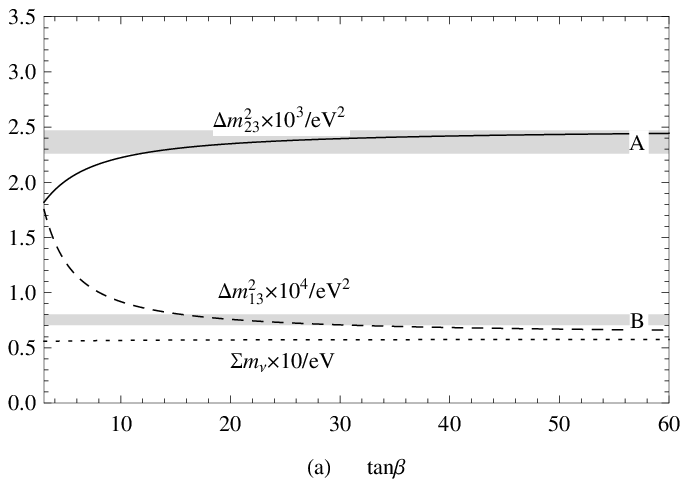}
\vspace{0.5cm}
\includegraphics[width=3.0in]{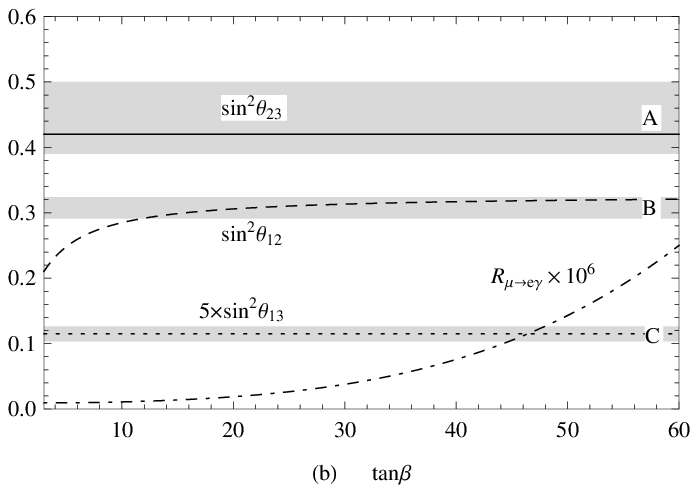}
\vspace{0cm}
\caption[]{Assuming neutrino mass spectrum with IO and taking
$m_{_{BL}}=1$ TeV, $\mu=2$ TeV, we plot the mass squared differences and mixing angles
of neutrinos versus $\tan\beta$. Where (a) the solid line stands $\Delta m_{_{31}}^2$ versus $\tan\beta$,
the dashed line stands $\Delta m_{_{23}}^2$ versus $\tan\beta$,
and the dotted line denotes $\sum m_{\nu}$ varying with $\tan\beta$, together with
the gray band A represents the points which deviate the experimental central value on
$\Delta m_{_A}^2=\Delta m_{_{13}}^2$ within 1 standard deviation,
the gray band B represents the points which deviate the experimental central value on
$\Delta m_{_\odot}^2=\Delta m_{_{21}}^2$ within 1 standard deviation;
and (b) solid line stands $\sin^2\theta_{_{23}}$
versus $\tan\beta$, the dashed line stands $\sin^2\theta_{_{12}}$ versus $\tan\beta$, the dotted
line stands $\sin^2\theta_{_{13}}$ versus $\tan\beta$,
the dashed-dotted line stands $R_{\mu\rightarrow e\gamma}\times10^6$ versus $\tan\beta$,
as well as the gray band A represents the points which deviate the experimental central value on
$\sin^2\theta_{_{23}}$ within 1 standard deviation,
the gray band B represents the points which deviate the experimental central value on
$\sin^2\theta_{_{12}}$ within 1 standard deviation,
the gray band C represents the points which deviate the experimental central value on
$\sin^2\theta_{_{13}}$ within 1 standard deviation, respectively.}
\label{fig3}
\end{figure}
%%%%%%%%%%%%%%%%%%%%%%%%%%%%%%%%%%%%%%%%%%%%%%%%%%%%%

In order to fit the experimental data on neutrino oscillations with
two solutions of the neutrino mass spectrum, we choose the VEVs of
left-handed sneutrinos and the Yukawa couplings of right-handed
neutrinos respectively as
\begin{itemize}
\item for the NO spectrum:
\begin{eqnarray}
&&\upsilon_{_{L_1}}=3.11 \times10^{-4}{\rm GeV},\;
\upsilon_{_{L_2}}=6.83\times10^{-4}{\rm GeV},\;
\upsilon_{_{L_3}}=3.43\times10^{-4}{\rm GeV},
\nonumber\\
&&(Y_{_1},\;Y_{_2},\;Y_{_3})=(0,\;2.24\times10^{-7},\;4.30\times10^{-7})\;,
\label{NH2}
\end{eqnarray}
and issue the theoretical predictions on neutrino masses and mixing
angles as
\begin{eqnarray}
&&m_{_{\nu_1}}\simeq0,\;\;m_{_{\nu_2}}\simeq8.71\times10^{-3}\;{\rm eV},\;\;m_{_{\nu_3}}\simeq4.85\times10^{-2}\;{\rm eV},
\nonumber\\
&&\sum\limits_i m_{_{\nu_i}}\simeq5.72\times10^{-2}\;{\rm eV},
\nonumber\\
&&\Delta m_{_{\odot}}^2=m_{_{\nu_2}}^2-m_{_{\nu_1}}^2\simeq7.58\times10^{-23}\;{\rm GeV^2},
\nonumber\\
&&\Delta m_{_A}^2=m_{_{\nu_3}}^2-m_{_{\nu_1}}^2\simeq2.35\times10^{-21}\;{\rm GeV^2},
\nonumber\\
&&\sin^2\theta_{_{12}}\simeq0.306,\;\;\sin^2\theta_{_{23}}\simeq0.420,\;\;
\sin^2\theta_{_{13}}\simeq0.023
\label{NH3}
\end{eqnarray}
when $\tan\beta=20$, $m_{_{BL}}=1$ TeV,  and $\mu=2$ TeV;

\item for the IO spectrum:
\begin{eqnarray}
&&\upsilon_{_{L_1}}=1.89 \times10^{-3}{\rm GeV},\;
\upsilon_{_{L_2}}=7.88\times10^{-4}{\rm GeV},\;
\upsilon_{_{L_3}}=1.05\times10^{-3}{\rm GeV},
\nonumber\\
&&(Y_{_1},\;Y_{_2},\;Y_{_3})=(1.61\times10^{-7},\;3.82\times10^{-8},\;0),
\label{IH2}
\end{eqnarray}
and issue the theoretical predictions on neutrino masses and mixing
angles as
\begin{eqnarray}
&&m_{_{\nu_1}}\simeq8.71\times10^{-3}\;{\rm eV},\;\;m_{_{\nu_2}}\simeq4.85\times10^{-2}\;{\rm eV},\;\;m_{_{\nu_3}}\simeq0,
\nonumber\\
&&\sum\limits_i m_{_{\nu_i}}\simeq5.72\times10^{-2}\;{\rm eV},
\nonumber\\
&&\Delta m_{_{\odot}}^2=m_{_{\nu_1}}^2-m_{_{\nu_3}}^2\simeq7.58\times10^{-23}\;{\rm GeV^2},
\nonumber\\
&&\Delta m_{_A}^2=m_{_{\nu_2}}^2-m_{_{\nu_3}}^2\simeq2.35\times10^{-21}\;{\rm GeV^2},
\nonumber\\
&&\sin^2\theta_{_{12}}\simeq0.306,\;\;\sin^2\theta_{_{23}}\simeq0.420,\;\;
\sin^2\theta_{_{13}}\simeq0.023
\label{IH3}
\end{eqnarray}
when $\tan\beta=20$, $m_{_{BL}}=1$ TeV,  and $\mu=2$ TeV.
\end{itemize}
Meanwhile, the theoretical predictions on the branching ratio of
$\mu\rightarrow3e$ are all about $10^{-20}$, far below the current
experimental bound in Eq.(\ref{exp-muto3e}) in both scenarios.

%%%%%%%%%%%%%%%%%%%%%%%%%%%%%%%%%%%%%%%%%%%%%%%%%%%%%
\begin{figure}[h]
\setlength{\unitlength}{1mm}
\centering
\includegraphics[width=3.0in]{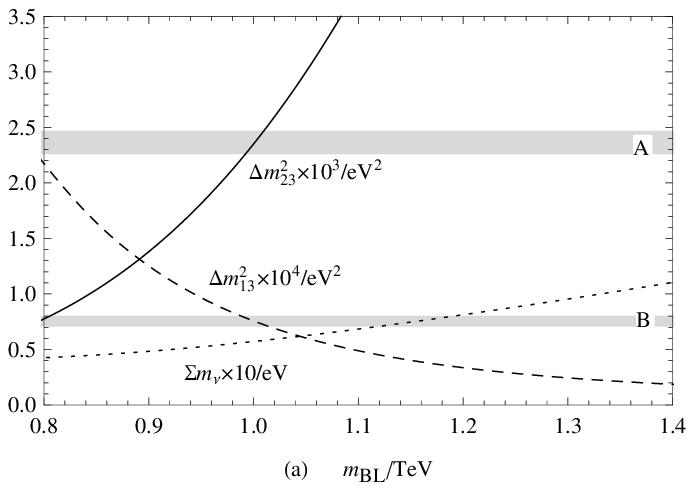}
\vspace{0.5cm}
\includegraphics[width=3.0in]{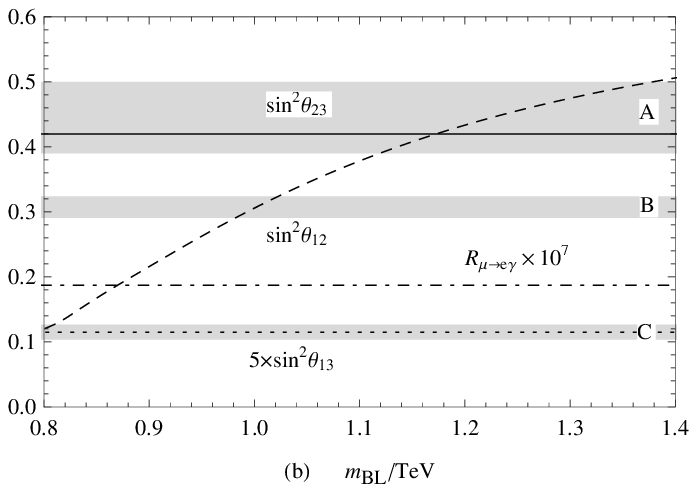}
\vspace{0cm}
\caption[]{Assuming neutrino mass spectrum with IO and taking
$\tan\beta=20$, $\mu=2$ TeV, we plot the mass squared differences and mixing angles
of neutrinos versus $U(1)_{_{(B-L)}}$ gauging mass $m_{_{BL}}$. Where (a) the solid line denotes
$\Delta m_{_{23}}^2$ versus $m_{_{BL}}$, the dashed line denotes $\Delta m_{_{13}}^2$ versus
$m_{_{BL}}$, and the dotted line denotes $\sum m_{\nu}$ versus $m_{_{BL}}$, together with
the gray band A represents the points which deviate the experimental central value on
$\Delta m_{_A}^2=\Delta m_{_{31}}^2$ within 1 standard deviation,
the gray band B represents the points which deviate the experimental central value on
$\Delta m_{_\odot}^2=\Delta m_{_{21}}^2$ within 1 standard deviation; and (b) solid line denotes $\sin^2\theta_{_{23}}$
versus $m_{_{BL}}$, the dashed line denotes $\sin^2\theta_{_{12}}$ versus $m_{_{BL}}$, the dotted
line denotes $\sin^2\theta_{_{13}}$ versus $m_{_{BL}}$,
the dashed-dotted line denotes $R_{\mu\rightarrow e\gamma}\times10^7$ versus $m_{_{BL}}$,
as well as the gray band A represents the points which deviate the experimental central value on
$\sin^2\theta_{_{23}}$ within 1 standard deviation,
the gray band B represents the points which deviate the experimental central value on
$\sin^2\theta_{_{12}}$ within 1 standard deviation,
the gray band C represents the points which deviate the experimental central value on
$\sin^2\theta_{_{13}}$ within 1 standard deviation,
respectively.}
\label{fig4}
\end{figure}
%%%%%%%%%%%%%%%%%%%%%%%%%%%%%%%%%%%%%%%%%%%%%%%%%%%%%
Additionally we can safely neglect the last two terms in
Eq.(\ref{chargino-lepton7}) which are originate from the mixing
between charginos and charged leptons, when we adopt the choices in
Eq.(\ref{NH2}) and Eq.(\ref{IH2}). Further assuming the $3\times3$ Yukawa couplings $Y_{_E}$
diagonally we obtain the solution as
$Y_{_E}\simeq{\rm diag}\;(\sqrt{2}m_{_e}/\upsilon_{_d},\;\sqrt{2}m_{_\mu}/\upsilon_{_d},\;
\sqrt{2}m_{_\tau}/\upsilon_{_d})$.

Assuming neutrino mass spectrum with NO and taking $m_{_{BL}}=1$ TeV, $\mu=2$ TeV
we depict the mass squared differences of neutrinos versus $\tan\beta$ in Fig.\ref{fig1}(a).
Where the solid line denotes $\Delta m_{_{31}}^2=\Delta m_{_A}^2$ varying with $\tan\beta$,
the dashed line denotes $\Delta m_{_{21}}^2=\Delta m_{_\odot}^2$ varying with $\tan\beta$, and
the dotted line denotes $\sum m_{\nu}$ varying with $\tan\beta$, respectively. With increasing of $\tan\beta$,
the theoretical evaluations on $\Delta m_{_{21}}^2,\;\Delta m_{_{31}}^2$ decrease steeply
as $\tan\beta\le10$, and diminish mildly as $\tan\beta>15$. Additional the theoretical evaluation
on sum of active neutrino masses satisfies the cosmological observations of Planck.
Using the same choice on parameter space, we also draw the mixing angles
of neutrinos versus $\tan\beta$ in Fig.\ref{fig1}(b). Where the solid line denotes $\sin^2\theta_{_{23}}$
versus $\tan\beta$, the dashed line denotes $\sin^2\theta_{_{12}}$ versus $\tan\beta$, the dotted
line denotes $\sin^2\theta_{_{13}}$ versus $\tan\beta$,
and the dashed-dotted line denotes $R_{\mu\rightarrow e\gamma}\times10^9$ versus $\tan\beta$,
respectively. Actually, those mixing angles $\theta_{_{12}},\;\theta_{_{23}},\;\theta_{_{13}}$
vary with $\tan\beta$ gently, and $R_{\mu\rightarrow e\gamma}$ depends on
$\tan\beta$ strongly.

As a 'brand new' parameter, the $U(1)_{B-L}$ gaugino mass $m_{_{BL}}$ also affects the finally
numerical results on neutrino sector in the MSSM with local $U(1)_{B-L}$ symmetry.
Supposing neutrino mass spectrum with NO and taking $\tan\beta=20$, $\mu=2$ TeV,
we plot the mass squared differences of neutrinos versus $m_{_{BL}}$ in
Fig.\ref{fig2}(a). Where the solid line represents $\Delta m_{_{31}}^2
=\Delta m_{_A}^2$ varying with $m_{_{BL}}$, the dashed line represents
$\Delta m_{_{21}}^2=\Delta m_{_\odot}^2$ varying with $m_{_{BL}}$, the dotted line denotes
$\sum m_\nu$ varying with $m_{_{BL}}$, respectively. With increasing of $m_{_{BL}}$,
the theoretical prediction on $\Delta m_{_{31}}^2$ diminishes steeply, and that on $\Delta m_{_{21}}^2$ raises
quickly, meanwhile the evaluation on sum of active neutrino masses is consistent with the cosmological
upper bound from Planck collaboration. Because the effective $3\times3$ mass matrix for light active neutrinos depends on $m_{_{BL}}$
through the term $\zeta_i\zeta_j/\Lambda_\zeta\simeq(4\Delta_{_{BL}}^2m_{_{BL}}/m_{_{Z_{BL}}}^4
-\tilde{m}\upsilon_{_d}^2/(4\tilde{\mu}^4))Y_iY_j\upsilon_{_N}^2$, the numerical evaluations on
$\Delta m_{_{31}}^2$, $\Delta m_{_{21}}^2$ and $\sum m_\nu$ vary with $m_{_{BL}}$ actually.
Using the same assumption on parameter space, we also present the mixing angles
of neutrinos versus $m_{_{BL}}$ in Fig.\ref{fig2}(b). Where the solid line denotes $\sin^2\theta_{_{23}}$
versus $m_{_{BL}}$, the dashed line denotes $\sin^2\theta_{_{12}}$ versus $m_{_{BL}}$, the dotted
line denotes $\sin^2\theta_{_{13}}$ versus $m_{_{BL}}$, and
the dashed-dotted line denotes $R_{\mu\rightarrow e\gamma}\times10^9$ versus $m_{_{BL}}$,
respectively. Actually, the mixing angle $\theta_{_{12}}$
depends on $m_{_{BL}}$ mildly, and other mixing angles $\theta_{_{23}},\;\theta_{_{13}}$
vary with $m_{_{BL}}$ acutely, and $R_{\mu\rightarrow e\gamma}$ varies with
$m_{_{BL}}$ slowly..

%%%%%%%%%%%%%%%%%%%%%%%%%%%%%%%%%%%%%%%%%%%%%%%%%%%%%
\begin{figure}[h]
\setlength{\unitlength}{1mm}
\centering
\includegraphics[width=3.0in]{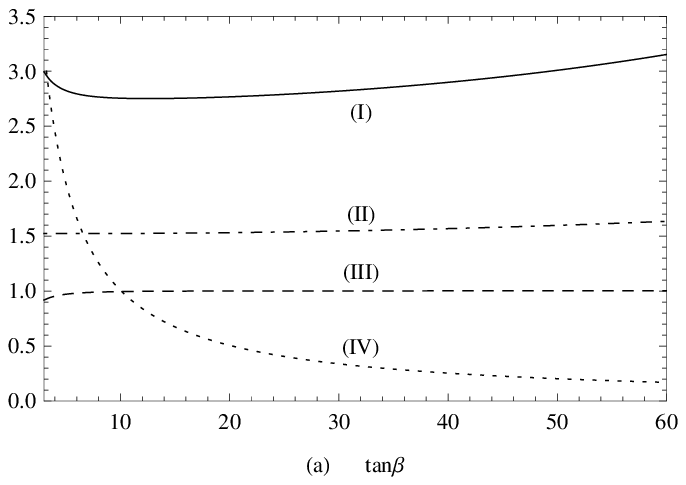}
\vspace{0.5cm}
\includegraphics[width=3.0in]{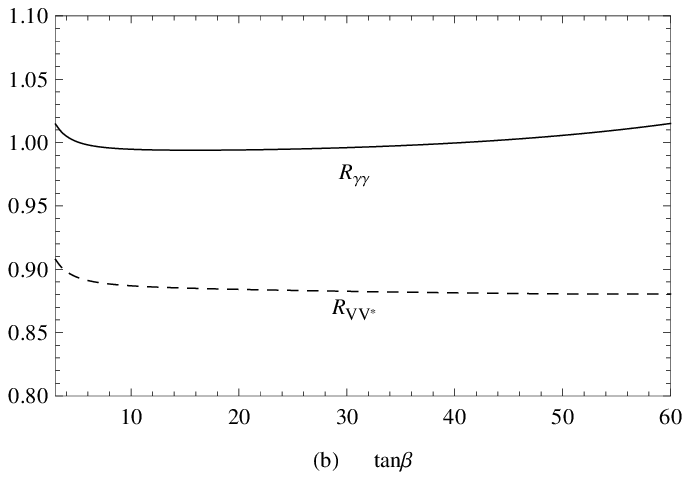}
\vspace{0cm}
\caption[]{Assuming neutrino mass spectrum with IO and taking $m_{_{\tilde{Q}_3}}=14$ TeV,
$m_{_{\tilde{L}_3}}=2$ TeV, $m_{_{\tilde{E}_3}}=1.72$ TeV,
$m_{_{BL}}=1$ TeV, $\mu=2$ TeV, $A_{_t}=1.8$ TeV, we plot $m_{_{h^0}}$,
$\Delta m_{_{h^0}}$, $\Big|\Big(Z_{_{H_0}}\Big)_{12}\Big|$, $Z_{_{h^0\tilde{N}^c}}$,
$R_{\gamma\gamma}$ and $R_{VV^*}\;(V=Z,\;W)$ versus $\tan\beta$.
Where (a) the solid line (I) denotes $\Delta m_{_{h^0}}/{\rm GeV}$ versus $\tan\beta$,
the dash-dot line (II) denotes $Z_{_{h^0\tilde{N}^c}}\times10^2$ versus $\tan\beta$,
the dash line (III) denotes $m_{_{h^0}}/125.9{\rm GeV}$ versus $\tan\beta$,
as well as the dot line (IV) denotes $\Big|\Big(Z_{_{H_0}}\Big)_{12}\Big|\times10$
versus $\tan\beta$; and (b)the solid line denotes $R_{\gamma\gamma}$ versus $\tan\beta$,
the dashed line denotes $R_{VV^*}$ versus  $\tan\beta$, respectively.}
\label{fig5}
\end{figure}
%%%%%%%%%%%%%%%%%%%%%%%%%%%%%%%%%%%%%%%%%%%%%%%%%%%%%

When the neutrino mass spectrum is IO, the manners of parameters $\tan\beta,\;m_{_{BL}}$
affecting the numerical results on neutrino sector differ from that of the neutrino mass spectrum with NO.
Assuming neutrino mass spectrum with IO and taking $m_{_{BL}}=1$ TeV, $\mu=2$ TeV,
we plot the mass squared differences of neutrinos versus $\tan\beta$ in Fig.\ref{fig3}(a).
Where the solid line denotes $\Delta m_{_{23}}^2=\Delta m_{_A}^2$ varying with $\tan\beta$,
the dashed line denotes $\Delta m_{_{13}}^2=\Delta m_{_\odot}^2$ varying with $\tan\beta$,
and the dotted line denotes $\sum m_{\nu}$ varying with $\tan\beta$, respectively.
Obviously the theoretical predictions on $\Delta m_{_{23}}^2,\;
\Delta m_{_{13}}^2$ vary with $\tan\beta$ steeply as $\tan\beta< 20$, and the theoretical predictions on
$\Delta m_{_{23}}^2,\;\Delta m_{_{13}}^2$ depend on $\tan\beta$ relatively mildly when $\tan\beta>25$.
Furthermore, the numerical evaluation on sum of active neutrino masses fulfills the upper limit
from Planck observations.
Adopting the same choice on parameter space, we also show the mixing angles
of neutrinos versus $\tan\beta$ in Fig.\ref{fig3}(b). Where the solid line denotes $\sin^2\theta_{_{23}}$
versus $\tan\beta$, the dashed line denotes $\sin^2\theta_{_{12}}$ versus $\tan\beta$, the dotted
line denotes $\sin^2\theta_{_{13}}$ versus $\tan\beta$,
and the dashed-dotted line denotes $R_{\mu\rightarrow e\gamma}\times10^7$ versus $\tan\beta$,
respectively.
Obviously the mixing angle $\theta_{_{12}}$ increases quickly with increasing
of $\tan\beta$, and the mixing angles
$\theta_{_{23}},\;\theta_{_{13}}$ vary with $\tan\beta$ slowly,
and $R_{\mu\rightarrow e\gamma}$ depends on
$\tan\beta$ strongly.

As mentioned above, the $U(1)_{_{(B-L)}}$ gaugino mass $m_{_{BL}}$ also affects
the numerical evaluations in neutrino sector when neutrino mass spectrum is IO.
Assuming neutrino mass spectrum with IO and taking $\tan\beta=20$, $\mu=2$ TeV,
we depict the mass squared differences of neutrinos versus the $U(1)_{_{(B-L)}}$ gaugino mass $m_{_{BL}}$ in
Fig.\ref{fig4}(a). Where the solid line represents $\Delta m_{_{23}}^2
=\Delta m_{_A}^2$ varying with $m_{_{BL}}$, the dashed line represents $\Delta m_{_{13}}^2=\Delta m_{_\odot}^2$
varying with $m_{_{BL}}$, respectively. With increasing of $m_{_{BL}}$,
the theoretical prediction on $\Delta m_{_{23}}^2$ raises steeply, and
that on $\Delta m_{_{13}}^2$ decreases quickly. For the aforementioned reason in NO neutrino spectrum,
those evaluations on $\Delta m_{_{23}}^2$, $\Delta m_{_{13}}^2$ and $\sum m_\nu$ depend on $m_{_{BL}}$
impressibly. Adopting the same assumption on parameter space, we also draw the mixing angles
of neutrinos versus $m_{_{BL}}$ in Fig.\ref{fig4}(b). Where the solid line stands $\sin^2\theta_{_{23}}$
versus $m_{_{BL}}$, the dashed line stands $\sin^2\theta_{_{12}}$ versus $m_{_{BL}}$, the dotted
line stands $\sin^2\theta_{_{13}}$ versus $m_{_{BL}}$,
and the dashed-dotted line denotes $R_{\mu\rightarrow e\gamma}\times10^7$
versus $m_{_{BL}}$, respectively.
Obviously the mixing angle $\theta_{_{12}}$ decreases quickly with increasing
of $m_{_{BL}}$, and the mixing angles
$\theta_{_{23}},\;\theta_{_{13}}$ and $R_{\mu\rightarrow e\gamma}$ vary with $m_{_{BL}}$ mildly.

%%%%%%%%%%%%%%%%%%%%%%%%%%%%%%%%%%%%%%%%%%%%%%%%%%%%%
\begin{figure}[h]
\setlength{\unitlength}{1mm}
\centering
\includegraphics[width=3.0in]{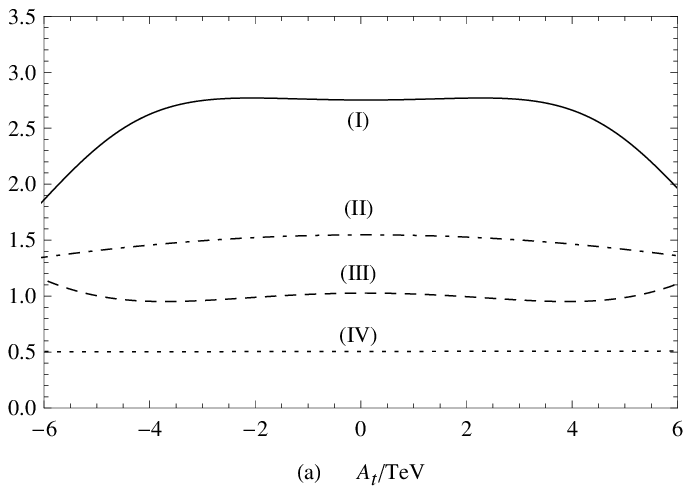}
\vspace{0.5cm}
\includegraphics[width=3.0in]{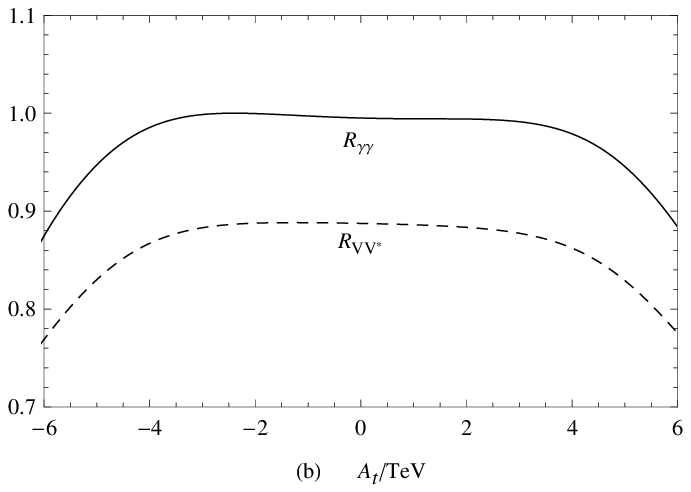}
\vspace{0cm}
\caption[]{Assuming neutrino mass spectrum with IO and taking $\tan\beta=20$,
$m_{_{\tilde{Q}_3}}=14$ TeV, $m_{_{\tilde{L}_3}}=2$ TeV, $m_{_{\tilde{E}_3}}=1.72$ TeV,
$m_{_{BL}}=1$ TeV, $\mu=2$ TeV,
we plot $m_{_{h^0}}$, $\Delta m_{_{h^0}}$, $\Big|\Big(Z_{_{H_0}}\Big)_{12}\Big|$, $Z_{_{h^0\tilde{N}^c}}$,
$R_{\gamma\gamma}$ and $R_{VV^*}\;(V=Z,\;W)$ versus $A_{_t}$.
Where (a) the solid line (I) denotes $\Delta m_{_{h^0}}/{\rm GeV}$ versus $A_{_t}$,
the dash-dot line (II) denotes $Z_{_{h^0\tilde{N}^c}}\times10^2$ versus $A_{_t}$,
the dash line (III) denotes $m_{_{h^0}}/125.9{\rm GeV}$ versus $A_{_t}$,
as well as the dot line (IV) denotes $\Big|\Big(Z_{_{H_0}}\Big)_{12}\Big|\times10$
versus $A_{_t}$; and (b)the solid line denotes $R_{\gamma\gamma}$ versus $A_{_t}$,
the dashed line denotes $R_{VV^*}$ versus  $A_{_t}$, respectively.}
\label{fig6}
\end{figure}
%%%%%%%%%%%%%%%%%%%%%%%%%%%%%%%%%%%%%%%%%%%%%%%%%%%%%

Choosing the assumptions presented in Eq.(\ref{NH2}) and Eq.(\ref{IH2}), one finds that
the radiative corrections from right-handed
neutrino sector to the CP-even Higgs mass squared matrix can be neglected safely\cite{Haber2}.
This fact implies that theoretical predictions on Higgs sector almost do not depend on
our hypothesis of neutrino mass spectrum. Assuming neutrino mass spectrum with IO and
taking  $\tan\beta=20$, $m_{_{\tilde{Q}_3}}=14$ TeV, $m_{_{\tilde{L}_3}}=2$ TeV, $m_{_{\tilde{E}_3}}=1.72$ TeV,
$m_{_{BL}}=1$ TeV, $\mu=2$ TeV, $A_{_t}=1.8$ TeV, we obtain the following numerical results in Higgs sector as
\begin{eqnarray}
&&m_{_{h^0}}\simeq125.9\;{\rm GeV},\;\;
\Delta m_{_{h^0}}=m_{_{h^0}}-m_{_{h^0}}^{MSSM}\simeq2.77\;{\rm GeV},
\nonumber\\
&&m_{_{H_2^0}}\simeq1.40\;{\rm TeV},\;\;\Big|\Big(Z_{_{H_0}}\Big)_{12}\Big|\simeq0.051,
\nonumber\\
&&Z_{_{h^0\tilde{N}^c}}=\sqrt{\sum\limits_{i=1}^3\Big|\Big(Z_{_{H_0}}\Big)_{1(3+i)}\Big|^2}
\simeq1.53\times10^{-2},
\nonumber\\
&&R_{_{\gamma\gamma}}=0.99,\;R_{_{VV^*}}=0.88\;.
\label{Higgs-sector1}
\end{eqnarray}
Here $m_{_{h^0}}^{MSSM}$ is theoretical evaluation on the lightest CP-even mass
including one-loop corrections and leading terms from two-loop contributions,
and $m_{_{h^0}}$ is corresponding evaluation in the MSSM with local $U(1)_{B-L}$ symmetry.
Furthermore $m_{_{H_2^0}}$ represents the heavier CP-even Higgs in the MSSM,
$\Big(Z_{_{H_0}}\Big)_{12}$ represents the mixing between $H_2^0$ and $h^0$,
$Z_{_{h^0\tilde{N}^c}}$ represents the mixing between the lightest CP-even Higgs
and real parts of right-handed sneutrinos, respectively.

%%%%%%%%%%%%%%%%%%%%%%%%%%%%%%%%%%%%%%%%%%%%%%%%%%%%%
\begin{figure}[h]
\setlength{\unitlength}{1mm}
\centering
\includegraphics[width=3.0in]{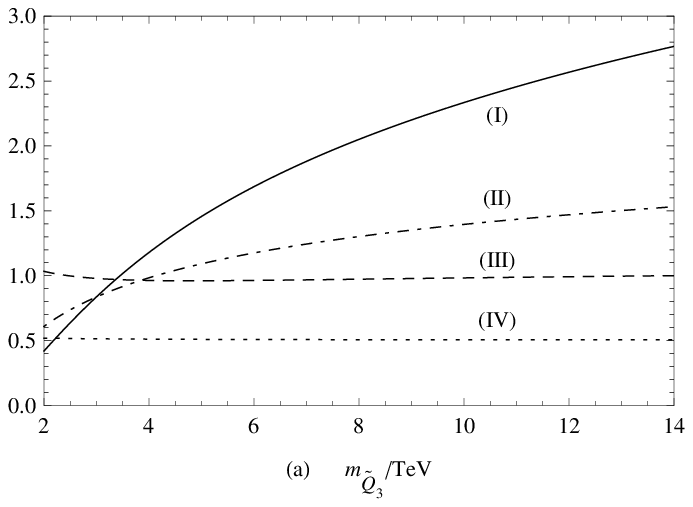}
\vspace{0.5cm}
\includegraphics[width=3.0in]{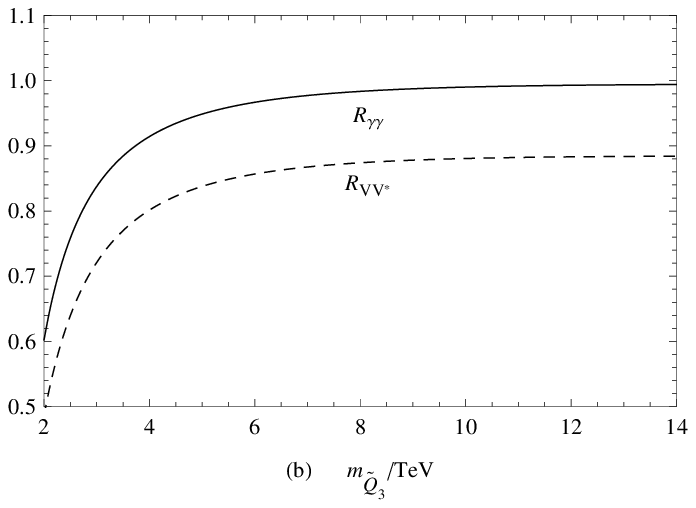}
\vspace{0cm}
\caption[]{Assuming neutrino mass spectrum with IO and taking $\tan\beta=20$,
$m_{_{\tilde{L}_3}}=2$ TeV, $m_{_{\tilde{E}_3}}=1.72$ TeV,
$m_{_{BL}}=1$ TeV, $\mu=2$ TeV, $A_{_t}=1.8$ TeV,
we plot $m_{_{h^0}}$, $\Delta m_{_{h^0}}$, $\Big|\Big(Z_{_{H_0}}\Big)_{12}\Big|$, $Z_{_{h^0\tilde{N}^c}}$,
$R_{\gamma\gamma}$ and $R_{VV^*}\;(V=Z,\;W)$ versus $m_{_{\tilde{Q}_3}}$.
Where (a) the solid line (I) denotes $\Delta m_{_{h^0}}/{\rm GeV}$ versus $m_{_{\tilde{Q}_3}}$,
the dash-dot line (II) denotes $Z_{_{h^0\tilde{N}^c}}\times10^2$ versus $m_{_{\tilde{Q}_3}}$,
the dash line (III) denotes $m_{_{h^0}}/125.9{\rm GeV}$ versus $m_{_{\tilde{Q}_3}}$,
as well as the dot line (IV) denotes $\Big|\Big(Z_{_{H_0}}\Big)_{12}\Big|\times10$
versus $m_{_{\tilde{Q}_3}}$; and (b)the solid line denotes $R_{\gamma\gamma}$ versus $m_{_{\tilde{Q}_3}}$,
the dashed line denotes $R_{VV^*}$ versus  $m_{_{\tilde{Q}_3}}$, respectively.}
\label{fig7}
\end{figure}
%%%%%%%%%%%%%%%%%%%%%%%%%%%%%%%%%%%%%%%%%%%%%%%%%%%%%

Assuming neutrino mass spectrum with IO and taking $m_{_{\tilde{Q}_3}}=14$ TeV,
$m_{_{\tilde{L}_3}}=1.72$ TeV, $m_{_{\tilde{E}_3}}=2$ TeV,
$m_{_{BL}}=1$ TeV, $\mu=2$ TeV, $A_{_t}=1.8$ TeV, we plot $m_{_{h^0}}$,
$\Delta m_{_{h^0}}$, $\Big|\Big(Z_{_{H_0}}\Big)_{12}\Big|$, and $Z_{_{h^0\tilde{N}^c}}$
versus $\tan\beta$ in Fig.\ref{fig5}(a).
Where the solid line (I) denotes $\Delta m_{_{h^0}}/{\rm GeV}$ versus $\tan\beta$,
the dash-dot line (II) denotes $Z_{_{h^0\tilde{N}^c}}\times10^2$ versus $\tan\beta$,
the dash line (III) denotes $m_{_{h^0}}/125.9{\rm GeV}$ versus $\tan\beta$,
as well as the dot line (IV) denotes $\Big|\Big(Z_{_{H_0}}\Big)_{12}\Big|\times10$
versus $\tan\beta$, respectively. The evaluation on the lightest CP-even Higgs mass $m_{_{h^0}}$
coincides with the ATLAS/CMS data in one standard deviation: $123.8\;{\rm GeV}\le
m_{_{h^0}}\le128.0\;{\rm GeV}$ as $3\le\tan\beta\le60$. Since the main corrections to
$\Delta m_{_{h^0}}$ originate from scalar top sector, the evaluation
on $\Delta m_{_{h^0}}$ decreases with increasing of $\tan\beta$ as $\tan\beta\lesssim10$.
Correspondingly, the evaluation on $\Delta m_{_{h^0}}$ increases with increasing of $\tan\beta$
as $\tan\beta\gtrsim30$ because the corrections to $\Delta m_{_{h^0}}$ from scalar bottom
and tau sectors are magnified. The mixing $Z_{_{h^0\tilde{N}^c}}$ between the lightest CP-even
Higgs and right-handed sneutinos is dominated by the loop corrections, and
is well above $10^{-2}$ with our assumptions on parameter space. The mixing between
$h^0$ and $H_2^0$ decreases steeply with increasing of $\tan\beta$.
Similarly, we also depict $R_{\gamma\gamma}$ and $R_{VV^*}$ versus $\tan\beta$
in Fig.\ref{fig5}(b). With increasing of $\tan\beta$, $R_{\gamma\gamma}$ decreases
slowly as $\tan\beta<5$ since the main corrections to $R_{\gamma\gamma}$ originate
from scalar top sector, and raises mildly as $\tan\beta>30$ since the corresponding corrections
mainly originate from the scalar tau sector. Meanwhile
$R_{VV^*}\;(V=Z,\;W)$ decreases smoothly with increasing of $\tan\beta$.
Although the loop induced mixing between $h^0$ and right-handed sneutrinos
exceeds $0.01$, this mixing cannot modify theoretical evaluation on $R_{\gamma\gamma}$ drastically.
%%%%%%%%%%%%%%%%%%%%%Begin Modification%%%%%%%%%%%%%%%%%%%%%%%
Actually the decay width of $h^0\rightarrow\gamma\gamma$ is enhanced by light scalar quarks originating from large mixing
between left- and right-hand scalar partners in the model considered here. However this effect is generally
offset by a suppression of the production rate for $gg\rightarrow h^0$, and the evaluations
on a Higgs gluon fusion production times photon decay rate is approximately equal to or slightly lower than
the corresponding one of the SM in the parameter region consistent with a $125$ GeV
Higgs. In most parameter space the corrections to the decay width of $h^0\rightarrow\gamma\gamma$
from charginos also tend to reduce corresponding evaluation of the SM, which
is consistent with the numerical results in the MSSM\cite{Dermisek,Carena2} qualitatively.
With our assumptions on the parameter space the corrections to $R_{\gamma\gamma}$
from charged scalar leptons cannot modify the corresponding SM evaluation drastically
since scalar taus all have TeV masses.

%%%%%%%%%%%%%%%%%%%%%%End Modification%%%%%%%%%%%%%%%%%%%%%%%

Assuming neutrino mass spectrum with IO and taking $\tan\beta=20$, $m_{_{\tilde{Q}_3}}=14$ TeV,
$m_{_{\tilde{L}_3}}=1.72$ TeV, $m_{_{\tilde{E}_3}}=2$ TeV, $m_{_{BL}}=1$ TeV, $\mu=2$ TeV,
we plot $m_{_{h^0}}$, $\Delta m_{_{h^0}}$, $\Big|\Big(Z_{_{H_0}}\Big)_{12}\Big|$, and
$Z_{_{h^0\tilde{N}^c}}$ versus $A_{_t}$ in Fig.\ref{fig6}(a).
The evaluation on the lightest CP-even Higgs mass $m_{_{h^0}}$
coincides with the ATLAS/CMS data in one standard deviation
as $|A_{_t}|\lesssim4\;{\rm TeV}$. With increasing of $|A_{_t}|$,
the evaluation on $m_{_{h^0}}$ raises mildly. Meanwhile the correction to $m_{_{h^0}}$ from
the mixing between the lightest CP-even Higgs and right-handed sneutrinos
exceeds $2\;{\rm GeV}$ around $|A_{_t}|\lesssim4\;{\rm TeV}$, and decreases
with increasing of $|A_{_t}|$. The mixing between the lightest CP-even
Higgs and right-handed sneutrinos $Z_{_{h^0\tilde{N}^c}}$ is dominated by the radiative corrections,
is well above $10^{-2}$ as $|A_{_t}|\lesssim6\;{\rm TeV}$. The mixing between
$h^0$ and $H_2^0$ changes mildly with $A_{_t}$.
Similarly, we also depict $R_{\gamma\gamma}$ and $R_{VV^*}$ versus $A_{_t}$
in Fig.\ref{fig6}(b). $R_{\gamma\gamma}$ and $R_{VV^*}\;(V=Z,\;W)$ vary gently as $|A_{_t}|\lesssim4$ TeV.
Similarly the theoretical evaluations on $R_{\gamma\gamma}$ and $R_{VV^*}\;(V=Z,\;W)$ are all
smaller than 1 since the reason mentioned above.

Assuming neutrino mass spectrum with IO and taking $\tan\beta=20$,
$m_{_{\tilde{L}_3}}=1.72$ TeV, $m_{_{\tilde{E}_3}}=2$ TeV,
$m_{_{BL}}=1$ TeV, $\mu=2$ TeV, $A_{_t}=1.8$ TeV, we plot $m_{_{h^0}}$,
$\Delta m_{_{h^0}}$, $\Big|\Big(Z_{_{H_0}}\Big)_{12}\Big|$, and $Z_{_{h^0\tilde{N}^c}}$
versus $m_{_{\tilde{Q}_3}}$ in Fig.\ref{fig7}(a).
The evaluation on the lightest CP-even Higgs mass $m_{_{h^0}}$
coincides with the ATLAS/CMS data in one standard deviation
as $m_{_{\tilde{Q}_3}}\gtrsim2\;{\rm TeV}$.
Because the one-loop corrections to the mixing between the lightest CP-even
Higgs and right-handed sneutinos are proportional to $m_{_{\tilde{Q}_3}}^2-m_{_{\tilde{U}_3}}^2$,
the correction to $m_{_{h^0}}$ from
the mixing between the lightest CP-even Higgs and right-handed sneutrinos
exceeds $1\;{\rm GeV}$ when $m_{_{\tilde{Q}_3}}\gtrsim4\;{\rm TeV}$, and raises
with increasing of $m_{_{\tilde{Q}_3}}$. The mixing between the lightest CP-even
Higgs and right-handed sneutinos $Z_{_{h^0\tilde{N}^c}}$ is well above $10^{-2}$ as
$m_{_{\tilde{Q}_3}}\gtrsim4\;{\rm TeV}$. The mixing between
$h^0$ and $H_2^0$ varies mildly with increasing of $m_{_{\tilde{Q}_3}}$.
Similarly, we also depict $R_{\gamma\gamma}$ and $R_{VV^*}$ versus $m_{_{\tilde{Q}_3}}$
in Fig.\ref{fig7}(b). Actually $R_{\gamma\gamma}$ and $R_{VV^*}\;(V=Z,\;W)$ vary
mildly with $m_{_{\tilde{Q}_3}}$ when $m_{_{\tilde{Q}_3}}\gtrsim4\;{\rm TeV}$.

%%%%%%%%%%%%%%%%%%%%%%%%%%%%%%%%%%%%%%%%%%%%%%%%%%%%%
\begin{figure}[h]
\setlength{\unitlength}{1mm}
\centering
\includegraphics[width=3.0in]{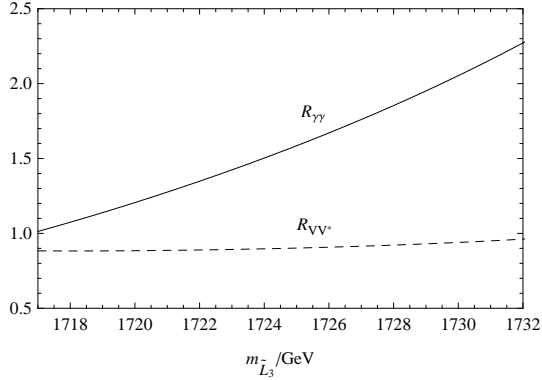}
\vspace{0.5cm}
\caption[]{$R_{\gamma\gamma}$ and $R_{VV^*}\;(V=Z,\;W)$ as a function of $m_{_{\tilde{L}_3}}$,
for $\tan\beta=60$ varying $\mu$ such that $m_{_{{\tilde\tau}_1}}=90$ GeV.
Where the Higgs mass $m_{_{h^0}}$ varies with $m_{_{\tilde{E}_3}}$, but satisfies with
Eq.(\ref{M-h0}).}
\label{fig8}
\end{figure}
%%%%%%%%%%%%%%%%%%%%%%%%%%%%%%%%%%%%%%%%%%%%%%%%%%%%%

%%%%%%%%%%%%%%%%%%%%%%%%%%%Begin Modification%%%%%%%%%%%%%%%%%%
Similar to scenarios in the MSSM, the possible correction to enhance theoretical evaluations
on $R_{\gamma\gamma}$ originates from light scalar tau leptons,
in which large mixing between left- and right-handed scalar tau leptons is evoked
by large values of $\mu$ parameter and $\tan\beta$ \cite{Carena2}.
In order to enhance the corrections from scalar tau sector to  $R_{\gamma\gamma}$,
we require the lightest scalar tau has a mass close to the lower bound
from LEP collaboration $m_{_{{\tilde\tau}_1}}\gtrsim90$ GeV.
In the MSSM, the elements of scalar tau mass squared matrix are
\begin{eqnarray}
%%%%%%%%%%%%%%%%%%%%%%%%%%%%%%%%%%%%%%%%%%%%%%%%%%%%
&&\Big(m_{_{\tilde\tau}}^2\Big)_{LL}\simeq m_{_{\tilde{L}_3}}^2
-{1\over2}m_{_{\rm Z}}^2(1-2\cos^2\beta)(1-2c_{_{\rm W}}^2)+m_\tau^2
\;,\nonumber\\
%%%%%%%%%%%%%%%%%%%%%%%%%%%%%%%%%%%%%%%%%%%%%%%%%%%%
&&\Big(m_{_{\tilde\tau}}^2\Big)_{RR}\simeq m_{_{\tilde{E}_3}}^2
+m_{_{\rm Z}}^2s_{_{\rm W}}^2(1-2\cos^2\beta)+m_\tau^2
\;,\nonumber\\
%%%%%%%%%%%%%%%%%%%%%%%%%%%%%%%%%%%%%%%%%%%%%%%%%%%%
&&\Big(m_{_{\tilde\tau}}^2\Big)_{LR}\simeq -m_\tau\Big(\mu\tan\beta-A_\tau\Big)
\;.\nonumber\\
%%%%%%%%%%%%%%%%%%%%%%%%%%%%%%%%%%%%%%%%%%%%%%%%%%%%
\label{stau-mass-squared-MSSM}
\end{eqnarray}
Assuming $m_{_{\tilde{L}_3}}=m_{_{\tilde{E}_3}}$ and varying the
$\mu$ parameter to guarantee the lightest scalar tau $m_{_{{\tilde\tau}_1}}=90$ GeV, we
can derive that the branching ratio of $h^0\rightarrow\gamma\gamma$ is larger than that of the SM
for large $\tan\beta$ scenarios in the MSSM.
In the MSSM with local $U(1)_{B-L}$ symmetry, the elements of scalar tau mass squared matrix are
modified as
\begin{eqnarray}
%%%%%%%%%%%%%%%%%%%%%%%%%%%%%%%%%%%%%%%%%%%%%%%%%%%%
&&\Big(m_{_{\tilde\tau}}^2\Big)_{LL}\simeq m_{_{\tilde{L}_3}}^2
-{1\over2}m_{_{\rm Z}}^2(1-2\cos^2\beta)(1-2c_{_{\rm W}}^2)+m_\tau^2-{1\over2}m_{_{Z_{BL}}}^2
\;,\nonumber\\
%%%%%%%%%%%%%%%%%%%%%%%%%%%%%%%%%%%%%%%%%%%%%%%%%%%%
&&\Big(m_{_{\tilde\tau}}^2\Big)_{RR}\simeq m_{_{\tilde{E}_3}}^2
+m_{_{\rm Z}}^2s_{_{\rm W}}^2(1-2\cos^2\beta)+m_\tau^2+{1\over2}m_{_{Z_{BL}}}^2
\;,\nonumber\\
%%%%%%%%%%%%%%%%%%%%%%%%%%%%%%%%%%%%%%%%%%%%%%%%%%%%
&&\Big(m_{_{\tilde\tau}}^2\Big)_{LR}\simeq -m_\tau\Big(\mu\tan\beta-A_\tau\Big)
\;.\nonumber\\
%%%%%%%%%%%%%%%%%%%%%%%%%%%%%%%%%%%%%%%%%%%%%%%%%%%%
\label{stau-mass-squared}
\end{eqnarray}
In order to get a lightest scalar tau $m_{_{{\tilde\tau}_1}}=90$ GeV for large $\tan\beta$,
we cannot adopt the assumption $m_{_{\tilde{L}_3}}\simeq m_{_{\tilde{E}_3}}$ any more
because of TeV scale mass of the neutral gauge boson $Z_{_{BL}}$. In addition, the positive
definite condition of $2\times2$ mass squared matrix $m_{_{\tilde\tau}}^2$ also requires
$m_{_{\tilde{L}_3}}^2\gtrsim m_{_{Z_{BL}}}^2/2$.
Choosing $\tan\beta=60$, $m_{_{\tilde{E}_3}}=2$ TeV, and varying $\mu$ such that $m_{_{{\tilde\tau}_1}}=90$ GeV,
we plot $R_{\gamma\gamma}$ and $R_{VV^*}\;(V=Z,\;W)$ versus $m_{_{\tilde{L}_3}}$
in Fig.(\ref{fig8}). $R_{\gamma\gamma}$ is magnified by the corrections from the lightest
scalar tau, and $R_{\gamma\gamma}\simeq2$ as $m_{_{\tilde{L}_3}}\simeq1.73$ TeV.
However, the fine tuning condition $m_{_{{\tilde\tau}_1}}=90$ GeV causes that the product $\mu \tan\beta$
turns large rapidly with increasing of $m_{_{\tilde{L}_3}}$. Under our assumptions on parameter
space, we find $\mu\simeq1.2$ TeV as  $m_{_{\tilde{L}_3}}=1.717$ TeV, and $\mu\simeq5.8$ TeV as
$m_{_{\tilde{L}_3}}=1.732$ TeV, respectively. Meanwhile, the heaviest scalar tau has a
TeV scalar mass $m_{_{{\tilde\tau}_2}}\simeq 2.6$ TeV.
%%%%%%%%%%%%%%%%%%%%%%%%%%%%End Modification%%%%%%%%%%%%%%%%%%

\section{Summary\label{sec8}}
\indent\indent
In the scenarios where sneutrinos all obtain nonzero VEVs, we study the constraints from
the observed Higgs signal and neutrino oscillation experimental data on parameter space
of the MSSM with local $U(1)_{B-L}$ symmetry\cite{Perez1,Perez2,Perez3,Perez4}.
Considering the constraints from neutrino oscillation, the mixing between real parts
of right-handed sneutrinos and the lightest CP-even Higgs is below $10^{-12}$ at tree level,
and the mixing between real parts of left-handed sneutrinos and the lightest CP-even Higgs is
about $10^{-6}$ at tree level. Including one-loop virtual corrections, we find
that the mixing between real parts of right-handed sneutrinos and the lightest CP-even
Higgs can reach $1.5\times10^{-2}$, and this mixing increases the MSSM theoretical
evaluation on the mass of the lightest CP-even Higgs exceeding $2.0\;{\rm GeV}$.
Meanwhile, we can safely neglect the one-loop corrections to the mixing
between real parts of left-handed sneutrinos and the lightest CP-even Higgs
which are proportional to the tiny nonzero VEVs of left-handed sneutrinos.
Numerically the MSSM with local $U(1)_{B-L}$ symmetry accommodates naturally
the observed Higgs signal from ATLAS/CMS collaborations
and the updated experimental data on neutrino oscillation simultaneously. In addition,
the model also predicts two sterile neutrinos with sub-eV masses\cite{Perez6,Senjanovic1} which are favored
by the BBN in cosmology\cite{Hamann}.

\begin{acknowledgments}
\indent\indent The work has been supported by the National Natural
Science Foundation of China (NNSFC) with Grant No. 11275243, No. 11275036,
No. 11047002, No. 11147001,
the open project of State Key Laboratory of Mathematics-Mechanization
with Grant No. Y3KF311CJ1, the Natural Science Foundation of
Hebei province with Grant No. A2013201277, and Natural Science Foundation of
Hebei University with Grant No. 2011JQ05, No. 2012-242.
\end{acknowledgments}

\appendix

\section{The mass squared matrices for squarks\label{app0}}
\indent\indent
With the minimal flavor violation assumption, the $2\times2$ mass squared matrix
for scalar tops is given as
\begin{eqnarray}
&&{\cal Z}_{_t}^\dagger\left(\begin{array}{cc}
m_{_{\tilde{t}_L}}^2\;\;&m_{_{\tilde{t}_X}}^2\\ \\
m_{_{\tilde{t}_X}}^2\;\;&m_{_{\tilde{t}_R}}^2
\end{array}\right){\cal Z}_{_t}=diag\Big(m_{_{\tilde{t}_1}}^2,\;m_{_{\tilde{t}_2}}^2\Big)\;,
\label{UP-squark1}
\end{eqnarray}
with
\begin{eqnarray}
&&m_{_{\tilde{t}_L}}^2={(g_1^2+g_2^2)\upsilon_{_{\rm EW}}^2\over24}
\Big(1-2\cos^2\beta\Big)\Big(1-4c_{_{\rm W}}^2\Big)
\nonumber\\
&&\hspace{1.4cm}
+{g_{_{BL}}^2\over6}\Big(\upsilon_{_N}^2-\upsilon_{_{\rm EW}}^2
+\upsilon_{_{\rm SM}}^2\Big)+m_{_t}^2+m_{_{\tilde{Q}_3}}^2
\;,\nonumber\\
&&m_{_{\tilde{t}_R}}^2=-{g_1^2\upsilon_{_{\rm EW}}^2\over6}
\Big(1-2\cos^2\beta\Big)
\nonumber\\
&&\hspace{1.4cm}
-{g_{_{BL}}^2\over6}\Big(\upsilon_{_N}^2-\upsilon_{_{\rm EW}}^2
+\upsilon_{_{\rm SM}}^2\Big)
+m_{_t}^2+m_{_{\tilde{U}_3}}^2\;,
\nonumber\\
&&m_{_{\tilde{t}_X}}^2=-{\upsilon_{_u}\over\sqrt{2}}A_{_t}Y_{_t}
+{\mu\upsilon_{_d}\over\sqrt{2}}Y_{_t}\;.
\label{UP-squark2}
\end{eqnarray}
Here $Y_{_t},\;A_{_t}$ denote Yukawa coupling
and trilinear soft-breaking parameters in top quark sector, respectively.
In a similar way, the mass-squared matrix for scalar bottoms is
\begin{eqnarray}
&&{\cal Z}_{_b}^\dagger\left(\begin{array}{cc}
m_{_{\tilde{b}_L}}^2\;\;&m_{_{\tilde{b}_X}}^2\\ \\
m_{_{\tilde{b}_X}}^2\;\;&m_{_{\tilde{b}_R}}^2
\end{array}\right){\cal Z}_{_b}=diag\Big(m_{_{\tilde{b}_1}}^2,\;m_{_{\tilde{b}_2}}^2\Big)\;,
\label{DOWN-squark1}
\end{eqnarray}
with
\begin{eqnarray}
&&m_{_{\tilde{b}_L}}^2={(g_1^2+g_2^2)\upsilon_{_{\rm EW}}^2\over24}
\Big(1-2\cos^2\beta\Big)\Big(1+2c_{_{\rm W}}^2\Big)
\nonumber\\
&&\hspace{1.4cm}
+{g_{_{BL}}^2\over6}\Big(\upsilon_{_N}^2-\upsilon_{_{\rm EW}}^2
+\upsilon_{_{\rm SM}}^2\Big)
+m_{_b}^2+m_{_{\tilde{Q}_3}}^2\;,
\nonumber\\
&&m_{_{\tilde{b}_R}}^2={g_1^2\upsilon_{_{\rm EW}}^2\over12}
\Big(1-2\cos^2\beta\Big)
\nonumber\\
&&\hspace{1.4cm}
-{g_{_{BL}}^2\over6}\Big(\upsilon_{_N}^2-\upsilon_{_{\rm EW}}^2
+\upsilon_{_{\rm SM}}^2\Big)
+m_{_b}^2+m_{_{\tilde{D}_3}}^2\;,
\nonumber\\
&&m_{_{\tilde{b}_X}}^2={\upsilon_{_d}\over\sqrt{2}}A_{_b}Y_{_b}
-{\mu\upsilon_{_u}\over\sqrt{2}}Y_{_b}\;,
\label{DOWN-squark2}
\end{eqnarray}
here $Y_{_b},\;A_{_b}$ denote Yukawa couplings
and trilinear soft-breaking parameters in b quark sector, respectively.

\section{The minimization conditions and mass squared matrices for Higgs\label{app1}}
\indent\indent
The tree level minimization conditions are formulated as
\begin{eqnarray}
&&T_{_u}^0=\upsilon_{_u}\Big\{\mu^2+m_{_{H_u}}^2+{g_1^2+g_2^2\over8}\Big(2\upsilon_{_u}^2-\upsilon_{_{\rm EW}}^2\Big)
+{1\over2}\sum\limits_{\alpha,\beta=1}^3\Big[\upsilon_{_{N_\alpha}}
\Big(Y_{_N}^TY_{_N}\Big)_{\alpha\beta}\upsilon_{_{N_\beta}}
\nonumber\\
&&\hspace{1.0cm}
+\upsilon_{_{L_\alpha}}\Big(Y_{_N}Y_{_N}^T\Big)_{\alpha\beta}\upsilon_{_{L_\beta}}\Big]\Big\}
+{1\over\sqrt{2}}\sum\limits_{\alpha,\beta=1}^3\upsilon_{_{L_\alpha}}\Big(A_{_N}\Big)_{\alpha\beta}
\upsilon_{_{N_\beta}}+B\mu\upsilon_{_d}
\;,\nonumber\\
%%%%%%%%%%%%%%%%%%%%%%%%%%%%%%%%%%%%%%%%%%%%%%%%%%%
&&T_{_d}^0=\upsilon_{_d}\Big\{\mu^2+m_{_{H_d}}^2-{g_1^2+g_2^2\over8}\Big(2\upsilon_{_u}^2
-\upsilon_{_{\rm EW}}^2\Big)\Big\}+{\mu\varepsilon_{_N}^2\over\sqrt{2}}+B\mu\upsilon_{_u}
\;,\nonumber\\
%%%%%%%%%%%%%%%%%%%%%%%%%%%%%%%%%%%%%%%%%%%%%%%%%%%
&&T_{_{\tilde{L}_I}}^0={1\over2}\sum\limits_{\alpha=1}^3\Big[\Big(m_{_{\tilde L}}^2\Big)_{I\alpha}
+\Big(m_{_{\tilde L}}^2\Big)_{\alpha I}\Big]\upsilon_{_{L_\alpha}}
+{\upsilon_{_u}\over\sqrt{2}}\sum\limits_{\alpha=1}^3\Big(A_{_N}\Big)_{I\alpha}\upsilon_{_{N_\alpha}}
+{\mu\upsilon_{_d}\over\sqrt{2}}\zeta_{_I}
\nonumber\\
&&\hspace{1.0cm}
+{\varepsilon_{_N}^2\zeta_{_I}\over2}
+{\upsilon_{_u}^2\over2}\sum\limits_{\alpha=1}^3\Big(Y_{_N}Y_{_N}^T\Big)_{I\alpha}\upsilon_{_{L_\alpha}}
-\upsilon_{_{L_I}}\Big\{{g_1^2+g_2^2\over8}\Big(2\upsilon_{_u}^2-\upsilon_{_{\rm EW}}^2\Big)
\nonumber\\
&&\hspace{1.0cm}
+{g_{_{BL}}^2\over2}\Big(\upsilon_{_N}^2-\upsilon_{_{\rm EW}}^2+\upsilon_{_{\rm SM}}^2\Big)\Big\}
\;,\nonumber\\
%%%%%%%%%%%%%%%%%%%%%%%%%%%%%%%%%%%%%%%%%%%%%%%%%%%
&&T_{_{\tilde{N}_I}}^0={1\over2}\sum\limits_{\alpha=1}^3\Big[\Big(m_{_{{\tilde N}^c}}^2\Big)_{I\alpha}
+\Big(m_{_{{\tilde N}^c}}^2\Big)_{\alpha I}\Big]\upsilon_{_{N_\alpha}}
+{\upsilon_{_u}\over\sqrt{2}}\sum\limits_{\alpha=1}^3\upsilon_{_{L_\alpha}}\Big(A_{_N}\Big)_{\alpha I}
\nonumber\\
&&\hspace{1.0cm}
+{\mu\upsilon_{_d}\over\sqrt{2}}\sum\limits_{\alpha=1}^3\upsilon_{_{L_\alpha}}\Big(Y_{_N}\Big)_{\alpha I}
+{\varepsilon_{_N}^2\over2}\sum\limits_{\alpha=1}^3\upsilon_{_{L_\alpha}}\Big(Y_{_N}\Big)_{\alpha I}
+{\upsilon_{_u}^2\over2}\sum\limits_{\alpha=1}^3\upsilon_{_{N_\alpha}}\Big(Y_{_N}^TY_{_N}\Big)_{\alpha I}
\nonumber\\
&&\hspace{1.0cm}
+{g_{_{BL}}^2\over2}\upsilon_{_{N_I}}\Big(\upsilon_{_N}^2-\upsilon_{_{\rm EW}}^2
+\upsilon_{_{\rm SM}}^2\Big)\;.
\label{minimizations1}
\end{eqnarray}
The radiative corrections from top, bottom and tau sectors to the minimization
conditions are
\begin{eqnarray}
%%%%%%%%%%%%%%%%%%%%%%%%%%%%%%%%%%%%%%%%%%%%%%%%%%%
&&\Delta T_{_u}={3\over(4\pi)^2}\Big\{\Big(Y_{_t}^2-{g_{_1}^2+g_{_2}^2\over8}\Big)\Big(
f(m_{_{{\tilde t}_1}}^2)+f(m_{_{{\tilde t}_2}}^2)\Big)
-2Y_{_t}^2f(m_{_t}^2)
\nonumber\\
&&\hspace{1.3cm}
+\Big(Y_{_t}^2A_{_t}(A_{_t}-\mu{\upsilon_{_d}\over\upsilon_{_u}})
-{3g_{_2}^2-5g_{_1}^2\over24}(m_{_{\tilde{t}_L}}^2-m_{_{\tilde{t}_R}}^2)\Big)
{f(m_{_{{\tilde t}_1}}^2)-f(m_{_{{\tilde t}_2}}^2)\over
m_{_{{\tilde t}_1}}^2-m_{_{{\tilde t}_2}}^2}
\nonumber\\
&&\hspace{1.3cm}
+{g_{_1}^2+g_{_2}^2\over8}\Big(f(m_{_{{\tilde b}_1}}^2)+f(m_{_{{\tilde b}_2}}^2)\Big)
\nonumber\\
&&\hspace{1.3cm}
-\Big(Y_{_b}^2\mu(A_{_b}{\upsilon_{_d}\over\upsilon_{_u}}-\mu)
-{3g_{_2}^2-g_{_1}^2\over24}(m_{_{\tilde{b}_L}}^2-m_{_{\tilde{b}_R}}^2)\Big)
{f(m_{_{{\tilde b}_1}}^2)-f(m_{_{{\tilde b}_2}}^2)\over
m_{_{{\tilde b}_1}}^2-m_{_{{\tilde b}_2}}^2}\Big\}
\nonumber\\
&&\hspace{1.3cm}
+{1\over(4\pi)^2}\Big\{{g_{_1}^2+g_{_2}^2\over8}\Big(f(m_{_{{\tilde\tau}_1}}^2)+f(m_{_{{\tilde\tau}_2}}^2)\Big)
-\Big(Y_{_\tau}^2\mu(A_{_\tau}{\upsilon_{_d}\over\upsilon_{_u}}-\mu)
\nonumber\\
&&\hspace{1.3cm}
-{g_{_2}^2-3g_{_1}^2\over8}(m_{_{\tilde{\tau}_L}}^2-m_{_{\tilde{\tau}_R}}^2)\Big)
{f(m_{_{{\tilde\tau}_1}}^2)-f(m_{_{{\tilde\tau}_2}}^2)\over
m_{_{{\tilde\tau}_1}}^2-m_{_{{\tilde\tau}_2}}^2}\Big\}
\;,\nonumber\\
%%%%%%%%%%%%%%%%%%%%%%%%%%%%%%%%%%%%%%%%%%%%%%%%%%%%%%%
&&\Delta T_{_d}={3\over(4\pi)^2}\Big\{{g_{_1}^2+g_{_2}^2\over8}\Big(
f(m_{_{{\tilde t}_1}}^2)+f(m_{_{{\tilde t}_2}}^2)\Big)
\nonumber\\
&&\hspace{1.3cm}
-\Big(Y_{_t}^2\mu(A_{_t}{\upsilon_{_u}\over\upsilon_{_d}}-\mu)
-{3g_{_2}^2-5g_{_1}^2\over24}(m_{_{\tilde{t}_L}}^2-m_{_{\tilde{t}_R}}^2)\Big)
{f(m_{_{{\tilde t}_1}}^2)-f(m_{_{{\tilde t}_2}}^2)\over
m_{_{{\tilde t}_1}}^2-m_{_{{\tilde t}_2}}^2}
\nonumber\\
&&\hspace{1.3cm}
+\Big(Y_{_b}-{g_{_1}^2+g_{_2}^2\over8}\Big)\Big(f(m_{_{{\tilde b}_1}}^2)+f(m_{_{{\tilde b}_2}}^2)\Big)
-2Y_{_b}^2f(m_{_b}^2)
\nonumber\\
&&\hspace{1.3cm}
+\Big(Y_{_b}^2A_{_b}(A_{_b}-\mu{\upsilon_{_u}\over\upsilon_{_d}})
-{3g_{_2}^2-g_{_1}^2\over24}(m_{_{\tilde{b}_L}}^2-m_{_{\tilde{b}_R}}^2)\Big)
{f(m_{_{{\tilde b}_1}}^2)-f(m_{_{{\tilde b}_2}}^2)\over
m_{_{{\tilde b}_1}}^2-m_{_{{\tilde b}_2}}^2}\Big\}
\nonumber\\
&&\hspace{1.3cm}
+{1\over(4\pi)^2}\Big\{\Big(Y_{_\tau}^2-{g_{_1}^2+g_{_2}^2\over8}\Big)
\Big(f(m_{_{{\tilde\tau}_1}}^2)+f(m_{_{{\tilde\tau}_2}}^2)\Big)-2Y_{_\tau}^2f(m_{_\tau}^2)
\nonumber\\
&&\hspace{1.3cm}
+\Big(Y_{_\tau}^2A_{_\tau}(A_{_\tau}-\mu{\upsilon_{_u}\over\upsilon_{_d}})
-{g_{_2}^2-3g_{_1}^2\over8}(m_{_{\tilde{\tau}_L}}^2-m_{_{\tilde{\tau}_R}}^2)\Big)
{f(m_{_{{\tilde\tau}_1}}^2)-f(m_{_{{\tilde\tau}_2}}^2)\over
m_{_{{\tilde\tau}_1}}^2-m_{_{{\tilde\tau}_2}}^2}\Big\}
\;,\nonumber\\
%%%%%%%%%%%%%%%%%%%%%%%%%%%%%%%%%%%%%%%%%%%%%%%%%%%%%%
&&\Delta T_{_{\tilde{L}}}={3\over(4\pi)^2}\Big\{\Big[{g_{_1}^2+g_{_2}^2\over8}\Big(
f(m_{_{{\tilde t}_1}}^2)+f(m_{_{{\tilde t}_2}}^2)\Big)
\nonumber\\
&&\hspace{1.3cm}
+\Big({3g_{_2}^2-5g_{_1}^2\over24}-{g_{_{BL}}^2\over3}\Big)
(m_{_{\tilde{t}_L}}^2-m_{_{\tilde{t}_R}}^2)
{f(m_{_{{\tilde t}_1}}^2)-f(m_{_{{\tilde t}_2}}^2)\over
m_{_{{\tilde t}_1}}^2-m_{_{{\tilde t}_2}}^2}\Big]
\nonumber\\
&&\hspace{1.3cm}
-\Big[{g_{_1}^2+g_{_2}^2\over8}\Big(
f(m_{_{{\tilde b}_1}}^2)+f(m_{_{{\tilde b}_2}}^2)\Big)
\nonumber\\
&&\hspace{1.3cm}
+\Big({3g_{_2}^2-g_{_1}^2\over24}+{g_{_{BL}}^2\over3}\Big)
(m_{_{\tilde{b}_L}}^2-m_{_{\tilde{b}_R}}^2)
{f(m_{_{{\tilde b}_1}}^2)-f(m_{_{{\tilde b}_2}}^2)\over
m_{_{{\tilde b}_1}}^2-m_{_{{\tilde b}_2}}^2}\Big]\Big\}
\nonumber\\
&&\hspace{1.3cm}
-{1\over(4\pi)^2}\Big\{{g_{_1}^2+g_{_2}^2\over8}\Big(
f(m_{_{{\tilde\tau}_1}}^2)+f(m_{_{{\tilde\tau}_2}}^2)\Big)
\nonumber\\
&&\hspace{1.3cm}
+\Big({g_{_2}^2-3g_{_1}^2\over8}+g_{_{BL}}^2\Big)
(m_{_{\tilde{\tau}_L}}^2-m_{_{\tilde{\tau}_R}}^2)
{f(m_{_{{\tilde\tau}_1}}^2)-f(m_{_{{\tilde\tau}_2}}^2)\over
m_{_{{\tilde\tau}_1}}^2-m_{_{{\tilde\tau}_2}}^2}\Big\}
\;,\nonumber\\
%%%%%%%%%%%%%%%%%%%%%%%%%%%%%%%%%%%%%%%%%%%%%%%%%%
&&\Delta T_{_{\tilde{N}}}={g_{_{BL}}^2\over16\pi^2}\Big\{
(m_{_{\tilde{t}_L}}^2-m_{_{\tilde{t}_R}}^2){f(m_{_{{\tilde t}_1}}^2)
-f(m_{_{{\tilde t}_2}}^2)\over m_{_{{\tilde t}_1}}^2-m_{_{{\tilde t}_2}}^2}
+(m_{_{\tilde{b}_L}}^2-m_{_{\tilde{b}_R}}^2){f(m_{_{{\tilde b}_1}}^2)
-f(m_{_{{\tilde b}_2}}^2)\over m_{_{{\tilde b}_1}}^2-m_{_{{\tilde b}_2}}^2}
\nonumber\\
&&\hspace{1.3cm}
+(m_{_{\tilde{\tau}_L}}^2-m_{_{\tilde{\tau}_R}}^2){f(m_{_{{\tilde\tau}_1}}^2)
-f(m_{_{{\tilde\tau}_2}}^2)\over m_{_{{\tilde\tau}_1}}^2-m_{_{{\tilde\tau}_2}}^2}\Big\}\;.
%%%%%%%%%%%%%%%%%%%%%%%%%%%%%%%%%%%%%%%%%%%%%%%%%%
\label{minimizations2}
\end{eqnarray}

In the interaction basis $H_{_{CH}}^T=(H_{_u}^-,\;H_{_d}^-,\;\tilde{L}_{_I}^-,\;\tilde{E}_{_J}^{c*}),\;\;(I,\;J=1,\;2,\;3)$,
elements of $A_{_{CH}}$ and the symmetric matrix $M_{_{\tilde E}}^2$ are written as
\begin{eqnarray}
%%%%%%%%%%%%%%%%%%%%%%%%%%%%%%%%%%%%%%%%%%%%%%%%%%%%
&&\Big[A_{_{CH}}\Big]_{1I^\prime}=
{1\over\sqrt{2}}\sum\limits_{\alpha=1}^3(A_{_N})_{I^\prime\alpha}\upsilon_{_{N_\alpha}}
+{\upsilon_{_u}\over2}\sum\limits_{\alpha=1}^3(Y_{_N}Y_{_N}^T)_{I^\prime\alpha}
\upsilon_{_{L_\alpha}}
-{g_2^2\over4}\upsilon_{_u}\upsilon_{_{L_{I^\prime}}}
\;,\nonumber\\
%%%%%%%%%%%%%%%%%%%%%%%%%%%%%%%%%%%%%%%%%%%%%%%%%%%%
&&\Big[A_{_{CH}}\Big]_{1(3+J^\prime)}={\mu\over\sqrt{2}}\sum\limits_{\alpha=1}^3
\Big(Y_{_E}\Big)_{J^\prime\alpha}\upsilon_{_{L_\alpha}}+{1\over2}\upsilon_{_d}\sum\limits_{\alpha=1}^3
\Big(Y_{_E}^TY_{_N}\Big)_{J^\prime\alpha}\upsilon_{_{N_\alpha}}
\;,\nonumber\\
%%%%%%%%%%%%%%%%%%%%%%%%%%%%%%%%%%%%%%%%%%%%%%%%%%%%
&&\Big[A_{_{CH}}\Big]_{2I^\prime}=-{g_2^2\over4}\upsilon_{_d}\upsilon_{_{L_{I^\prime}}}
-{\mu\zeta_{_{I^\prime}}\over\sqrt{2}}
-{1\over2}\upsilon_{_d}\sum\limits_{\alpha=1}^3\Big(Y_{_E}Y_{_E}^T\Big)_{I^\prime\alpha}\upsilon_{_{L_\alpha}}
\;,\nonumber\\
%%%%%%%%%%%%%%%%%%%%%%%%%%%%%%%%%%%%%%%%%%%%%%%%%%%%
&&\Big[A_{_{CH}}\Big]_{2(3+J^\prime)}={1\over\sqrt{2}}\sum\limits_{\alpha=1}^3\upsilon_{_{L_\alpha}}
\Big(A_{_E}Y_{_E}\Big)_{\alpha J^\prime}^*
+{\upsilon_{_u}\over2}\sum\limits_{\alpha=1}^3\upsilon_{_{N_\alpha}}\Big(Y_{_E}^TY_{_E}\Big)_{\alpha J^\prime}
\;,\nonumber\\
%%%%%%%%%%%%%%%%%%%%%%%%%%%%%%%%%%%%%%%%%%%%%%%%%%%%
&&\Big[M_{_{\tilde E}}^2\Big]_{II^\prime}=\Big(m_{_{\tilde L}}^2\Big)_{II^\prime}
-{g_1^2-g_2^2\over8}\Big(2\upsilon_{_u}^2-\upsilon_{_{\rm EW}}^2\Big)\delta_{II^\prime}
-{g_2^2\over4}\upsilon_{_{L_I}}\upsilon_{_{L_{I^\prime}}}
+{\upsilon_{_d}^2\over2}\Big(Y_{_E}Y_{_E}^T\Big)_{I^\prime I}
\nonumber\\
&&\hspace{2.2cm}
-{1\over2}\zeta_{_I}\zeta_{_{I^\prime}}
-{g_{_{BL}}^2\over2}\Big(\upsilon_{_N}^2-\upsilon_{_{\rm EW}}^2+\upsilon_{_{\rm SM}}^2\Big)\delta_{II^\prime}
+\Delta T_{_L}\delta_{II^\prime}
\;,\nonumber\\
%%%%%%%%%%%%%%%%%%%%%%%%%%%%%%%%%%%%%%%%%%%%%%%%%%%%
&&\Big[M_{_{\tilde E}}^2\Big]_{I(3+J^\prime)}={\upsilon_{_d}\over\sqrt{2}}\Big(A_{_E}Y_{_E}\Big)_{IJ^\prime}
-{\mu\upsilon_{_u}\over\sqrt{2}}\Big(Y_{_E}^T\Big)_{J^\prime I}
\;,\nonumber\\
%%%%%%%%%%%%%%%%%%%%%%%%%%%%%%%%%%%%%%%%%%%%%%%%%%%%
&&\Big[M_{_{\tilde E}}^2\Big]_{(3+J)(3+J^\prime)}=\Big(m_{_{\tilde E}}^2\Big)_{J^\prime J}
+{g_1^2\over4}\Big(2\upsilon_{_u}^2-\upsilon_{_{\rm EW}}^2\Big)\delta_{J^\prime J}
+{g_{_{BL}}^2\over2}\Big(\upsilon_{_N}^2-\upsilon_{_{\rm EW}}^2+\upsilon_{_{\rm SM}}^2\Big)\delta_{J^\prime J}
\nonumber\\
&&\hspace{3.2cm}
+{1\over2}\upsilon_{_d}^2\Big(Y_{_E}^TY_{_E}\Big)_{J^\prime J}
+{1\over2}\sum\limits_{\alpha,\beta=1}^3\Big(Y_{_E}^T\Big)_{J^\prime\alpha}\upsilon_{_{L_\alpha}}
\upsilon_{_{L_\beta}}\Big(Y_{_E}\Big)_{\beta J}\;.
\label{app1-CH}
\end{eqnarray}

In the interaction basis $P^{0,T}=(P_{_u}^0,\;P_{_d}^0,\;P_{_{\tilde{L}_I}}^0,\;P_{_{\tilde{N}_J}}^0),\;\;(I,\;J=1,\;2,\;3)$,
elements of $A_{_{CPO}}$  are written as
\begin{eqnarray}
%%%%%%%%%%%%%%%%%%%%%%%%%%%%%%%%%%%%%%%%%%%%%%%%%%%%
&&\Big[A_{_{CPO}}^{(0)}\Big]_{1I^\prime}=
{1\over\sqrt{2}}\sum\limits_{\alpha=1}^3(A_{_N})_{I^\prime\alpha}\upsilon_{_{N_\alpha}}
\;,\nonumber\\
%%%%%%%%%%%%%%%%%%%%%%%%%%%%%%%%%%%%%%%%%%%%%%%%%%%%
&&\Big[A_{_{CPO}}^{(0)}\Big]_{1(3+J^\prime)}=
{1\over\sqrt{2}}\sum\limits_{\alpha=1}^3(A_{_N})_{\alpha J^\prime}\upsilon_{_{L_\alpha}}
\;,\nonumber\\
%%%%%%%%%%%%%%%%%%%%%%%%%%%%%%%%%%%%%%%%%%%%%%%%%%%%
&&\Big[A_{_{CPO}}^{(0)}\Big]_{2I^\prime}=
-{\mu\over\sqrt{2}}\zeta_{_{I^\prime}}
\;,\nonumber\\
%%%%%%%%%%%%%%%%%%%%%%%%%%%%%%%%%%%%%%%%%%%%%%%%%%%%
&&\Big[A_{_{CPO}}^{(0)}\Big]_{2(3+J^\prime)}=
-{\mu\over\sqrt{2}}\sum\limits_{\alpha=1}^3\upsilon_{_{L_\alpha}}\Big(Y_{_N}\Big)_{\alpha J^\prime}\;,
%%%%%%%%%%%%%%%%%%%%%%%%%%%%%%%%%%%%%%%%%%%%%%%%%%%%
\label{app1-CP-odd1}
\end{eqnarray}
and elements of the symmetric matrix $M_{_P}^2$ are similarly given as
\begin{eqnarray}
%%%%%%%%%%%%%%%%%%%%%%%%%%%%%%%%%%%%%%%%%%%%%%%%%%%%
&&\Big[M_{_P}^2\Big]_{II^\prime}=\Big(m_{_{\tilde L}}^2\Big)_{II^\prime}
-{g_1^2+g_2^2\over8}\Big(2\upsilon_{_u}^2-\upsilon_{_{\rm EW}}^2\Big)\delta_{II^\prime}
-{g_{_{BL}}^2\over2}\Big(\upsilon_{_N}^2-\upsilon_{_{\rm EW}}^2+\upsilon_{_{\rm SM}}^2
\Big)\delta_{II^\prime}
\nonumber\\
&&\hspace{2.2cm}
+\Delta T_{_L}\delta_{II^\prime}+\Big[\delta^2m_{_{LL}}^{odd}\Big]_{II^\prime}
\;,\nonumber\\
%%%%%%%%%%%%%%%%%%%%%%%%%%%%%%%%%%%%%%%%%%%%%%%%%%%%
&&\Big[M_{_P}^2\Big]_{I(3+J^\prime)}=\Big[\delta^2m_{_{LR}}^{odd}\Big]_{IJ^\prime}
\;,\nonumber\\
%%%%%%%%%%%%%%%%%%%%%%%%%%%%%%%%%%%%%%%%%%%%%%%%%%%%
&&\Big[M_{_P}^2\Big]_{(3+J)(3+J^\prime)}=
\Big(m_{_{{\tilde N}^c}}^2\Big)_{JJ^\prime}+{g_{_{BL}}^2\over2}\Big(\upsilon_{_N}^2-\upsilon_{_{\rm EW}}^2
+\upsilon_{_{\rm SM}}^2\Big)\delta_{JJ^\prime}+\Delta T_{_N}\delta_{JJ^\prime}
\nonumber\\
&&\hspace{3.2cm}
+\Big[\delta^2m_{_{RR}}^{odd}\Big]_{JJ^\prime}\;,
%%%%%%%%%%%%%%%%%%%%%%%%%%%%%%%%%%%%%%%%%%%%%%%%%%%%
\label{app1-CP-odd2}
\end{eqnarray}
with
\begin{eqnarray}
%%%%%%%%%%%%%%%%%%%%%%%%%%%%%%%%%%%%%%%%%%%%%%%%%%%%
&&\Big[\delta^2m_{_{LL}}^{odd}\Big]_{II^\prime}=-{1\over2}\zeta_{_I}\zeta_{_{I^\prime}}
-{1\over2}\Big(Y_{_N}Y_{_N}^T\Big)_{II^\prime}\upsilon_{_u}^2
\;,\nonumber\\
%%%%%%%%%%%%%%%%%%%%%%%%%%%%%%%%%%%%%%%%%%%%%%%%%%%%
&&\Big[\delta^2m_{_{LR}}^{odd}\Big]_{IJ^\prime}=
{\upsilon_{_u}\over\sqrt{2}}\Big(A_{_N}\Big)_{IJ^\prime}+{\mu\upsilon_{_d}\over\sqrt{2}}\Big(Y_{_N}\Big)_{IJ^\prime}
+{\varepsilon_{_N}^2\over2}\Big(Y_{_N}\Big)_{IJ^\prime}
\nonumber\\
&&\hspace{2.5cm}
-{1\over2}\zeta_{_I}\sum\limits_{\alpha=1}^3\upsilon_{_{L_\alpha}}\Big(Y_{_N}^T\Big)_{J^\prime\alpha}
\;,\nonumber\\
%%%%%%%%%%%%%%%%%%%%%%%%%%%%%%%%%%%%%%%%%%%%%%%%%%%%
&&\Big[\delta^2m_{_{RR}}^{odd}\Big]_{JJ^\prime}=
-{1\over2}\upsilon_{_u}^2\Big(Y_{_N}^TY_{_N}\Big)_{JJ^\prime}
-{1\over2}\sum\limits_{\alpha,\beta=1}^3\Big(Y_{_N}^T\Big)_{J\alpha}\upsilon_{_{L_\alpha}}
\upsilon_{_{L_\beta}}\Big(Y_{_N}\Big)_{\beta J^\prime}\;.
%%%%%%%%%%%%%%%%%%%%%%%%%%%%%%%%%%%%%%%%%%%%%%%%%%%%
\label{app1-CP-odd2a}
\end{eqnarray}
Correspondingly the $6\times6$ matrix is approximated as
\begin{eqnarray}
%%%%%%%%%%%%%%%%%%%%%%%%%%%%%%%%%%%%%%%%%%%%%%%%%%%%
&&Z_{_P}\simeq\left(\begin{array}{ccc}1&\Big[{[\delta^2m_{_{HL}}^{odd}]_{I^\prime}\over m_{_{A_{3+I^\prime}^0}}^2
-m_{_{A_3^0}}^2}\Big]_{1\times3}&0_{1\times2}\\
\Big[{[\delta^2m_{_{HL}}^{odd}]_{I}\over m_{_{A_{3+I}^0}}^2
-m_{_{A_3^0}}^2}\Big]_{3\times1} & \left[(Z_{_{\tilde L}}^P)_{II^\prime}\right]_{3\times3} &
\Big[{[(\delta^2m_{_{LR}}^{odd}){\cal Z}_{_{{\tilde N}^c}}^P]_{I(i^\prime+1)}\over m_{_{A_{6+i^\prime}^0}}^2
-m_{_{A_{3+I}^0}}^2}\Big]_{3\times2}\\
0_{2\times1} &\Big[{[(\delta^2m_{_{LR}}^{odd}){\cal Z}_{_{{\tilde N}^c}}^P]_{I^\prime(i+1)}\over m_{_{A_{3+I^\prime}^0}}^2
-m_{_{A_{6+i}^0}}^2}\Big]_{2\times3} &\left[(Z_{_{\tilde R}}^P)_{ii^\prime}\right]_{2\times2}
\end{array}\right)\;,
%%%%%%%%%%%%%%%%%%%%%%%%%%%%%%%%%%%%%%%%%%%%%%%%%%%%
\label{app1-CP-odd2b}
\end{eqnarray}
where
\begin{eqnarray}
%%%%%%%%%%%%%%%%%%%%%%%%%%%%%%%%%%%%%%%%%%%%%%%%%%%%
&&Z_{_{\tilde L}}^P\simeq\left(\begin{array}{ccc}1&{[\delta^2m_{_{LL}}^{odd}]_{12}\over m_{_{A_5^0}}^2-m_{_{A_4^0}}^2}&
{[\delta^2m_{_{LL}}^{odd}]_{13}\over m_{_{A_6^0}}^2-m_{_{A_4^0}}^2}\\
{[\delta^2m_{_{LL}}^{odd}]_{12}\over m_{_{A_4^0}}^2-m_{_{A_5^0}}^2} &1&
{[\delta^2m_{_{LL}}^{odd}]_{23}\over m_{_{A_6^0}}^2-m_{_{A_5^0}}^2}\\
{[\delta^2m_{_{LL}}^{odd}]_{13}\over m_{_{A_4^0}}^2-m_{_{A_6^0}}^2} &
{[\delta^2m_{_{LL}}^{odd}]_{23}\over m_{_{A_5^0}}^2-m_{_{A_6^0}}^2}  &1
\end{array}\right)
\;,\nonumber\\
%%%%%%%%%%%%%%%%%%%%%%%%%%%%%%%%%%%%%%%%%%%%%%%%%%%%
&&Z_{_{\tilde R}}^P\simeq\left(\begin{array}{cc}1&{[{\cal Z}_{_{{\tilde N}^c}}^{P,T}(\delta^2m_{_{LR}}^{odd})
{\cal Z}_{_{{\tilde N}^c}}^P]_{23}\over m_{_{A_8^0}}^2-m_{_{A_7^0}}^2}\\
{[{\cal Z}_{_{{\tilde N}^c}}^{P,T}(\delta^2m_{_{LR}}^{odd}){\cal Z}_{_{{\tilde N}^c}}^P]_{32}\over m_{_{A_7^0}}^2
-m_{_{A_8^0}}^2} &1
\end{array}\right)\;.
%%%%%%%%%%%%%%%%%%%%%%%%%%%%%%%%%%%%%%%%%%%%%%%%%%%%
\label{app1-CP-odd2c}
\end{eqnarray}

In the interaction basis $H^{0,T}=(H_{_u}^0,\;H_{_d}^0,\;\tilde{\nu}_{_{L_I}},\;\tilde{\nu}_{_{R_J}}),\;\;(I,\;J=1,\;2,\;3)$,
elements of the matrix $A_{_{CPE}}^{(0)}$  are respectively written as
\begin{eqnarray}
%%%%%%%%%%%%%%%%%%%%%%%%%%%%%%%%%%%%%%%%%%%%%%%%%%%%
&&\Big(A_{_{CPE}}^{(0)}\Big)_{1I^\prime}=
-{g_1^2+g_2^2\over4}\upsilon_{_u}\upsilon_{_{L_{I^\prime}}}
-{1\over\sqrt{2}}\sum\limits_{\alpha=1}^3(A_{_N})_{I^\prime\alpha}\upsilon_{_{N_\alpha}}
-\upsilon_{_u}\sum\limits_{\alpha=1}^3\upsilon_{_{L_\alpha}}\Big(Y_{_N}Y_{_N}^T\Big)_{\alpha I^\prime}
\;,\nonumber\\
%%%%%%%%%%%%%%%%%%%%%%%%%%%%%%%%%%%%%%%%%%%%%%%%%%%%
&&\Big(A_{_{CPE}}^{(0)}\Big)_{1(3+J^\prime)}=
-{1\over\sqrt{2}}\sum\limits_{\alpha=1}^3(A_{_N})_{\alpha J^\prime}\upsilon_{_{L_\alpha}}
-\upsilon_{_u}\sum\limits_{\alpha=1}^3\upsilon_{_{N_\alpha}}\Big(Y_{_N}^TY_{_N}\Big)_{\alpha J^\prime}
\;,\nonumber\\
%%%%%%%%%%%%%%%%%%%%%%%%%%%%%%%%%%%%%%%%%%%%%%%%%%%%
&&\Big(A_{_{CPE}}^{(0)}\Big)_{2I^\prime}=
{g_1^2+g_2^2\over4}\upsilon_{_d}\upsilon_{_{L_{I^\prime}}}
-{\mu\zeta_{_{I^\prime}}\over\sqrt{2}}
\;,\nonumber\\
%%%%%%%%%%%%%%%%%%%%%%%%%%%%%%%%%%%%%%%%%%%%%%%%%%%%
&&\Big(A_{_{CPE}}^{(0)}\Big)_{2(3+J^\prime)}=
-{\mu\over\sqrt{2}}\sum\limits_{\alpha=1}^3\upsilon_{_{L_\alpha}}\Big(Y_{_N}\Big)_{\alpha J^\prime}
\;,\nonumber\\
%%%%%%%%%%%%%%%%%%%%%%%%%%%%%%%%%%%%%%%%%%%%%%%%%%%%
\label{app1-CP-even1}
\end{eqnarray}
and elements of the symmetric matrix $M_{_S}^2$ are similarly given as
\begin{eqnarray}
%%%%%%%%%%%%%%%%%%%%%%%%%%%%%%%%%%%%%%%%%%%%%%%%%%%%
&&\Big(M_{_{S}}^2\Big)_{II^\prime}=
\Big(m_{_{\tilde L}}^2\Big)_{II^\prime}-{g_1^2+g_2^2\over8}\Big(2\upsilon_{_u}^2-\upsilon_{_{\rm EW}}^2\Big)
\delta_{II^\prime}-{g_{_{BL}}^2\over2}\Big(\upsilon_{_N}^2-\upsilon_{_{\rm EW}}^2
+\upsilon_{_{\rm SM}}^2\Big)\delta_{II^\prime}
\nonumber\\
&&\hspace{2.2cm}
+\Delta T_{_L}\delta_{II^\prime}+\Big[\delta^2m_{_{LL}}^{even}\Big]_{II^\prime}
\;,\nonumber\\
%%%%%%%%%%%%%%%%%%%%%%%%%%%%%%%%%%%%%%%%%%%%%%%%%%%%
&&\Big(M_{_{S}}^2\Big)_{I(3+J^\prime)}=\Big[\delta^2m_{_{LR}}^{even}\Big]_{IJ^\prime}
\;,\nonumber\\
%%%%%%%%%%%%%%%%%%%%%%%%%%%%%%%%%%%%%%%%%%%%%%%%%%%%
&&\Big(M_{_{S}}^2\Big)_{(3+J)(3+J^\prime)}=
\Big(m_{_{{\tilde N}^c}}^2\Big)_{JJ^\prime}+{g_{_{BL}}^2\over2}\Big[\Big(\upsilon_{_N}^2-\upsilon_{_{\rm EW}}^2
+\upsilon_{_{\rm SM}}^2\Big)\delta_{JJ^\prime}+\upsilon_{_{N_J}}\upsilon_{_{N_{J^\prime}}}\Big]
\nonumber\\
&&\hspace{3.2cm}
+\Delta T_{_N}\delta_{JJ^\prime}+\Big[\delta^2m_{_{RR}}^{even}\Big]_{JJ^\prime}\;,
%%%%%%%%%%%%%%%%%%%%%%%%%%%%%%%%%%%%%%%%%%%%%%%%%%%%
\label{app1-CP-even2}
\end{eqnarray}
with
\begin{eqnarray}
%%%%%%%%%%%%%%%%%%%%%%%%%%%%%%%%%%%%%%%%%%%%%%%%%%%%
&&\Big[\delta^2m_{_{LL}}^{even}\Big]_{II^\prime}={g_1^2+g_2^2\over4}\upsilon_{_{L_I}}\upsilon_{_{L_{I^\prime}}}
-{1\over2}\zeta_{_I}\zeta_{_{I^\prime}}
+g_{_{BL}}^2\upsilon_{_{L_I}}\upsilon_{_{L_{I^\prime}}}
-{1\over2}\Big(Y_{_N}Y_{_N}^T\Big)_{II^\prime}\upsilon_{_u}^2
\;,\nonumber\\
%%%%%%%%%%%%%%%%%%%%%%%%%%%%%%%%%%%%%%%%%%%%%%%%%%%%
&&\Big[\delta^2m_{_{LR}}^{even}\Big]_{IJ^\prime}=
-g_{_{BL}}^2\upsilon_{_{L_I}}\upsilon_{_{N_{J^\prime}}}
-{\upsilon_{_u}\over\sqrt{2}}\Big(A_{_N}\Big)_{IJ^\prime}+{\mu\upsilon_{_d}\over\sqrt{2}}\Big(Y_{_N}\Big)_{IJ^\prime}
\nonumber\\
&&\hspace{2.8cm}
-{1\over2}\varepsilon_{_N}^2\Big(Y_{_N}\Big)_{IJ^\prime}
-{1\over2}\zeta_{_I}\sum\limits_{\alpha=1}^3\upsilon_{_{L_\alpha}}
\Big(Y_{_N}\Big)_{\alpha J^\prime}
\;,\nonumber\\
%%%%%%%%%%%%%%%%%%%%%%%%%%%%%%%%%%%%%%%%%%%%%%%%%%%%
&&\Big[\delta^2m_{_{RR}}^{even}\Big]_{JJ^\prime}=
-{1\over2}\sum\limits_{\alpha,\beta=1}^3\Big(Y_{_N}^T\Big)_{J\alpha}\upsilon_{_{L_\alpha}}
\upsilon_{_{L_\beta}}\Big(Y_{_N}\Big)_{\beta J^\prime}
-{1\over2}\upsilon_{_u}^2\Big(Y_{_N}^TY_{_N}\Big)_{JJ^\prime}\;.
%%%%%%%%%%%%%%%%%%%%%%%%%%%%%%%%%%%%%%%%%%%%%%%%%%%%
\label{app1-CP-even2a}
\end{eqnarray}
Furthermore,
\begin{eqnarray}
%%%%%%%%%%%%%%%%%%%%%%%%%%%%%%%%%%%%%%%%%%%%%%%%%%%%
&&Z_{_{\tilde L}}\simeq\left(\begin{array}{ccc}1&{[\delta^2m_{_{LL}}^{even}]_{12}\over m_{_{H_4^0}}^2-m_{_{H_3^0}}^2}&
{[\delta^2m_{_{LL}}^{even}]_{13}\over m_{_{H_5^0}}^2-m_{_{H_3^0}}^2}\\
{[\delta^2m_{_{LL}}^{even}]_{12}\over m_{_{H_4^0}}^2-m_{_{H_3^0}}^2} &1&
{[\delta^2m_{_{LL}}^{even}]_{23}\over m_{_{H_5^0}}^2-m_{_{H_4^0}}^2}\\
{[\delta^2m_{_{LL}}^{even}]_{13}\over m_{_{H_3^0}}^2-m_{_{H_5^0}}^2} &
{[\delta^2m_{_{LL}}^{even}]_{23}\over m_{_{H_4^0}}^2-m_{_{H_5^0}}^2}  &1
\end{array}\right)\;.
%%%%%%%%%%%%%%%%%%%%%%%%%%%%%%%%%%%%%%%%%%%%%%%%%%%%
\label{app1-CP-even2b}
\end{eqnarray}

\section{Radiative corrections to the CP-even Higgs mass squared matrix\label{app1a}}
\indent\indent
The radiative corrections from quark sector are formulated as
\begin{eqnarray}
%%%%%%%%%%%%%%%%%%%%%%%%%%%%%%%%%%%%%%%%%%%%%%%%%%%%
&&\Delta_{11}^B={3G_{_F}m_{_t}^4\over2\sqrt{2}\pi^2\sin^2\beta}
\Big\{\ln{m_{_{\tilde{t}_1}}^2m_{_{\tilde{t}_2}}^2\over m_{_t}^4}
+{2A_{_t}(A_{_t}-\mu\cot\beta)\over m_{_{\tilde{t}_1}}^2-m_{_{\tilde{t}_2}}^2}
\ln{m_{_{\tilde{t}_1}}^2\over m_{_{\tilde{t}_2}}^2}
\nonumber\\
&&\hspace{1.2cm}
+{A_{_t}^2(A_{_t}-\mu\cot\beta)^2\over(m_{_{\tilde{t}_1}}^2-m_{_{\tilde{t}_2}}^2)^2}
g(m_{_{\tilde{t}_1}}^2,m_{_{\tilde{t}_2}}^2)\Big\}
\nonumber\\
&&\hspace{1.2cm}
+{3G_{_F}m_{_b}^4\over2\sqrt{2}\pi^2\cos^2\beta}\cdot
{\mu^2(A_{_b}-\mu\tan\beta)^2\over(m_{_{\tilde{b}_1}}^2-m_{_{\tilde{b}_2}}^2)^2}
g(m_{_{\tilde{b}_1}}^2,m_{_{\tilde{b}_2}}^2)
\nonumber\\
&&\hspace{1.2cm}
+{3G_{_F}m_{_t}^4\over16\sqrt{2}\pi^4\sin^2\beta}\Big({3\pi\alpha m_{_t}^2\over4s_{_{\rm W}}^2m_{_{\rm W}}^2}
-8\pi\alpha_{_S}\Big)\ln{m_{_{\tilde{t}_1}}^2m_{_{\tilde{t}_2}}^2\over m_{_t}^4}
\nonumber\\
&&\hspace{1.2cm}\times
\Big\{\ln{m_{_{\tilde{t}_1}}^2m_{_{\tilde{t}_2}}^2\over m_{_t}^4}
+{4(A_{_t}-\mu\cot\beta)^2\over m_{_{\tilde{t}_1}}m_{_{\tilde{t}_2}}}
\Big[1-{(A_{_t}-\mu\cot\beta)^2\over12m_{_{\tilde{t}_1}}m_{_{\tilde{t}_2}}}\Big]\Big\}
\nonumber\\
&&\hspace{1.2cm}
+{G_{_F}m_{_b}^4\over64\sqrt{2}\pi^4\sin^2\beta\cos^4\beta}\Big({9\pi\alpha m_{_b}^2\over
s_{_{\rm W}}^2m_{_{\rm W}}^2\cos^2\beta}-{5\pi\alpha m_{_t}^2\over s_{_{\rm W}}^2m_{_{\rm W}}^2}
-32\pi\alpha_{_S}\Big)\ln{m_{_{\tilde{b}_1}}^2m_{_{\tilde{b}_2}}^2\over m_{_b}^4}
\;,\nonumber\\
%%%%%%%%%%%%%%%%%%%%%%%%%%%%%%%%%%%%%%%%%%%%%%%%%%%%%%%%%%
&&\Delta_{12}^B=-{3G_{_F}m_{_t}^4\over2\sqrt{2}\pi^2\sin^2\beta}\cdot
{\mu(A_{_t}-\mu\cot\beta)\over m_{_{\tilde{t}_1}}^2-m_{_{\tilde{t}_2}}^2}
\Big\{\ln{m_{_{\tilde{t}_1}}^2\over m_{_{\tilde{t}_2}}^2}
\nonumber\\
&&\hspace{1.2cm}
+{A_{_t}(A_{_t}-\mu\cot\beta)
\over m_{_{\tilde{t}_1}}^2-m_{_{\tilde{t}_2}}^2}g(m_{_{\tilde{t}_1}}^2,m_{_{\tilde{t}_2}}^2)\Big\}
\nonumber\\
&&\hspace{1.2cm}
-{3G_{_F}m_{_b}^4\over2\sqrt{2}\pi^2\cos^2\beta}\cdot
{\mu(A_{_b}-\mu\tan\beta)\over m_{_{\tilde{b}_1}}^2-m_{_{\tilde{b}_2}}^2}
\Big\{\ln{m_{_{\tilde{b}_1}}^2\over m_{_{\tilde{b}_2}}^2}
\nonumber\\
&&\hspace{1.2cm}
+{A_{_b}(A_{_b}-\mu\tan\beta)
\over m_{_{\tilde{b}_1}}^2-m_{_{\tilde{b}_2}}^2}g(m_{_{\tilde{b}_1}}^2,m_{_{\tilde{b}_2}}^2)\Big\}
\;,\nonumber\\
%%%%%%%%%%%%%%%%%%%%%%%%%%%%%%%%%%%%%%%%%%%%%%%%%%%%%%%%%%
&&\Delta_{22}^B=
{3G_{_F}m_{_t}^4\over2\sqrt{2}\pi^2\sin^2\beta}\cdot
{\mu^2(A_{_t}-\mu\cot\beta)^2\over(m_{_{\tilde{t}_1}}^2-m_{_{\tilde{t}_2}}^2)^2}
g(m_{_{\tilde{t}_1}}^2,m_{_{\tilde{t}_2}}^2)
\nonumber\\
&&\hspace{1.2cm}
+{3G_{_F}m_{_b}^4\over2\sqrt{2}\pi^2\cos^2\beta}
\Big\{\ln{m_{_{\tilde{b}_1}}^2m_{_{\tilde{b}_2}}^2\over m_{_b}^4}
+{2A_{_b}(A_{_b}-\mu\tan\beta)\over m_{_{\tilde{b}_1}}^2-m_{_{\tilde{b}_2}}^2}
\ln{m_{_{\tilde{b}_1}}^2\over m_{_{\tilde{b}_2}}^2}
\nonumber\\
&&\hspace{1.2cm}
+{A_{_b}^2(A_{_b}-\mu\tan\beta)^2\over(m_{_{\tilde{b}_1}}^2-m_{_{\tilde{b}_2}}^2)^2}
g(m_{_{\tilde{b}_1}}^2,m_{_{\tilde{b}_2}}^2)\Big\}\;.
%%%%%%%%%%%%%%%%%%%%%%%%%%%%%%%%%%%%%%%%%%%%%%%%%%%%
\label{app1a-rad-B}
\end{eqnarray}
Similarly the contributions from lepton sector to the mass-squared matrix of CP-even Higgs are
\begin{eqnarray}
%%%%%%%%%%%%%%%%%%%%%%%%%%%%%%%%%%%%%%%%%%%%%%%%%%%%
&&\Delta_{11}^\tau={G_{_F}m_{_\tau}^4\over2\sqrt{2}\pi^2\cos^2\beta}\cdot
{\mu^2(A_{_\tau}-\mu\tan\beta)^2\over(m_{_{\tilde{\tau}_1}}^2-m_{_{\tilde{\tau}_2}}^2)^2}
g(m_{_{\tilde{\tau}_1}}^2,m_{_{\tilde{\tau}_2}}^2)
\;,\nonumber\\
%%%%%%%%%%%%%%%%%%%%%%%%%%%%%%%%%%%%%%%%%%%%%%%%%%%%%%%%%%
&&\Delta_{12}^\tau=-{G_{_F}m_{_\tau}^4\over2\sqrt{2}\pi^2\cos^2\beta}\cdot
{\mu(A_{_\tau}-\mu\tan\beta)\over m_{_{\tilde{\tau}_1}}^2-m_{_{\tilde{\tau}_2}}^2}
\Big\{\ln{m_{_{\tilde{\tau}_1}}^2\over m_{_{\tilde{\tau}_2}}^2}
\nonumber\\
&&\hspace{1.2cm}
+{A_{_\tau}(A_{_\tau}-\mu\tan\beta)
\over m_{_{\tilde{\tau}_1}}^2-m_{_{\tilde{\tau}_2}}^2}g(m_{_{\tilde{\tau}_1}}^2,m_{_{\tilde{\tau}_2}}^2)\Big\}
\;,\nonumber\\
%%%%%%%%%%%%%%%%%%%%%%%%%%%%%%%%%%%%%%%%%%%%%%%%%%%%%%%%%%
&&\Delta_{22}^\tau={G_{_F}m_{_\tau}^4\over2\sqrt{2}\pi^2\cos^2\beta}
\Big\{\ln{m_{_{\tilde{\tau}_1}}^2m_{_{\tilde{\tau}_2}}^2\over m_{_\tau}^4}
+{2A_{_\tau}(A_{_\tau}-\mu\tan\beta)\over m_{_{\tilde{\tau}_1}}^2-m_{_{\tilde{\tau}_2}}^2}
\ln{m_{_{\tilde{\tau}_1}}^2\over m_{_{\tilde{\tau}_2}}^2}
\nonumber\\
&&\hspace{1.2cm}
+{A_{_\tau}^2(A_{_\tau}-\mu\tan\beta)^2\over(m_{_{\tilde{\tau}_1}}^2-m_{_{\tilde{\tau}_2}}^2)^2}
g(m_{_{\tilde{\tau}_1}}^2,m_{_{\tilde{\tau}_2}}^2)\Big\}\;,
%%%%%%%%%%%%%%%%%%%%%%%%%%%%%%%%%%%%%%%%%%%%%%%%%%%%
\label{app1a-rad-L}
\end{eqnarray}
where
\begin{eqnarray}
%%%%%%%%%%%%%%%%%%%%%%%%%%%%%%%%%%%%%%%%%%%%%%%%%%%%
&&g(x,y)=2-{x+y\over x-y}\ln{x\over y}\;.
%%%%%%%%%%%%%%%%%%%%%%%%%%%%%%%%%%%%%%%%%%%%%%%%%%%%
\label{app1a-rad-fun}
\end{eqnarray}

\section{The couplings between CP-even Higgs and charged scalars\label{app2}}
\indent\indent
The interaction between the CP-even Higgs and charged Higgs is written as
\begin{eqnarray}
&&{\cal L}_{_{H^0H^+H^-}}=\sum\limits_{i=1}^{10}\sum\limits_{\alpha,\beta=1}^8
\xi_{_{i\beta\alpha}}^{H^\pm}H_i^0H_\beta^-H_\alpha^+
\label{app2-1a}
\end{eqnarray}
with
\begin{eqnarray}
&&\xi_{_{i\beta\alpha}}^{H^\pm}={g_1^2+g_2^2\over4}\upsilon_{_{\rm EW}}{\cal R}_{1i}
A^{H^\pm}_{\alpha\beta}
+{g_2^2\upsilon_{_{\rm EW}}\over4}\Big\{\Big[\Big(Z_{_{H_0}}\Big)_{1i}\Big(Z_{_{CH}}\Big)_{2\alpha}
+\Big(Z_{_{H_0}}\Big)_{2i}\Big(Z_{_{CH}}\Big)_{1\alpha}\Big]{\cal V}_\beta^*
\nonumber\\
&&\hspace{1.4cm}
+\Big[\Big(Z_{_{H_0}}\Big)_{1i}\Big(Z_{_{CH}}^\dagger\Big)_{\beta2}
+\Big(Z_{_{H_0}}\Big)_{2i}\Big(Z_{_{CH}}^\dagger\Big)_{\beta1}\Big]{\cal V}_\alpha
+2{\cal R}_{1i}\sum\limits_{I=1}^3\Big(Z_{_{CH}}\Big)_{(2+I)\alpha}\Big(Z_{_{CH}}^\dagger\Big)_{\beta(2+I)}
\nonumber\\
&&\hspace{1.4cm}
+2\sum\limits_{I=1}^3{\upsilon_{_{L_I}}\over\upsilon_{_{\rm EW}}}\Big(Z_{_{H_0}}\Big)_{(2+I)i}
\Big[\Big(Z_{_{CH}}\Big)_{1\alpha}\Big(Z_{_{CH}}^\dagger\Big)_{\beta1}
-\Big(Z_{_{CH}}\Big)_{2\alpha}\Big(Z_{_{CH}}^\dagger\Big)_{\beta2}\Big]
\nonumber\\
&&\hspace{1.4cm}
+\sum\limits_{I=1}^3\Big[\sum\limits_{J=1}^3
{\upsilon_{_{L_I}}\over\upsilon_{_{\rm EW}}}\Big(Z_{_{H_0}}\Big)_{(2+J)i}
\Big[\Big(Z_{_{CH}}\Big)_{(2+I)\alpha}\Big(Z_{_{CH}}^\dagger\Big)_{\beta(2+J)}
+\Big(Z_{_{CH}}\Big)_{(2+J)\alpha}\Big(Z_{_{CH}}^\dagger\Big)_{\beta(2+I)}\Big]
\nonumber\\
&&\hspace{1.4cm}
+\Big[{\upsilon_{_{L_I}}\over\upsilon_{_{\rm EW}}}\Big(Z_{_{H_0}}\Big)_{1i}
+{\upsilon_{_u}\over\upsilon_{_{\rm EW}}}\Big(Z_{_{H_0}}\Big)_{(2+I)i}\Big]
\Big[\Big(Z_{_{CH}}\Big)_{(2+I)\alpha}\Big(Z_{_{CH}}^\dagger\Big)_{\beta1}
+\Big(Z_{_{CH}}\Big)_{1\alpha}\Big(Z_{_{CH}}^\dagger\Big)_{\beta(2+I)}\Big]
\nonumber\\
&&\hspace{1.4cm}
+\Big[{\upsilon_{_{L_I}}\over\upsilon_{_{\rm EW}}}\Big(Z_{_{H_0}}\Big)_{2i}
+{\upsilon_{_d}\over\upsilon_{_{\rm EW}}}\Big(Z_{_{H_0}}\Big)_{(2+I)i}\Big]
\Big[\Big(Z_{_{CH}}\Big)_{(2+I)\alpha}\Big(Z_{_{CH}}^\dagger\Big)_{\beta2}
+\Big(Z_{_{CH}}\Big)_{2\alpha}\Big(Z_{_{CH}}^\dagger\Big)_{\beta(2+I)}\Big]\Big]\Big\}
\nonumber\\
&&\hspace{1.4cm}
+{g_1^2\over2}\upsilon_{_{\rm EW}}{\cal R}_{1i}\sum\limits_{I=1}^3
\Big(Z_{_{CH}}\Big)_{(5+I)\alpha}\Big(Z_{_{CH}}^\dagger\Big)_{\beta(5+I)}
\nonumber\\
&&\hspace{1.4cm}
+g_{_{BL}}^2\upsilon_t{\cal R}_{2i}\sum\limits_{I=1}^3
\Big[\Big(Z_{_{CH}}\Big)_{(2+I)\alpha}\Big(Z_{_{CH}}^\dagger\Big)_{\beta(2+I)}
-\Big(Z_{_{CH}}\Big)_{(5+I)\alpha}\Big(Z_{_{CH}}^\dagger\Big)_{\beta(5+I)}\Big]
\nonumber\\
&&\hspace{1.4cm}
+\Big\{{\mu^*\over\sqrt{2}}\sum\limits_{I,J}^3\Big(Z_{_{CH}}\Big)_{(2+I)\alpha}
\Big(Y_{_N}\Big)_{IJ}\Big(Z_{_{H_0}}\Big)_{(5+J)i}\Big(Z_{_{CH}}^\dagger\Big)_{\beta2}
\nonumber\\
&&\hspace{1.4cm}
+{\mu\over\sqrt{2}}\sum\limits_{I,J}^3\Big(Z_{_{H_0}}^T\Big)_{i(5+I)}
\Big(Y_{_N}^\dagger\Big)_{IJ}\Big(Z_{_{CH}}^*\Big)_{(2+J)\beta}\Big(Z_{_{CH}}\Big)_{2\alpha}
\nonumber\\
&&\hspace{1.4cm}
+{1\over2}\sum\limits_{I,J,I^\prime}^3\Big[\zeta_{_{I^\prime}}^*
\Big(Z_{_{CH}}\Big)_{(2+I^\prime)\alpha}\Big(Z_{_{CH}}^\dagger\Big)_{\beta(2+J)}
\Big(Y_{_N}\Big)_{JI}\Big(Z_{_{H_0}}\Big)_{(5+I)i}
\nonumber\\
&&\hspace{1.4cm}
+\Big(Z_{_{H_0}}^T\Big)_{i(5+I)}\Big(Y_{_N}^\dagger\Big)_{IJ}\Big(Z_{_{CH}}\Big)_{(2+J)\alpha}
\Big(Z_{_{CH}}^\dagger\Big)_{\beta(2+I^\prime)}\zeta_{_{I^\prime}}\Big\}
\nonumber\\
&&\hspace{1.4cm}
+\Big\{-{\mu^*\over\sqrt{2}}\sum\limits_{I,J}^3\Big(Z_{_{H_0}}^T\Big)_{i1}
\Big(Z_{_{CH}}^\dagger\Big)_{\beta(2+I)}\Big(Y_{_E}\Big)_{IJ}\Big(Z_{_{CH}}\Big)_{(5+J)\alpha}
\nonumber\\
&&\hspace{1.4cm}
-{\mu\over\sqrt{2}}\sum\limits_{I,J}^3\Big(Z_{_{CH}}^\dagger\Big)_{\beta(5+I)}
\Big(Y_{_E}^\dagger\Big)_{IJ}\Big(Z_{_{CH}}\Big)_{(2+J)\alpha}\Big(Z_{_{H_0}}\Big)_{1i}
\nonumber\\
&&\hspace{1.4cm}
+{1\over2}\sum\limits_{I,J}^3\sum\limits_{I^\prime,J^\prime}^3\Big[\Big(Z_{_{CH}}^\dagger\Big)_{\beta(5+I)}
\Big(Y_{_E}^\dagger\Big)_{IJ}\upsilon_{_{L_J}}\Big(Z_{_{H_0}}^T\Big)_{i(2+J^\prime)}
\Big(Y_{_E}\Big)_{J^\prime I^\prime}\Big(Z_{_{CH}}\Big)_{(5+I^\prime)\alpha}
\nonumber\\
&&\hspace{1.4cm}
+\Big(Z_{_{CH}}^\dagger\Big)_{\beta(5+I)}
\Big(Y_{_E}^\dagger\Big)_{IJ}\Big(Z_{_{H_0}}\Big)_{(2+J)i}\upsilon_{_{L_{J^\prime}}}
\Big(Y_{_E}\Big)_{J^\prime I^\prime}\Big(Z_{_{CH}}\Big)_{(5+I^\prime)\alpha}\Big]\Big\}
\nonumber\\
&&\hspace{1.4cm}
+\sum\limits_{I,J}^3\Big\{\upsilon_{_u}\Big(Z_{_{CH}}^\dagger\Big)_{\beta(5+I)}
\Big(Y_{_E}^\dagger Y_{_E}\Big)_{IJ}\Big(Z_{_{CH}}\Big)_{(5+J)\alpha}\Big(Z_{_{H_0}}\Big)_{2i}
\nonumber\\
&&\hspace{1.4cm}
+{1\over2}\Big[\upsilon_{_{N_I}}\Big(Y_{_N}^\dagger Y_{_N}\Big)_{IJ}\Big(Z_{_{H_0}}\Big)_{(5+J)i}
+\Big(Z_{_{H_0}}^T\Big)_{i(5+I)}\Big(Y_{_N}^\dagger Y_{_N}\Big)_{IJ}\upsilon_{_{N_J}}\Big]
\Big(Z_{_{CH}}^\dagger\Big)_{\beta1}\Big(Z_{_{CH}}\Big)_{1\alpha}
\nonumber\\
&&\hspace{1.4cm}
+{1\over2}\Big(Z_{_{CH}}^\dagger\Big)_{\beta(5+I)}\Big(Y_{_E}^\dagger Y_{_N}\Big)_{IJ}\upsilon_{_{N_J}}
\Big(Z_{_{H_0}}\Big)_{2i}\Big[\Big(Z_{_{CH}}\Big)_{1\alpha}+\Big(Z_{_{CH}}\Big)_{2\alpha}\Big]
\nonumber\\
&&\hspace{1.4cm}
+{1\over2}\upsilon_{_{\rm EW}}\Big(Z_{_{CH}}^\dagger\Big)_{\beta(5+I)}\Big(Y_{_E}^\dagger Y_{_N}\Big)_{IJ}
\Big(Z_{_{H_0}}\Big)_{(5+J)i}{\cal V}_\alpha
\nonumber\\
&&\hspace{1.4cm}
+{1\over2}\Big[\Big(Z_{_{CH}}^\dagger\Big)_{\beta1}+\Big(Z_{_{CH}}^\dagger\Big)_{\beta2}\Big]
\upsilon_{_{N_I}}\Big(Y_{_N}^\dagger Y_{_E}\Big)_{IJ}\Big(Z_{_{CH}}^\dagger\Big)_{(5+J)\alpha}
\Big(Z_{_{H_0}}\Big)_{2i}
\nonumber\\
&&\hspace{1.4cm}
+{1\over2}\upsilon_{_{\rm EW}}{\cal V}_\beta^*\Big(Z_{_{H_0}}^T\Big)_{i(5+I)}
\Big(Y_{_N}^\dagger Y_{_E}\Big)_{IJ}\Big(Z_{_{CH}}^\dagger\Big)_{(5+J)\alpha}\Big\}
\nonumber\\
&&\hspace{1.4cm}
+\sum\limits_{I,J}^3\Big\{\upsilon_{_d}\Big(Z_{_{H_0}}\Big)_{2i}
\Big(Z_{_{CH}}^\dagger\Big)_{\beta(2+I)}\Big(Y_{_E}^*Y_{_E}^T\Big)_{IJ}\Big(Z_{_{CH}}\Big)_{(2+J)\alpha}
\nonumber\\
&&\hspace{1.4cm}
+{1\over2}\Big[\upsilon_{_{L_I}}\Big(Y_{_E}Y_{_E}^\dagger\Big)_{IJ}\Big(Z_{_{H_0}}\Big)_{(2+J)i}
+\Big(Z_{_{H_0}}^T\Big)_{i(2+I)}\Big(Y_{_E}Y_{_E}^\dagger\Big)_{IJ}\upsilon_{_{L_J}}\Big]
\Big(Z_{_{CH}}^\dagger\Big)_{\beta2}\Big(Z_{_{CH}}\Big)_{2\alpha}
\nonumber\\
&&\hspace{1.4cm}
-{1\over2}\upsilon_{_d}\Big(Z_{_{CH}}^\dagger\Big)_{\beta(2+I)}\Big(Y_{_E}Y_{_E}^\dagger\Big)_{IJ}
\Big(Z_{_{H_0}}\Big)_{(2+J)i}\Big(Z_{_{CH}}\Big)_{2\alpha}
\nonumber\\
&&\hspace{1.4cm}
-{1\over2}\upsilon_{_d}\Big(Z_{_{CH}}^\dagger\Big)_{\beta2}\Big(Z_{_{H_0}}^T\Big)_{i(2+I)}
\Big(Y_{_E}^*Y_{_E}^T\Big)_{IJ}\Big(Z_{_{CH}}\Big)_{(2+J)\alpha}
\nonumber\\
&&\hspace{1.4cm}
-{1\over2}\Big(Z_{_{H_0}}\Big)_{2i}\Big(Z_{_{CH}}^\dagger\Big)_{\beta(2+I)}\Big(Y_{_E}Y_{_E}^\dagger\Big)_{IJ}
\upsilon_{_{L_J}}\Big(Z_{_{CH}}\Big)_{2\alpha}
\nonumber\\
&&\hspace{1.4cm}
-{1\over2}\Big(Z_{_{H_0}}\Big)_{2i}\Big(Z_{_{CH}}^\dagger\Big)_{\beta2}\upsilon_{_{L_I}}
\Big(Y_{_E}^*Y_{_E}^T\Big)_{IJ}\Big(Z_{_{CH}}\Big)_{(2+J)\alpha}\Big\}
\nonumber\\
&&\hspace{1.4cm}
+\sum\limits_{I,J}^3\Big\{\upsilon_{_u}\Big(Z_{_{H_0}}\Big)_{1i}
\Big(Z_{_{CH}}^\dagger\Big)_{\beta(2+I)}\Big(Y_{_N}^*Y_{_N}^T\Big)_{IJ}\Big(Z_{_{CH}}\Big)_{(2+J)\alpha}
\nonumber\\
&&\hspace{1.4cm}
+{1\over2}\Big[\upsilon_{_{L_I}}\Big(Y_{_N}Y_{_N}^\dagger\Big)_{IJ}\Big(Z_{_{H_0}}\Big)_{(2+J)i}
+\Big(Z_{_{H_0}}^T\Big)_{i(2+I)}\Big(Y_{_N}Y_{_N}^\dagger\Big)_{IJ}\upsilon_{_{L_J}}\Big]
\Big(Z_{_{CH}}^\dagger\Big)_{\beta1}\Big(Z_{_{CH}}\Big)_{1\alpha}
\nonumber\\
&&\hspace{1.4cm}
-{1\over2}\upsilon_{_u}\Big(Z_{_{CH}}^\dagger\Big)_{\beta(2+I)}\Big(Y_{_N}Y_{_N}^\dagger\Big)_{IJ}
\Big(Z_{_{H_0}}\Big)_{(2+J)i}\Big(Z_{_{CH}}\Big)_{1\alpha}
\nonumber\\
&&\hspace{1.4cm}
-{1\over2}\upsilon_{_u}\Big(Z_{_{CH}}^\dagger\Big)_{\beta1}\Big(Z_{_{H_0}}^T\Big)_{i(2+I)}
\Big(Y_{_N}^*Y_{_N}^T\Big)_{IJ}\Big(Z_{_{CH}}\Big)_{(2+J)\alpha}
\nonumber\\
&&\hspace{1.4cm}
-{1\over2}\Big(Z_{_{H_0}}\Big)_{1i}\Big(Z_{_{CH}}^\dagger\Big)_{\beta(2+I)}\Big(Y_{_N}Y_{_N}^\dagger\Big)_{IJ}
\upsilon_{_{L_J}}\Big(Z_{_{CH}}\Big)_{1\alpha}
\nonumber\\
&&\hspace{1.4cm}
-{1\over2}\Big(Z_{_{H_0}}\Big)_{1i}\Big(Z_{_{CH}}^\dagger\Big)_{\beta1}\upsilon_{_{L_I}}
\Big(Y_{_N}^*Y_{_N}^T\Big)_{IJ}\Big(Z_{_{CH}}\Big)_{(2+J)\alpha}
\nonumber\\
&&\hspace{1.4cm}
+{1\over\sqrt{2}}\sum\limits_{I,J}^3\Big\{\Big(Z_{_{CH}}^\dagger\Big)_{\beta(2+I)}\Big(A_{_E}Y_{_E}\Big)_{IJ}
\Big(Z_{_{CH}}\Big)_{(5+J)\alpha}\Big(Z_{_{H_0}}\Big)_{2i}
\nonumber\\
&&\hspace{1.4cm}
+\Big(Z_{_{CH}}^\dagger\Big)_{\beta(5+I)}\Big(A_{_E}Y_{_E}\Big)_{IJ}^\dagger
\Big(Z_{_{CH}}\Big)_{(2+J)\alpha}\Big(Z_{_{H_0}}\Big)_{2i}
\nonumber\\
&&\hspace{1.4cm}
+\Big(Z_{_{CH}}^\dagger\Big)_{\beta(2+I)}\Big(A_{_N}Y_{_N}\Big)_{IJ}
\Big(Z_{_{H_0}}\Big)_{(5+J)i}\Big(Z_{_{CH}}\Big)_{1\alpha}
\nonumber\\
&&\hspace{1.4cm}
+\Big(Z_{_{CH}}^\dagger\Big)_{\beta1}\Big(Z_{_{H_0}}^T\Big)_{i(5+I)}\Big(A_{_N}Y_{_N}\Big)_{IJ}^\dagger
\Big(Z_{_{CH}}\Big)_{(2+J)\alpha}\Big\}\;.
\label{app2-1}
\end{eqnarray}
Where
\begin{eqnarray}
&&{\cal R}_{1i}={\upsilon_{_u}\over\upsilon_{_{\rm EW}}}\Big(Z_{_{H_0}}\Big)_{1i}-{\upsilon_{_d}\over\upsilon_{_{\rm EW}}}
\Big(Z_{_{H_0}}\Big)_{2i}-\sum\limits_{J=1}^3{\upsilon_{_{L_J}}\over\upsilon_{_{\rm EW}}}\Big(Z_{_{H_0}}\Big)_{(2+J)i}
\;,\nonumber\\
&&{\cal R}_{2i}=\sum\limits_{J=1}^3\Big[{\upsilon_{_{L_J}}\over\upsilon_t}\Big(Z_{_{H_0}}\Big)_{(2+J)i}
-{\upsilon_{_{N_J}}\over\upsilon_t}\Big(Z_{_{H_0}}\Big)_{(5+J)i}\Big]
\;,\nonumber\\
&&{\cal V}_\alpha={\upsilon_{_u}\over\upsilon_{_{\rm EW}}}\Big(Z_{_{CH}}\Big)_{2\alpha}
+{\upsilon_{_d}\over\upsilon_{_{\rm EW}}}\Big(Z_{_{CH}}\Big)_{1\alpha}
\;,\nonumber\\
&&A^{H^\pm}_{\alpha\beta}=\Big(Z_{_{CH}}\Big)_{1\alpha}\Big(Z_{_{CH}}\Big)_{1\beta}
-\Big(Z_{_{CH}}\Big)_{2\alpha}\Big(Z_{_{CH}}\Big)_{2\beta}
-\sum\limits_{I=1}^3\Big(Z_{_{CH}}\Big)_{(2+I)\alpha}\Big(Z_{_{CH}}\Big)_{(2+I)\beta}
\label{app2-1b}
\end{eqnarray}
and $Y_{_E}={\rm diag}(Y_{_e},\;Y_{_\mu},\;Y_{_\tau})$.

The couplings between CP-even Higgs and stops are formulated as
\begin{eqnarray}
&&{\cal L}_{_{H_i^0\tilde{t}_\alpha\tilde{t}_\beta^*}}=\sum\limits_{i=1}^{8}
\sum\limits_{\alpha,\beta}^2\xi_{i\beta\alpha}^{\tilde{t}}H_i^0\tilde{t}_\beta^*
\tilde{t}_\alpha\;,
\label{app2-2a}
\end{eqnarray}
with
\begin{eqnarray}
&&\xi_{i\beta\alpha}^{\tilde{t}}=
-{e^2\over4s_{_{\rm W}}^2c_{_{\rm W}}^2}\upsilon_{_{\rm EW}}{\cal R}_{1i}
\Big\{\Big[1-\Big(1+{\cal Y}_{_q}\Big)s_{_{\rm W}}^2\Big]
\Big(Z_{_{\tilde{U}^3}}^\dagger\Big)_{\beta 1}\Big(Z_{_{\tilde{U}^3}}\Big)_{1\alpha}
\nonumber\\
&&\hspace{1.4cm}
-{\cal Y}_{_u}s_{_{\rm W}}^2\Big(Z_{_{\tilde{U}^3}}^\dagger\Big)_{\beta2}
\Big(Z_{_{\tilde{U}^3}}\Big)_{2\alpha}\Big\}
\nonumber\\
&&\hspace{1.4cm}
-{g_{_{BL}}^2\over3}\upsilon_t{\cal R}_{2i}
\Big\{\Big(Z_{_{\tilde{U}^3}}^\dagger\Big)_{\beta1}\Big(Z_{_{\tilde{U}^3}}\Big)_{1\alpha}
-\Big(Z_{_{\tilde{U}^3}}^\dagger\Big)_{\beta2}\Big(Z_{_{\tilde{U}^3}}\Big)_{2\alpha}\Big\}
\nonumber\\
&&\hspace{1.4cm}
+\Big\{{\mu^*Y_{_t}\over\sqrt{2}}\Big(Z_{_{H_0}}\Big)_{2i}
\Big(Z_{_{\tilde{U}^3}}^\dagger\Big)_{\beta2}\Big(Z_{_{\tilde{U}^3}}\Big)_{1\alpha}
+{\mu Y_{_t}\over\sqrt{2}}\Big(Z_{_{H_0}}\Big)_{2i}
\Big(Z_{_{\tilde{U}^3}}^\dagger\Big)_{\beta 1}
\Big(Z_{_{\tilde{U}^3}}\Big)_{2\alpha}
\nonumber\\
&&\hspace{1.4cm}
+{Y_{_t}^*\over2}\sum\limits_{I,J}^3\Big[\upsilon_{_{L_{I}}}
\Big(Y_{_N}\Big)_{IJ}\Big(Z_{_{H_0}}\Big)_{(5+J)i}
+\Big(Z_{_{H_0}}^T\Big)_{i(5+I)}\Big(Y_{_N}\Big)_{IJ}
\upsilon_{_{N_{J}}}\Big]
\nonumber\\
&&\hspace{1.4cm}\times
\Big(Z_{_{\tilde{U}^3}}^\dagger\Big)_{\beta1}\Big(Z_{_{\tilde{U}^3}}\Big)_{2\alpha}
\nonumber\\
&&\hspace{1.4cm}
+{Y_{_t}\over2}\sum\limits_{I,J}^3\Big[\upsilon_{_{L_{I}}}
\Big(Y_{_N}^\dagger\Big)_{I J}\Big(Z_{_{H_0}}\Big)_{(5+J)i}
+\Big(Z_{_{H_0}}^T\Big)_{i(5+I)}\Big(Y_{_N}^\dagger\Big)_{IJ}
\upsilon_{_{N_{J}}}\Big]
\nonumber\\
&&\hspace{1.4cm}\times
\Big(Z_{_{\tilde{U}^3}}^\dagger\Big)_{\beta2}\Big(Z_{_{\tilde{U}^3}}\Big)_{1\alpha}\Big\}
+\upsilon_{_u}|Y_{_t}|^2\Big(Z_{_{H_0}}\Big)_{1i}\Big(Z_{_{\tilde{U}^3}}^\dagger\Big)_{\beta2}
\Big(Z_{_{\tilde{U}^3}}\Big)_{2\alpha}
\nonumber\\
&&\hspace{1.4cm}
-{A_{_t}Y_{_t}\over\sqrt{2}}\Big(Z_{_{H_0}}\Big)_{1i}
\Big(Z_{_{\tilde{U}^3}}^\dagger\Big)_{\beta2}\Big(Z_{_{\tilde{U}^3}}\Big)_{1\alpha}
-{A_{_t}^*Y_{_t}^*\over\sqrt{2}}\Big(Z_{_{H_0}}\Big)_{1i}
\Big(Z_{_{\tilde{U}^3}}^\dagger\Big)_{\beta1}\Big(Z_{_{\tilde{U}^3}}\Big)_{2\alpha}
\label{app2-2}
\end{eqnarray}
where ${\cal Y}_{_q}=1/3\;,\;\;{\cal Y}_{_u}=-4/3$.

The couplings between CP-even Higgs and sbottoms are formulated as
\begin{eqnarray}
&&{\cal L}_{_{H_i^0\tilde{b}_\alpha\tilde{b}_\beta^*}}=\sum\limits_{i=1}^{8}
\sum\limits_{\alpha,\beta}^2\xi_{i\beta\alpha}^{\tilde{b}}H_i^0
\tilde{b}_\beta^*\tilde{b}_\alpha\;,
\label{app2-3a}
\end{eqnarray}
with
\begin{eqnarray}
&&\xi_{i\beta\alpha}^{\tilde{b}}=
{e^2\over4s_{_{\rm W}}^2c_{_{\rm W}}^2}\upsilon_{_{\rm EW}}{\cal R}_{1i}
\Big\{\Big[1-\Big(1-{\cal Y}_{_q}\Big)s_{_{\rm W}}^2\Big]
\Big(Z_{_{\tilde{D}^3}}^\dagger\Big)_{\beta1}\Big(Z_{_{\tilde{D}^3}}\Big)_{1\alpha}
\nonumber\\
&&\hspace{1.4cm}
+{\cal Y}_{_d}s_{_{\rm W}}^2\Big(Z_{_{\tilde{D}^3}}^\dagger\Big)_{\beta2}
\Big(Z_{_{\tilde{D}^3}}\Big)_{2\alpha}\Big\}
\nonumber\\
&&\hspace{1.4cm}
-{g_{_{BL}}^2\over3}\upsilon_t{\cal R}_{2i}\Big\{\Big(Z_{_{\tilde{D}^3}}^\dagger\Big)_{\beta1}
\Big(Z_{_{\tilde{D}^3}}\Big)_{1\alpha}
-\Big(Z_{_{\tilde{D}^3}}^\dagger\Big)_{\beta2}\Big(Z_{_{\tilde{D}^3}}\Big)_{2\alpha}\Big\}
\nonumber\\
&&\hspace{1.4cm}
-\Big\{{\mu Y_{_b}^*\over\sqrt{2}}\Big(Z_{_{H_0}}\Big)_{1i}
\Big(Z_{_{\tilde{D}^3}}^\dagger\Big)_{\beta1}\Big(Z_{_{\tilde{D}^3}}\Big)_{2\alpha}
+{\mu^*Y_{_b}\over\sqrt{2}}\Big(Z_{_{H_0}}\Big)_{1i}
\Big(Z_{_{\tilde{D}^3}}^\dagger\Big)_{\beta2}\Big(Z_{_{\tilde{D}^3}}\Big)_{1\alpha}\Big\}
\nonumber\\
&&\hspace{1.4cm}
+\upsilon_{_d}|Y_{_b}|^2\Big(Z_{_{H_0}}\Big)_{2i}
\Big(Z_{_{\tilde{D}^3}}^\dagger\Big)_{\beta2}
\Big(Z_{_{\tilde{D}^3}}\Big)_{2\alpha}
\nonumber\\
&&\hspace{1.4cm}
+{A_{_b}Y_{_b}\over\sqrt{2}}\Big(Z_{_{H_0}}\Big)_{2i}
\Big(Z_{_{\tilde{D}^3}}^\dagger\Big)_{\beta2}\Big(Z_{_{\tilde{D}^3}}\Big)_{1\alpha}
+{A_{_b}^*Y_{_b}^*\over\sqrt{2}}\Big(Z_{_{H_0}}\Big)_{2i}
\Big(Z_{_{\tilde{D}^3}}^\dagger\Big)_{\beta1}\Big(Z_{_{\tilde{D}^3}}\Big)_{2\alpha}
\label{app2-3}
\end{eqnarray}
with ${\cal Y}_{_d}=2/3$.

The couplings among CP-even Higgs and charginos/charged leptons are written as
\begin{eqnarray}
&&{\cal L}_{_{H^0\chi^\pm e}}=\sum\limits_{i=1}^{2}\sum\limits_{\alpha=1}^2\sum\limits_{I=1}^3
[\delta\xi_1^{m}]_{_{iI\alpha}}H_i^0\overline{e_{_I}}\omega_-\chi_{_\alpha}^-
+\sum\limits_{i=1}^{2}\sum\limits_{\alpha=1}^2\sum\limits_{I=1}^3
[\delta\xi_2^{m}]_{_{i\alpha I}}H_i^0\overline{\chi_{_\alpha}^-}\omega_-e_{_I}
\nonumber\\
&&\hspace{1.6cm}
+\sum\limits_{K=1}^{3}\sum\limits_{\alpha=1}^2\sum\limits_{I=1}^3\Big\{
{\Big(U_-\Big)_{2\alpha}\over\sqrt{2}}\Big(Y_{_E}V_{_R}\Big)_{KI}
+[\delta^2\xi_1^{m}]_{_{(2+K)I\alpha}}\Big\}H_{_{2+K}}^0\overline{e_{_I}}\omega_-\chi_{_\alpha}^-
\nonumber\\
&&\hspace{1.6cm}
+\sum\limits_{K=1}^{3}\sum\limits_{\alpha=1}^2\sum\limits_{I=1}^3
\Big[{e\over\sqrt{2}s_{_{\rm W}}}\Big(U_+\Big)_{1\alpha}(V_{_L})_{KI}
+[\delta^2\xi_2^m]_{_{(2+K)\alpha I}}\Big]H_{_{2+K}}^0\overline{\chi_{_\alpha}^-}\omega_-e_{_I}
\nonumber\\
&&\hspace{1.6cm}
+\sum\limits_{K=1}^{3}\sum\limits_{\alpha=1}^2\sum\limits_{I=1}^3[\delta\xi_2^{m}]_{_{(5+K)\alpha I}}
H_{_{5+K}}^0\overline{\chi_{_\alpha}^-}\omega_-e_{_I}
\nonumber\\
&&\hspace{1.6cm}
-\sum\limits_{i=1}^{2}\sum\limits_{I,J}^3\Big\{{\sqrt{2}m_{_{e^I}}\over\upsilon_{_d}}\Big(Z_{_R}\Big)_{2i}\delta_{IJ}
-[\delta^2\xi_1^{e}]_{_{iIJ}}\Big\}H_i^0\overline{e_{_I}}\omega_-e_{_J}
\nonumber\\
&&\hspace{1.6cm}
+\sum\limits_{K=1}^{3}\sum\limits_{I,J}^3[\delta\xi_1^{e}]_{_{(2+K)IJ}}H_{_{2+K}}^0\overline{e_{_I}}\omega_-e_{_J}+h.c.
\label{chargino-lepton10}
\end{eqnarray}
with
\begin{eqnarray}
%%%%%%%%%%%%%%%%%%%%%%%%%%%%%%%%%%%%%%%%%%%%%%%%%%%%%%%%%%%%%%%%%%%%%%%%%%
&&[\delta\xi_1^{m}]_{_{iI\alpha}}\simeq{e\over\sqrt{2}s_{_{\rm W}}}\Big[
\Big(Z_{_{R}}\Big)_{2i}\Big(\xi_{_R}V_{_R}\Big)_{1I}\Big(U_-\Big)_{2\alpha}
+\Big(Z_{_{R}}\Big)_{1i}\Big(\xi_{_R}V_{_R}\Big)_{2I}\Big(U_-\Big)_{1\alpha}\Big]
\nonumber\\
&&\hspace{2.0cm}
+{1\over\sqrt{2}}\Big(Z_{_{R}}\Big)_{2i}\Big(U_-^T\xi_{_L}Y_{_E}V_{_R}\Big)_{\alpha I}
\nonumber\\
&&\hspace{2.0cm}
-{1\over\sqrt{2}}\Big(U_-\Big)_{2\alpha}\sum\limits_{I^\prime=1}^3
{(Z_{_R}^TA_{_{CPE}})_{iI^\prime}\Big(Y_{_E}V_{_R}\Big)_{I^\prime I}
\over m_{_{H_{2+I^\prime}^0}}^2-m_{_{H_{i}^0}}^2}
\;,\nonumber\\
%%%%%%%%%%%%%%%%%%%%%%%%%%%%%%%%%%%%%%%%%%%%%%%%%%%%%%%%%%%%%%%%%%%%%%%%%%
&&[\delta^2\xi_1^{m}]_{_{(2+K)I\alpha}}\simeq
-{1\over\sqrt{2}}\Big(U_-\Big)_{2\alpha}
\sum\limits_{I^\prime\neq K}^3{(\delta^2m_{_{LL}}^{even})_{I^\prime K}\over m_{_{H_{2+I^\prime}^0}}^2-m_{_{H_{_{2+K}}^0}}^2}
\Big(Y_{_E}V_{_R}\Big)_{I^\prime I}
\nonumber\\
&&\hspace{3.0cm}
+{e\over\sqrt{2}s_{_{\rm W}}}\sum\limits_{\beta=1}^2\Big[
{(Z_{_R})_{2\beta}(Z_{_R}A_{_{CPE}})_{\beta K}\over m_{_{H_{_{2+K}}^0}}^2-m_{_{H_{\beta}^0}}^2}
\Big(\xi_{_R}V_{_R}\Big)_{1I}\Big(U_-\Big)_{2\alpha}
\nonumber\\
&&\hspace{3.0cm}
+{(Z_{_R})_{1\beta}(Z_{_R}A_{_{CPE}})_{\beta K}\over m_{_{H_{_{2+K}}^0}}^2-m_{_{H_{\beta}^0}}^2}
\Big(\xi_{_R}V_{_R}\Big)_{2I}\Big(U_-\Big)_{1\alpha}\Big]
\nonumber\\
&&\hspace{3.0cm}
+{1\over\sqrt{2}}\Big(U_-^T\xi_{_L}Y_{_E}V_{_R}\Big)_{\alpha I}\sum\limits_{\beta=1}^2
{(Z_{_R})_{2\beta}(Z_{_R}A_{_{CPE}})_{\beta K}\over m_{_{H_{_{2+K}}^0}}^2-m_{_{H_{\beta}^0}}^2}
\;,\nonumber\\
%%%%%%%%%%%%%%%%%%%%%%%%%%%%%%%%%%%%%%%%%%%%%%%%%%%%%%%%%%%%%%%%%%%%%%%%%%
&&\Big[\delta\xi_2^{m}\Big]_{_{i\alpha I}}\simeq{e\over\sqrt{2}s_{_{\rm W}}}\Big[
\Big(Z_{_R}\Big)_{2i}\Big(\xi_{_L}V_{_L}\Big)_{2I}\Big(U_+\Big)_{1\alpha}
+\Big(Z_{_R}\Big)_{1i}\Big(\xi_{_L}V_{_L}\Big)_{1I}\Big(U_+\Big)_{2\alpha}
\nonumber\\
&&\hspace{2.0cm}
-\Big(U_+\Big)_{1\alpha}\sum\limits_{I^\prime=1}^3
{(Z_{_R}^TA_{_{CPE}})_{iI^\prime}(V_{_L})_{I^\prime I}\over m_{_{H_{2+I^\prime}^0}}^2-m_{_{H_{i}^0}}^2}\Big]
+{1\over\sqrt{2}}\Big(Z_{_R}\Big)_{2i}\Big(V_{_L}^TY_{_E}\xi_{_R}^TU_+\Big)_{I\alpha}
\;,\nonumber\\
%%%%%%%%%%%%%%%%%%%%%%%%%%%%%%%%%%%%%%%%%%%%%%%%%%%%%%%%%%%%%%%%%%%%%%%%%%
&&[\delta^2\xi_2^m]_{_{(2+K)\alpha I}}\simeq{e\over\sqrt{2}s_{_{\rm W}}}\Big\{
\Big(U_+\Big)_{1\alpha}{(\delta^2m_{LL}^{even})_{IK}\over m_{_{H_{2+K}^0}}^2
-m_{_{H_{2+I}^0}}^2}(1-\delta_{_{IK}})
\nonumber\\
&&\hspace{2.2cm}
-\cos\beta\Big(U_+\Big)_{1\alpha}\Big(\xi_{_L}V_{_L}\Big)_{2I}{\upsilon_{_{L_{K}}}\over\upsilon_{_{\rm EW}}}
+\sin\beta\Big(U_-\Big)_{1\alpha}\Big(\xi_{_R}V_{_R}\Big)_{2I}{\upsilon_{_{L_{K}}}\over\upsilon_{_{\rm EW}}}\Big\}
\nonumber\\
&&\hspace{2.2cm}
-{\cos\beta\over\sqrt{2}}{\upsilon_{_{L_{K}}}\over\upsilon_{_{\rm EW}}}\Big(V_{_L}^TY_{_E}\xi_{_R}^TU_+\Big)_{I\alpha}
-{1\over\sqrt{2}}\Big(U_+\Big)_{2\alpha}\sum\limits_{I^\prime=1}^3{\upsilon_{_{N_{I^\prime}}}\upsilon_{_{L_{K}}}\over
\upsilon_{_N}^2}\Big(Y_{_N}V_{_L}\Big)_{I^\prime I}
\;,\nonumber\\
%%%%%%%%%%%%%%%%%%%%%%%%%%%%%%%%%%%%%%%%%%%%%%%%%%%%%%%%%%%%%%%%%%%%%%%%%%
&&\Big[\delta\xi_2^{m}\Big]_{_{(5+K)\alpha I}}\simeq
{\Big(U_+\Big)_{2\alpha}\over\sqrt{2}}\Big(({\cal Z}_{_{\tilde N^c}}^T)Y_{_N}V_{_L}\Big)_{KI}
\;,\nonumber\\
%%%%%%%%%%%%%%%%%%%%%%%%%%%%%%%%%%%%%%%%%%%%%%%%%%%%%%%%%%%%%%%%%%%%%%%%%%
&&\Big[\delta^2\xi_1^{e}\Big]_{_{iIJ}}\simeq
{e\over\sqrt{2}s_{_{\rm W}}}\Big(\xi_{_R}V_{_R}\Big)_{1I}\sum\limits_{I^\prime=1}^3\Big(V_{_L}\Big)_{I^\prime J}
{(A_{_{CPE}}^TZ_{_R})_{I^\prime i}\over m_{_{H_{2+I^\prime}^0}}^2-m_{_{H_{i}^0}}^2}
\nonumber\\
&&\hspace{2.0cm}
+{1\over\sqrt{2}}\Big(\xi_{_L}V_{_L}\Big)_{2J}\sum\limits_{I^\prime,J^\prime}^3\Big(Y_{_E}V_{_R}\Big)_{I^\prime I}
{(A_{_{CPE}}^TZ_{_R})_{I^\prime i}\over m_{_{H_{2+I^\prime}^0}}^2-m_{_{H_{i}^0}}^2}
\;,\nonumber\\
%%%%%%%%%%%%%%%%%%%%%%%%%%%%%%%%%%%%%%%%%%%%%%%%%%%%%%%%%%%%%%%%%%%%%%%%%%
&&\Big[\delta\xi_1^{e}\Big]_{_{(2+K)IJ}}\simeq{e\over\sqrt{2}s_{_{\rm W}}}\Big(\xi_{_R}V_{_R}\Big)_{1I}\Big(V_{_L}\Big)_{KJ}
\nonumber\\
&&\hspace{2.5cm}
+{1\over\sqrt{2}}\Big(\xi_{_L}V_{_L}\Big)_{2J}\sum\limits_{J^\prime=1}^3
\Big(Y_{_E}\Big)_{KJ^\prime}\Big(V_{_R}\Big)_{J^\prime I}\;.
%%%%%%%%%%%%%%%%%%%%%%%%%%%%%%%%%%%%%%%%%%%%%%%%%%%%%%%%%%%%
\label{chargino-lepton11}
\end{eqnarray}

The couplings between CP-odd Higgs and charginos/charged leptons are
\begin{eqnarray}
%%%%%%%%%%%%%%%%%%%%%%%%%%%%%%%%%%%%%%%%%%%%%%%%%%%%%%%%%%%%
&&{\cal L}_{_{A^0\chi^\pm e}}=
\sum\limits_{i=1}^{3}\sum\limits_{\alpha=1}^2\sum\limits_{I=1}^3
A_i^0\Big\{[\delta\eta_1^m]_{_{iI\alpha}}\overline{e_{_I}}\omega_-\chi_{_\alpha}^-
-[\delta\eta_1^m]_{_{iI\alpha}}^*\overline{\chi_{_\alpha}^-}\omega_+e_{_I}\Big\}
\nonumber\\
&&\hspace{1.6cm}
+\sum\limits_{K=1}^{3}\sum\limits_{\alpha=1}^2\sum\limits_{I=1}^3
A_{_{3+K}}^0\Big\{\Big[{1\over\sqrt{2}}\Big(U_-\Big)_{2\alpha}\Big(Y_{_E}V_{_R}\Big)_{KI}
+[\delta^2\eta_1^m]_{_{(3+K)I\alpha}}\Big]\overline{e_{_I}}\omega_-\chi_{_\alpha}^-
\nonumber\\
&&\hspace{1.6cm}
-\Big[{1\over\sqrt{2}}\Big(U_-\Big)_{2\alpha}^*\Big(Y_{_E}V_{_R}\Big)_{KI}^*
+[\delta^2\eta_1^m]_{_{(3+K)I\alpha}}^*\Big]\overline{\chi_{_\alpha}^-}\omega_+e_{_I}\Big\}
\nonumber\\
&&\hspace{1.6cm}
+\sum\limits_{i=1}^{2}\sum\limits_{\alpha=1}^2\sum\limits_{I=1}^3
A_{6+i}^0\Big\{[\delta\eta_1^m]_{_{(6+i)I\alpha}}\overline{e_{_I}}\omega_-\chi_{_\alpha}^-
-\Big[\delta\eta_1^m\Big]_{_{(6+i)I\alpha}}^*\overline{\chi_{_\alpha}^-}\omega_+e_{_I}\Big\}
\nonumber\\
&&\hspace{1.6cm}
+\sum\limits_{i=1}^{3}\sum\limits_{\alpha=1}^2\sum\limits_{I=1}^3
A_i^0\Big\{[\delta\eta_2^m]_{_{i\alpha I}}\overline{\chi_{_\alpha}^-}\omega_-e_{_I}
-[\delta\eta_2^m]_{_{i\alpha I}}^*\overline{e_{_I}}\omega_+\chi_{_\alpha}^-\Big\}
\nonumber\\
&&\hspace{1.6cm}
+\sum\limits_{K=1}^{3}\sum\limits_{\alpha=1}^2\sum\limits_{I=1}^3
A_{_{3+K}}^0\Big\{\Big[{e\over\sqrt{2}s_{_{\rm W}}}\Big(U_+\Big)_{1\alpha}(V_{_L})_{KI}
+[\delta^2\eta_2^m]_{_{(3+K)\alpha I}}\Big]\overline{\chi_{_\alpha}^-}\omega_-e_{_I}
\nonumber\\
&&\hspace{1.6cm}
-\Big[{e\over\sqrt{2}s_{_{\rm W}}}\Big(U_+\Big)_{1\alpha}^*(V_{_L})_{KI}^*
+[\delta^2\eta_2^m]_{_{(3+K)\alpha I}}^*\Big]\overline{e_{_I}}\omega_+\chi_{_\alpha}^-\Big\}
\nonumber\\
&&\hspace{1.6cm}
+\sum\limits_{i=1}^{2}\sum\limits_{\alpha=1}^2\sum\limits_{I=1}^3
A_{_{6+i}}^0\Big\{[\delta\eta_2^m]_{_{(6+i)\alpha I}}\overline{\chi_{_\alpha}^-}\omega_-e_{_I}
-[\delta\eta_2^m]_{_{(6+i)\alpha I}}^*\overline{e_{_I}}\omega_+\chi_{_\alpha}^-\Big\}
\nonumber\\
&&\hspace{1.6cm}
+\sum\limits_{I,J}^3A_1^0\Big\{\Big[{\sqrt{2}m_{_{e^I}}\over\upsilon_{_d}}\cos\beta\delta_{IJ}+[\delta^2\eta^e]_{_{1IJ}}\Big]
\overline{e_{_I}}\omega_-e_{_J}
\nonumber\\
&&\hspace{1.6cm}
-\Big[{\sqrt{2}m_{_{e^I}}\over\upsilon_{_d}}\cos\beta\delta_{IJ}+[\delta^2\eta^e]_{_{1IJ}}^*\Big]\overline{e_{_J}}\omega_+e_{_I}\Big\}
\nonumber\\
&&\hspace{1.6cm}
+\sum\limits_{I,J}^3A_3^0\Big\{\Big[-{\sqrt{2}m_{_{e^I}}\over\upsilon_{_d}}\sin\beta\delta_{IJ}+[\delta^2\eta^e]_{_{3IJ}}\Big]
\overline{e_{_I}}\omega_-e_{_J}
\nonumber\\
&&\hspace{1.6cm}
-\Big[-{\sqrt{2}m_{_{e^I}}\over\upsilon_{_d}}\sin\beta\delta_{IJ}+[\delta^2\eta^e]_{_{3IJ}}^*\Big]\overline{e_{_J}}\omega_+e_{_I}\Big\}
\nonumber\\
&&\hspace{1.6cm}
+\sum\limits_{K=1}^{3}\sum\limits_{I,J}^3A_{_{3+K}}^0\Big\{[\delta\eta^e]_{_{(3+K)IJ}}\overline{e_{_I}}\omega_-e_{_J}
-[\delta\eta^e]_{_{(3+K)IJ}}^*\overline{e_{_J}}\omega_+e_{_I}\Big\}
%%%%%%%%%%%%%%%%%%%%%%%%%%%%%%%%%%%%%%%%%%%%%%%%%%%%%%%%%%%%
\label{chargino-lepton12}
\end{eqnarray}
with
\begin{eqnarray}
%%%%%%%%%%%%%%%%%%%%%%%%%%%%%%%%%%%%%%%%%%%%%%%%%%%%%%%%%%%%%%%%%%%%%%%%%%
&&\Big[\delta\eta_1^m\Big]_{1I\alpha}\simeq{e\over\sqrt{2}s_{_{\rm W}}}\Big[
-\cos\beta\Big(\xi_{_R}V_{_R}\Big)_{1I}\Big(U_-\Big)_{2\alpha}
+\sin\beta\Big(\xi_{_R}V_{_R}\Big)_{2I}\Big(U_-\Big)_{1\alpha}\Big]
\nonumber\\
&&\hspace{2.2cm}
-{1\over\sqrt{2}}\cos\beta\sum\limits_{I^\prime=1}^3\Big(Y_{_E}V_{_R}\Big)_{I^\prime I}
\Big(\xi_{_L}^TU_-\Big)_{I^\prime\alpha}
-{1\over\sqrt{2}}\Big(U_-\Big)_{2\alpha}\sum\limits_{I^\prime=1}^3{\upsilon_{_{L_{I^\prime}}}\over\upsilon_{_{\rm EW}}}
\Big(Y_{_E}V_{_R}\Big)_{I^\prime I}
\;,\nonumber\\
%%%%%%%%%%%%%%%%%%%%%%%%%%%%%%%%%%%%%%%%%%%%%%%%%%%%%%%%%%%%%%%%%%%%%%%%%%
&&\Big[\delta\eta_1^m\Big]_{2I\alpha}\simeq0
\;,\nonumber\\
%%%%%%%%%%%%%%%%%%%%%%%%%%%%%%%%%%%%%%%%%%%%%%%%%%%%%%%%%%%%%%%%%%%%%%%%%%
&&\Big[\delta\eta_1^m\Big]_{3I\alpha}\simeq{e\over\sqrt{2}s_{_{\rm W}}}\Big[
\sin\beta\Big(\xi_{_R}V_{_R}\Big)_{1I}\Big(U_-\Big)_{2\alpha}
+\cos\beta\Big(\xi_{_R}V_{_R}\Big)_{2I}\Big(U_-\Big)_{1\alpha}\Big]
\nonumber\\
&&\hspace{2.2cm}
+{1\over\sqrt{2}}\sin\beta\sum\limits_{I^\prime=1}^3\Big(Y_{_E}V_{_R}\Big)_{I^\prime I}
\Big(\xi_{_L}^TU_-\Big)_{I^\prime\alpha}
\nonumber\\
&&\hspace{2.2cm}
-{1\over\sqrt{2}}\Big(U_-\Big)_{2\alpha}\sum\limits_{I^\prime=1}^3
{(\delta^2m_{HL}^{odd})_{I^\prime}\over m_{_{A_{3+I^\prime}^0}}^2-m_{_{A_3^0}}^2}
\Big(Y_{_E}V_{_R}\Big)_{I^\prime I}
\;,\nonumber\\
%%%%%%%%%%%%%%%%%%%%%%%%%%%%%%%%%%%%%%%%%%%%%%%%%%%%%%%%%%%%%%%%%%%%%%%%%%
&&[\delta^2\eta_1^m]_{_{(3+K)I\alpha}}\simeq
{1\over\sqrt{2}}\Big(U_-\Big)_{2\alpha}
\sum\limits_{I^\prime\neq K}^3{(\delta^2m_{LL}^{odd})_{I^\prime K}\over m_{_{A_{3+K}^0}}^2-m_{_{A_{3+I^\prime}^0}}^2}
\Big(Y_{_E}V_{_R}\Big)_{I^\prime I}
\;,\nonumber\\
%%%%%%%%%%%%%%%%%%%%%%%%%%%%%%%%%%%%%%%%%%%%%%%%%%%%%%%%%%%%%%%%%%%%%%%%%%
&&[\delta\eta_1^m]_{_{(6+i)I\alpha}}\simeq
{1\over\sqrt{2}}\Big(U_-\Big)_{2\alpha}\sum\limits_{I^\prime=1}^3
{(\delta^2m_{LR}^{odd}{\cal Z}_{_{\tilde{N}^c}}^P)_{I^\prime i}\over
m_{_{A_{_{6+i}}^0}}^2-m_{_{A_{3+I^\prime}^0}}^2}\Big(Y_{_E}V_{_R}\Big)_{I^\prime I}\;,\;\;(i=1,\;2)
\;,\nonumber\\
%%%%%%%%%%%%%%%%%%%%%%%%%%%%%%%%%%%%%%%%%%%%%%%%%%%%%%%%%%%%%%%%%%%%%%%%%%
&&[\delta\eta_2^m]_{_{1\alpha I}}\simeq
{e\over\sqrt{2}s_{_{\rm W}}}\Big[-\cos\beta\Big(\xi_{_L}V_{_L}\Big)_{2I}\Big(U_+\Big)_{1\alpha}
+\sin\beta\Big(\xi_{_L}V_{_L}\Big)_{1I}\Big(U_+\Big)_{2\alpha}
\nonumber\\
&&\hspace{2.2cm}
-\Big(U_+\Big)_{1\alpha}\sum\limits_{I^\prime=1}^3{\upsilon_{_{L_{I^\prime}}}\over\upsilon_{_{\rm EW}}}
\Big(V_{_L}\Big)_{I^\prime I}\Big]
-{\cos\beta\over\sqrt{2}}\Big(V_{_L}^TY_{_E}\xi_{_R}^TU_+\Big)_{I\alpha}
\;,\nonumber\\
%%%%%%%%%%%%%%%%%%%%%%%%%%%%%%%%%%%%%%%%%%%%%%%%%%%%%%%%%%%%%%%%%%%%%%%%%%
&&[\delta\eta_2^m]_{_{2\alpha I}}\simeq
-{1\over\sqrt{2}\upsilon_{_N}}\Big(U_+\Big)_{2\alpha}\sum\limits_{I^\prime=1}^3\zeta_{_{I^\prime}}\Big(V_{_L}\Big)_{I^\prime I}
\;,\nonumber\\
%%%%%%%%%%%%%%%%%%%%%%%%%%%%%%%%%%%%%%%%%%%%%%%%%%%%%%%%%%%%%%%%%%%%%%%%%%
&&[\delta\eta_2^m]_{_{3\alpha I}}\simeq
{e\over\sqrt{2}s_{_{\rm W}}}\Big[\sin\beta\Big(\xi_{_L}V_{_L}\Big)_{2I}\Big(U_+\Big)_{1\alpha}
+\cos\beta\Big(\xi_{_L}V_{_L}\Big)_{1I}\Big(U_+\Big)_{2\alpha}
\nonumber\\
&&\hspace{2.2cm}
-\Big(U_+\Big)_{1\alpha}\sum\limits_{I^\prime=1}^3
{(\delta^2m_{HL}^{odd})_{(3+I^\prime)}\over m_{_{A_{I^\prime}^0}}^2-m_{_{A_3^0}}^2}
\Big(V_{_L}\Big)_{I^\prime I}\Big]+{\sin\beta\over\sqrt{2}}\Big(V_{_L}^TY_{_E}\xi_{_R}^TU_+\Big)_{I\alpha}
\;,\nonumber\\
%%%%%%%%%%%%%%%%%%%%%%%%%%%%%%%%%%%%%%%%%%%%%%%%%%%%%%%%%%%%%%%%%%%%%%%%%%
&&[\delta^2\eta_2^m]_{_{(3+K)\alpha I}}\simeq{e\over\sqrt{2}s_{_{\rm W}}}\Big\{
\Big(U_+\Big)_{1\alpha}{(\delta^2m_{LL}^{odd})_{I K}\over m_{_{A_{3+K}^0}}^2
-m_{_{A_{3+I}^0}}^2}(1-\delta_{IK})
\nonumber\\
&&\hspace{2.2cm}
-\cos\beta\Big(U_+\Big)_{1\alpha}\Big(\xi_{_L}V_{_L}\Big)_{2I}{\upsilon_{_{L_{K}}}\over\upsilon_{_{\rm EW}}}
+\sin\beta\Big(U_-\Big)_{1\alpha}\Big(\xi_{_R}V_{_R}\Big)_{2I}{\upsilon_{_{L_{K}}}\over\upsilon_{_{\rm EW}}}\Big\}
\nonumber\\
&&\hspace{2.2cm}
-{\cos\beta\over\sqrt{2}}{\upsilon_{_{L_{K}}}\over\upsilon_{_{\rm EW}}}\Big(V_{_L}^TY_{_E}\xi_{_R}^TU_+\Big)_{I\alpha}
-{1\over\sqrt{2}}\Big(U_+\Big)_{2\alpha}\sum\limits_{I^\prime=1}^3{\upsilon_{_{N_{I^\prime}}}\upsilon_{_{L_{K}}}\over
\upsilon_{_N}^2}\Big(Y_{_N}V_{_L}\Big)_{I^\prime I}
\;,\nonumber\\
%%%%%%%%%%%%%%%%%%%%%%%%%%%%%%%%%%%%%%%%%%%%%%%%%%%%%%%%%%%%%%%%%%%%%%%%%%
&&[\delta\eta_2^m]_{_{(6+i)\alpha I}}\simeq
{1\over\sqrt{2}}\Big(U_+\Big)_{2\alpha}\sum\limits_{I^\prime=1}^3
{(\delta^2m_{LR}^{odd}{\cal Z}_{_{\tilde{N}^c}}^P)_{I^\prime i}\over
m_{_{A_{6+i}^0}}^2-m_{_{A_{3+I^\prime}^0}}^2}\Big(V_{_R}\Big)_{I^\prime I}\;,\;\;(i=1,\;2)
\;,\nonumber\\
%%%%%%%%%%%%%%%%%%%%%%%%%%%%%%%%%%%%%%%%%%%%%%%%%%%%%%%%%%%%%%%%%%%%%%%%%%
&&[\delta^2\eta^e]_{_{1IJ}}\simeq-{e\over\sqrt{2}s_{_{\rm W}}}\Big(\xi_{_R}V_{_R}\Big)_{1I}\sum\limits_{I^\prime=1}^3
{\upsilon_{_{L_{I^\prime}}}\over\upsilon_{_{\rm EW}}}\Big(V_{_L}\Big)_{I^\prime J}
-{1\over\sqrt{2}}\Big(\xi_{_L}V_{_L}\Big)_{2J}\sum\limits_{I^\prime=1}^3
{\upsilon_{_{L_{I^\prime}}}\over\upsilon_{_{\rm EW}}}\Big(Y_{_E}V_{_R}\Big)_{I^\prime I}
\;,\nonumber\\
%%%%%%%%%%%%%%%%%%%%%%%%%%%%%%%%%%%%%%%%%%%%%%%%%%%%%%%%%%%%%%%%%%%%%%%%%%
&&[\delta^2\eta^e]_{_{3IJ}}\simeq
{e\over\sqrt{2}s_{_{\rm W}}}\Big(\xi_{_R}V_{_R}\Big)_{1I}\sum\limits_{I^\prime=1}^3
{(\delta^2m_{HL}^{odd})_{(I^\prime)}\over m_{_{A_{3+I^\prime}^0}}^2-m_{_{A_3^0}}^2}\Big(V_{_L}\Big)_{I^\prime J}
\nonumber\\
&&\hspace{2.2cm}
+{1\over\sqrt{2}}\Big(\xi_{_L}V_{_L}\Big)_{2J}\sum\limits_{I^\prime=1}^3
{(\delta^2m_{HL}^{odd})_{(I^\prime)}\over m_{_{A_{3+I^\prime}^0}}^2-m_{_{A_3^0}}^2}
\Big(Y_{_E}V_{_R}\Big)_{I^\prime I}\;.
%%%%%%%%%%%%%%%%%%%%%%%%%%%%%%%%%%%%%%%%%%%%%%%%%%%%%%%%%%%%
\label{chargino-lepton13}
\end{eqnarray}

The couplings between charged Higgs, charged leptons and neutralinos/neutrinos are
formulated as
\begin{eqnarray}
%%%%%%%%%%%%%%%%%%%%%%%%%%%%%%%%%%%%%%%%%%%%%%%%%%%%%%%%%%%%
&&{\cal L}_{_{H^-\overline{e}\chi^0}}=
\sum\limits_{I=1}^3\sum\limits_{\alpha=1}^{5}G^-\overline{e_I}\Big\{\Big[\cos\beta(Z_{_\nu}^TY_{_E}V_{_R})_{_{\alpha I}}
+\sum\limits_{b=1}^5[\delta^2\zeta_G^L]_{_{1Ib}}(Z_{_\nu})_{b\alpha}\Big]\omega_-
\nonumber\\
&&\hspace{1.5cm}
+\sum\limits_{b=1}^5[\delta^2\zeta_G^R]_{_{1Ib}}(Z_{_\nu})_{b\alpha}^*\omega_+\Big\}\nu_\alpha
\nonumber\\
&&\hspace{1.5cm}
+\sum\limits_{I=1}^3\sum\limits_{\alpha=1}^{5}H^-\overline{e_I}\Big\{\Big[-\sin\beta(Z_{_\nu}^TY_{_E}V_{_R})_{_{\alpha I}}
+\sum\limits_{b=1}^5[\delta^2\zeta_H^L]_{_{2Ib}}(Z_{_\nu})_{b\alpha}\Big]\omega_-
\nonumber\\
&&\hspace{1.5cm}
+\sum\limits_{b=1}^5[\delta^2\zeta_H^R]_{_{2Ib}}(Z_{_\nu})_{b\alpha}^*\omega_+\Big\}\nu_\alpha
\nonumber\\
&&\hspace{1.5cm}
+\sum\limits_{I,J}^3\sum\limits_{\alpha=1}^{5}\sum\limits_{b=1}^5\tilde{L}_J^-\overline{e_I}\Big\{
[\delta\zeta_{_{\tilde L}}^L]_{_{JIb}}(Z_{_\nu})_{b\alpha}\omega_-
+[\delta\zeta_{_{\tilde L}}^R]_{_{JIb}}(Z_{_\nu})_{b\alpha}^*\omega_+\Big\}\nu_\alpha
\nonumber\\
&&\hspace{1.5cm}
+\sum\limits_{I,J}^3\sum\limits_{\alpha=1}^{5}\sum\limits_{b=1}^5\tilde{R}_J^-\overline{e_I}\Big\{
[\delta\zeta_{_{\tilde R}}^L]_{_{JIb}}(Z_{_\nu})_{b\alpha}\omega_-
+[\delta\zeta_{_{\tilde R}}^R]_{_{JIb}}(Z_{_\nu})_{b\alpha}^*\omega_+\Big\}\nu_\alpha
\nonumber\\
&&\hspace{1.5cm}
+\sum\limits_{I=1}^3\sum\limits_{\beta=1}^{4}G^-\overline{e_I}\Big\{[\delta\zeta_G^L]_{_{1I\chi_\beta^0}}\omega_-
+[\delta\zeta_G^R]_{_{1I\chi_\beta^0}}\omega_+\Big\}\chi_\beta^0
\nonumber\\
&&\hspace{1.5cm}
+\sum\limits_{I=1}^3\sum\limits_{\beta=1}^{4}H^-\overline{e_I}\Big\{[\delta\zeta_H^L]_{_{2I\chi_\beta^0}}\omega_-
+[\delta\zeta_H^R]_{_{2I\chi_\beta^0}}\omega_+\Big\}\chi_\beta^0
\nonumber\\
&&\hspace{1.5cm}
+\sum\limits_{I,J}^3\sum\limits_{\beta=1}^{4}\tilde{L}_J^-\overline{e_I}\Big\{
\Big[{e\sqrt{2}\over c_{_{\rm W}}}(Z_{_{\tilde{E}_J}})_{21}\Big(V_{_R}\Big)_{JI}(U_\chi)_{1\beta}
\nonumber\\
&&\hspace{1.5cm}
+(Y_{_E}V_{_R})_{_{JI}}(Z_{_{\tilde{E}_J}})_{11}(U_\chi)_{3\beta}
+[\delta^2\zeta_{_{\tilde L}}^L]_{_{JI\chi_\beta^0}}\Big]\omega_-
\nonumber\\
&&\hspace{1.5cm}
+\Big[-{e\over\sqrt{2}s_{_{\rm W}}c_{_{\rm W}}}(Z_{_{\tilde{E}_J}})_{11}\Big(V_{_L}\Big)_{JI}^*\Big[
c_{_{\rm W}}(U_\chi)_{2\beta}^*+s_{_{\rm W}}(U_\chi)_{1\beta}^*\Big]
\nonumber\\
&&\hspace{1.5cm}
+(Z_{_{\tilde{E}_J}})_{21}(V_{_L}^\dagger Y_{_E})_{_{IJ}}(U_\chi)_{3\beta}^*
+[\delta^2\zeta_{_{\tilde L}}^R]_{_{JI\chi_\beta^0}}\Big]\omega_+\Big\}\chi_\beta^0
\nonumber\\
&&\hspace{1.5cm}
+\sum\limits_{I,J}^3\sum\limits_{\beta=1}^{4}\tilde{R}_J^-\overline{e_I}\Big\{\Big[
{e\sqrt{2}\over c_{_{\rm W}}}(Z_{_{\tilde{E}_J}})_{22}\Big(V_{_R}\Big)_{JI}(U_\chi)_{1\beta}
\nonumber\\
&&\hspace{1.5cm}
+(Y_{_E}V_{_R})_{_{JI}}(Z_{_{\tilde{E}_J}})_{12}(U_\chi)_{3\beta}
+[\delta^2\zeta_{_{\tilde R}}^L]_{_{JI\chi_\beta^0}}\Big]\omega_-
\nonumber\\
&&\hspace{1.5cm}
+\Big[-{e\over\sqrt{2}s_{_{\rm W}}c_{_{\rm W}}}(Z_{_{\tilde{E}_J}})_{12}\Big(V_{_L}\Big)_{JI}^*\Big(
c_{_{\rm W}}(U_\chi)_{2\beta}^*+s_{_{\rm W}}(U_\chi)_{1\beta}^*\Big)
\nonumber\\
&&\hspace{1.5cm}
+(V_{_L}^\dagger Y_{_E})_{_{IJ}}(Z_{_{\tilde{E}_J}})_{21}(U_\chi)_{3\beta}^*
+[\delta^2\zeta_{_{\tilde R}}^R]_{_{JI\chi_\beta^0}}\Big]\omega_+\Big\}\chi_\beta^0+h.c.
%%%%%%%%%%%%%%%%%%%%%%%%%%%%%%%%%%%%%%%%%%%%%%%%%%%%%%%%%%%%
\label{chargino-lepton14}
\end{eqnarray}
with
\begin{eqnarray}
%%%%%%%%%%%%%%%%%%%%%%%%%%%%%%%%%%%%%%%%%%%%%%%%%%%%%%%%%%%%
&&[\delta^2\zeta_G^L]_{_{1Ib}}\simeq
-{e\sin\beta\over s_{_{\rm W}}c_{_{\rm W}}}
\Big[{c_{_{\rm W}}\over\sqrt{2}}(\xi_{_R}V_{_R})_{2I}\Big(\delta U_\chi\Big)_{2b}
+{s_{_{\rm W}}\over\sqrt{2}}(\xi_{_R}V_{_R})_{2I}\Big(\delta U_\chi\Big)_{1b}
\nonumber\\
&&\hspace{2.2cm}
+c_{_{\rm W}}(\xi_{_R}V_{_R})_{1I}\Big(\delta U_\chi\Big)_{4b}\Big]
\nonumber\\
&&\hspace{2.2cm}
-{\cos\beta\over2}\sum\limits_{I^\prime=1}^3\Big[(Y_{_E}\xi_{_R}^T\xi_{_R}V_{_R})_{I^\prime I}
\delta_{I^\prime b}+(Y_{_E}V_{_R})_{I^\prime I}\Big(\delta^2\chi_{_D}\Big)_{I^\prime b}\Big]
\;,\nonumber\\
%%%%%%%%%%%%%%%%%%%%%%%%%%%%%%%%%%%%%%%%%%%%%%%%%%%%%%%%%%%%
&&[\delta^2\zeta_G^R]_{_{1Ib}}\simeq
{e\cos\beta\over s_{_{\rm W}}c_{_{\rm W}}}
\Big[{c_{_{\rm W}}\over\sqrt{2}}(\xi_{_L}V_{_L})_{2I}^*\Big(\delta U_\chi\Big)_{2b}
+{s_{_{\rm W}}\over\sqrt{2}}(\xi_{_L}V_{_L})_{2I}^*\Big(\delta U_\chi\Big)_{1b}
\nonumber\\
&&\hspace{2.2cm}
-c_{_{\rm W}}(\xi_{_L}V_{_L})_{1I}^*\Big(\delta U_\chi\Big)_{3b}\Big]
\;,\nonumber\\
%%%%%%%%%%%%%%%%%%%%%%%%%%%%%%%%%%%%%%%%%%%%%%%%%%%%%%%%%%%%
&&[\delta^2\zeta_H^L]_{_{2Ib}}\simeq
{e\over s_{_{\rm W}}c_{_{\rm W}}}\Big\{
-\cos\beta\Big[(\xi_{_R}V_{_R})_{2I}\Big({c_{_{\rm W}}\over\sqrt{2}}\Big(\delta U_\chi\Big)_{2b}
+{s_{_{\rm W}}\over\sqrt{2}}\Big(\delta U_\chi\Big)_{1b}\Big)
\nonumber\\
&&\hspace{2.2cm}
+c_{_{\rm W}}(\xi_{_R}V_{_R})_{1I}\Big(\delta U_\chi\Big)_{4b}\Big]
+\sqrt{2}s_{_{\rm W}}\sum\limits_{J=1}^3(\varepsilon_{_E}^\prime)_{_{3+J}}(V_{_R})_{JI}
\Big(\delta U_\chi\Big)_{1b}\Big\}
\nonumber\\
&&\hspace{2.2cm}
+{1\over2}\sum\limits_{I^\prime=1}^3\Big\{\sin\beta\Big[
(Y_{_E}\xi_{_R}^T\xi_{_R}V_{_R})_{I^\prime I}\delta_{I^\prime b}
+(Y_{_E}V_{_R})_{I^\prime I}\Big(\delta^2\chi_{_D}\Big)_{I^\prime b}\Big]
\nonumber\\
&&\hspace{2.2cm}
+2(\varepsilon_{_E}^\prime)_{_{I^\prime}}(Y_{_E}V_{_R})_{I^\prime I}
\Big(\delta U_\chi\Big)_{3b}\Big\}
\;,\nonumber\\
%%%%%%%%%%%%%%%%%%%%%%%%%%%%%%%%%%%%%%%%%%%%%%%%%%%%%%%%%%%%
&&[\delta^2\zeta_H^R]_{_{2Ib}}\simeq
{e\over s_{_{\rm W}}c_{_{\rm W}}}\Big\{
\sin\beta\Big[(\xi_{_L}V_{_L})_{2I}^*\Big({c_{_{\rm W}}\over\sqrt{2}}(\delta U_\chi)_{2b}
+{s_{_{\rm W}}\over\sqrt{2}}(\delta U_\chi)_{1b}\Big)
\nonumber\\
&&\hspace{2.2cm}
-c_{_{\rm W}}(\xi_{_L}V_{_L})_{1I}^*(\delta U_\chi)_{3b}\Big]
-{1\over\sqrt{2}}\sum\limits_{J^\prime=1}^3(\varepsilon_{_E}^\prime)_{_{J^\prime}}(V_{_L})_{J^\prime I}^*
\Big(c_{_{\rm W}}(\delta U_{\chi})_{2b}
\nonumber\\
&&\hspace{2.2cm}
+s_{_{\rm W}}(\delta U_{\chi})_{3b}\Big)\Big\}
-{e\over s_{_{\rm W}}}(\xi_{_L}V_{_L})_{1I}^*\sum\limits_{J^\prime=1}^3(\varepsilon_{_E}^\prime)_{J^\prime}
\delta_{J^\prime b}
\nonumber\\
&&\hspace{2.2cm}
+\sum\limits_{I^\prime,J^\prime}^3(Y_{_E})_{_{I^\prime J^\prime}}(\varepsilon_{_E}^\prime)_{_{3+J^\prime}}\Big\{
(V_{_L})_{I^\prime I}^*(\delta U_{\chi})_{3b}+(\xi_{_L}V_{_L})_{1I}^*\delta_{I^\prime b}\Big\}
\;,\nonumber\\
%%%%%%%%%%%%%%%%%%%%%%%%%%%%%%%%%%%%%%%%%%%%%%%%%%%%%%%%%%%%
&&[\delta\zeta_{_{\tilde L}}^L]_{_{JIb}}\simeq
{\sqrt{2}e\over c_{_{\rm W}}}(Z_{_{\tilde{E}_{J}}})_{21}
(V_{_R})_{JI}(\delta U_{\chi})_{1b}
-\sum\limits_{I^\prime=1}^3(Y_{_E}V_{_R})_{_{I^\prime I}}\Big\{
\Big[\sin\beta(\varepsilon_{_E}^\prime)_{_{3+J}}
\nonumber\\
&&\hspace{2.0cm}
-\cos\beta{\upsilon_{_{L_J}}\over\upsilon_{_{\rm EW}}}\Big]\delta_{I^\prime b}
+\delta_{I^\prime J}(Z_{_{\tilde{E}_{I^\prime}}})_{11}(\delta U_{\chi})_{3b}\Big\}
\;,\nonumber\\
%%%%%%%%%%%%%%%%%%%%%%%%%%%%%%%%%%%%%%%%%%%%%%%%%%%%%%%%%%%%
&&[\delta\zeta_{_{\tilde L}}^R]_{_{JIb}}\simeq
{e\over s_{_{\rm W}}c_{_{\rm W}}}(Z_{_{\tilde{E}_{J}}})_{11}\Big\{c_{_{\rm W}}
(\xi_{_L}V_{_L})^*_{1I}\delta_{Jb}+{1\over\sqrt{2}}(V_{_L})^*_{JI}
\Big[s_{_{\rm W}}(\delta U_{\chi})_{1b}
\nonumber\\
&&\hspace{2.0cm}
+c_{_{\rm W}}(\delta U_{\chi})_{2b}\Big]\Big\}
+(Z_{_{\tilde{E}_J}})_{21}\Big\{\sum\limits_{I^\prime=1}^3(Y_{_E})_{_{I^\prime I}}
(\xi_{_L}V_{_L})^*_{1I}\delta_{I^\prime b}
\nonumber\\
&&\hspace{2.0cm}
+(V_{_L}^\dagger Y_{_E})_{IJ}(\delta U_{\chi})_{3b}\Big\}
\;,\nonumber\\
%%%%%%%%%%%%%%%%%%%%%%%%%%%%%%%%%%%%%%%%%%%%%%%%%%%%%%%%%%%%
&&[\delta\zeta_{_{\tilde R}}^L]_{_{JIb}}\simeq
{\sqrt{2}e\over c_{_{\rm W}}}\sum\limits_{I^\prime=1}^3(Z_{_{\tilde{E}_{I^\prime}}})_{22}
(V_{_R})_{JI}(\delta U_{\chi})_{1b}
\nonumber\\
&&\hspace{2.0cm}
-\sum\limits_{I^\prime=1}^3(Y_{_E}V_{_R})_{_{I^\prime I}}\Big[\cos\beta
(\varepsilon_{_E}^\prime)_{_{3+J}}\delta_{I^\prime b}
+\delta_{I^\prime J}(Z_{_{\tilde{E}_{I^\prime}}})_{12}(\delta U_{\chi})_{3b}\Big]
\;,\nonumber\\
%%%%%%%%%%%%%%%%%%%%%%%%%%%%%%%%%%%%%%%%%%%%%%%%%%%%%%%%%%%%
&&[\delta\zeta_{_{\tilde R}}^R]_{_{JIb}}\simeq
{e\over s_{_{\rm W}}c_{_{\rm W}}}(Z_{_{\tilde{E}_{J}}})_{12}\Big\{c_{_{\rm W}}(\xi_{_L}V_{_L})^*_{1I}\delta_{Jb}
+{1\over\sqrt{2}}(V_{_L})^*_{JI}\Big[s_{_{\rm W}}(\delta U_{\chi})_{1b}
+c_{_{\rm W}}(\delta U_{\chi})_{2b}\Big]\Big\}
\nonumber\\
&&\hspace{2.0cm}
+\sum\limits_{I^\prime,J^\prime}^3(Y_{_E})_{_{I^\prime J^\prime}}(Z_{_{\tilde{E}_{J^\prime}}})_{22}
\Big\{\delta_{IJ^\prime}(\xi_{_L}V_{_L})^*_{1I}\delta_{I^\prime b}+(V_{_L})_{I^\prime I}^*(\delta U_{\chi})_{3b}\Big\}
\;,\nonumber\\
%%%%%%%%%%%%%%%%%%%%%%%%%%%%%%%%%%%%%%%%%%%%%%%%%%%%%%%%%%%%
&&[\delta\zeta_G^L]_{_{1I\chi_\beta^0}}\simeq
{e\sin\beta\over s_{_{\rm W}}c_{_{\rm W}}}\Big[{c_{_{\rm W}}\over\sqrt{2}}(\xi_{_R}V_{_R})_{2I}(U_\chi)_{2\beta}
+{s_{_{\rm W}}\over\sqrt{2}}(\xi_{_R}V_{_R})_{2I}(U_\chi)_{1\beta}
\nonumber\\
&&\hspace{2.2cm}
+c_{_{\rm W}}(\xi_{_R}V_{_R})_{1I}(U_\chi)_{3\beta}\Big]
+\sum\limits_{I^\prime,J^\prime}^3(Y_{_E}V_{_R})_{_{I^\prime I}}
\Big[\cos\beta{(m_{_D})_{I^\prime(2+\beta)}\over m_{_{\chi_\beta^0}}}
-{\upsilon_{_{L_I^\prime}}\over\upsilon_{_{\rm EW}}}(U_\chi)_{3\beta}\Big]
\;,\nonumber\\
%%%%%%%%%%%%%%%%%%%%%%%%%%%%%%%%%%%%%%%%%%%%%%%%%%%%%%%%%%%%
&&[\delta\zeta_G^R]_{_{1I\chi_\beta^0}}\simeq
{e\cos\beta\over s_{_{\rm W}}c_{_{\rm W}}}\Big[{c_{_{\rm W}}\over\sqrt{2}}(\xi_{_L}V_{_L})_{2I}^*(U_\chi)_{2\beta}^*
+{s_{_{\rm W}}\over\sqrt{2}}(\xi_{_L}V_{_L})_{2I}^*(U_\chi)_{1\beta}^*
\nonumber\\
&&\hspace{2.2cm}
-c_{_{\rm W}}(\xi_{_L}V_{_L})_{1I}^*(U_\chi)_{3\beta}^*\Big]
+\sum\limits_{I^\prime=1}^3{\upsilon_{_{L_I^\prime}}\over\sqrt{2}\upsilon_{_{\rm EW}}}(V_{_L})_{I^\prime I}^*
\Big[c_{_{\rm W}}(U_\chi)_{2\beta}+s_{_{\rm W}}(U_\chi)_{1\beta}\Big]
\;,\nonumber\\
%%%%%%%%%%%%%%%%%%%%%%%%%%%%%%%%%%%%%%%%%%%%%%%%%%%%%%%%%%%%
&&[\delta\zeta_H^L]_{_{2I\chi_\beta^0}}\simeq
{e\over s_{_{\rm W}}c_{_{\rm W}}}\Big\{\cos\beta\Big[{c_{_{\rm W}}\over\sqrt{2}}(\xi_{_R}V_{_R})_{2I}(U_\chi)_{2\beta}
+{s_{_{\rm W}}\over\sqrt{2}}(\xi_{_R}V_{_R})_{2I}(U_\chi)_{1\beta}
\nonumber\\
&&\hspace{2.2cm}
+c_{_{\rm W}}(\xi_{_R}V_{_R})_{1I}(U_\chi)_{3\beta}\Big]
+\sqrt{2}s_{_{\rm W}}\sum\limits_{I^\prime=1}^3(\varepsilon_{_E}^\prime)_{_{3+I^\prime}}(V_{_R})_{I^\prime I}
(U_\chi)_{1\beta}\Big\}
\nonumber\\
&&\hspace{2.2cm}
-\sum\limits_{I^\prime=1}^3(Y_{_E}V_{_R})_{_{I^\prime I}}
\Big[\sin\beta{(m_{_D})_{I^\prime(2+\beta)}\over m_{_{\chi_\beta^0}}}
+(\varepsilon_{_E}^\prime)_{_{I^\prime}}(U_\chi)_{3\beta}\Big]
\;,\nonumber\\
%%%%%%%%%%%%%%%%%%%%%%%%%%%%%%%%%%%%%%%%%%%%%%%%%%%%%%%%%%%%
&&[\delta\zeta_H^R]_{_{2I\chi_\beta^0}}\simeq
-{e\over s_{_{\rm W}}c_{_{\rm W}}}\Big\{\sin\beta\Big[{c_{_{\rm W}}\over\sqrt{2}}(\xi_{_L}V_{_L})_{2I}^*(U_\chi)_{2\beta}^*
+{s_{_{\rm W}}\over\sqrt{2}}(\xi_{_L}V_{_L})_{2I}^*(U_\chi)_{1\beta}^*
\nonumber\\
&&\hspace{2.2cm}
-c_{_{\rm W}}(\xi_{_L}V_{_L})_{1I}^*(U_\chi)_{3\beta}^*\Big]
-\sum\limits_{I^\prime=1}^3{1\over\sqrt{2}}(\varepsilon_{_E}^\prime)_{_{I^\prime}}(V_{_L})_{I^\prime I}^*
\Big[c_{_{\rm W}}(U_\chi)_{2\beta}^*
\nonumber\\
&&\hspace{2.2cm}
+s_{_{\rm W}}(U_\chi)_{1\beta}^*\Big]\Big\}
-\sum\limits_{I^\prime=1}^3(V_{_L}^\dagger Y_{_E})_{_{II^\prime}}
(\varepsilon_{_E}^\prime)_{_{3+I^\prime}}(U_\chi)_{3\beta}
\;,\nonumber\\
%%%%%%%%%%%%%%%%%%%%%%%%%%%%%%%%%%%%%%%%%%%%%%%%%%%%%%%%%%%%
&&[\delta^2\zeta_{_{\tilde L}}^L]_{_{JI\chi_\beta^0}}\simeq
-{e\over\sqrt{2}c_{_{\rm W}}}(Z_{_{\tilde{E}_J}})_{21}\Big[\Big(\xi_{_R}^T\xi_{_R}V_{_R}\Big)_{JI}(U_\chi)_{1\beta}
+(V_{_R})_{JI}\Big(\delta^2U_\chi\Big)_{1\beta}\Big]
\nonumber\\
&&\hspace{2.2cm}
-{1\over2}(Z_{_{\tilde{E}_J}})_{11}\Big(Y_{_E}\xi_{_R}^T\xi_{_R}V_{_R}\Big)_{JI}(U_\chi)_{3\beta}
-\sum\limits_{I^\prime=1}^3(Y_{_E}V_{_R})_{_{I^\prime I}}\Big\{
\Big[\sin\beta(\varepsilon_{_E}^\prime)_{_{3+J}}
\nonumber\\
&&\hspace{2.2cm}
-\cos\beta{\upsilon_{_{L_J}}\over\upsilon_{_{\rm EW}}}\Big]
{(m_{_D})_{I^\prime(2+\beta)}\over m_{_{\chi_\beta^0}}}
-(Z_{_{\tilde{E}_J}})_{11}\delta_{I^\prime J}\Big(\delta^2U_\chi\Big)_{1\beta}\Big\}
\;,\nonumber\\
%%%%%%%%%%%%%%%%%%%%%%%%%%%%%%%%%%%%%%%%%%%%%%%%%%%%%%%%%%%%
&&[\delta^2\zeta_{_{\tilde L}}^R]_{_{JI\chi_\beta^0}}\simeq
{e\over2\sqrt{2}s_{_{\rm W}}c_{_{\rm W}}}(Z_{_{\tilde{E}_J}})_{11}\Big(\xi_{_L}^T\xi_{_L}V_{_L}\Big)_{JI}^*\Big[
c_{_{\rm W}}(U_\chi)_{2\beta}^*+s_{_{\rm W}}(U_\chi)_{1\beta}^*\Big]
\nonumber\\
&&\hspace{2.2cm}
-{1\over2}(V_{_L}^\dagger\xi_{_L}^T\xi_{_L}Y_{_E})_{_{IJ}}(Z_{_{\tilde{E}_J}})_{21}(U_\chi)_{3\beta}^*
\nonumber\\
&&\hspace{2.2cm}
-{e\over\sqrt{2}s_{_{\rm W}}c_{_{\rm W}}}\Big\{\Big[\sin\beta(\varepsilon_{_E}^\prime)_{_{3+J}}
-\cos\beta{\upsilon_{_{L_J}}\over\upsilon_{_{\rm EW}}}\Big]\Big[c_{_{\rm W}}(\xi_{_L}V_{_L})_{2I}^*(U_\chi)_{2\beta}^*
\nonumber\\
&&\hspace{2.2cm}
+s_{_{\rm W}}(\xi_{_L}V_{_L})_{2I}^*(U_\chi)_{1\beta}^*-\sqrt{2}c_{_{\rm W}}(\xi_{_L}V_{_L})_{1I}^*(U_\chi)_{3\beta}^*\Big]
\nonumber\\
&&\hspace{2.2cm}
-(Z_{_{\tilde{E}_J}})_{11}
\Big[c_{_{\rm W}}(V_{_L})_{JI}^*(\delta^2U_\chi)_{2\beta}^*+s_{_{\rm W}}(V_{_L})_{JI}^*(\delta^2U_\chi)_{1\beta}^*\Big]
\nonumber\\
&&\hspace{2.2cm}
-\sqrt{2}c_{_{\rm W}}(Z_{_{\tilde{E}_J}})_{11}(\xi_{_L}V_{_L})_{1I}^*{(m_{_D})_{J(2+\beta)}\over m_{_{\chi_\beta^0}}}\Big\}
\nonumber\\
&&\hspace{2.2cm}
-\sum\limits_{I^\prime=1}^3(Y_{_E})_{_{I^\prime J}}(Z_{_{\tilde{E}_J}})_{21}\Big[
(\xi_{_L}V_{_L})_{1I}^*{(m_{_D})_{I^\prime(2+\beta)}\over m_{_{\chi_\beta^0}}}
+(V_{_L})_{I^\prime I}^*(\delta^2U_\chi)_{3\beta}^*\Big]
\;,\nonumber\\
%%%%%%%%%%%%%%%%%%%%%%%%%%%%%%%%%%%%%%%%%%%%%%%%%%%%%%%%%%%%
&&[\delta^2\zeta_{_{\tilde R}}^L]_{_{JI\chi_\beta^0}}\simeq
-{e\over\sqrt{2}c_{_{\rm W}}}(Z_{_{\tilde{E}_J}})_{22}\Big(\xi_{_R}^T\xi_{_R}V_{_R}\Big)_{JI}(U_\chi)_{1\beta}
-{1\over2}(Z_{_{\tilde{E}_J}})_{12}\Big(Y_{_E}\xi_{_R}^T\xi_{_R}V_{_R}\Big)_{JI}(U_\chi)_{3\beta}
\nonumber\\
&&\hspace{2.2cm}
+{e\over\sqrt{2}s_{_{\rm W}}c_{_{\rm W}}}\Big\{\cos\beta(\varepsilon_{_E}^\prime)_{_{3+J}}
\Big[c_{_{\rm W}}(\xi_{_R}V_{_R})_{2I}(U_\chi)_{2\beta}+s_{_{\rm W}}(\xi_{_R}V_{_R})_{2I}(U_\chi)_{1\beta}
\nonumber\\
&&\hspace{2.2cm}
-\sqrt{2}c_{_{\rm W}}(\xi_{_R}V_{_R})_{1I}(U_\chi)_{4\beta}\Big]
-c_{_{\rm W}}(Z_{_{\tilde{E}_J}})_{22}(V_{_R})_{JI}(\delta^2 U_\chi)_{1\beta}\Big\}
\nonumber\\
&&\hspace{2.2cm}
-\sum\limits_{I^\prime=1}^3(Y_{_E}V_{_R})_{_{I^\prime I}}
\Big[\sin\beta(\varepsilon_{_E}^\prime)_{_{3+J}} {(m_{_D})_{I^\prime(2+\beta)}\over m_{_{\chi_\beta^0}}}
+\delta_{I^\prime J}(Z_{_{\tilde{E}_J}})_{12}(\delta^2U_\chi)_{3\beta}^*\Big]
\;,\nonumber\\
%%%%%%%%%%%%%%%%%%%%%%%%%%%%%%%%%%%%%%%%%%%%%%%%%%%%%%%%%%%%
&&[\delta^2\zeta_{_{\tilde R}}^R]_{_{JI\chi_\beta^0}}\simeq
{e\over2\sqrt{2}s_{_{\rm W}}c_{_{\rm W}}}(Z_{_{\tilde{E}_J}})_{12}\Big(\xi_{_L}^T\xi_{_L}V_{_L}\Big)_{JI}^*\Big[
c_{_{\rm W}}(U_\chi)_{2\beta}^*+s_{_{\rm W}}(U_\chi)_{1\beta}^*\Big]
\nonumber\\
&&\hspace{2.2cm}
-{1\over2}(Z_{_{\tilde{E}_J}})_{22}(V_{_L}^\dagger\xi_{_L}^T\xi_{_L}Y_{_E})_{_{I^\prime J}}(U_\chi)_{3\beta}^*
\nonumber\\
&&\hspace{2.2cm}
-{e\over\sqrt{2}s_{_{\rm W}}c_{_{\rm W}}}\Big\{\sin\beta(\varepsilon_{_E}^\prime)_{_{3+J}}
\Big[c_{_{\rm W}}(\xi_{_L}V_{_L})_{2I}^*(U_\chi)_{2\beta}^*
\nonumber\\
&&\hspace{2.2cm}
+s_{_{\rm W}}(\xi_{_L}V_{_L})_{2I}^*(U_\chi)_{1\beta}^*-\sqrt{2}c_{_{\rm W}}(\xi_{_L}V_{_L})_{1I}^*(U_\chi)_{3\beta}^*\Big]
\nonumber\\
&&\hspace{2.2cm}
-(Z_{_{\tilde{E}_J}})_{12}\Big[c_{_{\rm W}}(V_{_L})_{JI}^*(\delta^2U_\chi)_{2\beta}^*
+s_{_{\rm W}}(V_{_L})_{JI}^*(\delta^2U_\chi)_{1\beta}^*\Big]
\nonumber\\
&&\hspace{2.2cm}
-\sqrt{2}c_{_{\rm W}}(Z_{_{\tilde{E}_J}})_{12}(\xi_{_L}V_{_L})_{1I}^*{(m_{_D})_{J(2+\beta)}\over m_{_{\chi_\beta^0}}}\Big\}
\nonumber\\
&&\hspace{2.2cm}
-\sum\limits_{I^\prime=1}^3(Y_{_E})_{_{I^\prime J}}(Z_{_{\tilde{E}_J}})_{22}\Big[
(\xi_{_L}V_{_L})_{1I}^*{(m_{_D})_{I^\prime(2+\beta)}\over m_{_{\chi_\beta^0}}}
+(V_{_L})_{I^\prime I}^*(\delta^2U_\chi)_{3\beta}^*\Big]
%%%%%%%%%%%%%%%%%%%%%%%%%%%%%%%%%%%%%%%%%%%%%%%%%%%%%%%%%%%%
\label{chargino-lepton15}
\end{eqnarray}
Here the abreviations are
\begin{eqnarray}
%%%%%%%%%%%%%%%%%%%%%%%%%%%%%%%%%%%%%%%%%%%%%%%%%%%%%%%%%%%%
&&\Big(\delta U_\chi\Big)_{\alpha b}=\sum\limits_{\beta=1}^4(U_\chi)_{\alpha\beta}{(m_{_D})_{b(2+\beta)}\over m_{_{\chi_\beta^0}}}\;,
\;\;(\alpha=1,\;\cdots,\;4,\;\;\;b=1,\;\cdots,\;5)
\;,\nonumber\\
%%%%%%%%%%%%%%%%%%%%%%%%%%%%%%%%%%%%%%%%%%%%%%%%%%%%%%%%%%%%
&&\Big(\delta^2\chi_{_D}\Big)_{ab}={(m_{_D})_{a1}(m_{_D})_{b1}\over(\Delta_{_{BL}}-m_{_{BL}})^2}
+{(m_{_D})_{a2}(m_{_D})_{b2}\over(\Delta_{_{BL}}+m_{_{BL}})^2}
\nonumber\\
&&\hspace{2.2cm}
+\sum\limits_{\alpha=1}^4{(m_{_D})_{a(2+\alpha)}(m_{_D})_{b(2+\alpha)}\over m_{_{\chi_\alpha^0}}^2}
\;\;(a,\;b=1,\;\cdots,\;5)
\;,\nonumber\\
%%%%%%%%%%%%%%%%%%%%%%%%%%%%%%%%%%%%%%%%%%%%%%%%%%%%%%%%%%%%
&&\Big(\delta^2U_\chi\Big)_{\alpha\beta}=
\sum\limits_{\beta^\prime=1}^4\sum\limits_{\alpha^\prime=1}^5(U_\chi)_{\alpha\beta^\prime}
{(m_{_D})_{\alpha^\prime(2+\beta^\prime)}(m_{_D})_{4(2+\beta)}\over m_{_{\chi_\beta^0}}m_{_{\chi_{\beta^\prime}^0}}}\;,
\;\;(\alpha,\;\beta=1,\;\cdots,\;4)
\;,\nonumber\\
%%%%%%%%%%%%%%%%%%%%%%%%%%%%%%%%%%%%%%%%%%%%%%%%%%%%%%%%%%%%
&&(\varepsilon_{_E}^\prime)_{_{J}}=\sum\limits_{i=1}^2{(\cos\beta(A_{_{CH}})_{_{1(3i-3+J)}}
+\sin\beta(A_{_{CH}})_{_{2(3i-3+J)}})(Z_{_{\tilde{E}_J}})_{i1}\over m_{_{H_{_{2+J}}^\pm}}^2-m_{_{H_2^\pm}}^2}
\;,\nonumber\\
%%%%%%%%%%%%%%%%%%%%%%%%%%%%%%%%%%%%%%%%%%%%%%%%%%%%%%%%%%%%
&&(\varepsilon_{_E}^\prime)_{_{3+J}}=\sum\limits_{i=1}^2{(\cos\beta(A_{_{CH}})_{_{1(3i-3+J)}}
+\sin\beta(A_{_{CH}})_{_{2(3i-3+J)}})(Z_{_{\tilde{E}_J}})_{i2}\over m_{_{H_{_{5+J}}^\pm}}^2-m_{_{H_2^\pm}}^2}\;.
%%%%%%%%%%%%%%%%%%%%%%%%%%%%%%%%%%%%%%%%%%%%%%%%%%%%%%%%%%%%
\label{chargino-lepton15a}
\end{eqnarray}

\begin{eqnarray}
%%%%%%%%%%%%%%%%%%%%%%%%%%%%%%%%%%%%%%%%%%%%%%%%%%%%%%%%%%%%
&&{\cal L}_{_{B\bar{\chi}\chi}}={e\over2s_{_{\rm W}}c_{_{\rm W}}}Z_{_\mu}\sum\limits_{I,J}^3
\overline{e_I^-}\Big\{\Big[-\Big(1-2s_{_{\rm W}}^2\Big)\delta_{IJ}+[\delta^2{\cal C}^L]_{_{IJ}}\Big]\gamma^\mu\omega_-
\nonumber\\
&&\hspace{1.5cm}
+\Big[2s_{_{\rm W}}^2\delta_{IJ}+[\delta^2{\cal C}^R]_{_{IJ}}\Big]\gamma^\mu\omega_+\Big\}e_J^-
\nonumber\\
&&\hspace{1.5cm}
+{e\over2s_{_{\rm W}}c_{_{\rm W}}}Z_{_\mu}\sum\limits_{I=1}^3\sum\limits_{\beta=1}^2
\overline{e_I^-}\Big\{[\delta{\cal C}^L]_{_{I\chi_\beta^-}}\gamma^\mu\omega_-
+[\delta{\cal C}^R]_{_{I\chi_\beta^-}}\gamma^\mu\omega_+\Big\}\chi_\beta^-
\nonumber\\
&&\hspace{1.5cm}
+g_{_{BL}}Z_{_{BL,\mu}}\sum\limits_{I,J}^3\overline{e_I^-}\Big\{\Big[\delta_{IJ}+[\delta^2{\cal B}^L]_{_{IJ}}\Big]
\gamma^\mu\omega_-+\Big[\delta_{IJ}+[\delta^2{\cal B}^R]_{_{IJ}}\Big]\gamma^\mu\omega_+\Big\}e_J^-
\nonumber\\
&&\hspace{1.5cm}
+g_{_{BL}}Z_{_{BL,\mu}}\sum\limits_{I=1}^3\sum\limits_{\beta=1}^2
\overline{e_I^-}\Big\{[\delta{\cal B}^L]_{_{I\chi_\beta^-}}\gamma^\mu\omega_-
+[\delta{\cal B}^R]_{_{I\chi_\beta^-}}\gamma^\mu\omega_+\Big\}\chi_\beta^-
\nonumber\\
&&\hspace{1.5cm}
+{e\over s_{_{\rm W}}}\Big\{W_{_\mu}^-\sum\limits_{I=1}^3\sum\limits_{\alpha=1}^{5}
\overline{e_I^-}\Big[\Big(-{1\over\sqrt{2}}(V_{_L}^\dagger Z_{_\nu})_{I\alpha}
+\sum\limits_{b=1}^5[\delta^2{\cal V}^L]_{_{Ib}}(Z_{_\nu})_{b\alpha}\Big)\gamma^\mu\omega_-
\nonumber\\
&&\hspace{1.5cm}
+\sum\limits_{b=1}^5[\delta^2{\cal V}^R]_{_{Ib}}(Z_{_\nu})_{b\alpha}^*\gamma^\mu\omega_+\Big]\nu_\alpha^0
\nonumber\\
&&\hspace{1.5cm}
+W_{_\mu}^-\sum\limits_{I=1}^3\sum\limits_{\beta=1}^{4}
\overline{e_I^-}\Big([\delta{\cal V}^L]_{_{I\chi_\beta^0}}\gamma^\mu\omega_-
+[\delta{\cal V}^R]_{_{I\chi_\beta^0}}\gamma^\mu\omega_+\Big)\chi_\beta^0+h.c.\Big\}
%%%%%%%%%%%%%%%%%%%%%%%%%%%%%%%%%%%%%%%%%%%%%%%%%%%%%%%%%%%%
\label{chargino-lepton16}
\end{eqnarray}
with
\begin{eqnarray}
%%%%%%%%%%%%%%%%%%%%%%%%%%%%%%%%%%%%%%%%%%%%%%%%%%%%%%%%%%%%
&&[\delta^2{\cal C}^L]_{_{IJ}}=-(\xi_{_L}V_{_L})_{1I}^*(\xi_{_L}V_{_L})_{1J}
\;,\nonumber\\
%%%%%%%%%%%%%%%%%%%%%%%%%%%%%%%%%%%%%%%%%%%%%%%%%%%%%%%%%%%%
&&[\delta^2{\cal C}^R]_{_{IJ}}=
-2(\xi_{_R}V_{_R})_{1I}(\xi_{_R}V_{_R})_{1J}^*-(\xi_{_R}V_{_R})_{2I}(\xi_{_R}V_{_R})_{2J}^*
\;,\nonumber\\
%%%%%%%%%%%%%%%%%%%%%%%%%%%%%%%%%%%%%%%%%%%%%%%%%%%%%%%%%%%%
&&[\delta{\cal C}^L]_{_{I\chi_\beta^-}}=-(\xi_{_L}V_{_L})_{1I}^*(U_-)_{1\beta}
\;,\nonumber\\
%%%%%%%%%%%%%%%%%%%%%%%%%%%%%%%%%%%%%%%%%%%%%%%%%%%%%%%%%%%%
&&[\delta{\cal C}^R]_{_{I\chi_\beta^-}}=-2(\xi_{_R}V_{_R})_{1I}(U_+)_{1\beta}^*-(\xi_{_R}V_{_R})_{2I}(U_+)_{2\beta}^*
\;,\nonumber\\
%%%%%%%%%%%%%%%%%%%%%%%%%%%%%%%%%%%%%%%%%%%%%%%%%%%%%%%%%%%%
&&[\delta^2{\cal B}^L]_{_{IJ}}=-(\xi_{_L}V_{_L})_{1I}^*(\xi_{_L}V_{_L})_{1J}-(\xi_{_L}V_{_L})_{2I}^*(\xi_{_L}V_{_L})_{2J}
\;,\nonumber\\
%%%%%%%%%%%%%%%%%%%%%%%%%%%%%%%%%%%%%%%%%%%%%%%%%%%%%%%%%%%%
&&[\delta^2{\cal B}^R]_{_{IJ}}=
-(\xi_{_R}V_{_R})_{1I}(\xi_{_R}V_{_R})_{1J}^*-(\xi_{_R}V_{_R})_{2I}(\xi_{_R}V_{_R})_{2J}^*
\;,\nonumber\\
%%%%%%%%%%%%%%%%%%%%%%%%%%%%%%%%%%%%%%%%%%%%%%%%%%%%%%%%%%%%
&&[\delta{\cal B}^L]_{_{I\chi_\beta^-}}=-(\xi_{_L}V_{_L})_{1I}^*(U_-)_{1\beta}-(\xi_{_L}V_{_L})_{2I}^*(U_-)_{2\beta}
\;,\nonumber\\
%%%%%%%%%%%%%%%%%%%%%%%%%%%%%%%%%%%%%%%%%%%%%%%%%%%%%%%%%%%%
&&[\delta{\cal B}^R]_{_{I\chi_\beta^-}}=-(\xi_{_R}V_{_R})_{1I}(U_+)_{1\beta}^*-(\xi_{_R}V_{_R})_{2I}(U_+)_{2\beta}^*
\;,\nonumber\\
%%%%%%%%%%%%%%%%%%%%%%%%%%%%%%%%%%%%%%%%%%%%%%%%%%%%%%%%%%%%
&&[\delta^2{\cal V}^L]_{Ib}={1\over2\sqrt{2}}\sum\limits_{I^\prime=1}^3
\Big\{(V_{_L}^\dagger\xi_{_L}^T\xi_{_L}))_{II^\prime}\delta_{I^\prime b}
-{1\over2}(V_{_L})_{I^\prime I}^*\Big(\delta^2\chi_{_D}\Big)_{I^\prime b}\Big\}
\nonumber\\
&&\hspace{1.4cm}
+\Big[{1\over\sqrt{2}}(\xi_{_L}V_{_L})_{2I}^*(\delta U_{\chi})_{3b}+(\xi_{_L}V_{_L})_{1I}^*(\delta U_{\chi})_{2b}\Big]
\;,\nonumber\\
%%%%%%%%%%%%%%%%%%%%%%%%%%%%%%%%%%%%%%%%%%%%%%%%%%%%%%%%%%%%
&&[\delta^2{\cal V}^R]_{_{Ib}}=
-\Big[{1\over\sqrt{2}}(\xi_{_R}V_{_R})_{2I}(\delta U_{\chi})_{3b}^*-(\xi_{_R}V_{_R})_{1I}(\delta U_{\chi})_{2b}^*\Big]
\;,\nonumber\\
%%%%%%%%%%%%%%%%%%%%%%%%%%%%%%%%%%%%%%%%%%%%%%%%%%%%%%%%%%%%
&&[\delta{\cal V}^L]_{_{I\chi_\beta^0}}=-{1\over\sqrt{2}}\Big[\sum\limits_{I^\prime=1}^3(V_{_L})_{I^\prime I}^*
{(m_{_D})_{I^\prime(2+\beta)}\over m_{_{\chi_\beta^0}}}+(\xi_{_L}V_{_L})_{2I}^*(U_{\chi})_{3\beta}\Big]
-(\xi_{_L}V_{_L})_{1I}^*(U_{\chi})_{2\beta}
\;,\nonumber\\
%%%%%%%%%%%%%%%%%%%%%%%%%%%%%%%%%%%%%%%%%%%%%%%%%%%%%%%%%%%%
&&[\delta{\cal V}^R]_{_{I\chi_\beta^0}}={1\over\sqrt{2}}(\xi_{_R}V_{_R})_{2I}(U_{\chi})_{4\beta}^*
-(\xi_{_R}V_{_R})_{1I}(U_{\chi})_{2\beta}^*
%%%%%%%%%%%%%%%%%%%%%%%%%%%%%%%%%%%%%%%%%%%%%%%%%%%%%%%%%%%%
\label{chargino-lepton17}
\end{eqnarray}

\section{The inverse matrix of ${\cal M}$ \label{app3}}
\indent\indent
The inverse mass matrix of six heavy majorana fermions is
\begin{eqnarray}
&&{\cal M}^{-1}=\left(\begin{array}{cc}
\Big[{\cal M}_{_N}^{(1)}\Big]^{-1}_{2\times2}\;\;\;&\Big[{\cal M}_{_N}^{(2)T}\Big]^{-1}_{2\times4}\;\;\;\\\\
\Big[{\cal M}_{_N}^{(2)}\Big]^{-1}_{4\times2}\;\;\;&\Big[{\cal M}_{_N}\Big]^{-1}_{4\times4}\;\;\;
\end{array}\right)
\label{app3-1}
\end{eqnarray}
with
\begin{eqnarray}
&&\Big[{\cal M}_{_N}\Big]^{-1}=\left(\begin{array}{cccc}
{4m_2\mu^2+g_2^2\mu\upsilon_{_u}\upsilon_{_d}\over2\tilde{\mu}^4}\;\;\;
&{g_1g_2\mu\upsilon_{_u}\upsilon_{_d}\over2\tilde{\mu}^4}\;\;\;&-{g_1m_2\mu\upsilon_{_u}\over\tilde{\mu}^4}\;\;\;
&{g_1m_2\mu\upsilon_{_d}\over\tilde{\mu}^4}\;\;\;\\
{g_1g_2\mu\upsilon_{_u}\upsilon_{_d}\over2\tilde{\mu}^4}\;\;\;
&{4m_1\mu^2+g_1^2\mu\upsilon_{_u}\upsilon_{_d}\over2\tilde{\mu}^4}\;\;\;
&{g_2m_1\mu\upsilon_{_u}\over\tilde{\mu}^4}\;\;\;&-{g_2m_1\mu\upsilon_{_d}\over\tilde{\mu}^4}\;\;\;\\
-{g_1m_2\mu\upsilon_{_u}\over\tilde{\mu}^4}\;\;\;&{g_2m_1\mu\upsilon_{_u}\over\tilde{\mu}^4}\;\;\;&
{\tilde{m}\upsilon_{_u}^2\over2\tilde{\mu}^4}\;\;\;&{8\mu m_1m_2+\tilde{m}\upsilon_{_u}\upsilon_{_d}\over2\tilde{\mu}^4}\;\;\;\\
{g_1m_2\mu\upsilon_{_d}\over\tilde{\mu}^4}\;\;\;&-{g_2m_1\mu\upsilon_{_d}\over\tilde{\mu}^4}\;\;\;
&{8\mu m_1m_2+\tilde{m}\upsilon_{_u}\upsilon_{_d}\over2\tilde{\mu}^4}\;\;\;&
{\tilde{m}\upsilon_{_d}^2\over2\tilde{\mu}^4}\;\;\;
\end{array}\right)\;,\nonumber\\\nonumber\\
&&\Big[{\cal M}_{_N}^{(1)}\Big]^{-1}=\left(\begin{array}{cc}
{1\over\Delta_{_{BL}}-m_{_{BL}}}\;\;\;&0\;\;\;\\
0\;\;\;&{1\over\Delta_{_{BL}}+m_{_{BL}}}\;\;\;
\end{array}\right)\;,\nonumber\\\nonumber\\
&&\Big[{\cal M}_{_N}^{(2)}\Big]^{-1}=\left(\begin{array}{cccc}
{ig_1m_2\mu\upsilon_{_d}\varepsilon_-\over\tilde{\mu}^4(\Delta_{_{BL}}-m_{_{BL}})}\;\;\;&
-{g_1m_2\mu\upsilon_{_d}\varepsilon_+\over\tilde{\mu}^4(\Delta_{_{BL}}+m_{_{BL}})}\;\;\;\\
-{ig_2m_1\mu\upsilon_{_d}\varepsilon_-\over\tilde{\mu}^4(\Delta_{_{BL}}-m_{_{BL}})}\;\;\;&
{g_2m_1\mu\upsilon_{_d}\varepsilon_+\over\tilde{\mu}^4(\Delta_{_{BL}}+m_{_{BL}})}\;\;\;\\
{i(8\mu m_1m_2+\tilde{m}\upsilon_{_u}\upsilon_{_d}\varepsilon_-)\over2\tilde{\mu}^4(\Delta_{_{BL}}-m_{_{BL}})}\;\;\;&
-{8\mu m_1m_2+\tilde{m}\upsilon_{_u}\upsilon_{_d}\varepsilon_+\over2\tilde{\mu}^4(\Delta_{_{BL}}+m_{_{BL}})}\;\;\;\\
{i\tilde{m}\upsilon_{_d}^2\varepsilon_-\over2\tilde{\mu}^4(\Delta_{_{BL}}-m_{_{BL}})}\;\;\;&
-{\tilde{m}\upsilon_{_d}^2\varepsilon_+\over2\tilde{\mu}^4(\Delta_{_{BL}}+m_{_{BL}})}\;\;\;
\end{array}\right)\;.
\label{app3-2}
\end{eqnarray}
Where the abbreviations are
\begin{eqnarray}
&&\tilde{m}=g_1^2m_2+g_2^2m_1\;,\nonumber\\
&&\tilde{\mu}^4=4m_1m_2\mu^2+\tilde{m}\mu\upsilon_{_u}\upsilon_{_d}\;.
\label{app3-3}
\end{eqnarray}

\end{document}